\UseRawInputEncoding

\documentclass[aip,graphicx,reprint,pop,amsmath,amssymb]{revtex4-2}

\usepackage{graphicx}
\usepackage{dcolumn}
\usepackage{bm}
\usepackage[hidelinks]{hyperref}
\usepackage[utf8]{inputenc}
\usepackage[T1]{fontenc}
\usepackage{mathptmx}
\usepackage{etoolbox}
\usepackage{physics}
\usepackage{cleveref}
\usepackage{subcaption}
\usepackage{bibunits}

\DeclareMathOperator{\sgn}{sgn} 
\DeclareMathOperator{\erfc}{erfc} 
\DeclareUnicodeCharacter{2009}{\,}

\makeatletter
\def\@email#1#2{%
 \endgroup
 \patchcmd{\titleblock@produce}
  {\frontmatter@RRAPformat}
  {\frontmatter@RRAPformat{\produce@RRAP{*#1\href{mailto:#2}{#2}}}\frontmatter@RRAPformat}
  {}{}
}%
\makeatother

\crefname{equation}{eqn.}{eqns.}
\Crefname{equation}{Eqn.}{Eqns.}

\let\bmog\bm 
\renewcommand{\bm}[1]{{\mathbf{#1}}}

\defaultbibliographystyle{unsrt} 

\draft 

\begin{document}

\begin{bibunit}

\title{Fast electrostatic microinstability evaluation in arbitrary toroidal magnetic geometry using a variational approach}

\author{M.C.L. Morren}
    \email{m.c.l.morren@tue.nl}

\author{P. Mulholland}
\affiliation{Department of Applied Physics and Science Education, Eindhoven University of Technology, 5600 MB Eindhoven, The Netherlands}

\author{J.H.E. Proll}
\affiliation{Department of Applied Physics and Science Education, Eindhoven University of Technology, 5600 MB Eindhoven, The Netherlands}
\affiliation{Max-Planck-Institut für Plasmaphysik, 17491 Greifswald, Germany}

\author{M.J. Pueschel}
\affiliation{Department of Applied Physics and Science Education, Eindhoven University of Technology, 5600 MB Eindhoven, The Netherlands}
\affiliation{Dutch Institute for Fundamental Energy Research, 5612 AJ Eindhoven, The Netherlands}
\affiliation{Department of Physics \& Astronomy, Ruhr-Universität Bochum, 44780 Bochum, Germany}

\author{L. Podavini}
\affiliation{Max-Planck-Institut für Plasmaphysik, 17491 Greifswald, Germany}

\author{D.D. Kiszkiel}

\author{J.A. Schuurmans}
\affiliation{Department of Applied Physics and Science Education, Eindhoven University of Technology, 5600 MB Eindhoven, The Netherlands}

\author{A. Zocco}
\affiliation{Max-Planck-Institut für Plasmaphysik, 17491 Greifswald, Germany}

\date{5 December 2025}

\begin{abstract}
Small-scale turbulence originating from microinstabilities limits the energy confinement time in magnetic confinement fusion. Here we develop a semi-analytical dispersion relation based on lowest-order solutions to the gyrokinetic equations in an asymptotic expansion in the ratio of transit (bounce) frequency to the mode frequency for ions (electrons), capable of describing two common instabilities: the ion temperature gradient (ITG) mode and trapped-electron mode (TEM), in the electrostatic limit. The dispersion relation, which is valid in arbitrary toroidal geometry, takes into account resonances with the magnetic ion and bounce-averaged electron drifts, incorporates non-local effects along the magnetic field line, is valid for arbitrary sign of the growth rate and magnetic curvature, and is shown to satisfy a variational property. Several common approximation models are introduced for both the magnetic drift and finite Larmor radius (FLR) damping, with the Padé approximation for FLR effect in particular resulting in remarkable agreement with the baseline dispersion relation model at significantly reduced costs. The baseline model is verified by comparing solutions of the dispersion relation model to high-fidelity linear gyrokinetic simulations, where the exact eigenfunction of the electrostatic potential from simulations is used as a trial function, showing good quantitative agreement for ITGs and TEMs in (shaped) tokamaks as well as low-magnetic-shear stellarators. 
\end{abstract}


\maketitle 

\section{Introduction}
For the successful exploitation of magnetic confinement fusion as an energy source, it is important that the extreme plasma conditions persist for a sufficiently long time. In both tokamaks and modern stellarators optimised for low neoclassical transport \cite{Canik2007ExperimentalSymmetry,Beidler2021}, the confinement losses are dominated by the anomalous transport channel. This transport is due to turbulent fluctuations, which are driven by microinstabilities in the plasma\cite{Stroth1998,Connor1994}. In typical cases, most of the turbulent transport is driven by the ion-temperature gradient (ITG) mode and trapped-electron mode (TEM)\cite{Jenko2000,Merz2010,Tang1978MicroinstabilityTokamaks,Horton1999DriftTransport}, though in reactor-relevant scenarios, finite-pressure-gradient-driven magnetic instabilities such as kinetic ballooning modes (KBMs) \cite{Mulholland2025Finite-Modes,Pueschel2008GyrokineticBeta} or microtearing (MT)\cite{Doerk2011GyrokineticTurbulence} modes may also contribute to turbulence-induced transport. Nevertheless, in the present work, we focus on the former class of electrostatic instabilities. \par
These microinstabilities can be investigated using (linear) gyrokinetics \cite{Rutherford1968,Catto1981,Jr.1980,Brizard2007}, which reveals a strong susceptibility to the magnetic geometry \cite{Rewoldt2005ComparisonGeometries,Alcuson2020,Proll2022,Belli2008EffectsTurbulence,NAKATA2014LocalEquilibria,Morren2024InfluenceStellarators,Mynick2009GeometryTurbulence,Landreman2025HowLearning}. In particular, TEMs, driven by resonances with the toroidal-precession drift of trapped electrons\cite{Adam1976}, may be avoided altogether in so-called maximum-$J$ configurations\cite{Proll2012,Rosenbluth1968LowFrequencyInstability}, where all trapped particles experience a net-favourable bounce-averaged drift. While it is possible to design stellarators that provide a close approximation to such a maximum-$J$ property \cite{Goodman2024Quasi-IsodynamicCandidates,CamachoMata2022DirectConfigurations,Sanchez2023A}, in tokamaks\cite{Wang2022GlobalPlasmas} and quasi-symmetric stellarators\cite{Gerard2024OnStellarators} (both of which cannot be made maximum-$J$ \cite{Rodriguez2024TheStellarators}), the stability of TEMs may alternatively be adjusted by reducing the trapped-particle fraction and modifying the cross-field drifts through flux-surface shaping. Likewise, the (toroidal) ITG instability is directly affected by the magnetic geometry through the arrangement of bad-curvature regions and shear-enhanced finite-Larmor-radius (FLR) suppression \cite{Plunk2014,Rodriguez2025TheLocalisation}. \par
It it thus clear that there is potential for optimising the magnetic geometry for increased resilience against these microinstabilities, especially for stellarators due to their substantial flexibility in magnetic-configuration space. For realistic geometries, detailed microstability investigations require direct numerical simulations using gyrokinetic codes. Whilst the involved linear calculations have relatively low computational cost in comparison to nonlinear simulations for the heat losses, they are still too computationally intense to be directly used inside optimisation routines, where instead simpler and cheaper geometry-based proxies for microstability have been opted for opted for in the past \cite{Proll2015,Mynick2010OptimizingTransport,Duff2025SuppressingShaping,Mynick2014TurbulentConfigurations,Roberg-Clark2023CriticalConcept}. Noticeable exceptions include the work done by Refs.~\onlinecite{Jorge2024DirectDevices,Kim2024OptimizationStellarators} where a quasilinear mixing-length estimate and nonlinear heatflux of ITG, respectively, are directly included in the optimisation target, though both approaches necessitate an adiabatic treatment of electrons for numerical feasibility.  Hence, there is a need for the development of fast reduced models to assess the stability of microturbulence. \par
Whilst in the past strides have been made to analytically investigate stability, in particular of ITG modes, these often rely on using ``toy geometries'' like a square drift-well embedded in a shearless constant magnetic field\cite{Rodriguez2025TheLocalisation}, or by invoking the so-called ``local limit''\cite{Ivanov2023AnField,Zocco2018ThresholdPlasmas,Biglari1998b,Calvo2025ModificationPlasmas} where the variation of the geometry and eigenfunction along the field line are neglected. Only by invoking a non-resonant limit can effects of the geometry variation be retained in an analytically tractable form, though typically still requiring a simplification of the magnetic geometry parameterisation\cite{Zocco2018ThresholdPlasmas,Plunk2014,Biglari1998b,Zocco2016GeometricSystems}. For TEMs, the salient geometric features are of much more importance due to the strong interplay between geometry and arrangement of trapped-particle populations, hence non-local treatments are rendered a necessity. Whilst binary statements regarding the existence of unstable TEMs in arbitrary toroidal geometry can be made using energy-budget calculations\cite{Proll2012,Helander2015b,Helander2013}, obtaining quantitative growth rates in arbitrary geometry usually requires invoking a non-resonant expansion as well\cite{Helander2013,Plunk2017}. \par
Here, we investigate the feasibility of a reduced-physics model for the dispersion relation of ITG and TEM, based on a lowest-order expansion of the collisionless electrostatic gyrokinetic system in the ratio of transit (bounce) frequency to the mode frequency for ions (electrons). The model retains gyrokinetic passing-ion and drift-kinetic (DK) trapped-electron responses to fluctuations in the electrostatic potential, includes impurities\footnote[3333]{ \label{fn:imp_assump} In order for a gyrokinetic treatment to be sensible $\omega/\Omega_{s} \ll 1$ and $k_\perp \rho_{Ts} \sim \order{1}$ need to be respected by each species $s$, otherwise requiring a fully kinetic description. Additionally, we have assumed $\omega_{ts}/\omega \ll 1$ for the approximate treatment of the non-adiabatic part of the impurity distribution in \Cref{eq:approx-gi-sol} with index $i\rightarrow s$. As we assume these requirements to be met by the main (hydrogenic) ion species, and $\Omega_{s} = (Z_{s} m_i/m_{s}) \Omega_i, \ \rho_{Ts} = (\sqrt{T_s/T_i} \sqrt{m_s/m_i} /Z_s)\rho_{Ti}, \ \omega_{ts} = \sqrt{T_s/T_i} \sqrt{m_i/m_s} \omega_{ti}$, it follows that the treatment of impurities is only potentially limited by the $\omega/\Omega_{s} \ll 1$ assumption, as the mass of the impurity species increases faster than the ionisation state}
, is valid in arbitrary toroidal geometry, and accounts for non-local effects of the geometry along the magnetic field line. As a result of these approximations, the model lacks the physics associated with transit resonances, and essentially considers the passing electrons to be adiabatic, and therefore cannot describe the slab branch of the ITG mode\cite{Kadomtsev1970TurbulenceSystems,Plunk2014}, nor instabilities predominantly driven by passing electrons such as the electron-temperature gradient (ETG) mode\cite{Kadomtsev1970TurbulenceSystems,Horton1988ToroidalModes,Liu1971InstabilitiesCurrent,Lee1987CollisionlessInstability,Hirose1990ElectronTokamaks,Jenko2000} or the universal instability (UI) \cite{Landreman2015,Helander2015b,Costello2023TheStellarators}. However, the model does allow for analytical treatment of kinetic resonances for both ions and electrons. Most importantly, we extend the validity range of these analytical expressions to arbitrary growth rate and curvature, where previous treatments typically implicitly assumed these to be positive. The model is \textit{a posteriori} validated by comparing the growth rates and frequencies with those obtained from linear simulation with the \textsc{Gene} code\cite{Jenko2000}, using the exact gyrokinetic eigenfunctions as model input. To highlight the versatile geometry capabilities of the model, we consider a variety of realistic magnetic geometries based on experimental devices. These include the DIII-D tokamak\cite{Luxon2002}, both positive- and negative-triangularity configurations of the TCV tokamak\cite{Hofmann1994CreationTCV} (with equilibria corresponding to last closed-flux surface triangularities of $\delta_{\mathrm{LCFS}}\approx\pm 0.4$), the standard quasi-symmetric configuration of the Helically Symmetric eXperiment (HSX) stellarator\cite{Anderson1995}, and both the high- and negative-mirror configurations of the Wendelstein 7-X (W7-X) stellarator \cite{Beidler1990}. Here we will only present results for DIII-D, HSX and the high-mirror configuration of W7-X, with results for the remainder of configurations shown in the Supplementary Material. \par
The paper is organised as follows; in \Cref{sec:disp-theory}, the collisionless electrostatic gyrokinetic system is briefly introduced and the dispersion relation model is derived. Next, in \Cref{sec:numerics}, we describe the numerical solution strategy and normalisations used for solving the dispersion relation, as well as the setup of simulation parameters. We then present the results of the gyrokinetic simulations in \Cref{sec:GENE-application}, comparing these with the eigenfrequencies obtained by the dispersion model separately in the cases of adiabatic electrons (\Cref{sec:ae}) and kinetic electrons (\Cref{sec:ke}). Afterwards, in \Cref{sec:redmod} we explore the feasibility of using further approximations for the FLR stabilisation and magnetic-drift resonance, aimed at alleviating the need of numerical integration of the ion terms in the dispersion relation. These approximations allow for a fully analytical treatment of the ion terms, yielding simplified reduced versions of the dispersion relation with significantly lower computational costs. These reduced models are then subsequently tested against the baseline dispersion model to probe for the essential underlying physics necessary to resolve toroidal-instability drive. Lastly, in \Cref{sec:summary} we provide a summary and brief outlook.

\section{Gyrokinetic formulation of dispersion relation} \label{sec:disp-theory}
The ITG and TEM microinstabilities may be described as electrostatic drift-wave fluctuations, hence we consider the $\beta \rightarrow 0$ limit of the gyrokinetic framework\cite{Catto1981,Brizard2007}. Here $\beta = p/(B^2/2\mu_0)$, with $p$ the  total plasma pressure, $\mu_0$ the vacuum permeability and $B$ the magnetic field strength, expresses the ratio between kinetic- and magnetic pressures, such that magnetic fluctuations are neglected altogether in the electrostatic approximation ($\bm{\delta B}\approx 0$). Additionally, as we focus on microinstabilities responsible for core plasma turbulence we consider the collisionless limit. This collisionless electrostatic limit is commonly employed to describe both instabilities\cite{Helander2013,Plunk2014,Plunk2017,Rodriguez2025TheLocalisation,Zocco2018ThresholdPlasmas}. However, both finite-$\beta$ and finite collisionality  tend to have a stabilising influence on the ITG and TEM \cite{Belli2017ImplicationsSimulation,Pueschel2008GyrokineticBeta}, such that the modes governed by the collisionless electrostatic gyrokinetic system may be considered as upper bounds of instability. \par
The distribution function of each species $s$ is split into equilibrium and fluctuating parts as
\begin{align}
    f_s(\bm{x},\bm{v},t) =  F_{Ms}(\bm{x},\bm{v}) \left(1-\frac{q_s\phi(\bm{x},t)}{T_s}\right) + g_s(\bm{R},\bm{v},t)
    \label{eq:distribution-split}
\end{align}
where the first term includes both the equilibrium Maxwellian distribution 
\begin{align}
  F_{Ms}({\bf{x}},{\bf{v}}) = \frac{n_{s}(\bm{x})}{\left(\sqrt{2\pi v_{Ts}}\right)^3} \exp(-\frac{v^2}{2v_{Ts}^2})   
  \label{eq:Maxwellian}
\end{align}
and the adiabatic response. 
In \Cref{eq:distribution-split,eq:Maxwellian} $n_{s}, T_s, q_s, m_s,$ and $ v_{Ts}=\sqrt{T_s/m_s}$ denote the equilibrium density, equilibrium temperature, charge, mass and thermal velocity of species $s$, respectively, while $\bm{x},\bm{v},t$ denote position, velocity and time, respectively. Additionally, the last term in \Cref{eq:distribution-split} is the non-adiabatic response, which describes the distribution function of particle gyrocenters $\bm{R} = \bm{x} - \bm{e_b}\cross\bm{v}/\Omega_s$, where $\bm{e_b}=\bm{B}/B$, with $\bm{B}$ denoting the (equilibrium) magnetic flux density, $\Omega_s = q_s B/m_s$ the cyclotron frequency and $B=\norm{\bm{B}}$. The gyrocenter distribution evolves according to the (linearised) gyrokinetic Vlasov equation\cite{Rutherford1968}
\begin{align}
    v_{\parallel} \nabla_{\parallel} \hat{g}_s - i \left(\omega-\omega_{ds}\right) \hat{g}_s = - i \frac{q_s \hat{\phi}}{T_s} J_0(k_\perp \rho_s) \left(\omega-\omega_{\ast s}^{T}\right) F_{Ms}
    \label{eq:lin-GKE}
\end{align}
where it has been assumed that we are dealing with low-frequency drift waves characterised by eigenfrequencies $\omega/\Omega_s \ll 1$, and cross-field spatial scales $\lambda_\perp$ much shorter than equilibrium length scales, $\lambda_\perp/a \ll 1$, with $a$ the minor radius. With those conditions satisfied, each fluctuating quantity can be written as $\xi(\bm{x},t) = \hat{\xi}(l) \exp(iS(\bm{x}) - i\omega t)$, where the eikonal $e^{iS}$ describes the short-scale variation across the magnetic field, such that $\grad{S}\vdot\bm{B}=0$ and $\hat{\xi}(l)$ describes the long-scale variation of fluctuations along the magnetic field. Furthermore, in \Cref{eq:lin-GKE}, $\nabla_{\parallel} = \bm{e_b}\vdot\grad{}$ is the differential operator along the field-line coordinate, $\omega_{ds} = \bm{k_\perp} \vdot \bm{v_{ds}}$ the magnetic drift frequency, where $\bm{k_\perp} = \grad{S}$ is the perpendicular wavenumber, and $\bm{v_{ds}} = \bm{e_b} \cross \left(v_\parallel^2 \bm{\kappa} + v_\perp^2 \grad{\ln{B}}/2\right)/\Omega_s$ is the magnetic drift velocity, with $\bm{\kappa} = (\bm{e_b}\vdot\grad)\bm{e_b}$ the magnetic curvature vector, and $v_\parallel,v_\perp$ denote the velocity components decomposed with respect to their alignment to the magnetic field. Additionally, $J_0$ is the zeroth-order Bessel function of the first kind, $k_\perp = \norm{\bm{k_\perp}}$ is the magnitude of the perpendicular wavenumber (such that $k_\perp \sim 1/\lambda_\perp$), $\rho_{s} = \norm{\bm{e_b}\cross\bm{v}/\Omega_{s}}$ the Larmor radius, such that $J_0(k_\perp \rho_s)$ describes the effect of gyroaveraging the fluctuations over the unperturbed particle orbit, and $\omega_{\ast s}^{T} = \omega_{\grad{n_s}} + \omega_{\grad{T_s}} \left(v^2/(2v_{Ts}^2) - 3/2\right)$ is the temperature-dependent diamagnetic frequency in terms of the two components of the pressure gradient $\omega_{\grad{\{n_s,T_s\}}} = T_s \left(\bm{k_\perp}\cross\bm{e_b}\right)\vdot\grad{\ln \{n_s,T_s\}}/(q_s B)$, where $\omega_{\grad{n_s}}$ is commonly referred to as the diamagnetic frequency. In interpreting \Cref{eq:lin-GKE} as a one-dimensional differential equation in $\hat{g}_s(l)$ by introducing a perpendicular wavevector $\bm{k_\perp}$, the ballooning transform\cite{Connor1978,Dewar1998,Candy2004SmoothnessSurface} has been employed to reconcile the presence of sheared toroidal magnetic field lines with the periodicity constraint of Fourier modes. \par
The electrostatic potential is obtained self-consistently from Poisson's equation, which under reactor-relevant conditions where the length scales of interest far exceed the Debye length, reduces to a quasi-neutrality constraint on the density fluctuations $\delta n_{s} = \int (f_s-F_{Ms}) \dd^3{\bm{v_s}}$
\begin{align}
    \sum_{s} \frac{q_s^2 n_s \phi(\bm{x},t)}{T_s} = \sum_{s} q_s \int \expval{g_s(\bm{R},\bm{v},t)}_{\bm{x}} \dd^3{\bm{v_s}}
    \label{eq:gen-QN}
\end{align}
where $\expval{\cdots}_{\bm{x}}$ denotes a ring-average over the gyrophase, performed at constant particle position \cite{Howes2006}. For the linear eigenmodes with strong spatial anisotropy with respect to the magnetic-field orientation considered here, \Cref{eq:gen-QN} reduces to a constitutive relation between the slowly varying parallel profiles (denoted by hats)
\begin{align}
    \sum_{s} \frac{q_s^2 n_s \hat{\phi}}{T_s} = \sum_{s} q_s \int J_0(k_\perp \rho_s)\hat{g}_s \dd^3{\bm{v_s}}.
    \label{eq:lin-QN}
\end{align}

\subsection{Approximate solutions and local dispersion relation}
Whilst it is possible to find a general solution to \Cref{eq:lin-GKE} in terms of an integrating factor\cite{Taylor1968StabilityTheory}, the resulting quasi-neutrality condition becomes expressed in terms of charge-density kernels\cite{Connor1980StabilityIII}, which tends to obscure some of the more salient physical features underlying the instability mechanisms due to their non-tractability (see e.g.~Ref.~\onlinecite{Romanelli1989IonTokamaks} for such a general description of the ITG mode). \par
Instead here we focus on modes with characteristic frequencies ordered as
\begin{align}
   \omega_{b\mathrm{i}} \ll \omega_{t\mathrm{i}} \ll \omega \ll \omega_{b\mathrm{e}} \ll \omega_{t\mathrm{e}}
   \label{eq:freq-ordering}
\end{align}
where $\omega_{bs}$ and respectively $\omega_{ts}$ represent the characteristic bounce- and transit frequencies of species $s$. The former is defined in terms of the time it takes a thermal trapped particle to bounce between regions of magnetic maxima on a flux surface, while the latter is defined in terms of the time it takes a thermal passing particle to traverse the connection length, i.e.~the region between locations of good and bad magnetic curvature\cite{Plunk2014}. In a tokamak, this simply corresponds to the distance along the field line between the inboard and outboard side, but can be considerably shorter in a stellarator. The frequency ordering of \Cref{eq:freq-ordering} is equivalent to the condition for the mode to avoid strong Landau damping, and has also been considered in e.g.~ Refs.~\onlinecite{Helander2013,Plunk2014,Plunk2017,Romanelli1989IonTokamaks,Morren2024InfluenceStellarators,Connor1980StabilityIII,Terry1982KineticMode,Cowley1998ConsiderationsTurbulence,Biglari1998b}. Applying the frequency ordering of \Cref{eq:freq-ordering} to the gyrokinetic Vlasov equation [\Cref{eq:lin-GKE}], we can obtain asymptotic solutions for $\hat{g}_\mathrm{i}$ and $\hat{g}_\mathrm{e}$ in the limit of $\omega_{t\mathrm{i}}/\omega \sim \omega/\omega_{b\mathrm{e}} \ll 1$, which effectively weights the significance of the two terms of the left-hand sight, as the streaming term $v_\parallel \nabla_{\parallel}$ can be associated with the bounce- and transit motion. For the ions by neglecting the streaming term to lowest order we straightforwardly obtain 
\begin{align}
    \hat{g}_\mathrm{i} \approx \frac{Z_\mathrm{i} e \hat{\phi}}{T_\mathrm{i}} \frac{\omega-\omega_{\ast \mathrm{i}}^T}{\omega-\omega_{d\mathrm{i}}} J_0(k_\perp \rho_\mathrm{i}) F_{M\mathrm{i}},
    \label{eq:approx-gi-sol}
\end{align}
where $Z_\mathrm{i} = q_\mathrm{i}/e$ is the charge number and $e$ denotes the elementary charge.
Meanwhile, for the electrons the streaming term dominates, and to lowest order the distribution $\hat{g}_\mathrm{e}$ is constant along the field line. By taking a bounce-average of the gyrokinetic Vlasov equation [\Cref{eq:lin-GKE}], the electron gyrocenter distribution is found as\cite{Dominguez1992}
\begin{align}
    \hat{g}_\mathrm{e} \approx \frac{-e}{T_\mathrm{e}} \frac{\omega - \omega_{\ast \mathrm{e}}^T}{\omega - \overline{\omega_{d\mathrm{e}}}} \overline{J_0(k_\perp \rho_\mathrm{e})\hat{\phi}} F_{M\mathrm{e}}
    \label{eq:approx-ge-sol}
\end{align}
where $\overline{\cdots}$ denotes the bounce-average (transit-average) operator
\begin{align*}
    \overline{\cdots} = \frac{\oint \frac{\dd{l} \left(\cdots\right)}{\abs{v_\parallel}}}{\oint \frac{\dd{l}}{\abs{v_\parallel}}},
\end{align*}
where the integral is taken along subsequent bounce-points (along the entire field line), defined as the positions along the field line where $\abs{v_\parallel} = \sqrt{2(E-\mu B(l))}=0$, with $E=mv^2/2$ the particle kinetic energy and $\mu = mv_\perp^2/(2B)$ the magnetic moment, for trapped (passing) particles. We invoke one final approximation with regard to the spatial scales, where we consider modes characterised by $k_\perp \rho_{T\mathrm{i}} \sim \order{0.1}\mathrm{-}\order{1}$ where $\rho_{T\mathrm{i}} = v_{T\mathrm{i}}/\Omega_\mathrm{i}$ is the characteristic ion Larmor radius at thermal speed, which corresponds to the scales of interests where the bulk of the transport is typically observed in nonlinear simulations \cite{Merz2010,Pueschel2008GyrokineticBeta,Faber2015,Dannert2005,Garcia-Regana2021TurbulentPlasmas,Mulholland2025Finite-Modes}. At these scales, we may approximate $k_\perp \rho_\mathrm{e} \approx 0$ in the argument of the Bessel function in \Cref{eq:approx-ge-sol}, thereby effectively treating the electrons drift-kinetically, as also considered in e.g.~Refs.~\onlinecite{Helander2013,Connor1980StabilityIII,Kadomtsev1970TurbulenceSystems}. \par 
As a result of these simplifying assumptions on the parameter regime, we effectively treat all ions as passing particles, thus ignoring the possibility of trapped-ion modes, though such instabilities would require much larger spatial scales than supported by our $k_\perp \rho_{T\mathrm{i}} \sim \order{0.1}\mathrm{-}\order{1}$ ordering \cite{Kadomtsev1970TurbulenceSystems,Hahm1996NonlinearModes}. Moreover, we effectively treat passing electrons adiabatically, as passing particles will experience a net vanishing perturbation $\overline{\hat{\phi}} \rightarrow 0$ due to the finite extent of the modes along the field lines. Therefore, within the drift-kinetic (DK) treatment for electrons, the non-adiabatic response of electrons [\Cref{eq:approx-ge-sol}] vanishes for passing particles. As a consequence, short-wavelength (i.e.~electron-scale) instabilities associated with drift-resonances from passing particles like ETG modes\cite{Kadomtsev1970TurbulenceSystems,Horton1988ToroidalModes,Liu1971InstabilitiesCurrent,Lee1987CollisionlessInstability,Hirose1990ElectronTokamaks,Jenko2000} 
are absent from the model we present here. 
With these simplifications in mind, substituting \Cref{eq:approx-gi-sol,eq:approx-ge-sol} into the quasi-neutrality constraint \Cref{eq:lin-QN} yields the following local dispersion relation
\begin{widetext}
\begin{align}
    D_\mathrm{loc}(\omega,\bm{k_\perp},\{\mathrm{geo}\},\{\mathrm{plasma}\},l) = & \left(1+\sum_{j} \frac{Z_j^2 n_j}{n_\mathrm{e}} \frac{T_\mathrm{e}}{T_j} - \sum_j \frac{Z_j^2 n_j}{n_\mathrm{e}}\frac{T_\mathrm{e}}{T_j} \int \frac{\omega - \omega_{\ast j}^{T}}{\omega - \omega_{dj}(l)} \mathcal{F}_{Mj} J_0(k_\perp \rho_j(l))^2 \dd^3{\bm{v_j}}\right) \hat{\phi}(l)  \nonumber \\ 
    &-\int\limits_{\mathrm{trap}(l)} \frac{\omega-\omega_{\ast \mathrm{e}}^{T}}{\omega-\overline{\omega_{d\mathrm{e}}}} \overline{\hat{\phi}} \mathcal{F}_{M\mathrm{e}} \dd^3{\bm{v_\mathrm{e}}} = 0
    \label{eq:Dloc}
\end{align}
\end{widetext}
where the sum over $j$ accounts for all ion species present in the plasma. In obtaining \Cref{eq:Dloc} it has been assumed that the non-adiabatic part of the perturbed distribution function for all ion species can accurately be described by the gyrokinetic Vlasov equation [\Cref{eq:lin-GKE}] and its approximate solution of \Cref{eq:approx-gi-sol}, as e.g. also considered in the impurity model of Ref.~\onlinecite{Calvo2025ModificationPlasmas}, which are benign assumptions for all but the heaviest of impurities (see\footnotemark[3333]). Henceforth, the subscript $i$ shall exclusively refer to the main ion species. In \Cref{eq:Dloc} we have introduced the single-particle Maxwellian as $\mathcal{F}_{Ms} = F_{Ms}/n_s$. This dispersion relation depends on the geometric parameterisation through the magnetic drifts, as well as the FLR damping terms of the ions, as the magnitude of the perpendicular wavevector $\bm{k_\perp} = k_\psi \nabla{\psi} + k_{\alpha} \nabla{\alpha}$ is determined by
\begin{align}
    k_\perp = \sqrt{g^{\psi\psi}k_\psi^2 + g^{\alpha\alpha}k_{\alpha}^2 + 2 g^{\psi\alpha}k_\psi k_{\alpha}}
    \label{eq:kperp-clebsh}
\end{align}
where $g^{mn} = \grad{x^m}\vdot\grad{x^n}$ are the components of the contravariant metric tensor, $\psi$ the (poloidal) magnetic flux function and $\alpha=q\theta-\zeta$ the Clebsch angle, corresponding to the magnetic flux density $\bm{B}=\grad{\psi}\cross\grad{\alpha}$. Additionally, $\theta,\zeta$ denote the poloidal- and toroidal Boozer angles, respectively, while $q(\psi)$ is the inverse of the rotational transform, commonly known as the safety factor in tokamak literature \cite{Dhaeseleer1991}. In \Cref{eq:Dloc}, the dependency on parameters like the plasma composition and kinetic profiles are combined in $\{\mathrm{plasma}\}$, which like the magnetic-configuration dependency $\{\mathrm{geo}\}$ are considered as external parameters set by macroscopic considerations. The trapped region of velocity space is delimited by pitch angles $\lambda = \mu/E = v_\perp^2/(v^2B)$ in the range of $1/B_{\mathrm{max}} \leq \lambda \leq 1/B(l)$, where $B_{\mathrm{max}}$ is the global maximum of the magnetic field strength on a flux surface, and depends on the position along the field line, hence the nomer ``local'' for \Cref{eq:Dloc} despite the appearance of the bounce-averages. \par
By virtue of the approximations made above, we now gain physical insight regarding the underlying physics of the instabilities, for \Cref{eq:Dloc} provides two singularities in the non-adiabatic density fluctuations of ions ($\omega\approx\omega_{dj}$) and electrons ($\omega\approx\overline{\omega_{d\mathrm{e}}^{T}}$), leading to the resonance conditions for the (toroidal) ITG and TEM, respectively. This, however, results in an oversimplified picture, where it is assumed that the mode frequency is purely real-valued $\omega = \omega_R$ (corresponding to a stable drift wave), while 
such electrostatic fluctuations must have a finite growth ($\gamma>0$) or damping rate ($\gamma<0$), per $\omega = \omega_R + i \gamma$, producing Landau damping in homogeneous plasmas\cite{Landau1965ONPLASMA}, which can further be driven unstable at sufficiently large gradients\cite{Kadomtsev1970TurbulenceSystems}. \par 
In order to solve the kinetic integrals analytically, the drift resonances need to be carefully resolved. We propose here a generalisation of the method introduced in Ref.~\onlinecite{Biglari1998b} to resolve the ITG resonance in the DK limit, by writing the frequency denominators as
\begin{align}
    \frac{1}{\omega-\omega_{ds}} = \frac{1}{i\sigma_\gamma} \int\limits_{0}^{\infty}\dd{\xi} e^{i\sigma_\gamma (\omega-\omega_{ds})\xi} \quad \mathrm{iff} \ \Im[\omega] \neq 0
    \label{eq:resonce-denom-trick}
\end{align}
where $\sigma_\gamma = \sgn{\gamma} = \pm 1$ ensures absolute convergence of the auxiliary integral, which is strictly evaluated along $\mathbb{R}^{+}$, and therefore integrability of the resonance. A fundamental limitation of \Cref{eq:resonce-denom-trick}, however, is that the right-hand side is undefined for pure drift waves with $\omega=\omega_R$, though the left hand-side is, with exception of the point $\omega_R = \omega_{ds}$. For a given $\omega_R$, however, such a point will always occur while evaluating \Cref{eq:Dloc}, such that pure drift waves cannot constitute a valid solution to the dispersion relation. As conjectured above, solutions to the dispersion relation, which we seek to obtain using \Cref{eq:resonce-denom-trick}, must thus have finite $\gamma$. \par
Typically, this auxiliary integral approach to integrate the resonance is only considered for unstable modes\cite{Frei2022LocalMode,Beer1995,Gaur2024OmnigenousStability,Parisi2020ToroidalPedestals}, in which case it reduces to the approach taken by Biglari, Diamond and Rosenbluth\cite{Biglari1998b}, hereafter referred to as BDR. We note that \Cref{eq:resonce-denom-trick} closely resembles the form used in Eqn 3.5 from Ref.~\onlinecite{Ivanov2023AnField} to similarly extend the applicability of the BDR approach to damped modes, where instead the upper bound is considered to be $\sigma_\gamma \infty$ for robust numerical integration regardless of whether modes are damped or growing. Though, as we shall see in \Cref{app:disp-integral-deriv-general}, maintaining the integration path along the positive real line is beneficial when solving the resulting velocity integrals from the dispersion relation analytically. After applying \Cref{eq:resonce-denom-trick} to \Cref{eq:Dloc}, the remainder of the derivation involves exchanging the order of integration between $\xi$ and velocity-space variables, followed by a series of changes of integration variables, each tailored to maintain the absolute convergence that is guaranteed by \Cref{eq:resonce-denom-trick}, and is deferred to \Cref{app:disp-integral-deriv-general}, resulting in

\begin{widetext}
\begin{align}
    & D_\mathrm{loc}(\omega,k_\psi,k_{\alpha},\{\mathrm{geo}\},\{\mathrm{plasma}\},l) = \left(1+\sum_{j} \frac{Z_j^2 n_j}{n_\mathrm{e}} \frac{T_\mathrm{e}}{T_j} - \sum_j \frac{Z_j^2 n_j}{n_\mathrm{e}}\frac{T_\mathrm{e}}{T_j} \left[\Gamma_0(b_j) - \frac{\omega_{\grad{n_j}}-\frac{3}{2}\omega_{\grad{T_j}}}{\omega} J_{j}^{(0)}(\omega,\{\mathrm{geo}\},\bm{k_\perp},l) \right. \right. \nonumber \\ 
    & \left. \left. + \frac{\omega_{j,\grad{B}}-\omega_{\grad{T_j}}}{2\omega} J_{j,\perp}^{(2)}(\omega,\{\mathrm{geo}\},\bm{k_\perp},l) + \frac{\omega_{j,\kappa}-\frac{\omega_{\grad{T_j}}}{2}}{\omega} J_{j,\parallel}^2(\omega,\{\mathrm{geo}\},\bm{k_\perp},l)\right]\right) \hat{\phi}(l) - B(l) \int\limits_{1/B_\mathrm{max}}^{1/B(l)} \dd{\lambda} \frac{\overline{\hat{\phi}}(\lambda)}{\sqrt{1-\lambda B(l)}} \times \nonumber \\
    &  
    \left[\frac{1}{2}+\frac{3}{4}\frac{\overline{\omega_{d\mathrm{e}}^{T}}(\lambda)-\omega_{\grad{T_\mathrm{e}}}}{\omega} J_{\textrm{tr-el}}^{(2)}(\omega,\lambda,\{\mathrm{geo}\},\bm{k_\perp}) - \frac{\omega_{\grad{n_\mathrm{e}}}-\frac{3}{2}\omega_{\grad{T_\mathrm{e}}}}{2\omega} J_{\textrm{tr-el}}^{(1)}(\omega,\lambda,\{\mathrm{geo}\},\bm{k_\perp}) \right] = 0.
    \label{eq:loc-disp-solve}
\end{align}
\end{widetext}

In \Cref{eq:loc-disp-solve} we have absorbed the variation of $\bm{k_\perp}$ due to the metric into the geometric dependency $\{\mathrm{geo}\}$, while $\omega_{s,\kappa} = \bm{k_\perp}\vdot \left(\bm{e_b} \cross \bm{\kappa}\right) v_{Ts}^2/\Omega_s$ and $\omega_{s,\grad{B}} = \bm{k_\perp}\vdot \left(\bm{e_b} \cross \grad{\ln B}\right)v_{Ts}^2/\Omega_s$ are the usual thermal curvature and $\grad{B}$ components of the magnetic drift frequency, $b_j = \left(k_\perp \rho_{Tj}\right)^2$ represents the magnitude of the perpendicular wavenumber normalised to thermal ion Larmor radius, and  $\Gamma_n(z)=\exp(-z)I_n(z)$ denotes the exponentially scaled $n^{\mathrm{th}}$-order modified Bessel functions of the first kind. Additionally, we have used an energy and pitch angle representation of velocity space for electrons, such that the bounce-average operator 

\begin{align}
    \overline{\cdots} = \frac{\oint \frac{\dd{l}}{{\sqrt{1-\lambda B(l)}}} (\cdots)}{\oint \frac{\dd{l}}{{\sqrt{1-\lambda B(l)}}}}
    \label{eq:bav-operator-lambda}
\end{align}
becomes a function of pitch angle only, and $\omega_{d\mathrm{e}}^T(\lambda,l) = 2\omega_{e,\kappa}(l) \left(1-\lambda B(l)\right) + \omega_{e,\grad{B}}(l) \lambda B(l)$ is the orbit-modulated total drift frequency of an electron at thermal energy $E=T_\mathrm{e}$ with pitch angle $\lambda$. In \Cref{eq:loc-disp-solve}, the following ion integrals appear associated with the drift resonance:

\begin{widetext}
\begin{align}
    \begin{split}
            J_{j}^{(0)} &= \frac{\omega}{i\sigma_\gamma} \int\limits_{0}^{\infty} \dd{\xi} \frac{e^{i\sigma_\gamma \omega \xi}\Gamma_0(\hat{b_j}(\xi))}{\sqrt{1+2i\sigma_\gamma \omega_{j,\kappa} \xi}(1+i\sigma_\gamma \omega_{j,\grad{B}}\xi)} \\  
            J_{j,\perp}^{(2)} &=\frac{2\omega}{i\sigma_\gamma} \int\limits_{0}^{\infty} \dd{\xi} \frac{e^{i\sigma_\gamma \omega \xi} \left(\Gamma_0(\hat{b_j}(\xi)) + \hat{b_j}(\xi) \left[\Gamma_1(\hat{b_j}(\xi))-\Gamma_0(\hat{b_j}(\xi))\right]\right)}{\sqrt{1+2i\sigma_\gamma \omega_{j,\kappa} \xi}(1+i\sigma_\gamma \omega_{j,\grad{B}}\xi)^2} \\
            J_{j,\parallel}^{(2)} &= \frac{\omega}{i\sigma_\gamma} \int\limits_{0}^{\infty} \dd{\xi} \frac{e^{i\sigma_\gamma \omega \xi}\Gamma_0(\hat{b_j}(\xi))}{\left(1+2i\sigma_\gamma \omega_{j,\kappa} \xi\right)^{3/2}(1+i\sigma_\gamma \omega_{j,\grad{B}}\xi)} 
        \end{split}
        \label{eq:ion-1Dres-integrals}
    \end{align}
\end{widetext}
where $\hat{b_j}(\xi) = b_j/(1+i\sigma_\gamma \omega_{j,\grad{B}} \xi)$. Likewise, in \Cref{eq:loc-disp-solve} we have introduced the resonant electron integrals
\begin{align}
    J_{\textrm{tr-el}}^{(1)} &= -2\left(a_e + a_e^{3/2}\sigma_\gamma \sigma_{de}Z(\sigma_\gamma \sigma_{de} \sqrt{a_e})\right), \label{eq:trel-genplasmaZ-integrals} \\
    J_{\textrm{tr-el}}^{(2)} &= -2\left(\frac{a_e}{3}\left(1+2a_e\right) + \frac{2}{3} a_e^{5/2} \sigma_\gamma \sigma_{de}Z(\sigma_\gamma  \sigma_{de} \sqrt{a_e})\right), \nonumber
\end{align}
where $a_e  = \omega/\overline{\omega_{d\mathrm{e}}^{T}}$, and $\sigma_{de} = \sgn{\overline{\omega_{d\mathrm{e}}^{T}}}$, and the plasma dispersion function $Z(\zeta)=i\sqrt{\pi}\mathcal{W}(\zeta)$ here appears explicitly in the form of its fundamental definition in terms of the Faddeeva function\cite{Abramowitz1968HandbookTables} % Eqn 7.1.3.
$\mathcal{W}(\zeta) = \exp(-\zeta^2)\erfc(-i\zeta)$. We note that the kinetic-ion-related terms from \Cref{eq:loc-disp-solve,eq:ion-1Dres-integrals} match with the expressions given in Appendix A of Ref.~\onlinecite{Zocco2018ThresholdPlasmas} -- corrected for the typo $J_{j,\parallel}^{2} \leftrightarrow J_{j,\perp}^{2}$ that was recently also identified by Ref.~\onlinecite{Podavini2025EnergeticLandscape} -- in the case of $\sigma_\gamma = +1$, consistent with their use of the original BDR approach to integrate the resonant denominator. Meanwhile, the trapped-electron-related terms in \Cref{eq:loc-disp-solve,eq:trel-genplasmaZ-integrals} match with Eqn.~(25) of Ref.~\onlinecite{Connor1980StabilityIII} in the case of $\sigma_{\gamma} \sigma_{de} = +1$. Here, this discrepancy is caused by the use in Ref.~\onlinecite{Connor1980StabilityIII} of the integral representation of the plasma dispersion function
\begin{align}
    Z(\zeta) = \frac{1}{\sqrt{\pi}} \int\limits_{-\infty}^{\infty}\frac{\exp(-z^2)}{z-\zeta} \dd{z} \quad \mathrm{iff} \ \Im[\zeta] > 0
    \label{eq:plasmaZ-int}
\end{align}
as defined by Fried and Conte\cite{FriedContePlasmaZ}, though this representation suffers from the restriction that it is only valid in the upper half-plane, unlike the definition in terms of a scaled Faddeeva function used throughout this work, which is valid throughout the complex plane  \cite{Abramowitz1968HandbookTables}.In the context of Laplace-Fourier transform for the distribution function and potential fluctuations the plasma dispersion function is commonly analytically continued to extend application of \Cref{eq:plasmaZ-int} to damped fluctuations \cite{Ivanov2023AnField,Gultekin2020GeneralizedITG,Gurcan2014NumericalCurvature}. However, we omit such an approach in this work as the resulting damped fluctuations do not correspond to pure eigenmodes of the linear gyrokinetic system, which contradicts our prior fluctuation ansatz of $\xi(\bm{x},t)\propto\xi(\bm{x})e^{-i\omega t}$ for all fluctuating quantities. Meanwhile, pure damped eigenmodes of the linear gyrokinetic system do exist and are known to play an important role in turbulence saturation through nonlinear interactions \cite{Hatch2011SaturationEigenmodes,Pueschel2016}, and are hence of interest to model. Regardless of the approach taken to model damped fluctuations, the validity constraint of \Cref{eq:plasmaZ-int} is nonetheless restrictive in the accurate modelling of unstable modes -- which is the main use case of the model in \Cref{sec:GENE-application} -- for among the various trapping wells along the field line the bounce-averaged trapped-electron drift, appearing in the argument of the plasma dispersion function in \Cref{eq:trel-genplasmaZ-integrals}, will change its sign in some regions. This influence of the magnetic drift on the validity plasma dispersion function is often disregarded, though noticeable exceptions are the work from Ref.~ \onlinecite{Gurcan2014NumericalCurvature}, which considers a similar sign generalisation of $Z(\zeta)=\sigma Z(\sigma\zeta)$ where $\sigma=\sgn{\Im[\zeta]}$ in the context of ITG modes, and the TEM model recently developed by Ref.~\onlinecite{Garbet2024TheModel} where the role of $\sgn{\overline{\omega_{d\mathrm{e}}^{T}}}$ (equivalent to $\epsilon$ in their notation) has been accounted for in the evaluation of the plasma dispersion function, though their TEM model neglects the role of kinetic ions and is limited to geometry of shaped tokamaks.\par 
Additionally, we note that using our form of the plasma dispersion function as in \Cref{eq:trel-genplasmaZ-integrals}, is crucial to satisfy the fundamental property that the trapped-electron energy integral $I_{\textrm{tr-el}}$ [see \Cref{eq:Itrel-def}] is complex-conjugate symmetric, i.e.~it has the property that $I_{\textrm{tr-el}}(\omega^*) = [I_{\textrm{tr-el}}(\omega)]^{*} $, with $*$ denoting the complex conjugate, as can readily be proven using the symmetry property of the Faddeeva function\cite{Abramowitz1968HandbookTables} $\mathcal{W}(\zeta^*) = [\mathcal{W}(-\zeta)]^*$. % Eqn 7.1.12
Similar considerations regarding complex-conjugate symmetry also hold for the resonant ion velocity integrals (i.e. the first term in square brackets in \Cref{eq:loc-disp-solve}). \par

Lastly, before continuing with presenting the global dispersion model, we briefly remark about the influence of transit resonances due to the streaming term in the gyrokinetic Vlasov equation [\Cref{eq:lin-GKE}]. One particular analytically tractable case, is that of a plasma in a constant and unsheared magnetic field. In such a slab geometry fluctuations do not resonate with the magnetic drift frequency, but may resonate with the transit frequency $\omega\approx\omega_{ts}=k_\parallel v_{Ts}$ where $k_\parallel=2\pi/L_\parallel$ is the wavenumber of fluctuations aligned with the magnetic field. For completeness the derivation of the dispersion relation belonging to such slab-ITG modes is presented in \Cref{app:generalised-slab-derivation}, and can be similarly expressed in terms of the sign generalised plasma dispersion function as occur in our prescription of the trapped-electron resonant integral \Cref{eq:trel-genplasmaZ-integrals}. In literature the effect of transit resonances on the curvature-driven, i.e. toroidal, ITG mode is often modelled by modifying the resonant denominator in \Cref{eq:approx-gi-sol} to $1/(\omega - \omega_{ds} - k_\parallel v_\parallel)$ creating quasi-local models\cite{Frei2022LocalMode,Beer1995,Gultekin2020GeneralizedITG,Ivanov2023AnField,Gao2005ShortPlasmas,Parisi2020ToroidalPedestals,Xie2016GlobalGradient}. While forming a perfect analogy to the slab-branch of ITG instability, it should be noted that the use of a parallel wavenumber is formally ill-founded as the geometry is varying along the field line, and such a $k_\parallel$ should be considered as an ad-hoc effective parallel wavenumber meant to heuristically reincorporate the effects of Landau resonances in a toroidal system, while a proper inclusion of streaming effects in a toroidal system requires $k_\parallel$ to be considered as the streaming operator.

\subsection{Global dispersion relation}
\begin{figure}[hb]
    \centering
    \includegraphics[width=0.95\linewidth]{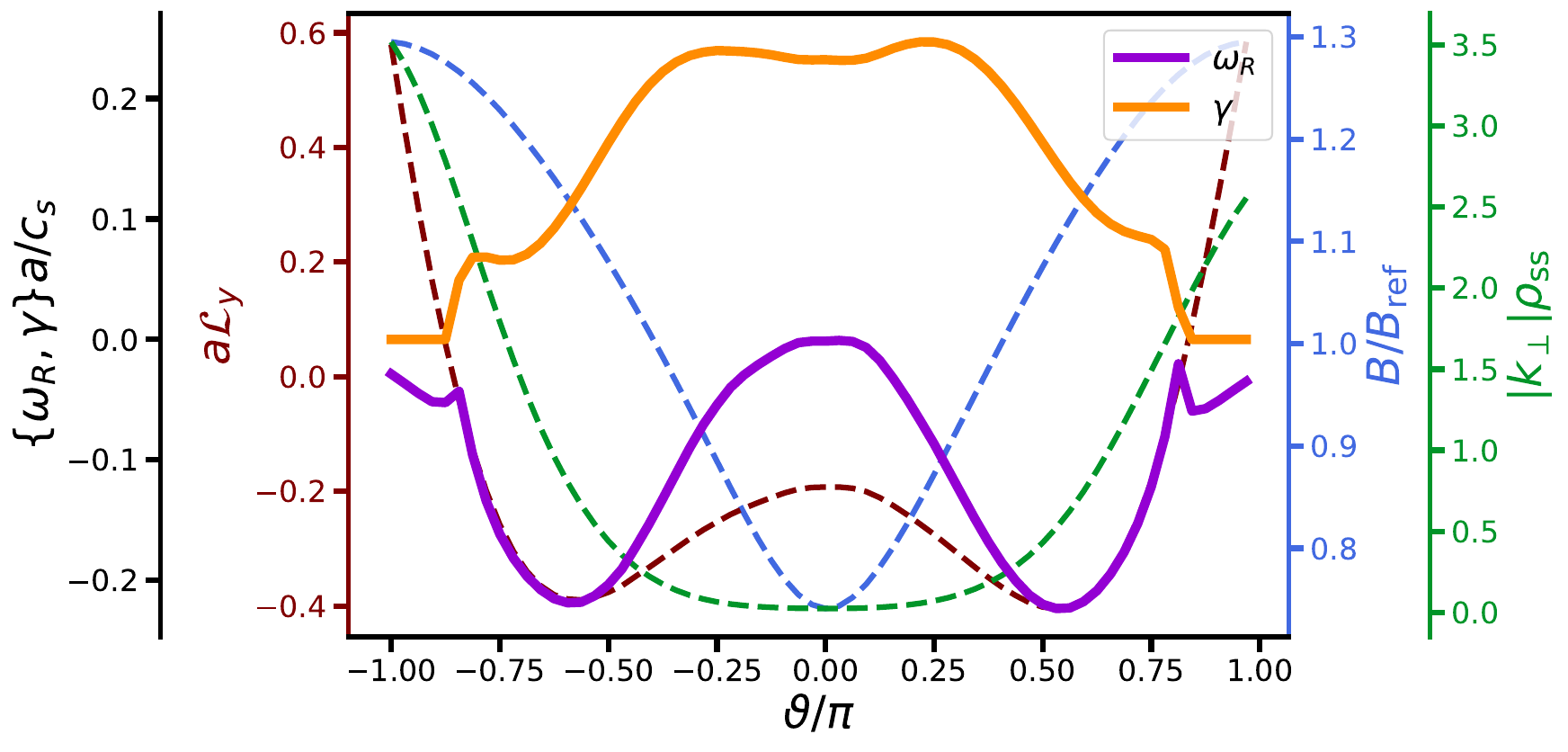}
    \caption{Variation along the magnetic field line of the normalised local eigenfrequency $\omega_{\mathrm{loc}}(l)$ solution to \Cref{eq:Dloc} in the adiabatic-electron limit, for a toroidal ITG instability characterised by radial and bi-normal wavenumbers $k_x \rho_{\mathrm{ss}} = 0, k_y \rho_{\mathrm{ss}}=0.3$ driven by density and temperature gradients of $a/L_{ni} = 2$ and $a/L_{T\mathrm{i}}=4$, respectively, in the flux tube of the DIII-D tokamak considered in this work (see \Cref{sec:numerics}). The plasma consists of only a single ion species with $Z_\mathrm{i} = 1$ and $T_\mathrm{i}=T_\mathrm{e}$. Contrasted against the eigenfrequency (solid lines; split into growth rate and propagation frequency) are the variation of the magnetic geometry (dashed lines) focussing on the magnetic field strength in blue, the magnitude of the (normalised) perpendicular wavevector in green and the bi-normal component of the $\grad{B}$ drift operator (see \Cref{sec:fluxtubes}) in maroon. For reference, the corresponding eigenfrequency obtained by \textsc{Gene} is $\omega a/c_{s}=-0.048+0.1820i$, being within the extremes of the local solutions.}
    \label{fig:loc-freq-sol-visual}
\end{figure}

We now turn our attention to a problematic issue in the local dispersion relation, which makes \Cref{eq:loc-disp-solve} inadequate for obtaining meaningful eigenfrequencies in toroidal geometry. If we consider a location along a field line in close proximity of $B_{\mathrm{max}}$, the trapped-electron response will vanish and \Cref{eq:loc-disp-solve} reduces to the form $D_\mathrm{loc}  = h(\omega,\{\mathrm{geo}\},\{\mathrm{plasma}\},l_\mathrm{max}) \times \hat{\phi}(l_{\mathrm{max}}) = 0$, where $l_{\mathrm{max}}$ is (one of) the position(s) along the field line where the $B(l)=B_{\mathrm{max}}$, which has either the trivial solution that the eigenmode must vanish at $l_{\mathrm{max}}$ consistent with an instability driven solely by trapped electrons, or we obtain a family of purely local eigenmode solutions $\omega(l_{\mathrm{max}})$ determined by the plasma composition and local geometry at $l_{\mathrm{max}}$. In the class of (approximately) omnigeneous devices of interest for fusion applications, there will be multiple equivalent magnetic maxima, all spread equidistantly along the field line\cite{Landreman2012OmnigenityQuasisymmetry,Cary1997OmnigenitySystems}. However, in the presence of finite (global) magnetic shear $\dv*{q}{\psi}$, the perpendicular wavevector $\bm{k_\perp}$ is secular (i.e.~increasing along the field line without bound) \cite{Plunk2014,Waltz1993}, resulting in neither the FLR damping nor the magnetic drift being equivalent between these magnetic maxima. Consequently, this local family of eigenfrequency solutions becomes disjoint between the various magnetic maxima. This problem is further exacerbated in the limit of the adiabatic electron model, in which case the trapped-electron contribution to \Cref{eq:loc-disp-solve} vanishes altogether, propagating the locality issue to all points on the field line as $D_{\mathrm{loc}} = 0$ reduces to $h(\omega,\{\mathrm{geo}\},\{\mathrm{plasma}\},l)=0$, yielding purely local eigenfrequency solutions $\omega_{\mathrm{loc}}(l)$ determined solely by the local geometry. The variation of these local solutions can be quite significant, even in the axisymmetric limit where the connection length (proportional to the characteristic length scale of variation in the geometry along the field line) is much larger than in a typical stellarator\cite{Plunk2014}, see \Cref{fig:loc-freq-sol-visual}. The interplay between drive (magnetic drift) and damping factors (FLR effects) is clear, with the (local) propagation frequency $\omega_R$ adjusting itself to the profile of the magnetic drift to maximise the resonance $\omega_R \approx \omega_{d\mathrm{i}}$ in the thermal bulk of the distribution, whilst the (local) growth rate decreases as $\norm{\bm{k_\perp}}$ increases away from $\theta = 0$ (corresponding to the outboard midplane) due to FLR damping. The growth rate eventually vanishes (asymptotically) as the magnetic drift changes sign (around $\theta \approx \pm0.85\pi$), thus locally rendering ion diamagnetic drift waves ($\omega_R \sim \omega_{\grad{n_{\mathrm{i}}}} < 0$ under conventional radially peaked profiles) stable against magnetic drift resonances.  \par

Clearly, such a collection of local solutions to the eigenfrequency are incompatible with the fundamental notion of an eigenmode, whose characteristics should be a global system property set by the overall magnetic configuration. 
This discrepancy between local and global solutions is a consequence of the frequency ordering \Cref{eq:freq-ordering}, which results in the streaming term of the gyrokinetic Vlasov equation [\Cref{eq:lin-GKE}] being neglected for ions, resulting in purely local solutions for the non-adiabatic part of the perturbed ion distribution function \Cref{eq:approx-gi-sol}. Without the effect of particle streaming, the adiabatic electron dispersion relation $D_{\mathrm{loc}}^{\mathrm{ae}}=h(\omega,\{\mathrm{geo}\},\{\mathrm{plasma}\},l)\hat{\phi}(l)$ formally has an eigenfunction basis consisting of Dirac delta functions ($\hat{\phi}\sim\delta(l-l_0)$)\cite{Connor1980StabilityIII}, which violate the underlying assumption that the eigenmodes are characterised by a long-scale variation along the field lines. The streaming term provides the necessary regularising effect on the admissible eigenfunctions, as it introduces non-local (along the field line) physics to the dispersion relation. When kinetic electrons are reintroduced, such regularisation is naturally provided by the bounce-average operator, resulting in an integral problem necessitating self-consistent solutions for $\hat{\phi}(l)$ as well-behaved smooth functions. In the adiabatic-electron limit, however, non-local effects can be re-introduced by not completely neglecting the influence of particle streaming, but considering an asymptotic expansion in $\omega_{ti}/\omega \ll 1$ for the ion gyrocenter distribution (to which \Cref{eq:approx-gi-sol} would be the leading zeroth-order term), resulting to lowest order in an $\order{(\omega_{ti}/\omega)^2}$ correction to the dispersion relation which introduces differential operators acting on $\hat{\phi}(l)$, providing the necessary regularising non-local effects\cite{Plunk2014,Romanelli1989IonTokamaks}. As this ion sound term is formally only $\order{(\omega_{ti}/\omega)^2}$, its effects are negligible compared with the regularlising influence introduced by the bounce-averaged trapped electrons when kinetic electrons are retained\cite{Connor1980StabilityIII}.\par

Here we propose an alternative approach to reintroduce non-local effects to the electrostatic dispersion relation, which equally applies to the adiabatic and kinetic treatments of the electrons. We construct a global (along the field line) dispersion relation by multiplying by $\hat{\phi}/B(l)$ and integrating along the entire field line, which facilitates a variational principle as is shown below, yielding
\begin{widetext}
    \begin{align}
        D_\mathrm{glob}(\omega,k_\psi,k_{\alpha},\{\mathrm{geo}\},\{\mathrm{plasma}\}) = \int \dd{l} \frac{D_\mathrm{loc}(\omega,\bm{k_\perp},\{\mathrm{geo}\},\{\mathrm{plasma}\},l) \hat{\phi}}{B} =\left(1+\sum_{j} \frac{Z_j^2 n_j}{n_e} \frac{T_e}{T_j}\right) \int \dd{l} \frac{\hat{\phi}^2}{B} \nonumber\\
        - \sum_j \frac{Z_j^2 n_j}{n_e}\frac{T_e}{T_j} \int \dd{l} \frac{h_{\mathrm{ion},j}(\omega,\bm{k_\perp},\{\mathrm{geo}\},\{\mathrm{plasma}\},l)\hat{\phi}^2}{B} 
        -  \int\limits_{1/B_\mathrm{max}}^{1/B_\mathrm{min}} \dd{\lambda} \sum_{\mathrm{wells}(\lambda)} K_{\textrm{tr-el},\mathrm{well}}(\omega,\bm{k_\perp},\lambda,\{\mathrm{geo}\},\{\mathrm{plasma}\}) \overline{\hat{\phi}}_{\mathrm{well}}^2 L_{\mathrm{well}}^{\mathrm{eff}}
        \label{eq:glob-disp}
    \end{align}
\end{widetext}
where we have reversed the order of integration over pitch angle and field line for the trapped-electron contribution. The summation takes into account the different disconnected trapping wells for a given pitch angle $\lambda$ along the field line, which corresponds to the various (non-equivalent) populations of trapped particles, each with their intrinsic bounce-average
\begin{align*}
    \overline{\hat{\phi}}_{\mathrm{well}} = \frac{\int\limits_{l_{a,\mathrm{well}}}^{l_{b,\mathrm{well}}} \frac{\dd{l}\hat{\phi}(l)}{\sqrt{1-\lambda B(l)}}}{\int\limits_{l_{a,\mathrm{well}}}^{l_{b,\mathrm{well}}} \frac{\dd{l}}{\sqrt{1-\lambda B(l)}}} \equiv \frac{\int\limits_{l_{a,\mathrm{well}}}^{l_{b,\mathrm{well}}} \frac{\dd{l}\hat{\phi}(l)}{\sqrt{1-\lambda B(l)}}}{L_{\mathrm{well}}^{\mathrm{eff}}},
\end{align*}
where $l_{\{a,b\},\mathrm{well}}$ are the successive bounce-points of a magnetic well, now formally defined as each simply connected region along the field line where $B(l)\leq 1/\lambda$ is satisfied, and we have introduced $L_{\mathrm{well}}^{\mathrm{eff}}$ as the effective arc-length of each well, which defines the bounce time as $\tau_{b}^\mathrm{well} = L_{\mathrm{well}}^{\mathrm{eff}}/\norm{\bm{v}}$. In writing \Cref{eq:glob-disp}, we have introduced the short-hand notations for the resonant integrals (with explicit dependencies suppressed)
\begin{align}
    \begin{split}
        h_{\mathrm{ion},j} =& \Gamma_0(b_j) - \frac{\omega_{\grad{n_j}}-\frac{3}{2}\omega_{\grad{T_j}}}{\omega} J_{j}^{(0)} \\ &+ \frac{\omega_{j,\grad{B}}-\omega_{\grad{T_j}}}{2\omega} J_{j,\perp}^{(2)}+ \frac{\omega_{j,\kappa}-\frac{1}{2}\omega_{\grad{T_j}}}{\omega} J_{j,\parallel}^{(2)}, \\
        K_{\textrm{tr-el}} =& \frac{1}{2}+\frac{3}{4}\frac{\overline{\omega_{d\mathrm{e}}^{T}}(\lambda)-\omega_{\grad{T_\mathrm{e}}}}{\omega} J_{\textrm{tr-el}}^{(2)} - \frac{\omega_{\grad{n_\mathrm{e}}}-\frac{3}{2}\omega_{\grad{T_\mathrm{e}}}}{2\omega} J_{\textrm{tr-el}}^{(1)},
    \end{split}
    \label{eq:ion-el-glob-kernels}
\end{align}
which respectively provide the kernels for the normalised fluctuating ion number density and trapped-electron pitch-angle density. \par 
Analogous approaches for considering global geometric effects have been considered ranging from perturbative approaches to include the effect of the TEM precession resonance on the eigenmode frequency\cite{Morren2024InfluenceStellarators,Dominguez1992,Rafiq2006,Hastie1971StabilityMultipoles,Rutherford1968,Gang1991}, energy budget analyses to determine stability properties\cite{Proll2012,Helander2015b,Ivanov2025ThePlasmas,Helander2013} and general instability limits\cite{Helander2022,Plunk2022EnergeticGrowth}, to obtaining quadratic forms of the dispersion relation which closely resemble a traditional variational formalism\cite{Rutherford1968,Coppi1977,Garbet1990VariationalTokamaks,Plunk2017,Helander2013,Aleynikova2017} as commonly used in e.g. reduced transport models based on quasilinear mixing-length approaches in tokamaks\cite{Bourdelle2015,Bourdelle2007APlasmas}. Here, however, in light of recent findings from Ref.~\onlinecite{Stephens2025AInstabilities}, we deviate from the aforementioned works by multiplying with $\hat{\phi}$ rather than $\hat{\phi}^{*}$, as is formally required to render the non-Hermitian gyrokinetic system variational, as is proven momentarily. Therefore, our work can be regarded as both a correction and an extension to the work from Refs.~\onlinecite{Helander2013,Plunk2017}, where such approaches for electrostatic drift waves have been generalised to stellarator geometry, though only applied in the high-frequency limit where the drift resonances are expanded, and are obtained by multiplying with $\hat{\phi}^{*}$ instead. Adopting the notation of the aforementioned works, we consider \Cref{eq:glob-disp} as a functional $D_{\mathrm{glob}}(\omega) \rightarrow S[\omega,\hat{\phi}]$ where we suppress the dependence on other fixed quantities. Within the framework of the trapped-electron model, where admissible self-consistent eigenmodes are characterised by $D_{\mathrm{loc}} = 0$, it follows that $S[\omega_0,\hat{\phi}_0]=0$ for the same frequency-eigenmode pair $(\omega_0,\hat{\phi}_0)$. If the mode structure were to be perturbed $\hat{\phi} \rightarrow \hat{\phi}_0 + \delta \hat{\phi}$, then the mode frequency will necessarily change according to $\omega \rightarrow \omega_0+\delta\omega$ so that $S[\omega_0+\delta\omega,\hat{\phi}_0+\delta\hat{\phi}]=0$, and consequently $\Delta S=0$ to all orders in $\delta$. Let us now consider some arbitrary $\omega$ and $\hat{\phi}$, and require the functional $S$ to be stationary under small perturbations such that 
\begin{widetext}
    \begin{align}
         \delta S=0 =& \left(1+\sum_{j} \frac{Z_j^2 n_j T_e}{n_e T_j}\right) \int \frac{\dd{l}}{B} 2\hat{\phi}\delta\hat{\phi} - 
         \sum_{j} \frac{Z_j^2 n_j T_e}{n_e T_j} \int \frac{\dd{l}}{B} \left( 2 h_{\mathrm{ion},j}(\omega,l) \hat{\phi}\delta\hat{\phi}+ \pdv{h_{\mathrm{ion},j}}{\omega} \delta \omega \hat{\phi}^2\right) \nonumber \\
         &-\int\limits_{1/B_\mathrm{max}}^{1/B_\mathrm{min}} \dd{\lambda} \sum_{\mathrm{wells}(\lambda)}  L_{\mathrm{well}}^{\mathrm{eff}} \left( 2K_{\textrm{tr-el},\mathrm{well}}(\omega,\lambda)\overline{\hat{\phi}}_{\mathrm{well}} \ \overline{\delta\hat{\phi}}_{\mathrm{well}} + \pdv{K_{\textrm{tr-el},\mathrm{well}}}{\omega} \delta \omega \overline{\hat{\phi}}_{\mathrm{well}}^2 \right)
         \label{eq:deltaS_var_form}
    \end{align}
\end{widetext}
and consequently $\Delta S \sim \order{\delta^2}$. The vanishing of the right hand side of \Cref{eq:deltaS_var_form} requires that the derivatives of the kernels \Cref{eq:ion-el-glob-kernels} exist despite sharp resonances in the velocity integrals. Both $\pdv*{h_{\mathrm{ion},j}}{\omega}$ and $\pdv*{K_{\mathrm{trel},\mathrm{well}}}{\omega}$ are well defined and finite provided that $\omega$ is sufficiently far away from marginality ($\gamma=0$), such that $\sigma_{\gamma} = \sgn{\Im[\omega]}$ remains fixed during the variation $\omega\rightarrow \omega + \delta \omega$, which is slightly stricter than the validity condition for \Cref{eq:resonce-denom-trick}, which guarantees the existence of the ion- [\Cref{eq:ion-1Dres-integrals}] and electron velocity integrals [\Cref{eq:trel-genplasmaZ-integrals}]. The frequency modification is then obtained from \Cref{eq:deltaS_var_form} and reads
\begin{widetext}
    \begin{align}
        \delta \omega = \frac{2\times\int \frac{\dd{l}}{B} \delta\hat{\phi} \left[ \left(1+\sum_{j} \frac{Z_j^2 n_j T_e}{n_e T_j} \left(1-h_{\mathrm{ion},j}\right)\right) \hat{\phi} - B \int\limits_{1/B_{\mathrm{max}}}^{1/B} \dd{\lambda} \frac{\overline{\hat{\phi}}}{\sqrt{1-\lambda B}} K_{\textrm{tr-el}} \right]}{\sum_{j} \frac{Z_j^2 n_j T_e}{n_e T_j} \int \frac{\dd{l}}{B} \pdv{h_{\mathrm{ion},j}}{\omega} \hat{\phi}^2+ \int\limits_{1/B_\mathrm{max}}^{1/B_\mathrm{min}} \dd{\lambda} \sum\limits_{\mathrm{wells}(\lambda)}  L_{\mathrm{well}}^{\mathrm{eff}} \pdv{K_{\mathrm{trel},\mathrm{well}}}{\omega} \overline{\hat{\phi}}_{\mathrm{well}}^2} \equiv \frac{2 A[\omega,\delta{\hat{\phi}},\hat{\phi}]}{C[\omega,\hat{\phi}]}
        \label{eq:omega_var_statement}
    \end{align}
\end{widetext}
where we have unravelled the sum over the various magnetic wells back into a field line integral by changing the order of integration in the numerator and introduced two new functionals for notational convenience. Note that these emergent functionals can be written succinctly in terms of the local dispersion relation [\Cref{eq:loc-disp-solve}] as $C[\omega,\hat{\phi}]=-\int \dd{l} (\hat{\phi} \fdv*{D_{\mathrm{loc}}[\omega,\hat{\phi}]}{\omega})/B = - \fdv*{D_{\mathrm{glob}}[\omega,\hat{\phi}]}{\omega}$ and $A[\omega,\delta\hat{\phi},\hat{\phi}] = \int \dd{l} \delta\hat{\phi} D_{\mathrm{loc}}[\omega,\hat{\phi}]/B$, where $\fdv*{f}{g}$ denotes the functional derivative and we require $C[\omega,\hat{\phi}]\neq 0$. We note that upon replacing the kernel functions [\Cref{eq:ion-el-glob-kernels}] by their non-resonant expansions [\Cref{eq:fluid-ion-kernel,eq:fluid-TEM-kernel} for $h_{\mathrm{ion},j}$ and $K_{\textrm{tr-el}}$, respectively] and assuming the absence of impurities, \Cref{eq:omega_var_statement} almost reduces to the variational form in Ref.~\onlinecite{Helander2013} where $\delta \omega = 2 A[\omega,\delta \hat{\phi},\hat{\phi}]/C[\omega,\abs{\hat{\phi}}]$ is obtained. In the aforementioned work, however, not only was the dispersion relation multiplied by $\hat{\phi}^{*}$, but $\hat{\phi}$ and $\hat{\phi}^{*}$ were not considered as separate variables when constructing $\delta S$, as formally required in e.g.~analogous variational formulations of quantum mechanics, where the Hermitian nature of the Schrödinger equation facilitates the construction of a Lagrangian density in terms of the wavefunction and its complex conjugate\cite{Ohta2000,Wagner1994,Sanz2012}. If one would propagate this rationale to the non-Hermitian gyrokinetic system under consideration here, one would instead have obtained $\delta \omega = (A[\omega,\delta \hat{\phi}^{*},\hat{\phi}]+A[\omega,\delta\hat{\phi},\hat{\phi}^{*}])/C[\omega,\abs{\phi}]$ for the frequency variation if the functional were constructed by multiplying the (local) dispersion relation [\Cref{eq:Dloc}] with $\hat{\phi}^{*}$ instead of $\hat{\phi}$. \par
We now state our variational principle: it follows that $\delta \omega = 0$ when $A[\omega,\delta\hat{\phi},\hat{\phi}] = 0$, whilst simultaneously $A[\omega,\delta\hat{\phi},\hat{\phi}]=0$ if $(\omega,\hat{\phi}) \rightarrow (\omega_0,\hat{\phi}_0)$ are a proper frequency-eigenmode pair solving the (local) dispersion relation [\Cref{eq:loc-disp-solve}]. Conversely, if $\delta \omega = 0$ for all variations $\delta{\phi}$, then \Cref{eq:loc-disp-solve} must be satisfied by the ($\omega,\hat{\phi}$) pair under consideration. Hence, in a liberal sense, this resembles a conventional variational principle.

This variational formalism of \Cref{eq:glob-disp} is only applicable when the effect of kinetic electrons is included such that \Cref{eq:loc-disp-solve} is a properly regularised integral problem. If, however, we consider \Cref{eq:glob-disp} in the context of the adiabatic-electron model, it follows that a root of $D_{\mathrm{glob}}$ provides an frequency-eigenmode pair $(\tilde{\omega},\tilde{\hat{\phi}})$ with a suitable eigenmode $\tilde{\hat{\phi}}\neq\delta(l-l_0)$ and a global eigenmode frequency $\tilde{\omega}$. Whilst this frequency-eigenmode pair does not solve $D_{\mathrm{loc}}=0$ (and therefore $\tilde{\omega}$ deviates from $\omega_{\mathrm{loc}}(l)$), it provides a measure of the overall instability drive to the mode as a result of FLR damping and the distribution of good- and bad-curvature regions along the field line. In the adiabatic-electron limit, one may therefore instead interpret \Cref{eq:glob-disp} not as being variational, but as an eigenmode-weighted average of $D_{\mathrm{loc}}$ along the field line. Electrostatic instabilities within the intermediate-frequency regime considered here, see \Cref{eq:freq-ordering}, always have eigenmodes of the electrostatic potential localised within regions of bad curvature\cite{Ivanov2025ThePlasmas}. Weighting the local dispersion relation by an appropriate $\hat{\phi}$ informed about mode localisation physics (as e.g.~obtained from first principles by considering the effect of ion sound dynamics or data-driven proxies) will thus result in a stronger influence of the bad-curvature regions on the global dispersion relation, resulting in global eigenfrequencies with $\Im[\tilde{\omega}]>0$. Therefore, for now, we shall consider \Cref{eq:glob-disp} as an ad-hoc means to introduce non-local effects to the dispersion relation in the adiabatic-electron limit, and verify \textit{a posteriori} that the obtained eigenfrequencies from this method are in agreement with linear gryokinetic simulations.\par

\section{Gyrokinetic simulation setup and dispersion relation solution strategy} \label{sec:numerics}
In order to test the validity of the global dispersion relation \Cref{eq:glob-disp}, we perform linear gyrokinetic simulations with the flux-tube version of the \textsc{Gene} code\cite{Jenko2000}. These simulations allow us to assess the ability of \Cref{eq:glob-disp} to describe (toroidal) ITG and TEM by using the \textsc{Gene} linear eigenfunctions as a guess for $\hat{\phi}$. As these eigenfunctions are exact (to within convergence precision), formally $\delta \hat{\phi}=0$, and ergo from \Cref{eq:omega_var_statement}, it follows that $\delta\omega = 0$. Consequently, any deviation between the eigenfrequency of \textsc{Gene} simulations and the (roots of) global dispersion relation [\Cref{eq:glob-disp}] should therefore be directly attributable to missing physics in our model. Whilst the variational principle only strictly holds for TEMs, the same approach allows us to erify the ad-hoc applicability of \Cref{eq:glob-disp} to describe adiabatic-electron ITGs. A similar approach for verifying the variational principle has been considered in Ref.~\onlinecite{Aleynikova2017} for KBMs. \par 

\subsection{Geometric considerations for flux-tubes} \label{sec:fluxtubes}
In the flux-tube approach \cite{Beer1995Field-alignedTurbulence}, the equilibrium gradients are considered fixed (valid for systems with small $\rho^{*} = \rho/a$), as the simulation domain is typically limited to a perpendicular extent of a few tens of (ion) gyroradii whilst extending along a magnetic field line for $N_{\mathrm{pol}}$ poloidal turns. As a consequence, the geometry (predominantly determined by $\grad{\bm{B}}$) varies along the field line, as the salient geometric features only change over macroscopic length-scales. In order to capture the anisotropy of the fluctuations with respect to the alignment of the a magnetic field, a (non-Cartesian) coordinate system $(x,y,z)$ is adopted
\begin{align}
    x=L_{\mathrm{ref}} \sqrt{s}, \quad y=L_{\mathrm{ref}}\sqrt{s_0}\alpha/q_0, \quad z=\theta
    \label{eq:flux-tube-coo}
\end{align}
where $s=\Phi/\Phi_{\mathrm{edge}}$ is the normalised (toroidal) magnetic flux function, $L_{\mathrm{ref}}$ denotes the macroscopic normalisation length scale (typically taken to be the minor radius), $s_0$ indicates the radial position position at the centre of the flux tube, $q = \dv*{\Phi}{\psi}$ and $q_0$ is a short-hand notation for $q(s_0)$. As a result of \Cref{eq:flux-tube-coo}, the magnetic field within the flux tube is characterised by $\bm{B} = B_{\mathrm{ref}}(x) \grad{x}\cross\grad{y}$, with $B_{\mathrm{ref}}$ a reference magnetic field strength as determined by the MHD equilibrium such that $\psi_{\mathrm{edge}} = B_{\mathrm{ref}} a^2/2$, therefore rendering $z$ as the field-aligned coordinate, which is taken to be the poloidal angle. Periodic boundary conditions are enforced across the perpendicular domain, while a twist-and-shift boundary condition\cite{Martin2018,Faber2018} is applied along $z$:
\begin{align}
    \tilde{\phi}_{k_x,k_y}(z,t) = \tilde{\phi}_{k_x+p\Delta k_x,k_y}(z+2\pi p N_{\mathrm{pol}},t) \tilde{C}_{ky}^p
    \label{eq:twist-and-shift}
\end{align}
where $\tilde{\phi}_{k_x,k_y}$ represents a perpendicular Fourier mode of the electrostatic potential, $p\in\mathbb{Z}$, $\Delta k_x = 2\pi N_{\mathrm{pol}} \hat{s} k_{y}$ with $\hat{s} = \left.\dv*{\ln{q}}{\ln{x}}\right|_{x=x_0}$ being the global magnetic shear, and $\tilde{C}_{k_y}$ is a phase factor with $\abs{\tilde{C}_{k_y}}=1$. Whilst \Cref{eq:twist-and-shift} explicitly shows the application of the twist-and-shift boundary condition for the electrostatic potential, the same rule is applied to all fluctuations. 
Although this twist-and-shift boundary condition is formally incorrect in stellarators due to the general lack of periodicity in the magnetic geometry after $N_{\mathrm{pol}}$ poloidal turns, reasonably accurate results can be obtained if the parallel extent of the flux tube is large enough such that the modes experience the proper drive and damping mechanisms as set by the real geometry over a turbulent correlation length \cite{Martin2018,Faber2018}. \par
In terms of these flux-tube coordinates, the relevant geometric quantities in the (linear) gyrokinetic Vlasov equation [\Cref{eq:lin-GKE}] can all be computed from the elements of the corresponding contravariant metric tensor\cite{Xanthopoulos2006,Xanthopoulos2009,Gorler2011} $g^{nm} = \grad{x^n}\vdot\grad{x^m}$, where $\{x^1,x^2,x^3\}=\{x,y,z\}$. In particular, the perpendicular wavenumber $\norm{\bm{k_\perp}}$ can be computed from \Cref{eq:kperp-clebsh} under the substitution $\psi \rightarrow x$ and $\alpha \rightarrow y$, while the parallel derivative follows as
$\nabla_{\parallel} = 1/(\sqrt{g} B_N) \pdv{}{z}$, where $B_N = B/B_{\mathrm{ref}}$ represents the normalised magnetic field strength and $\sqrt{g} = \left( \left(\grad{x}\cross\grad{y}\right)\vdot\grad{z}\right)^{-1}$ is the determinant of the metric tensor. Meanwhile the geometric component of the $\grad{B}$ drift is given by $\left(\bm{e_b}\cross\ln{B}\right)\vdot \bm{k_\perp} = k_x \mathcal{L}_x + k_y \mathcal{L}_x $ in terms of the drift-operators
\begin{align}
    \begin{split}
       \mathcal{L}_x =& - \left(\left.\pdv{B_N}{y}\right|_{x,z} + \frac{\gamma^{2}}{\gamma^{1}} \left.\pdv{B_N}{z}\right|_{x,y}\right) \\
        \mathcal{L}_y =& \left.\pdv{B_N}{x}\right|_{y,z} + \frac{\gamma^{3}}{\gamma^{1}} \left.\pdv{B_N}{z}\right|_{x,y}
    \end{split}
    \label{eq:gradB-operators-GENE}
\end{align}
where $\gamma^{1} = g^{xx}g^{yy}-(g^{xy})^2$, $\gamma^{2} = g^{yz}g^{xx}-g^{xz}g^{xy}$ and $\gamma^{3} = g^{xz}g^{yy}-g^{xy}g^{yz}$. Analogously, the geometric component of the curvature drift can be written as $\left(\bm{e_b}\cross\bm{\kappa}\right)\vdot \bm{k_\perp} = k_x \mathcal{K}_x + k_y \mathcal{K}_y$. Since the curvature vector can be expressed as\cite{Helander2014}
\begin{align}
    \bm{\kappa} = \grad_{\perp}{\ln B} + 2\beta \grad{\ln p},
    \label{eq:curv-gradp-rel}
\end{align}
it follows, due to the (assumed) existence of (isobaric) flux surfaces, that the drift-operators for the curvature drift are $\mathcal{K}_x = \mathcal{L}_x$ and $\mathcal{K}_y = \mathcal{L}_y + (\dv*{\beta_{\mathrm{ref}}}{x})/(2B_N)$, where $\beta_{\mathrm{ref}}$ is the normalised total plasma pressure with respect to magnetic pressure of $B_{\mathrm{ref}}$.

\subsection{Description of geometries and simulation parameters} \label{sec:num_settings}
In order to assess the capabilities of the global dispersion model \Cref{eq:glob-disp} to describe ITG and TEM instabilities in arbitrary toroidal geometry, we compare against gyrokinetic simulations in a variety of realistic configurations of existing devices. The flux-tube geometry information is generated by the Geometry Interface for Tokamaks and Stellarators (GIST) code\cite{Xanthopoulos2009} from the magnetic equilibrium of each configuration. For each device we perform \textsc{Gene} simulations using a single flux-tube, which is sufficient to capture all salient geometric features on a tokamak flux surface. For the stellarator configurations, we choose the so-called \textit{bean flux tube} -- corresponding to the flux tube where the bean-shaped poloidal cross-section is intersected at the out-board midplane -- which for both HSX\cite{Faber2015} and W7-X \cite{Sanchez2021,Proll2013} tends to be the most unstable flux tube. The global (i.e. MHD) geometric parameters determining the flux-tube domain for each of the considered magnetic configurations are shown in \Cref{tab:FT_global_header_params}. Note that all flux tubes are centred at a normalised radius of $s_0 = 0.5$, chosen to be sufficiently far radially outward to have both appreciable gradients as well as an appreciable trapped-electron fraction, with the exception of the TCV flux tubes. These are placed more radially outward at $s_0 = 0.774$ where the effects of triangular shaping of the last closed-flux surface are more discernable\cite{Marinoni2009TheSimulations,Merlo2015InvestigatingTransport}, resulting in flux tubes characterised by local triangularities of $\delta(s_0) \approx \pm 0.2$.  
For consistency with the electrostatic approximation, all flux tubes correspond to vacuum configurations of their respective devices.

\begin{table*}
    \centering
    \caption{Global geometric parameters that determine the flux-tube coordinate system \Cref{eq:flux-tube-coo} for the various geometries.}
    \begin{ruledtabular}
        \begin{tabular}{ccccccc}
            & DIII-D & TCV $\delta^+$ & TCV $\delta^-$ & HSX & W7-X (hm) & W7-X (nm) \\ \hline
            $s_0$ & 0.5 & 0.774 & 0.774 & 0.5 & 0.5 & 0.5 \\ 
            $q_0$ & 2.566 & 1.052 & 1.052 & 0.941 & 1.102 & 1.135 \\ 
            $\hat{s}$ & 1.579 & 0.705 & 0.705 & -0.046 & -0.129 & -0.141 \\ 
            $N_{\mathrm{pol}}$ & 1 & 1 & 1 & 4 & 1 & 1 \\
        \end{tabular}
    \end{ruledtabular}
    \label{tab:FT_global_header_params}
\end{table*}
As discussed in \Cref{sec:fluxtubes}, a flux tube spanning a single poloidal turn contains all essential geometry information for a tokamak, however, for stellarators, $N_{\mathrm{pol}}$ has to be chosen carefully such that the modes experience a sufficient amount of the physical driving and damping forces set by the real geometry, as opposed to the artificial geometry generated by the twist-and-shift boundary condition (see \Cref{sec:Dglob-method} on this matter). In previous studies, flux tubes spanning $N_{\mathrm{pol}}=1$ and $N_\mathrm{pol}=4$ poloidal turns have been found to be sufficiently long for W7-X\cite{Martin2018,Sanchez2021} and HSX\cite{Faber2018}, respectively, as are also considered for the flux tubes used in this work. \par
For both ITG and TEM we perform linear initial-value \textsc{Gene} simulations for a various scenarios. First, to focus on the relevant ITG physics, we perform adiabatic-electron simulations at fixed density gradient $a/L_{n} = 2$ while varying the temperature gradient $a/L_{T\mathrm{i}} = \left[3,4,5,6\right]$, where $1/L_{Ts}= -\dv*{\ln T_s}{x}$ is the inverse temperature gradient scale length of species $s$ (with an analogous definition for $1/L_{ns}$), defined such that $L_{Ts},L_{ns}>0$ for centrally peaked profiles. 
At low concentrations, the presence of impurities in the plasma results in dilution effects that weaken the ITG instability, see  Ref.~\onlinecite{Angioni2021ImpurityExperiments} and references therein.  These dilution effects are due to ambipolarity (i.e. quasi-neutrality for the equilibrium density profiles), which, assuming the electron density profile to be fixed to focus on impact of impurities, limits the main ion density and its gradient to
\begin{align}
    \begin{split}
        \frac{n_\mathrm{i}}{n_\mathrm{e}} &= 1 - \sum_{\mathrm{imp}} \frac{Z_{\mathrm{imp}}}{Z_i} \frac{n_{\mathrm{imp}}}{n_\mathrm{e}}  \\
        \frac{L_{n\mathrm{e}}}{L_{n\mathrm{i}}} &= \frac{1 - \sum_{\mathrm{imp}} Z_{\mathrm{imp}} \frac{n_{\mathrm{imp}}}{n_\mathrm{e}} \frac{L_{n\mathrm{e}}}{L_{n\mathrm{imp}}}}{1 - \sum_{\mathrm{imp}} Z_{\mathrm{imp}} \frac{n_{\mathrm{imp}}}{n_\mathrm{e}}}
    \end{split}
    \label{eq:impurity-ambipolarity}
\end{align}
where we have explicitly split the sum over ion quantities into the main species and impurities. From \Cref{eq:impurity-ambipolarity} it is evident that the ion concentration decreases as the impurity concentration increases, leading to a decrease in resonant population for ITGs. The presence of impurities can also modify the ITG drive by either enhancing or reducing the ion density gradient with respect to the (nominal) electron density gradient, depending on the shape of the impurity density profile (with respect to the electron density profile). This trend in enhancement/reduction of linear drive depending on the shape of the impurity density profile has also recently been reported to carry over to turbulent heat fluxes in nonlinear simulations\cite{Garcia-Regana2024ReductionImpurities}. Lastly, when the impurity concentration becomes significantly large, an impurity-driven ITG may emerge as the dominant instability. For all simulations in this paper, we consider fully ionised carbon $\mathrm{C}^{6+}$ as the only impurity species, such that the effective charge $Z_{\mathrm{eff}} = \sum_j Z_j^2 n_j/n_\mathrm{e}$ together with the impurity (and electron) density gradient(s) fully determines the ion quantities through \Cref{eq:impurity-ambipolarity}. We perform two sets of simulations at fixed nominal gradients of $a/L_{n\mathrm{e}} = 2, \ a/L_{T\mathrm{i}}=a/L_{T\mathrm{C}}=4$, one scanning over $Z_{\mathrm{eff}} = [1.2,1.4,\ldots,2]$ while considering $a/L_{n\mathrm{C}} = 2$ (such that \Cref{eq:impurity-ambipolarity} implies $a/L_{n\mathrm{i}}= 2$ as well) to isolate the density dilution effect, and one scanning over $a/L_{n\mathrm{C}} = [-4,-2,\ldots, 6]$ at fixed value of $Z_{\mathrm{eff}} = 1.4$ to isolate the effect of density gradient dilution. In present-day experiments, values of $Z_{\mathrm{eff}}\approx 1.4\textrm{-}1.6$ are commonly observed in devices with a carbon wall\cite{Mariani2018IdentifyingFluxes,Romba2023EvaluationW7-X,Grierson2018,Wolf2019}, though higher values of $Z_{\mathrm{eff}}>2$ are achievable through deliberate impurity seeding, as will be the case for future reactor-scale devices\cite{Siccinio2020,Siccinio2022}. Additionally, realistic impurity density profiles typically tend to be flat ($a/L_{n\textrm{C}}= 0$)\cite{Romba2023EvaluationW7-X}, though hollow impurity profiles with an off-axis peak have also been observed in both tokamaks\cite{Manas2017GyrokineticPlasmas,Grierson2018} and the Large Helical Device stellarator\cite{Ida2009ObservationDevice}. \par
Secondly, in simulations including kinetic electrons we obtain TEMs by considering artificial profiles with finite density and (optionally) electron temperature gradient, but a vanishing ion temperature gradient $\grad{T_\mathrm{i}} =0$ to eliminate the possibility of ITGs. This generates two datasets, one for density-gradient-driven TEMs (where in addition we assume $\grad{T_\mathrm{e}}=0$) scanning over $a/L_{n} = [1,2,3,4]$, and one where we scan over $a/L_{T\mathrm{e}}=[1,2,\ldots,6]$ at fixed $a/L_{n} = 3$ to probe the predictive capabilities of the global dispersion model with respect to electron-temperature-gradient-driven TEMs. Furthermore, the inclusion of kinetic electrons also has an effect on the ITG instability, hence we consider a third set where we scan over $a/L_{T\mathrm{i}} = [1,2,\ldots,6]$ in presence of a fixed density gradient $a/L_{n} = 3$, while assuming a vanishing electron temperature gradient to limit the drive of TEMs in favour of ITGs as the ion temperature gradient is increased. Lastly, a fourth kinetic-electron dataset is considered based on realistic profiles where both ion and electron temperature gradient are finite and are separately modified to $a/L_{Ts} \rightarrow \{1,5\}$ with respect to a nominal scenario of $a/L_{n} =3,\ a/L_{T\mathrm{i}} = 3, a/L_{T\mathrm{e}} = 3$. For all simulations with kinetic electrons, impurities are excluded. \par
For all simulations, we consider a deuterium plasma, with a mass ratio of $m_\mathrm{i}/m_\mathrm{e} = 3670$ when electrons are treated kinetically, and assume $T_\mathrm{i} = T_\mathrm{e}$, corresponding to reactor-relevant fusion conditions. Whenever impurities are considered, we assume thermal equilibrium between ion species, $T_\mathrm{i} = T_\mathrm{C}$, due to the fast collisional relaxation timescale between species with similar mass\cite{Romba2023EvaluationW7-X},
which implies likewise that $L_{T\mathrm{i}} = L_{T\mathrm{C}}$, as has been considered in all simulations including impurities. Similar assumptions regarding the impurity temperature have been applied by e.g.~Refs.~\cite{Garcia-Regana2024ReductionImpurities,Belli2017ImplicationsSimulation}. For each scenario, we scan over the bi-normal wavenumber $k_y\rho_{\mathrm{ss}} = [0.1,0.2,\ldots,2]$ and the driving gradient (or $Z_{\mathrm{eff}}$ in case of impurities). The ballooning space of modes has been centred at $k_x=0$, which typically corresponds to the most unstable mode in axisymmetric configurations\cite{Duff2022EffectTurbulence} as well as the bean flux tube in stellarators\cite{Sanchez2021,Proll2013,Faber2015}. 
For each scenario, we have performed convergence tests in the number of grid points used to discretise the velocity-space variables $v_\parallel$ and $\mu$ as well as the total extent of ballooning space as set by $2\pi N_{\mathrm{pol}} n_{kx}$ at both extremities of the driving gradient to ensure that the eigenfrequencies are converged within $\leq5$\%. The resulting resolutions for each configuration and scenario are shown in \Cref{tab:GENE_sim_resolutions}.

\begin{table*}
    \centering
    \caption{Numerical resolutions for the \textsc{Gene} simulations for various geometries and drive-scenarios. The numbers in the table correspond to the tuple $(n_{kx},n_{z},n_{v},n_{\mu})$ representing the number of radial Fourier modes included in the extended eigenmode envelope, and the number of grid points used for the field-line following coordinate, the parallel velocity space and the magnetic moment. Below ``ae'' denotes an adiabatic electron simulation, ``ke'' refers to a simulation with kinetic electrons, whilst ``full gradient'' indicates that the effect of finite ion and electron temperature gradient drive are both included.}
    \begin{ruledtabular}
        \begin{tabular}{ccccccc}
            & DIII-D & TCV $\delta^+$ & TCV $\delta^-$ & HSX & W7-X (hm) & W7-X (nm) \\ \hline
            ae ITG & (11,64,24,8) & (21,64,32,12) & (21,64,24,8) & (21,512,24,8) & (43,96,32,12) & (43,128,32,12) \\ 
            ae ITG + impurity & (11,64,24,8) & (21,64,48,16) & (21,64,36,18) & (21,512,24,8) & (43,96,32,12) & (43,128,32,12) \\ 
            $\grad{n}$-TEM & (21,64,48,12) & (33,64,64,12) & (33,64,64,16) & (31,512,48,18) & (43,96,64,24) & (33,128,64,24) \\ 
            ke ITG & (21,64,64,18) & (33,64,48,12) & (33,64,64,16) & (21,512,48,18) & (43,96,64,24) & (43,128,48,18) \\ 
            $\grad{T_e}$-TEM & (21,64,36,8) & (33,64,36,8) & (33,64,48,12) & (41,512,48,18) & (43,96,48,18) & (43,128,48,18) \\ 
            Full gradient & (21,64,64,18) & (33,64,48,12) & (33,64,48,12) & (31,512,48,18) & (43,96,64,24) & (43,128,48,18) \\ 
        \end{tabular}
    \end{ruledtabular}
    \label{tab:GENE_sim_resolutions}
\end{table*}

\subsection{Solving the global dispersion relation} \label{sec:Dglob-method}
In order to facilitate a direct comparison between the eigenfrequencies obtained by the global dispersion relation model and the linear gyrokinetic simulations we consider a similar normalisation scheme as used in \textsc{Gene}. 
This entails that equilibrium (gradient) length scales are normalised to $L_{\mathrm{ref}}$, whilst the perpendicular fluctuation length scales are normalised to a reference gyroradius  $\rho_{\mathrm{ref}} = c_{\mathrm{ref}}/\Omega_{\mathrm{ref}}$, where $c_{\mathrm{ref}}=\sqrt{T_{\mathrm{ref}}/m_{\mathrm{ref}}}$ and $\Omega_{\mathrm{ref}}=eB_{\mathrm{ref}}/m_{\mathrm{ref}}$. Here we choose $m_{\mathrm{ref}}=m_\mathrm{i}$ and $T_{\mathrm{ref}}=T_\mathrm{e}$ such that $\rho_{\mathrm{ref}}$ is the ion gyroradius at sound speed $\rho_{\mathrm{ss}}$ (note the distinction from $\rho_{s}$ -- as commonly used in literature to denote the sound ion Larmor radius\cite{Mynick2009GeometryTurbulence,Proll2013,Gerard2024OnStellarators,Costello2023TheStellarators} -- defined in \Cref{sec:disp-theory}). Subsequently, all temperatures are normalised by $T_{\mathrm{ref}}$, masses by $m_{\mathrm{ref}}$, and densities by $n_{\mathrm{ref}}$, the latter taken to be the electron density. 
Consequently, time scales are subsequently normalised by $L_{\mathrm{ref}}/c_{\mathrm{ref}}$, while the electrostatic potential is normalised to $e/T_{\mathrm{ref}}$. By choosing $L_{\mathrm{ref}}$ to coincide with minor radius $a$, compatibility with the GIST code providing normalised flux-tube geometry information is ensured. In terms of these normalisations and the flux-tube coordinates from \Cref{eq:flux-tube-coo}, the expressions for the various quantities appearing in \Cref{eq:glob-disp,eq:ion-el-glob-kernels} are summarised in \Cref{tab:normalised-geo-quantities}. \par

\begin{table*}
    \centering
    \caption{Overview of normalised terms appearing in the global dispersion relation. Whenever applicable, the terms are expressed in quantities related to the flux-tube coordinates \Cref{eq:flux-tube-coo}, and in terms of various geometric quantities defined in \Cref{sec:fluxtubes}.}
    \begin{ruledtabular}
        \begin{tabular}{cc}
            $\dd{l}$ & $L_{\mathrm{ref}} \sqrt{g} B\dd{z}/B_{\mathrm{ref}}$ \\
            $B$ & $B_{\mathrm{ref}} \sqrt{g^{xx}g^{yy}-(g^{xy})^2}$ \\
            $b_j$ & $\left[(k_x \rho_{\mathrm{ref}})^2 g^{xx} + (k_y \rho_{\mathrm{ref}})^2 g^{yy} + 2 k_x k_y \rho_{\mathrm{ref}}^2 g^{xy}\right]m_jT_j B_{\mathrm{ref}}^2/(Z_j^2 B^2 T_{\mathrm{ref}}m_{\mathrm{ref}})$  \\
            $\omega_{\grad{n_j}}$ & $-\left[k_y \rho_{\mathrm{ref}} T_jL_{\mathrm{ref}}/(Z_j T_{\mathrm{ref}}L_{n_j})\right] c_{\mathrm{ref}}/L_{\mathrm{ref}}$ \\
            $\omega_{\grad{T_j}}$ & $-\left[k_y \rho_{\mathrm{ref}} T_jL_{\mathrm{ref}}/(Z_j T_{\mathrm{ref}}L_{T_j})\right] c_{\mathrm{ref}}/L_{\mathrm{ref}}$ \\
            $\omega_{j,\grad{B}}$ & $\left[T_j B_{\mathrm{ref}}\left(k_x \rho_{\mathrm{ref}} \mathcal{L}_{x} L_{\mathrm{ref}}+k_y \rho_{\mathrm{ref}} \mathcal{L}_{y} L_{\mathrm{ref}}\right)/(Z_j T_{\mathrm{ref}} B)\right] c_{\mathrm{ref}}/L_{\mathrm{ref}}$ \\
            $\omega_{j,\kappa}$ & $\left[T_j B_{\mathrm{ref}}\left(k_x \rho_{\mathrm{ref}} \mathcal{K}_{x} L_{\mathrm{ref}}+k_y \rho_{\mathrm{ref}} \mathcal{K}_{y} L_{\mathrm{ref}}\right)/(Z_j T_{\mathrm{ref}} B)\right] c_{\mathrm{ref}}/L_{\mathrm{ref}}$
        \end{tabular}
    \end{ruledtabular}
    \label{tab:normalised-geo-quantities}
\end{table*}

One key concern for the global dispersion relation model is the geometry representation. Since we obtain the electrostatic potential from \textsc{Gene} simulations using the twist-and-shift boundary condition, see \Cref{eq:twist-and-shift}, we consider the extended ballooning representation of the eigenmode beyond the original $2\pi N_{\mathrm{pol}}$ extent of the flux tube where geometric quantities are defined, allowing us to consider only the effect of a $k_x=0$ mode with a vanishing amplitude $\abs{\hat{\phi}(\theta)}\rightarrow 0$ as $\abs{\theta} \rightarrow \infty$, which is a required boundary condition on the fluctuations in the ballooning formalism\cite{Connor1980StabilityIII}. As this formalism was used to obtain the gyrokinetic Vlasov equation [\Cref{eq:lin-GKE}] from which we derived our dispersion relation, it should be respected, while the eigenmode amplitude is typically not vanishing at the flux-tube boundaries of $\abs{\theta} = \pi N_{\mathrm{pol}}$ unless $N_{\mathrm{pol}}$ is large, hence necessitating a description of the geometry beyond its original domain. 
The geometry-related terms appearing in \Cref{tab:normalised-geo-quantities} are therefore expanded in ballooning space (see Supplemental Material) as 

\begin{align}
    \sqrt{g}(\theta+2\pi p N_{\mathrm{pol}}) =& \sqrt{g}(\theta) \nonumber \\
    g^{xx}(\theta+2\pi p N_{\mathrm{pol}})=& g^{xx}(\theta) \nonumber \\
    g^{xy}(\theta + 2\pi p N_{\mathrm{pol}}) =& g^{xy}(\theta)+ 2\pi p N_{\mathrm{pol}} \hat{s}g^{xx}(\theta) \nonumber \\
    g^{yy}(\theta + 2\pi p N_{\mathrm{pol}}) =& g^{yy}(\theta)+ 4\pi p N_{\mathrm{pol}} \hat{s} g^{xy}(\theta) \nonumber \\ 
    & +(2\pi p N_{\mathrm{pol}}\hat{s})^2 g^{xx}(\theta) \nonumber \\
    B(\theta+2\pi pN_{\mathrm{pol}}) =& B(\theta) \nonumber \\
    \mathcal{L}_x(\theta+2\pi p N_{\mathrm{pol}})=& \mathcal{L}_x(\theta) \nonumber \\
    \mathcal{K}_x(\theta+2\pi p N_{\mathrm{pol}}) = & \mathcal{K}_x(\theta) \nonumber \\
    \mathcal{L}_y(\theta+2\pi p N_{\mathrm{pol}}) = & \mathcal{L}_y(\theta) + 2\pi p N_{\mathrm{pol}} \hat{s} \mathcal{L}_x(\theta) \nonumber \\
    \mathcal{K}_y(\theta+2\pi p N_{\mathrm{pol}}) = & \mathcal{K}_y(\theta) + 2\pi p N_{\mathrm{pol}} \hat{s} \mathcal{L}_x(\theta).
    \label{eq:geometry-extend-formulae}
\end{align}
where $p\in \mathbb{Z}$ defines the extended ballooning angle as $\theta_p=\theta+2\pi pN_{\mathrm{pol}}$ as related to the parallel boundary condition [\Cref{eq:twist-and-shift}]. It should be noted that \Cref{eq:geometry-extend-formulae} is formally incorrect for stellarators, as the toroidal variation of the geometry has been neglected, resulting in discontinuities in e.g. the drift-operators and $\norm{\bm{k_\perp}}$ proportional to the global shear, though the consequences of this discrepancy on the linear eigenmode properties are fairly benign \cite{Martin2018}. 
A last geometric consideration is the calculation of the bounce-averages, since the involved bounce-integrals are formally divergent as one approaches the bounce-points. As we consider realistic 3D geometries, we cannot make use of analytical expressions for trapped-particle precession as exist for tokamaks\cite{Garbet2024TheModel,Stephens2021b}. Rather we use the specialised trapezoidal method developed by Ref.~\onlinecite{Mackenbach2023Bounce-averagedCases} for these bounce-integrals with a $1/\sqrt{f(x)}$ singularity.
\par
In order to evaluate $D_{\mathrm{glob}}(\omega|k_x,k_y,\{\mathrm{geo}\},\{\mathrm{plasma}\})$, the ion integrals in \Cref{eq:ion-1Dres-integrals} are evaluated numerically
, whilst the analytical expressions for \Cref{eq:trel-genplasmaZ-integrals} are used, except in magnetic wells where the bounce-averaged drift is close to vanishing. In the latter case, the argument of the plasma dispersion function diverges, which makes the analytical expressions ill-defined. Hence, whenever $\abs{\omega/\overline{\omega_{d\mathrm{e}}^T}} > 250$ we use a cubic asymptotic expansion [see \Cref{eq:TEM-asymp-int})] for $J_{\textrm{tr-el}}^{(1)},J_{\textrm{tr-el}}^{(2)}$ instead. Finally, in order to obtain the roots of \Cref{eq:glob-disp}, we make use of a quasi-Newton method employing a combination of the so-called ``good'' Broyden algorithm\cite{Broyden1965AEquations,Broyden2000OnMethod} with line search\cite{Armijo1966MinimizationDerivatives} to allow for adaptive step-sizing. In order to guarantee convergence of this method, both a good initial guess for both the mode frequency and the Jacobian are necessary \cite{More1976OnMethod,Demeester2022OnMethod,Decker1985BROYDENSROOT.}. 
The initial guess for the eigenfrequency is obtained by considering a high-frequency expansion of the ion- and electron resonant integrals of \Cref{eq:ion-1Dres-integrals,eq:trel-genplasmaZ-integrals} to approximate their respective density kernels [\Cref{eq:ion-el-glob-kernels}]. Such a high-frequency expansion can be made physically valid in the limit of strong drive, which, as shown in \Cref{app:fluid-limit-derivs}, yields to first order in $\omega_{ds}/\omega \ll 1$ a quadratic dispersion relation in the mode frequency
\begin{widetext}
    \begin{align}
        D_{\mathrm{glob}}(\omega,k_x,k_y,\{\mathrm{geo}\},\{\mathrm{plasma}\}) \approx & \left(1+\sum_{j} \frac{Z_j^2 n_j}{n_\mathrm{e}} \frac{T_\mathrm{e}}{T_j}\right) \int \dd{l} \frac{\hat{\phi}^2}{B} - \sum_j \frac{Z_j^2 n_j}{n_\mathrm{e}} \frac{T_\mathrm{e}}{T_j}\int \dd{l} \frac{\hat{\phi}^2}{B} \biggl(\Gamma_0(b_j)\biggl[1-\frac{\omega_{\grad{n_j}}}{\omega} +\frac{\omega_{j,\kappa}+\omega_{j,\grad{B}}}{\omega} \nonumber \\
        & -\frac{\left(\omega_{\grad{n_j}}+\omega_{\grad{T_j}}\right)\left(\omega_{j,\grad{B}}+\omega_{j,\kappa}\right)}{\omega^2}+b_j\frac{\omega_{\grad{T_j}}\omega_{j,\grad{B}}}{\omega^2}\biggr] 
        + b_j \left(\Gamma_1(b_j)-\Gamma_{0}(b_j) \right) \times \nonumber \\ 
        & \left[\frac{\omega_{j,\grad{B}}}{\omega}-\frac{\omega_{\grad{T_j}}}{\omega} -\frac{\omega_{j,\grad{B}}\left(\omega_{\grad{n_j}}+\omega_{\grad{T_j}}\right)}{\omega^2}-\frac{\omega_{\grad{T_j}}\left(\omega_{j,\kappa}+\omega_{j,\grad{B}}\right)}{\omega^2}+2b_j \frac{\omega_{\grad{T_j}}\omega_{j,\grad{B}}}{\omega^2}\right] \biggr) \nonumber \\
        & -\int\limits_{1/B_{\mathrm{max}}}^{1/B_{\mathrm{min}}} \dd{\lambda} \sum_{\mathrm{wells}(\lambda)} \left[\frac{1}{2} \left(1-\frac{\omega_{\grad{n_\mathrm{e}}}}{\omega}\right) + \frac{3}{4} \frac{\overline{\omega_{d\mathrm{e},\mathrm{well}}^{T}}}{\omega} \left(1-\frac{\omega_{\grad{n_\mathrm{e}}}+\omega_{\grad{T_\mathrm{e}}}}{\omega}\right)\right] \overline{\hat{\phi}}_{\mathrm{well}}^2 L_{\mathrm{well}}^{\mathrm{eff}},
        \label{eq:glob-nonres-quadratic-disprel}
    \end{align}
\end{widetext}
where we have reflected the change to the flux-tube coordinate system discussed in \Cref{sec:fluxtubes} by denoting the radial and bi-normal wavenumbers as $k_\psi \rightarrow k_x$ and $k_{\alpha} \rightarrow k_y$, respectively. In the limit of a single ion species, \Cref{eq:glob-nonres-quadratic-disprel} reduces to the dispersion relation found in e.g.~Refs.~\onlinecite{Helander2013,Plunk2017}, where the high-frequency expansion was directly applied to the resonant denominators, i.e.~$1/(\omega-\omega_{ds}) \approx (1+\omega_{ds}/\omega)/\omega$. Additionally, within the adiabatic-electron approximation, upon taking the DK limit $b_j\approx0$ and assuming the ions to be far from marginality ($\omega_{\nabla T_j}/\omega_{\nabla n_j} \gg 1)$, the local equivalent of \Cref{eq:glob-nonres-quadratic-disprel} (i.e. neglecting the role of the $\int \dd{l} \hat{\phi}^2/B$ operator) further reduces to the dispersion relation model used to describe the role of impurities on the ITG instability developed in Ref.~\onlinecite{Calvo2025ModificationPlasmas}.
Precomputing the geometry-dependent field-line and pitch-angle integrals in \Cref{eq:glob-nonres-quadratic-disprel} allows for an analytical solution for the approximate eigenfrequency $\omega$, which maintains the essential physics of instability drive by (unfavourable) magnetic drifts being balanced by FLR damping, as well as frequency scalings proportional to the driving gradients. The approximate eigenfrequency obtained from the quadratic dispersion relation [\Cref{eq:glob-nonres-quadratic-disprel}] corresponding to an unstable mode is used as an initial guess for root-finding. Meanwhile the initial guess for the Jacobian is obtained using second-order finite differences. It has been verified that using a regular Newton method yields similar results at increased computational cost due to the need to re-estimate derivatives at every iteration. In cases where root finding algorithm does not reach convergence, a conventional grid-search approach is used instead to solve the dispersion relation. 

\section{Dispersion model versus linear gyrokinetic simulations} \label{sec:GENE-application}
In this section we present the main findings of the work comparing the (complex) eigenmode frequency obtained by the global dispersion model developed in \Cref{sec:disp-theory} to the eigenfrequency obtained from \textsc{Gene} simulations. For all results presented in this section, the \textsc{Gene} eigenfunction of the electrostatic potential is used as input to the model to test its variational nature according to \Cref{eq:omega_var_statement}. The variational aspect of the model (with regard to the eigenfunction) is further assessed in \Cref{sec:redmod} where rudimentary trial functions are used instead. In all cases, a radial wavenumber of $k_x = 0$ is considered, consistent with centering of ballooning space of eigenmodes at $k_x = 0$ in the simulations. Here we will focus on (a subset of) results for the DIII-D tokamak, HSX stellarator and high-mirror configuration of the W7-X stellarator only, with the results for the remainder of geometries and scenarios listed in \Cref{tab:GENE_sim_resolutions} available in the Supplementary Material, which further extend our validation to include geometric sensitivity effects due to differences in triangularity (for TCV cases) and degree of maximum-$J$-ness (for W7-X cases), as well as additional scalings with respect to instability drive parameters.

\begin{figure*}
    \hspace*{\fill}
    \begin{subfigure}{.45\linewidth}
        \centering
        \includegraphics[width=\linewidth]{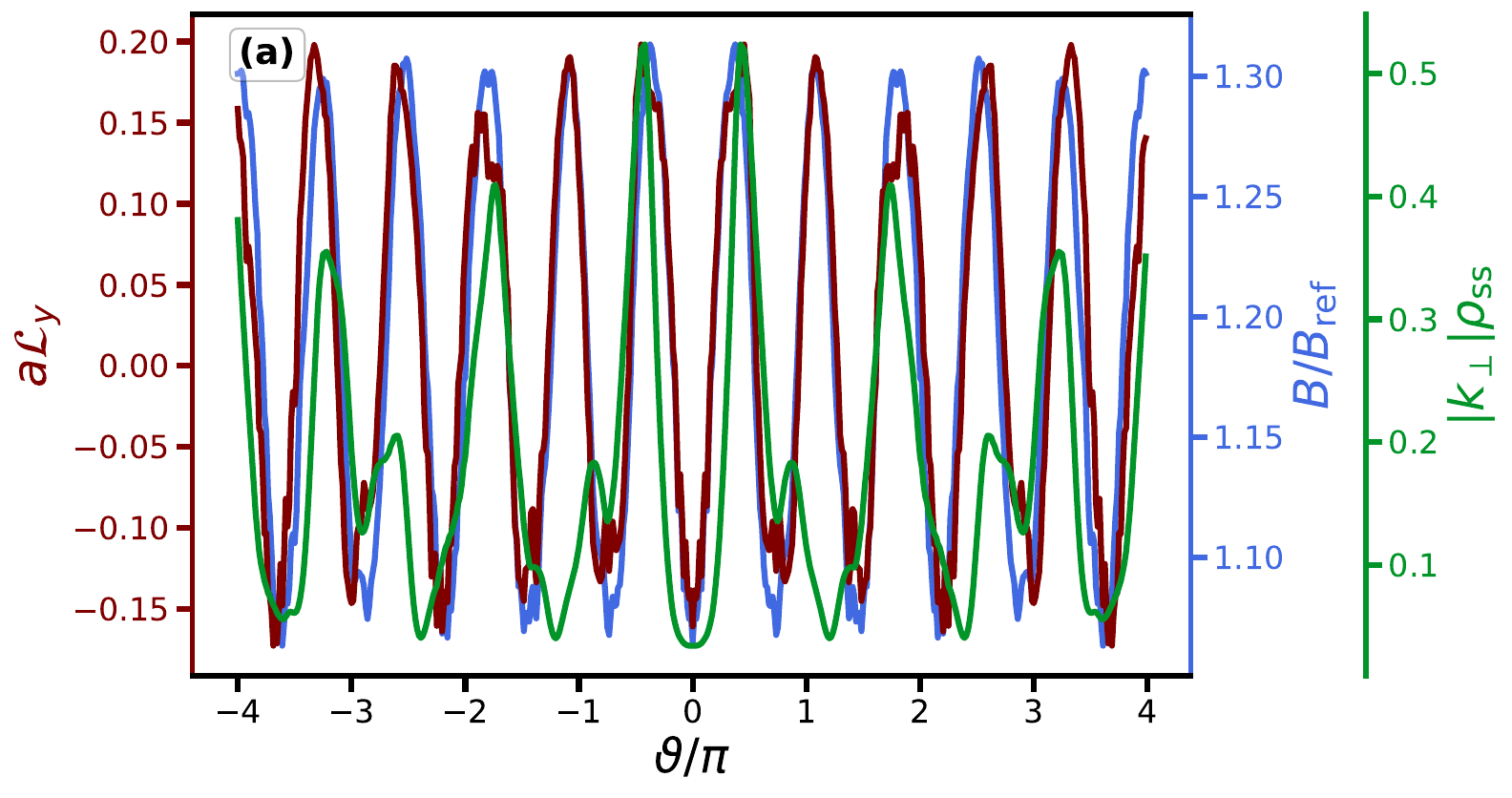}
    \end{subfigure}
    \begin{subfigure}{.45\linewidth}
        \centering
        \includegraphics[width=\linewidth]{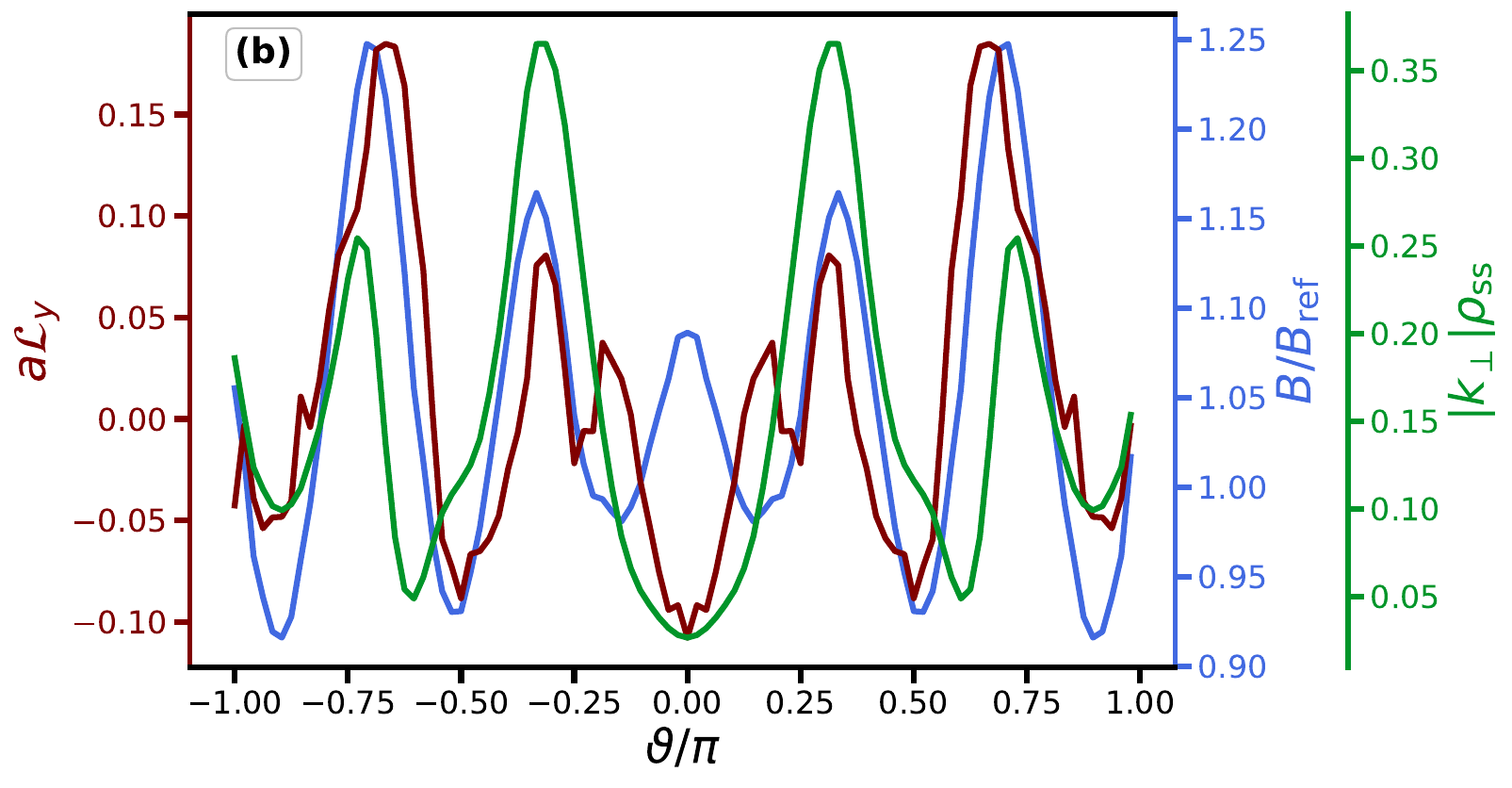}
    \end{subfigure}
    \hspace*{\fill}
    \caption{Flux-tube geometries for (a) the HSX stellarator and (b) high-mirror configuration of the W7-X stellarator showing the variation of the magnetic field strength (blue), bi-normal component of the $\grad{B}$ drift (maroon) and magnitude of the perpendicular wavenumber (green) along the field line. In the latter we have taken $k_x \rho_{\mathrm{ss}}=0 , \ k_y \rho_{\mathrm{ss}}= 0.3$, such that a direct comparison to $\norm{\bm{k_\perp}} \rho_{s}$ in the DIII-D geometry from \Cref{fig:loc-freq-sol-visual} is facilitated. Note that the W7-X flux-tube spans only a single poloidal turn, whereas the HSX flux-tube contains the geometric data of four poloidal turns.}
    \label{fig:FTgeo_stell}
\end{figure*}

The focus on these configurations allows us to cover both realistic axisymmetric geometry with (moderate) shaping beyond circular cross-sections as relevant for exploitation of higher fusion yields in ITER and beyond\cite{Siccinio2022}, as well as different classes of stellarators (quasi-helically symmetric and approximately quasi-isodynamic for HSX and W7-X, respectively), hence offering a wide range of salient geometric features that affect microinstability. Here, quasi-isodynamicity refers to configurations where the contours of $\norm{\bm{B}}$ are poloidally closed, and the second adiabatic invariant $J$ is constant on a flux surface, a key requirement for achieving a maximum-$J$ configuration\cite{Goodman2023ConstructingFields,Sanchez2023A}. The most relevant geometric quantities in the HSX and the high-mirror configuration W7-X flux tubes used in this work are shown in \Cref{fig:FTgeo_stell}, while those in the DIII-D flux tube geometry are displayed in \Cref{fig:loc-freq-sol-visual}. Comparing geometries, we find qualitative similarity between the arrangement of magnetic wells and regions of unfavourable magnetic curvature (characterised by $\mathcal{L}_y<0$) between DIII-D and HSX, exemplifying axisymmetry and quasi-helical symmetry, respectively\footnote[9999]{\label{fn:QIQS_cmt} It should be noted that HSX was designed and constructed at the turn of the millennium, while through advancements in optimisation techniques in recent years magnetic fields achieving significantly higher degree of quasi-symmetry can be achieved\cite{Landreman2022MagneticConfinement,Jorge2020ConstructionApproach,Giuliani2024DirectDevices,Giuliani2025ASets}. Likewise for W7-X, magnetic fields with much higher degree of quasi-isodynamicity and maximum-$J$-ness even at low $\beta$ can be constructed through modern optimisation targets\cite{Goodman2023ConstructingFields,CamachoMata2022DirectConfigurations,Sanchez2023A}.}, whilst such alignments are minimised in W7-X, corresponding to the key feature of a maximum-$J$ configuration\footnotemark[9999]. While this feature is only partially achieved by the vacuum field considered in this work, the degree of maximum-$J$-ness generally increases at higher $\beta$ \cite{Nuhrenberg2010DevelopmentStellarators,Rodriguez2024TheStellarators,Subbotin2006IntegratedStrength}. 
\nocite{Landreman2022MagneticConfinement,Jorge2020ConstructionApproach,Giuliani2024DirectDevices,Giuliani2025ASets}
The most striking difference between the stellarator configurations and DIII-D is the magnitude of the perpendicular wavenumber, defined as in \Cref{eq:kperp-clebsh} under substitutions $\{\psi,\alpha\}\rightarrow \{x,y\}$ to match flux-tube coordinates used by \textsc{Gene}, which is an order of magnitude larger in the latter case at the same bi-normal wavenumber. This can be attributed to the significantly lower global shear $\hat{s}$ for both stellarator configurations, resulting in a much weaker effect of the secular term in the $g^{yy}$ component of the magnetic tensor. \par
We note that in our analytical theory, we prescribed a time dependence of $\exp(-i\omega t)$, which results in a sign discrepancy for the propagation frequency $\omega_R$ between the analytical theory from \Cref{sec:disp-theory} and high-fidelity \textsc{Gene} simulations. In what follows, we shall always adopt the \textsc{Gene} convention, where $\omega_R>0$ corresponds to propagation in the ion diamagnetic direction, thereby inverting the sign of $\Re[\omega]$ obtained by the global dispersion model.

\subsection{Adiabatic-electron density-dilution scenario} \label{sec:ae}
Here we present the scenario probing for the density dilution effect for the adiabatic-electron simulations including $\mathrm{C}^{6+}$ as an impurity while varying $Z_{\mathrm{eff}}$, therefore simultaneously verifying the multi-species capabilities of the model and testing the validity of the ad-hoc method of introducing non-local effects along the field line by operating on the local dispersion relation [\Cref{eq:Dloc}] with $\int \dd{l} \hat{\phi}/B$ in the absence of a regularising trapped-electron term. Results for the driving gradient scans in $a/L_{T\mathrm{i}}$ (no impurity) and $a/L_{n\mathrm{C}}$ can be found in the Supplementary Material and show similar degrees of agreement. \par
\begin{figure}
    \centering  
    \begin{subfigure}{.85\linewidth}
        \centering
        \includegraphics[width=\linewidth]{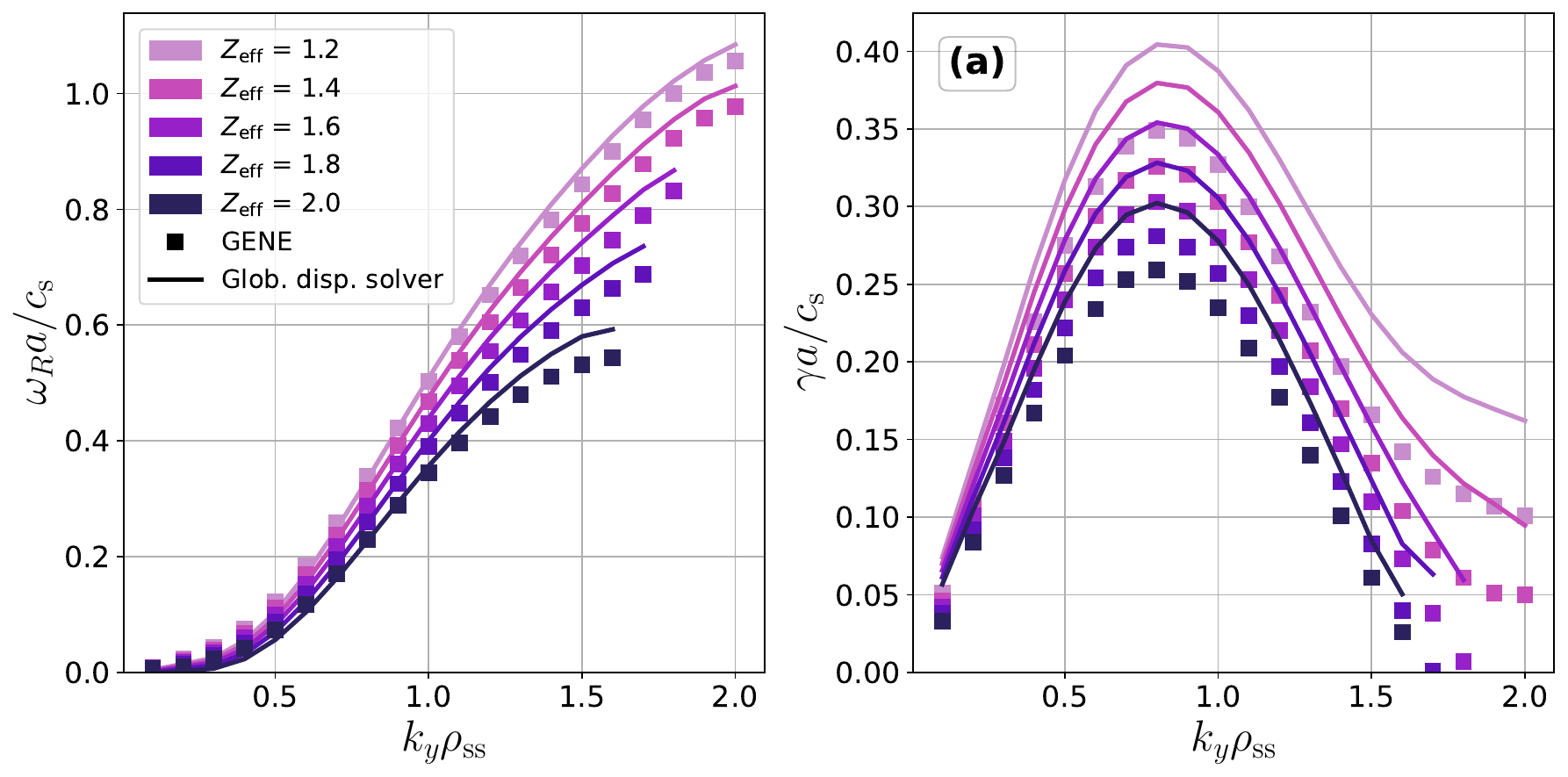}
    \end{subfigure}
    \begin{subfigure}{.85\linewidth}
        \centering
        \includegraphics[width=\linewidth]{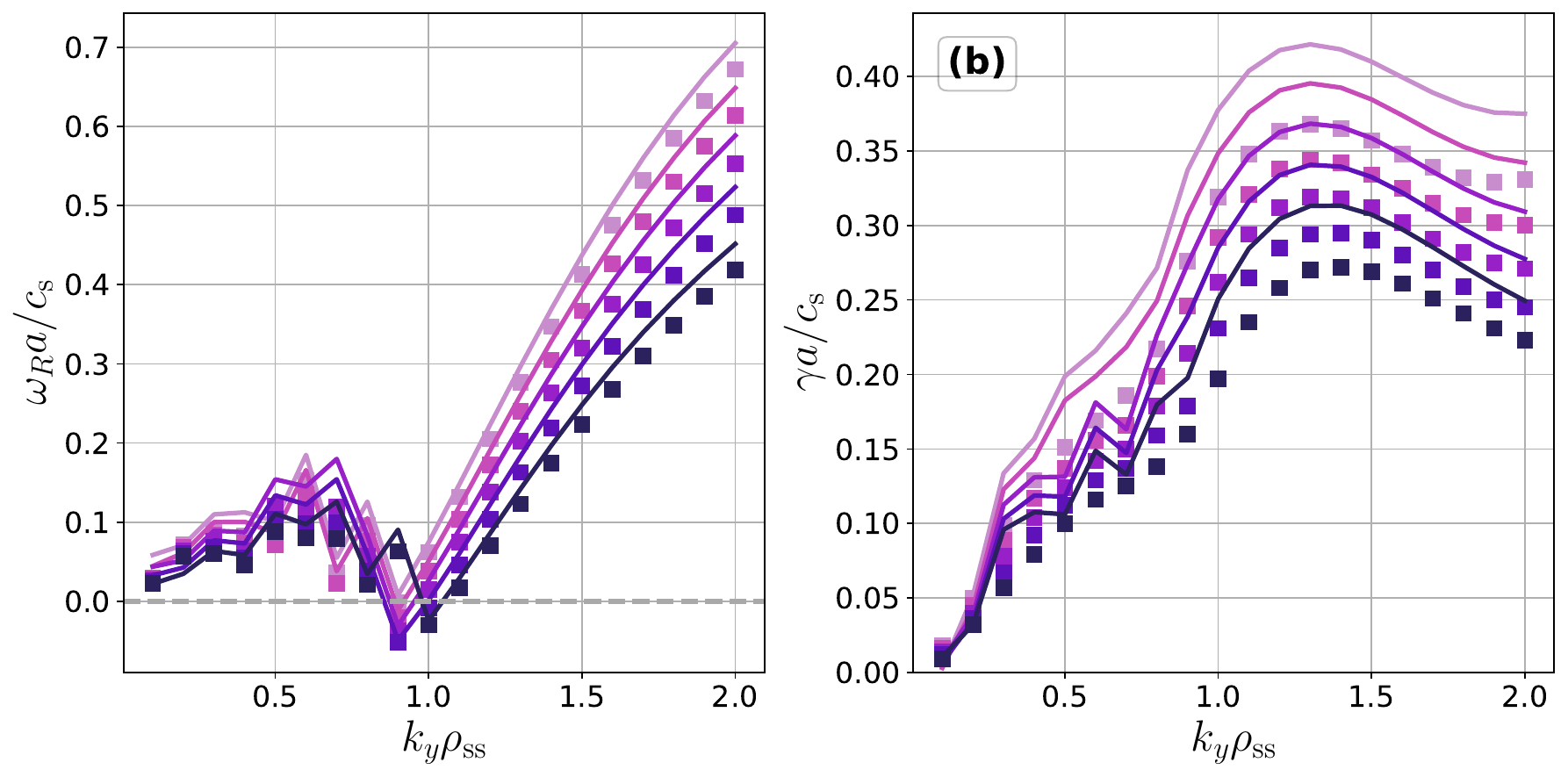}
    \end{subfigure}
    \begin{subfigure}{.85\linewidth}
        \centering
        \includegraphics[width=\linewidth]{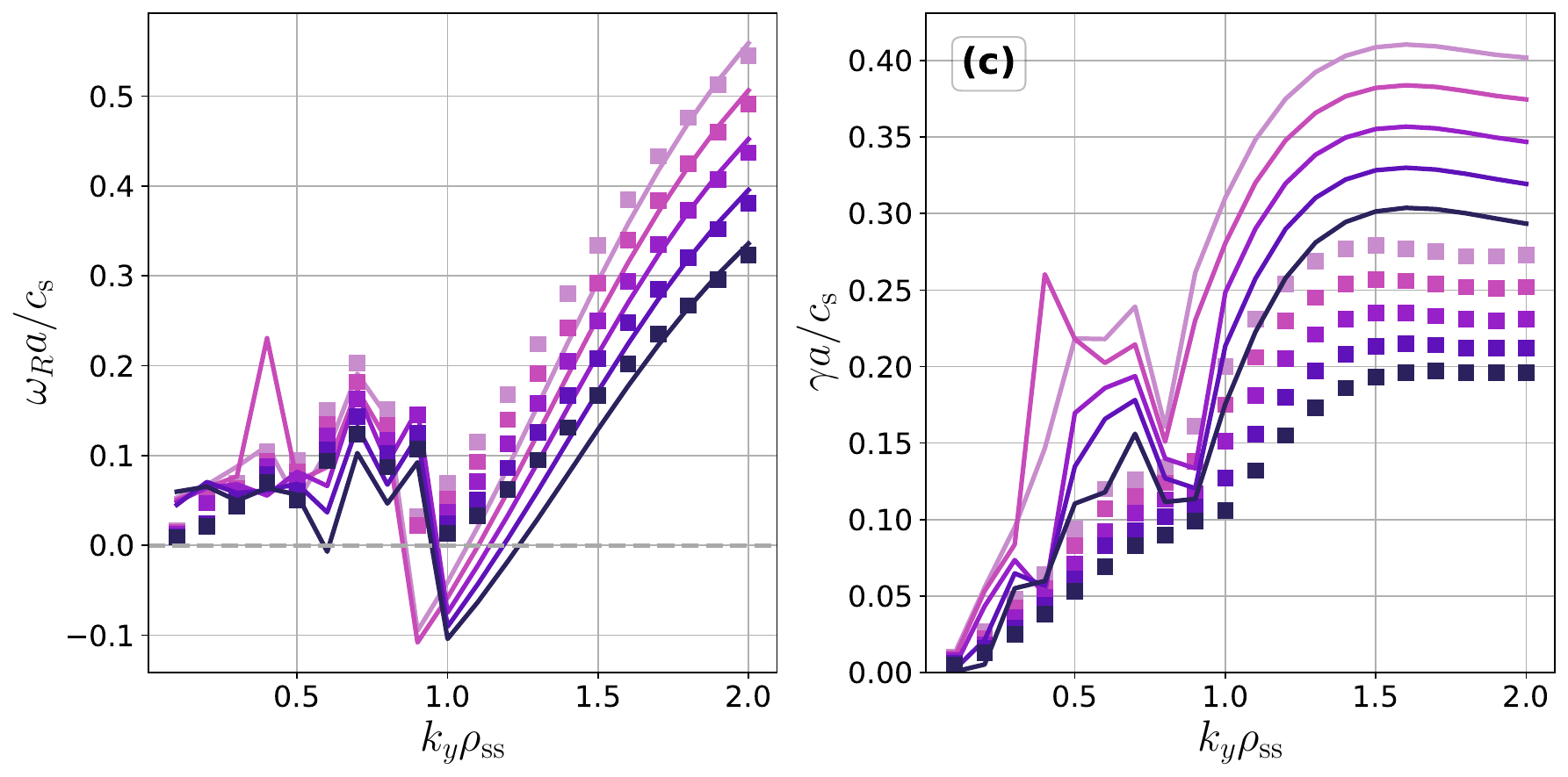}
    \end{subfigure}
    \caption{Eigenfrequency solutions of the global dispersion relation model (solid lines) contrasted with \textsc{Gene} simulations (symbols) for adiabatic-electron ITG in the presence of $\mathrm{C}^{6+}$ impurities, while varying the impurity concentration by modifying $Z_{\mathrm{eff}}$ (darker colours indicate larger impurity concentration). Shown are results for (a) the DIII-D tokamak, (b) the HSX stellarator and (c) the high-mirror configuration of the W7-X stellarator. In all cases, a density- and temperature gradient of $a/L_{n_s}=2$ and $\ a/L_{T_s}=4$, respectively, are considered for both deuterium ions and the carbon impurity.}
    \label{fig:Zeff_result}
\end{figure}

The results obtained for the $Z_{\mathrm{eff}}$ scan obtained with the global dispersion model are contrasted with \textsc{Gene} eigenfrequencies, and are shown in \Cref{fig:Zeff_result}.
A general trend observable across all geometries is that both the growth rate and the propagation frequency monotonically decrease as $Z_{\mathrm{eff}}$ increases, consistent with the dilution of the main-ion density. The global dispersion model universally predicts larger growth rates, which can be attributed to the lack of finite-transit effects which provide Landau damping. Since, at a given bi-normal wavenumber $k_y$, this overestimation is roughly constant, it suggests that Landau damping could be heuristically included as $\gamma \rightarrow \gamma - v_{T\mathrm{i}}/(\sqrt{2}L_\parallel)$ where $L_{\parallel}$ is the connection length\cite{Plunk2014}. The larger discrepancies in the growth rate observed in W7-X can therefore be attributed to the shorter connection length ($L_\parallel/(\pi q R) \approx 0.12$, taking $R=5.5$m for W7-X) compared with typical tokamaks, where one commonly estimates $L_{\parallel}\approx \pi q R$\cite{Plunk2014,Tang1978MicroinstabilityTokamaks}. 
The propagation frequency varies monotonically in DIII-D, whilst several mode transitions (indicated by non-monotonous behaviour in $\omega_R$) occur in \textsc{Gene} simulations for both stellarator geometries, where multiple unstable eigenmodes co-exist partly due to the presence of multiple non-equivalent magnetic and drift wells. These mode transitions are reproduced by the global dispersion model, though with greater accuracy in the propagation frequency in HSX compared with W7-X. The corresponding eigenmodes range from being broad in ballooning space at the lowest wavenumbers to strongly ballooned around the outboard midplane ($\theta = 0$) for $k_y \rho_{\mathrm{ss}} \gtrapprox 1$. These extended eigenmodes are a typical feature of low-shear stellarators, where FLR suppression is weaker due to the limited role of the secularity in $g^{yy}$, keeping $\norm{\bm{k_\perp}}$ small. These findings illustrate the utility of adopting an extension of the flux-tube geometry beyond its original $2\pi N_{\mathrm{pol}}$ domain using \Cref{eq:geometry-extend-formulae} to describe the drive of these broad eigenmodes by magnetic drift resonances. Additionally, the results from the $Z_{\mathrm{eff}}$ scan in \Cref{fig:Zeff_result} and the $a/L_{n\mathrm{C}}$ scan in the Supplementary Material show that the global dispersion solver is capable of predicting how the presence of impurities directly affects the absolute ITG growth rates, which can be considered as an extension of the recent perturbative impurity model from Ref,~\onlinecite{Calvo2025ModificationPlasmas} which predicts the relative modification to the ITG growth rates due to impurities.

\subsection{Kinetic-electron scenarios} \label{sec:ke}
Here we present the result for both the $a/L_{T\mathrm{e}}$ and $a/L_{T\mathrm{i}}$ gradient scans including kinetic electrons, while results for the $a/L_{n}$ scan (no temperature gradients) and non-synthetic profiles with non-zero gradients for both species are shown in the Supplementary Material. As the variational nature of the model is retained when kinetic electrons are included, these scenarios allow us to verify the variational property of \Cref{eq:glob-disp}, in both electron-dominated and ion-dominated scenarios. \par 
\subsubsection*{Application to pure TEM case}
\begin{figure}
    \centering  
    \begin{subfigure}{.85\linewidth}
        \centering
        \includegraphics[width=\linewidth]{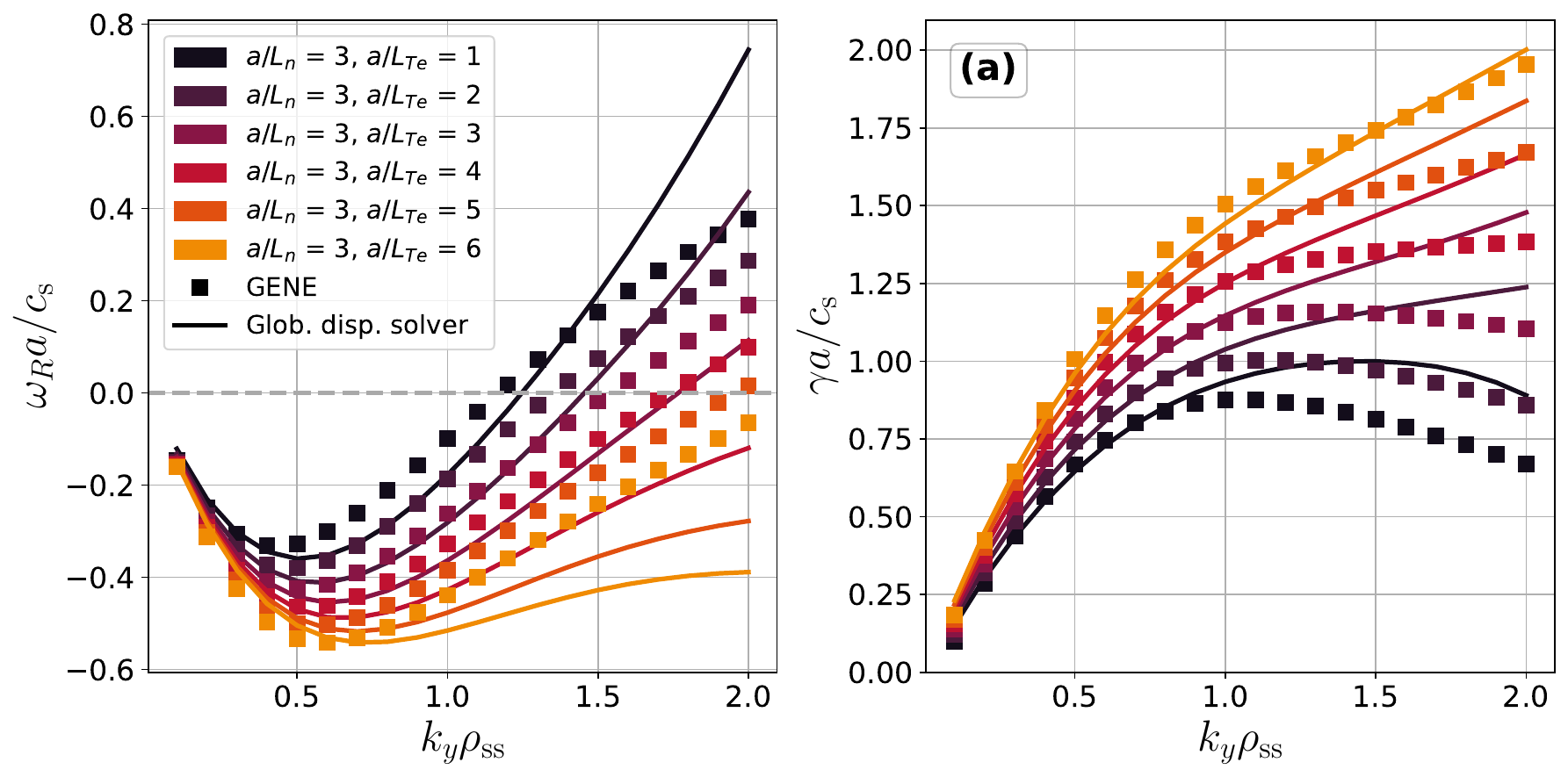}
    \end{subfigure}
    \begin{subfigure}{.85\linewidth}
        \centering
        \includegraphics[width=\linewidth]{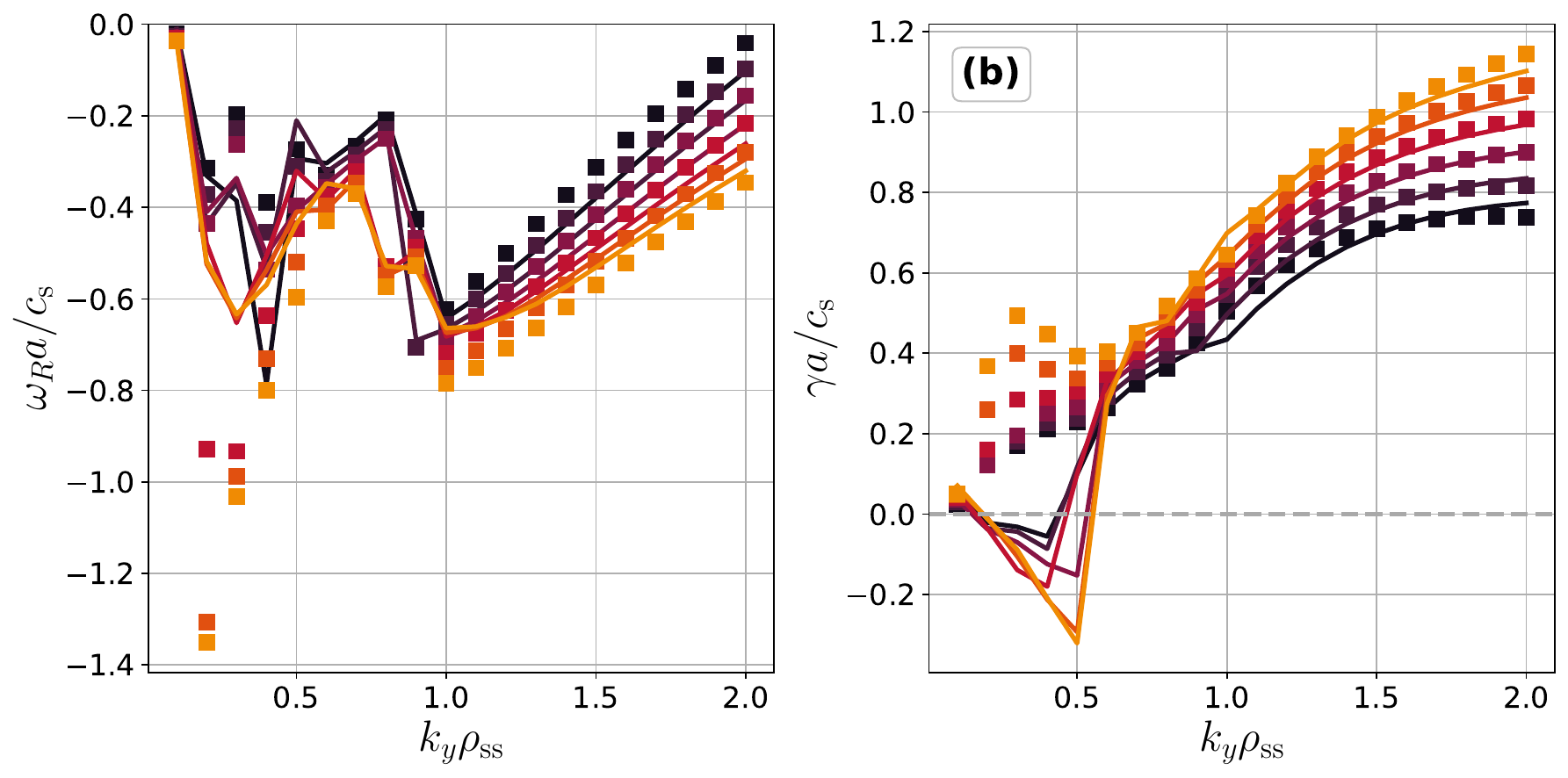}
    \end{subfigure}
    \begin{subfigure}{.85\linewidth}
        \centering
        \includegraphics[width=\linewidth]{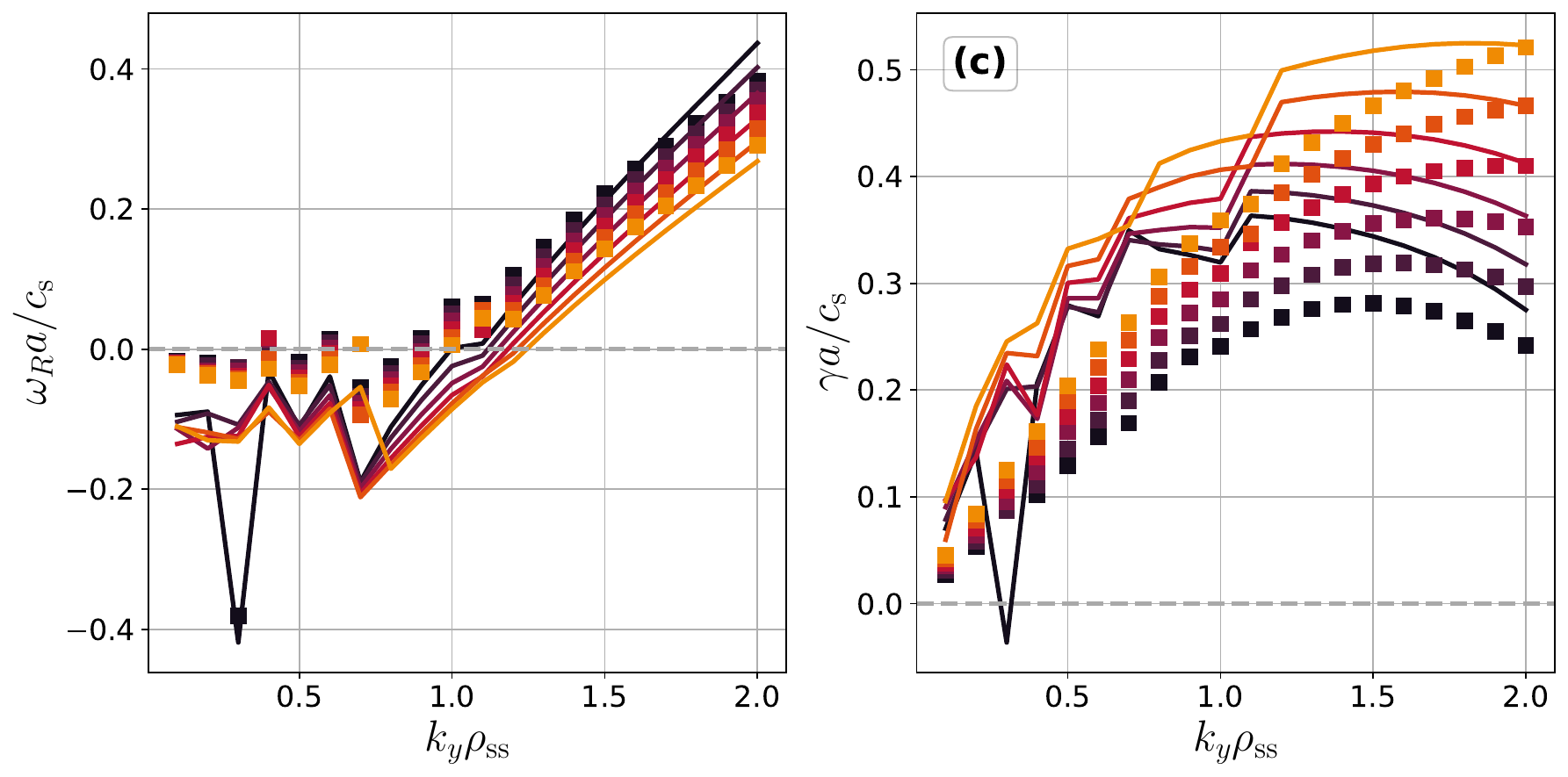}
    \end{subfigure}
    \caption{Eigenfrequency solutions of the global dispersion relation model (solid lines) contrasted with \textsc{Gene} simulations (symbols) including kinetic electrons while varying the electron temperature gradient $a/L_{T_e}$ (lighter colours indicate a stronger temperature gradient). Shown are results for (a) the DIII-D tokamak, (b) the HSX stellarator and (c) the high-mirror configuration of the W7-X stellarator. In all cases, the density gradient is fixed at $a/L_n=3$ whilst the ion temperature gradient is suppressed ($a/L_{T\mathrm{i}} = 0$).}
    \label{fig:LTe_scan_result}
\end{figure}

First, we focus on the physics associated with the precession resonance by testing the model against \textsc{Gene} simulations without an ion temperature gradient. Here, only TEMs are expected to emerge, and results for the $a/L_{T\mathrm{e}}$ scan are shown in \Cref{fig:LTe_scan_result}. The results for the density gradient scan of density-gradient-driven TEMs in absence of both temperature gradients are shown in the Supplementary Material, which show similar levels of agreement at high $a/L_n$, with agreement further improving at smaller $a/L_n$.

Unlike in the adiabatic-electron case, there is no universal overprediction of the growth rates across devices: Growth rates are being slightly overpredicted in DIII-D for higher wavenumbers ($k_y \rho_{\mathrm{ss}} \gtrapprox 1.0$) -- though this overprediction decreases with increasing electron-temperature gradient -- while growth rates in HSX mostly match those from \textsc{Gene} simulations, with exception of the cluster of modes appearing in the low-wavenumber regime of $k_y \rho_{\mathrm{ss}} = 0.2\textrm{-}0.5$. Meanwhile a consistent overprediction of growth rates persists in W7-X, though the latter is less severe compared with the adiabatic-electron ITG scenario. With regard to the propagation frequency, the inclusion of kinetic electrons has lead to a deterioration in the quantitative agreement between the global dispersion model and \textsc{Gene} simulations, with the disagreement being most pronounced in DIII-D (especially at larger wavenumber), while the disagreement in both stellarator configurations has only increased in low-$k_y$ region. 
Nonetheless, the pertinent qualitative trends of increasing destabilisation and downshift of propagation frequencies into the electron diamagnetic direction as the electron temperature gradient increases are reproduced by the global dispersion model. Obtaining perfect agreement between the model and simulation is, despite the use of exact eigenfunctions for the electrostatic potential in the variational formalism, not expected since the passing electrons have been assumed to be adiabatic in the model. \par
Analogous to the adiabatic-electron ITG case, we observe a continuous propagation frequency for DIII-D, indicating that the same eigenmode remains the dominant instability at all scales, whereas mode transitions again occur in both HSX and W7-X. The qualitative behaviour of $\omega_R$ differs between configurations, which is to be expected as the response of trapped electrons to electrostatic fluctuations is strongly influenced by the details of the magnetic configuration in which they are embedded. Note that despite the fact that in DIII-D $\omega_R$ changes from the electron diamagnetic direction to the ion diamagnetic direction as $k_y$ is increased, the underlying instability is not in fact an ITG as temperature gradients are absent, but corresponds to a ubiquitous mode (UM)\cite{Coppi1977}, which is an instability non-resonantly driven by a combination of the ion magnetic drift and bounce-averaged electron drift. As the electron temperature gradient is increased, the destabilising contribution from the electrons to this mode is enhanced (with the latter scaling as\cite{Plunk2017} $\gamma \propto \sqrt{1+\eta_{\textrm{e}}}$, with $\eta_s = \norm{\grad{T_s}}/\norm{\grad{n_s}}$ the gradient-length ratio of kinetic profiles for species $s$) causing the shift of $\omega_R$ to the electron diamagnetic direction. The TEM-UM transition is characterised by the existence of a growth-rate extremum between the drive from resonantly driven TEMs and the non-resonantly driven instability (which coincides with the sign change of $\omega_R$ in the absence of temperature gradients\cite{Coppi1977}). As such a growth-rate extremum is absent for the high-wavenumber modes for $a/L_{T\mathrm{e}} \geq 4$, these modes likely correspond to temperature-gradient-driven TEMs, as supported by the stronger scaling of the growth rate with the electron temperature gradient and propagation frequencies that are persistently in the electron diamagnetic direction, where resonances with the precession drift are possible. \par 
Much of the above discussion applies to the high-mirror configuration of W7-X as well, where the majority of the modes are found to propagate in the ion diamagnetic direction, corresponding to the so-called ion-driven trapped electron mode (iTEM)\cite{Plunk2017}, which may be considered as a general-geometry equivalent to UMs observed in tokamaks. The existence of this instability, originally discovered through gyrokinetic simulations\cite{Proll2013}, is attributable to the (approximate) maximum-$J$ property of this configuration, which makes classical TEMs driven by resonances with the trapped-electron precession drift (near) impossible. Though, originally this resilience of maximum-$J$ configurations to instabilities driven by the trapped-electron precession drift was shown to hold up to at least\cite{Proll2012} $\eta_{\textrm{e}} = 2/3$, it was later proven to persist for arbitrary electron temperature gradients\cite{Plunk2017}. This finding supports the fairly weak scaling (in comparison to DIII-D) of the propagation frequency with $a/L_{T\mathrm{e}}$ in W7-X, despite the $\eta_{\textrm{e}} = 2/3$ threshold being exceeded. The vacuum high-mirror configuration does not achieve exact maximum-$J$, with the magnetic wells closer to the inboard side having an overlap between regions of unfavourable curvature (see \Cref{fig:FTgeo_stell}), 
and indeed, the low-wavenumber modes ($k_y \rho_{\mathrm{ss}} \leq 0.6$) are found to propagate in the electron diamagnetic direction and correspond to conventional TEMs -- with the exception of the $k_y = 0.3$ mode for $a/L_{T\mathrm{e}} = 1$ which corresponds to a different instability as discussed below -- though their destabilisation is weak. Likewise, as the configuration is not exactly maximum-$J$, the electrons are expected to contribute to the drive of the iTEM, explaining the stronger scaling of the eigenfrequency with $a/L_{T\mathrm{e}}$ for high-wavenumber modes. The fact that these qualitative trends are observed in both the high-fidelity simulations as well as in the global dispersion relation model, demonstrates the substantial predictive capabilities of the latter. \par
The most striking differences between the model and \textsc{Gene} simulations are observed in HSX. Whilst good agreement is achieved at high wavenumber and the mode at $k_y \rho_{\mathrm{ss}}=0.1$, there is a strong discrepancy for the modes at $k_y \rho_{\mathrm{ss}}=0.2\textrm{-}0.5$ which show a significantly stronger scaling of the eigenfrequency with $a/L_{T\mathrm{e}}$. The $k_y \rho_{\mathrm{ss}}=0.1$ and high-$k_y$ modes correspond to conventional TEMs, as expected from the driving gradients and the isomorphism of the equilibrium for a quasi-symmetric stellarator and a tokamak, and scale similarly with $a/L_{T\mathrm{e}}$ as in DIII-D. The modes within the $k_y \rho_{\mathrm{ss}} = 0.2\textrm{-}0.5$ interval are characterised by a unique range in $\omega_R$ much larger than those corresponding to TEMs at comparable wavenumbers, and are subject to strong destabilisation by the electron temperature gradient. These modes correspond to the so-called universal instability (UI)\cite{Landreman2015,Helander2015b}, which are characterised by broad eigenmodes along the field line without preferential localisation to any geometric features, occurring typically at long wavelength and low global shear $\hat{s}$, and are predominantly driven by transit resonances of passing electrons\footnote{In toroidal geometry the presence of the magnetic drifts provides another potential driving mechanism for UI modes, though the contribution from drift effects to the growth rate is much smaller\cite{Costello2023TheStellarators,Chowdhury2010}.}. While this instability thrives in weakly sheared slab-like geometries, in other geometries the existence of a shear threshold tends to prevent their presence (with the critical parameter being\cite{Chowdhury2010,Landreman2015} $L_s/L_n$, with $L_s \propto \hat{s}$).
Therefore, by a combination of the low shear in HSX as well as the high density gradient of $a/L_n = 3$, an ideal situation for the emergence of the UI is created, and in Ref.~\onlinecite{Costello2023TheStellarators} such modes have been shown to exist in HSX at similar wavenumbers and frequencies. Despite the adiabatic treatment of passing electrons, the global dispersion model manages to qualitatively reproduce the trends in the propagation frequency $\omega_R$ for these UI modes, at least for $\eta_{\textrm{e}} \leq 1$, though erroneously predicting these UI modes to be damped rather than unstable. It should be mentioned, however, that we used the damped approximate eigenfrequency solution to the quadratic dispersion relation [\Cref{eq:glob-nonres-quadratic-disprel}] instead of the unstable one as an initial guess for root finding to obtain this matching solution in $\omega_R$. These results are consistent with the findings of Ref.~\onlinecite{Helander2015b}, where it is argued that in the limit of steep density gradients (equivalent to $\eta_{\textrm{e}} \ll 1$) the UI mode frequency is, to lowest order, purely real and determined by a balance of ion and trapped-electron density fluctuations. To next order, the destabilisation of this mode would subsequently be dominated by the non-adiabatic response of passing electrons. As the latter is absent in our model, we do not expect agreement in the growth rate, and indeed the model predicts that these modes cannot be driven unstable at all. The subsequent deterioration of agreement in the propagation frequency of these UI modes as $a/L_{T\mathrm{e}}$ is increased may be attributed to the increasing non-adiabaticity of passing electrons, which no longer perturbatively affect the eigenfrequency like in the steep-density-gradient limit. 
The emergence of the UI is not a feature unique to HSX, and has also been observed in the high-mirror configuration of W7-X\cite{Costello2023TheStellarators,Morren2024InfluenceStellarators}, where despite the higher shear, the TEM drive is sufficiently weakened by the maximum-$J$ property for the UI to become the dominant instability. In fact, the aforementioned non-TEM mode at $k_y \rho_{\mathrm{ss}}=0.3$ observed at the lowest temperature gradient in W7-X in \Cref{fig:LTe_scan_result} also corresponds to a UI, explaining the erroneously predicted damping rate of that mode by the model. \par

\subsubsection*{Application to mixed ITG-TEM case}
\begin{figure}
    \centering  
    \begin{subfigure}{.85\linewidth}
        \centering
        \includegraphics[width=\linewidth]{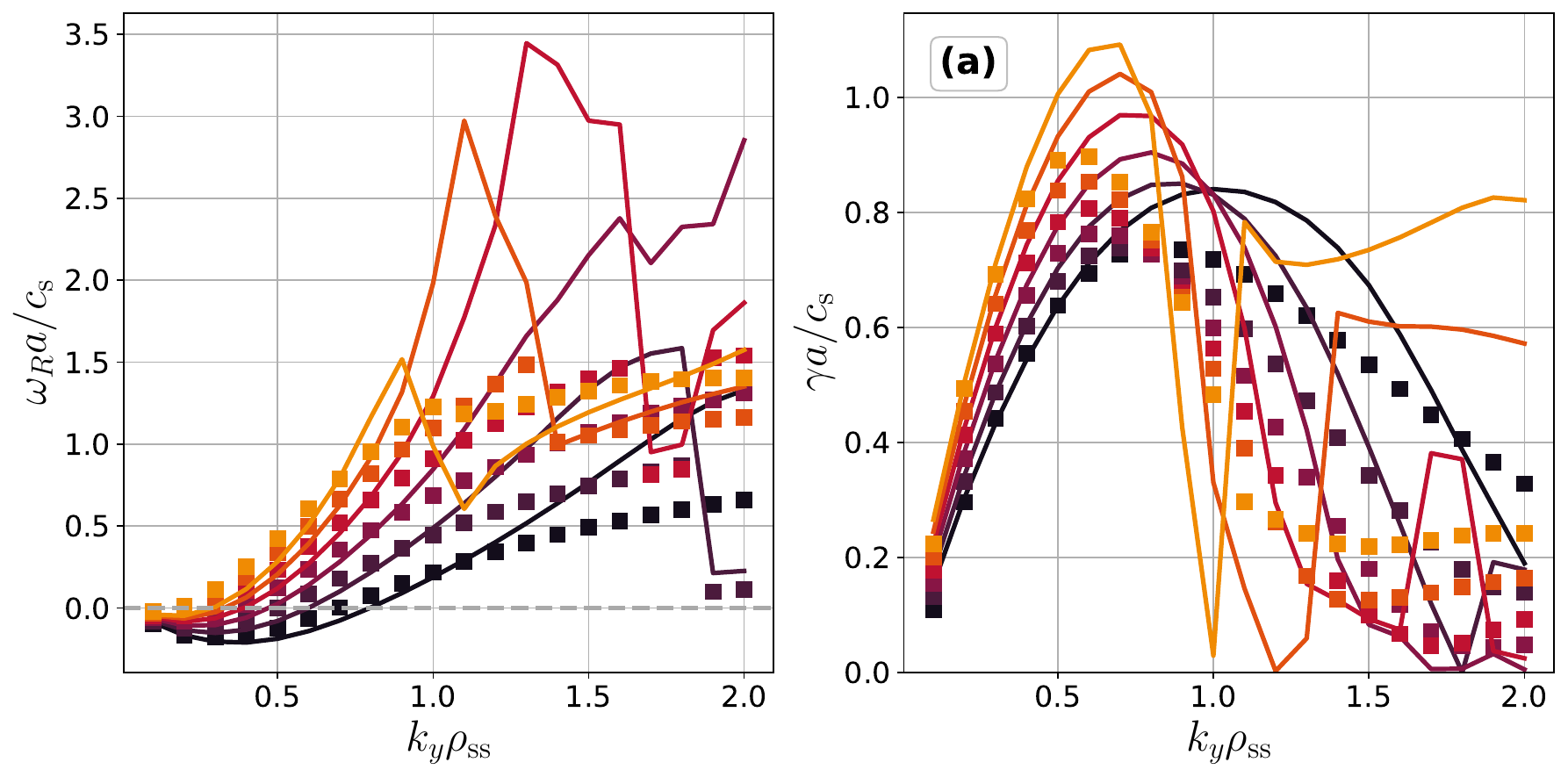}
    \end{subfigure}
    \begin{subfigure}{.85\linewidth}
        \centering
        \includegraphics[width=\linewidth]{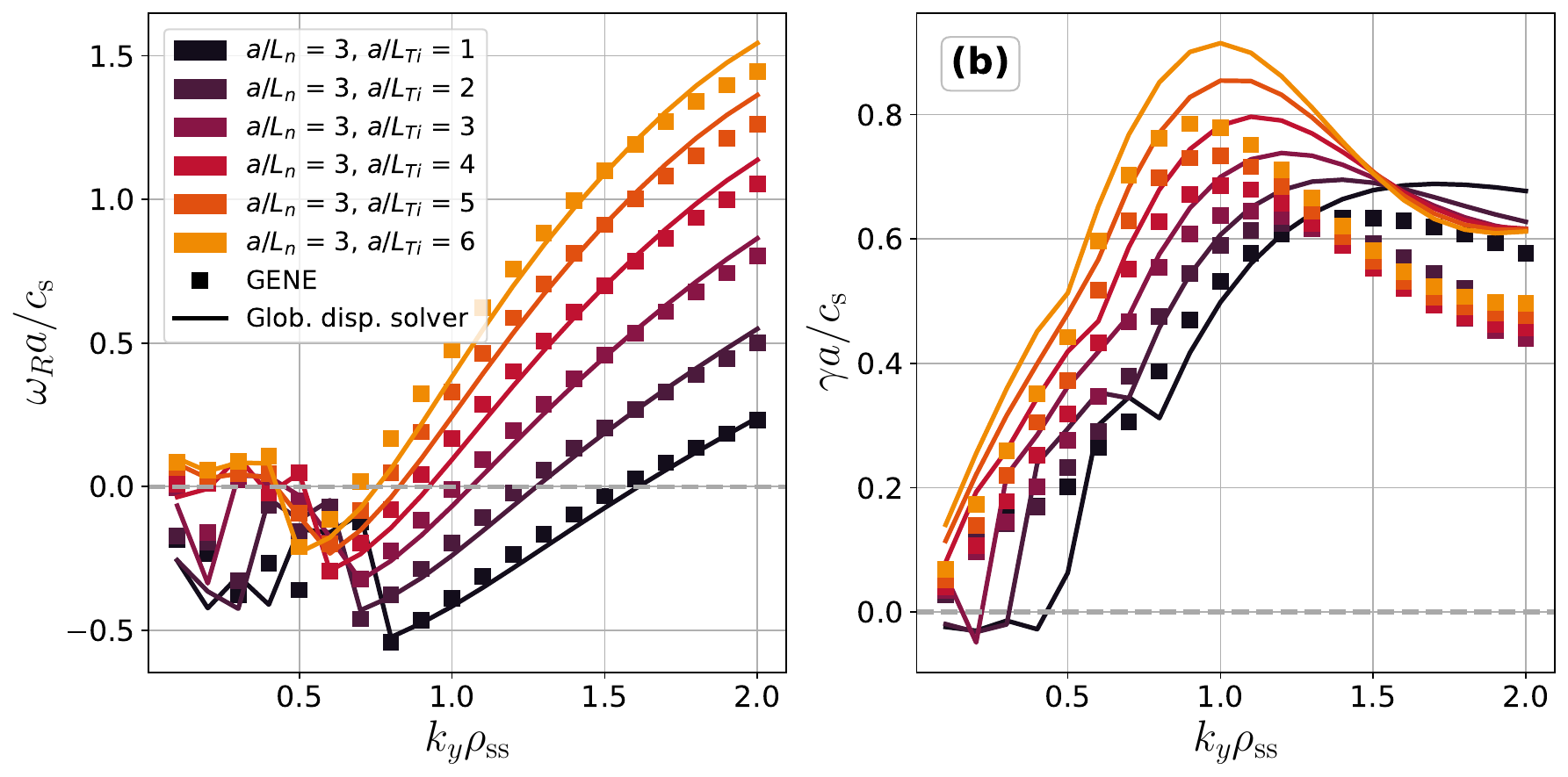}
    \end{subfigure}
    \begin{subfigure}{.85\linewidth}
        \centering
        \includegraphics[width=\linewidth]{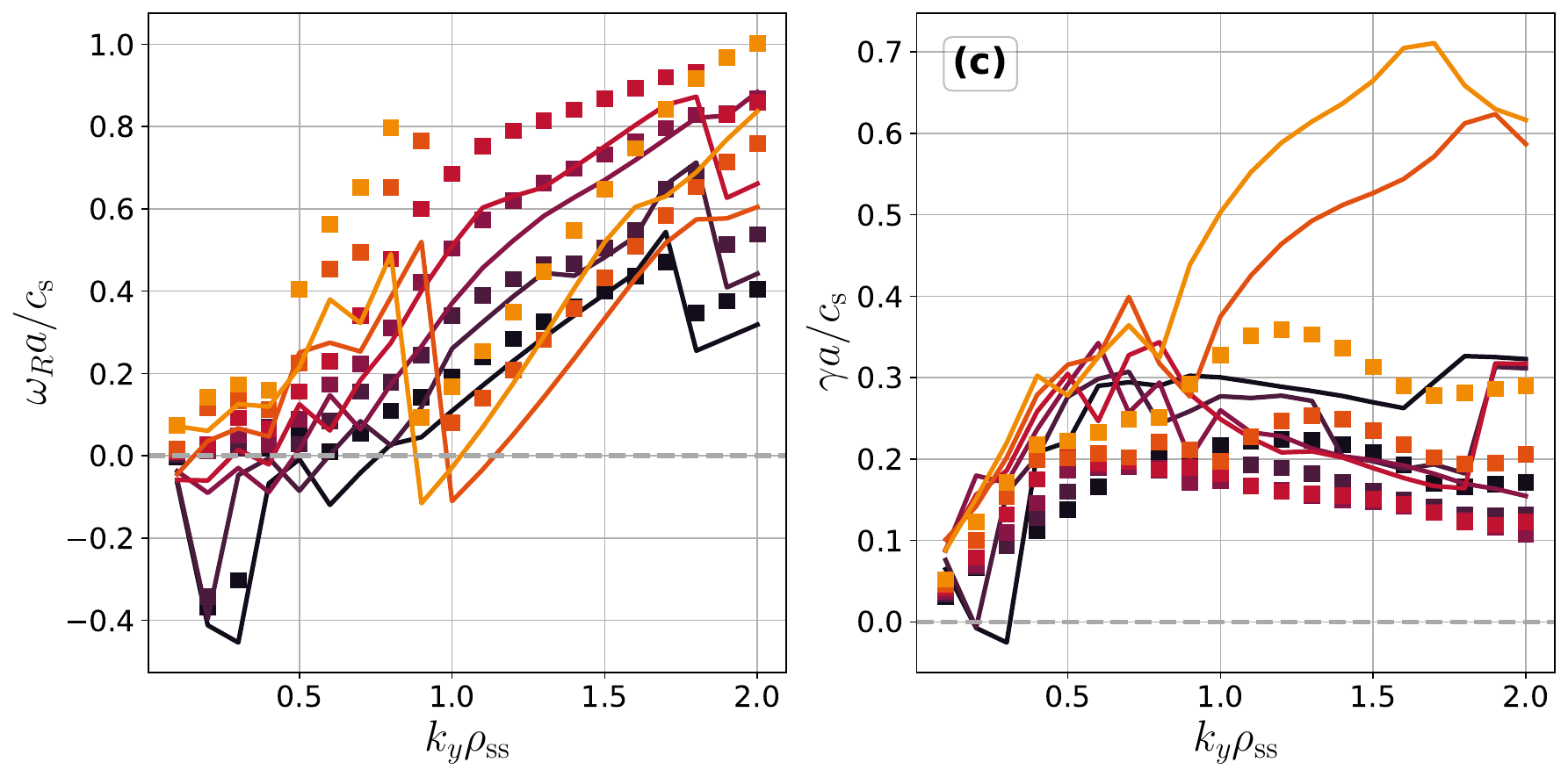}
    \end{subfigure}
    \caption{Eigenfrequency solutions of the global dispersion relation model (solid lines) contrasted with \textsc{Gene} simulations (symbols) including kinetic electrons while varying the ion temperature gradient $a/L_{T\mathrm{i}}$ (lighter colours indicate a stronger temperature gradient). Shown are results for (a) the DIII-D tokamak, (b) the HSX stellarator and (c) the high-mirror configuration of the W7-X stellarator. In all cases, the density gradient is fixed at $a/L_n=3$ whilst the electron temperature gradient is suppressed ($a/L_{T\mathrm{e}} = 0$).}
    \label{fig:ke_LTi_scan_result}
\end{figure}

When a finite ion temperature gradient is considered, the presence of trapped electrons will modify the ITG eigenfrequency with respect to an adiabatic-electron ITG mode at the same gradients, and ITGs and TEMs may now coexist, leading to potential mode transitions between the two instabilities. Whilst the global dispersion model has been shown to adequately resolve the physics of drift resonances for ITG and TEM individually, it needs to be validated in mixed-mode scenarios as well. Here we focus on the results for the $a/L_{T\mathrm{i}}$ scan in absence of electron temperature gradients, shown in \Cref{fig:ke_LTi_scan_result}, where 
TEMs are driven only by the fixed density gradient. Results for simulations with a finite temperature gradient for both species are shown in the Supplementary Material, in which case with the inclusion of an electron temperature gradient the model generally exhibits even better agreement.

In DIII-D, at moderate temperature gradients ($\eta_{\textrm{i}} \leq 1$) only a single mode transition is found to occur between $k_y \rho_{\mathrm{ss}}=1.8\textrm{-}1.9$ for the $a/L_{T\mathrm{i}}=2$ simulation, which is reproduced by the dispersion model. At those lower temperature gradients the TEM generally transitions to a UM as the wavenumber is increased, with the transition wavenumber decreasing with $a/L_{T\mathrm{i}}$, as is consistent with analytical theory of Ref.~\onlinecite{Coppi1977}. However, as the TEM-to-UM transition occurs close to the growth rate peak, this renders UM increasingly stable at shorter wavelengths. At higher temperature gradients of $a/L_{T\mathrm{i}} \geq 4$ a transition from UM to ITG modes is observed at these shorter wavelengths -- with these ITG modes persisting up to $k_y \rho_{\mathrm{ss}} =2$ for $a/L_{T\mathrm{i}} \geq 5$, though transitioning back into UM at $k_y\rho_{\mathrm{ss}} = 1.9$ for the $a/L_{T\mathrm{i}}=4$ case  -- close to the wavenumber where the growth rate attains a local minimum. While the qualitative trends of these mode transitions and the dispersion of the eigenfrequency in DIII-D are captured by the dispersion model, the quantitative agreement with \textsc{Gene} simulations has significantly deteriorated, especially at large $a/L_{T\mathrm{i}}$, compared with the TEM-dominated case of \Cref{fig:LTe_scan_result}. Remarkably, the model predicts the propagation frequency for these high-wavenumber ITG modes more accurately than for the intermediate-wavenumber UM modes at large $a/L_{T\mathrm{i}}$, though at the cost of a significantly overestimated growth rate. The eigenmodes of these high-wavenumber high-gradient ITG modes were found to be characterised by more finer-scale substructure (in both $\Re{\hat{\phi}},\Im{\hat{\phi}}$) beneath the mode envelope (i.e. $\abs{\phi}$), thereby classifying them as higher-excitation state ITG. Such higher-excitation state ITG modes are characterised by higher critical gradients for destabilisation, as well as increased stiffness of growth rate with respect to driving gradient once destabilised compared to those of conventional ground-state ITG eigenmodes (i.e. without finer-scale substructure) \cite{Pueschel2019OnDevices,Han2017MultipleBarriers,Xie2016GlobalGradient}. As a result from this finer-scale substructure in the eigenfunction, these higher-excitation state ITGs are subjected to stronger Landau damping from parallel dynamics\cite{Rodriguez2025TheLocalisation,Pueschel2019OnDevices}, which is explicitly absent from our model and may explain the strong overprediction of the growth rate for these modes. By contrast, for the intermediate-wavenumber UMs observed for these higher gradients, the model significantly underpredicts the growth rates and overpredicts the propagation frequency, particularly so close to the mode transitions to ITGs. It should be emphasized, however, that these observations about the model's reduced performance in the intermediate-to-high wavenumber range with increasing $a/L_{T\mathrm{i}}$ are specific to the DIII-D flux-tube geometry, rather than a generic property of the model for axisymmetric configurations. Indeed, as shown in the Supplementary Material, the model results are in close agreement with \textsc{Gene} simulations in both of the considered triangularity configurations of the TCV tokamak over the full range of wavenumbers and temperature gradients. As the most significant geometric differences between TCV and DIII-D are found in the safety factor $q_0$ and magnetic shear $\hat{s}$ [see \Cref{tab:FT_global_header_params}], with both being $\approx 2.2-2.4$ times larger in DIII-D, respectively, the reduction of the model's performance may be attributed to the stronger FLR damping -- represented by the modified Bessel functions in the ion resonant integrals of \Cref{eq:ion-1Dres-integrals} -- due to the stronger secular increase in $\norm{\bm{k_\perp}}$. Additionally, the combination of high propagation frequency and low growth rate renders the numerical integration of these resonant integrals increasingly more challenging to accurately resolve, while simultaneously these integrals have a more dominant effect on the dispersion relation, as they are weighted by $\omega_{\grad{T_i}} \propto a/L_{T\mathrm{i}}$ in the ion-density kernel [\Cref{eq:ion-el-glob-kernels}]. Both of these explanations do not contradict the high accuracy achieved by the model in the low wavenumber range, where both $\norm{\bm{k_\perp}}$ as well as the ratio $\omega_R/\gamma$ are much smaller. \par
For HSX, there is a much clearer distinction between ITG modes and trapped-particle modes compared with DIII-D, with low-wavenumber modes ($k_y \rho_{\mathrm{ss}} \leq 0.5$) changing from UI modes to ITG modes, as clearly indicated by the sign reversal of the propagation frequency when the ion temperature gradient exceeds $\eta_{\textrm{i}} \geq 1$, with the exception of the $k_y \rho_{\mathrm{ss}} = 0.2$ mode for $a/L_{T\mathrm{i}}=3$. In contrast, at higher wavenumber a monotonic change of propagation frequency between electron and ion diamagnetic frequencies occurs, with the associated transition wavenumber and the growth-rate maximum gradually shifting to lower $k_y$, with the latter also increasing with the ion temperature gradient. By convention, these modes are labelled as iTEMs due to the increasing influence of ions to the instability drive. Remarkably the global dispersion model accurately predicts both growth rates and propagation frequencies -- with the exception of the UI modes at $k_y \rho_{\mathrm{ss}} \leq 0.4$ being damped, and the $k_y\rho_{\mathrm{ss}} = 0.5$ mode at $a/L_{T\mathrm{i}}=1$ which shows a hybridization between TEM and UI features -- for the full temperature gradient and wavenumber range, unlike what has been observed in DIII-D. The growth rates for the intermediate-to-high wavenumber iTEM modes are, however, consistently overestimated, resulting in a small mismatch for the wavenumber at which the growth rate maximum is attained between the high-fidelity simulations and the model, though the trend of growth rates with $k_y$ are generally preserved for each gradient. As HSX shares many of the main qualitative geometric features with DIII-D due to its quasi-helical symmetry, the main differences being only the lack of exact physical symmetry and a significantly lower magnetic shear [see \Cref{tab:FT_global_header_params}]. This further enhances the argument of the model's reduced performance in DIII-D being a numerical issue caused by large $\norm{\bm{k_\perp}}$. \par
Lastly, for the high-mirror configuration of W7-X we find similar trends as observed in HSX. Aside from the modes at $k_y \rho_{\mathrm{ss}}=0.2,0.3$ for $a/L_{T\mathrm{i}} \leq 2$ where UIs appear, the presence of a finite ion temperature gradient is sufficient to shift the propagation frequency of all trapped-particle modes that appeared for the $a/L_{T\mathrm{e}}$ scan in \Cref{fig:LTe_scan_result} into the ion diamagnetic direction. By convention this classifies the majority of these modes as iTEMs at lower temperature gradients. At higher temperature gradients $a/L_{T\mathrm{i}} \geq 5$, however, two mode transitions occur around $k_y \rho_{\mathrm{ss}} \approx 0.5$ and $k_y \rho_{\mathrm{ss}} \approx 0.9$ corresponding to an ITG-to-iTEM and iTEM-to-ITG transition, respectively. At intermediate temperature gradient of $a/L_{T\mathrm{i}}=4$ ITG modes occur only at $k_y \rho_{\mathrm{ss}}=0.3\textrm{-}0.4$, with remaining modes being iTEM. Whilst these mode transitions between ITG and iTEMs and the qualitative trends among them are well captured by the model, the growth rates are again significantly overestimated, especially for the high-wavenumber ($k_y \rho_{\mathrm{ss}} \gtrapprox 1$) ITG modes observed for $a/L_{T\mathrm{i}} \geq 5$. The eigenmodes of these high-wavenumber high-gradient ITGs also show higher-excitation state features, similar to what was observed in the DIII-D simulations, explaining the strong overestimation of growth rates for these modes due to the lack of parallel ion dynamics. Unlike the DIII-D case, however, the model does not suffer from a significant overprediction of propagation frequency and underprediction of growth rate of the intermediate wavenumber iTEMs close to the transition point to these higher-excitation state ITG modes, This may again be attributed to the smaller global shear of W7-X, preventing numerical issues from arising due to very large $\norm{\bm{k_\perp}}$. The global dispersion relation model, however, does predict both propagation frequency and growth rate for the low-wavenumber ITG modes with greater accuracy than the low-wavenumber iTEMs. This may be attributed to the lesser role (compared with iTEMs) of trapped electrons in determining the ITG mode properties, such that the influence of the maximum-$\mathcal{J}$ feature of the geometry -- which tends to increase the relative importance of non-adiabatic passing electron response with respect to the trapped-electron response\cite{Helander2015b} -- is not as prominent for these instabilities. 

\section{Reduced fidelity physics model} \label{sec:redmod}
\subsection{Description of reduced models}
Despite the success of the global dispersion relation \Cref{eq:glob-disp} in predicting eigenfrequencies (given a correct eigenmode for $\hat{\phi}$ is provided), its main drawback is the lack of an analytical solution for the resonant ion integrals [\Cref{eq:ion-1Dres-integrals}], requiring evaluation through numerical integration instead. As this integral contains an oscillatory component, it is prone to either high computational cost or risk of not being well-resolved, which as we have seen in \Cref{sec:ke} is most likely to occur in high-shear devices. By contrast, the trapped-electron resonant integrals [\Cref{eq:trel-genplasmaZ-integrals}] are fully analytic and can thus be computed at relatively low computational cost to machine precision; the resulting pitch angle integral is thus fairly robust to numerical integration, containing no singularities or oscillatory components. Indeed, solving for the eigenfrequency requires no significant additional cost when switching between an adiabatic- and a kinetic-electron description in the model. As \Cref{eq:glob-disp} contains most of the linear collisionless electrostatic physics embedded in the Vlasov-Poisson system, aside from finite ion-transit-effects and a kinetic model for passing electrons, it may be advantageous to consider some reduced-fidelity physics models which allow for analytical solutions to the ion integrals. \par 
\subsubsection*{FLR effect approximations}
The main issue preventing analytical solutions to \Cref{eq:ion-1Dres-integrals} is the coupling between the drift resonance and FLR physics through the Bessel function argument $\hat{b_j}(\xi)=b_j/\left(1+i\omega_{j,\grad{B}}\xi\right)$, which arises as a result of the velocity-space asymmetry between the magnetic drift components $\omega_{ds} = \omega_{s,\kappa} (v_{\parallel}/v_{Ts})^2 + \omega_{s,\grad{B}} \left(v_{\perp}/v_{Ts}\right)^2 / 2$, and the Larmor radius dependence on $v_\perp$. Hence, by leveraging adequate approximations for the magnetic drift and FLR effects, the desired complexity reduction of the ion integral may be obtained. Aside from providing a reduction in computational load for the global dispersion model, the inclusion of such approximations for the drift and FLR physics also allows for probing the sensitivity of a particular instability to the various driving and damping mechanisms, enabling one to make decisions on which physics should be retained when constructing a reduced model to yield eigenmode frequencies within a desired accuracy. \par
The FLR and toroidal-resonance physics are most straightforwardly decoupled by invoking the DK approximation for the ions as well, therefore taking the $b_j\rightarrow 0$ limit\cite{Biglari1998b}. However, the complete suppression of FLR effects, together with the assumption of streamer-like instabilities with a radial wavenumber $k_x = 0$, implies the eigenfrequency spectrum to be roughly linear in $k_y$ [compare \Cref{eq:Dloc} and \Cref{tab:normalised-geo-quantities}], which only applies up to $k_y \rho_{\mathrm{ss}} \approx 0.5$ for most of the cases considered in \Cref{sec:GENE-application}. Hence, this approximation will likely be too egregious -- as e.g. recently shown by Ref.~\onlinecite{Merlo2023OnTokamaks} where growth rates beyond $k_y\rho_{\mathrm{ss}} \approx 0.3$ were significantly overpredicted in the DK approximation compared with simulations retaining full FLR effects -- and FLR effects should be retained to some degree. In previous works, a Taylor series approach to the FLR damping has been considered, expanding either $\Gamma_0$ after integration\cite{Gaur2024OmnigenousStability} or expanding $J_0^2$ before integrating\cite{Ivanov2023AnField,Terry1982KineticMode}. While these methods yield analytically tractable solutions for the ion integral, they are restrictively limited in their validity to small $b$, as these (truncated) power series expansions introduce artificial zero crossings at $b\sim\order{1}$ when expanded to any odd order, and rapidly grow without bound beyond $b\sim\order{1}$ for both even and odd order expansions, whilst their unapproximated counterparts are strictly positive and bounded between $[0,1]$. As $b\propto g^{yy}k_y^2$ and the metric component contains a secular term, such approximations will introduce erroneous physics in our dispersion model even at low $k_y$ when the full effects along the field line are taken into account, especially in high magnetic shear configurations. An alternative approximation to account for FLR damping effects is the Padé approximation $\Gamma_0(b_j)\approx1/(1+b_j)$, originally considered to remove the deleterious effects introduced by Taylor series approach for intermediate-wavelength modes in gyrofluid models\cite{Dorland1998}. Unlike their Taylor series counterpart, this Padé approximation does not introduce spurious zero crossings, closely captures the behaviour of the modified Bessel function even at $b_j\sim\order{1}$, and smoothly decays at decreasing wavelengths like $\Gamma_0(b_j)$. Although, at very short wavelengths $b_j\gg 1$ the effect of FLR damping is overestimated in the Padé approximaiton as the proper asymptotic form\cite{Abramowitz1968HandbookTables} $\Gamma_0(b_j)\sim1/\sqrt{b_j}$ decays more slowly. % Eqn 9.7.1.
As the transport-relevant scales for ITG-TEM driven turbulence are typically long-wavelength, the Padé approximation is thus expected to be adequate to describe the behaviour of the instabilities at the length-scales of interest, since it alleviates the issues that render the Taylor-series approach unsuitable for our model. \par 
The application of the Padé approximation to include FLR effects in the toroidal ITG dispersion relation has recently been considered by Ref.~\onlinecite{Zocco2018ThresholdPlasmas}, where for unstable modes ($\sigma_{\gamma} = +1$) and regions of positive curvature it was shown that the ion integrals [\Cref{eq:ion-1Dres-integrals}] have analytical solutions. In \Cref{app:PadéZocco-derivation}, we extend the validity of this treatment to arbitrary growth rate and arbitrary sign of the magnetic curvature and derive an alternative derivative-free expression for $J_{j,\parallel}^{(2)},J_{j,\perp}^{(2)}$. Applying the Padé approximation to the resonant ion integrals of \Cref{eq:ion-1Dres-integrals} as shown in \Cref{app:PadéZocco-derivation} results in the following analytical expressions for the resonant ion integrals:
\begin{widetext}
    \begin{align}
        J_{j}^{\textrm{Padé},0} &= \frac{a_{j,\grad{B}}}{\sqrt{g_{\kappa,\grad{B},b}}} \left(i\sqrt{\pi}e^{-\frac{\omega}{2\omega_{j,\kappa}}}Z\left(\sigma_{\gamma}\sigma_{\kappa}\sqrt{\frac{a_{j,\kappa}}{2}}\sqrt{g_{\kappa,\grad{B},b}}\right)
        + 4\pi e^{-a_{j,\grad{B}}\left(1+b_j\right)}T\left[-i\sigma_{\gamma}\sigma_{\kappa}\sqrt{a_{j,\kappa}}\sqrt{g_{\kappa,\grad{B},b}},\frac{1}{\sqrt{g_{\kappa,\grad{B},b}}}\right]\right), \nonumber \\
        J_{j,\parallel}^{\textrm{Padé},2} &= -\frac{1}{g_{\kappa,\grad{B},b}} \left( J_{j}^{\textrm{Padé},0} + 2a_{j,\grad{B}}\left[1+\sigma_{\gamma}\sigma_{\kappa}\sqrt{a_{j,\kappa}}Z\left(\sigma_{\gamma}\sigma_{\kappa}\sqrt{\frac{a_{j,\kappa}}{2}}\right)\right]\right), \nonumber \\
        J_{j,\perp}^{\textrm{Padé},2} &= 2a_{j,\grad{B}} \left(J_{j}^{\textrm{Padé},0}-\frac{1}{1+b_j}\right) - 2\frac{\omega_{j,\kappa}}{\omega_{j,\grad{B}}} J_{j,\parallel}^{\textrm{Padé},2},
        \label{eq:Padé-integrals}
    \end{align} 
\end{widetext}
where we have introduced $a_{j,\kappa} = \omega/\omega_{j,\kappa}, \ a_{j,\grad{B}} = \omega/\omega_{j,\grad{B}}$ and $\sigma_{\kappa}=\sgn{\omega_{j,\kappa}}$, use the short-hand notation $g_{\kappa,\grad{B},b}=2\omega_{j,\kappa}(1+b_j)/\omega_{j,\grad{B}}-1$, and $T[h,a]$ is the Owen's T-function\cite{Owen1956TablesProbabilities}. The only remaining validity limit to \Cref{eq:Padé-integrals} is that the expression for $J_{j}^{\textrm{Padé},0}$ was obtained under the assumption that $\omega_{j,\kappa}/\omega_{j,\grad{B}} > 0$ [though the expressions for $J_{j,\parallel}^{\textrm{Padé},2}$ and $J_{j,\perp}^{\textrm{Padé},2}$ are generally valid if one computes $J_{j}^{\textrm{Padé},0}$ numerically from \Cref{eq:ion-Padé-integrals} in those cases instead]. Using \Cref{eq:curv-gradp-rel}, the validity of this assumption can be estimated to hold up to plasma pressures of $\beta^{\mathrm{crit}} \sim \order{a/R}$, with $R$ the major radius. For the devices considered here, the tokamaks are characterised by $\beta^{\mathrm{crit}}\approx0.3$, while for the stellarators $\beta^{\mathrm{crit}}\approx 0.1$, far above the thresholds where electromagnetic effects have a non-negligible influence on the ITG\cite{Pueschel2008GyrokineticBeta,Mulholland2025Finite-Modes}, such that \Cref{eq:Padé-integrals} is adequate within the electrostatic approximation considered in this work. In the limit of $\sigma_{\kappa}=\sigma_{\gamma}=+1$, the expression of $J_{j}^{\textrm{Padé},0}$ reduces to that obtained in Appendix~B of Ref.~\onlinecite{Zocco2018ThresholdPlasmas}. The DK limit is then straightforwardly obtained from \Cref{eq:Padé-integrals} by setting $b_j = 0$, and in \Cref{app:DK-derivation} we show if one further assumes $\omega_{j,\kappa}\approx\omega_{j,\grad{B}}$, consistent with $\beta \approx 0$, that $J^{\textrm{Padé},0}_{j}$ reduces to
\begin{align*}
    J_{j}^{\mathrm{DK},0} = \lim_{\omega_{\kappa}\rightarrow \omega_{\grad{B}}}\lim_{b\rightarrow0} J_{j}^{\textrm{Padé},0} = \frac{a_{j,\kappa}}{2} Z\left(\sigma_{\gamma}\sigma_{\kappa}\sqrt{\frac{a_{j,\kappa}}{2}}\right)^2,
\end{align*}
which when substituted in \Cref{eq:Padé-integrals} to obtain expressions for $J_{j,\perp}^{(2)}$ and $J_{j,\parallel}^{(2)}$ in the DK limit, reduces the adiabatic-electron dispersion relation for ITG modes [\Cref{eq:loc-disp-solve} with the trapped-particle contribution suppressed] to the well-known BDR result\cite{Biglari1998b} [$D_0$ in Eqn.~(3) therein] under the conventional assumptions of $\sigma_{\gamma}=\sigma_{\kappa}=+1$, see \Cref{eq:DK-Biglari-dispersion}. \par
\subsubsection*{Magnetic drift approximations}
Aside from approximating the FLR effects, another method to decrease the complexity of the resonant ion integrals in \Cref{eq:ion-1Dres-integrals} is to reduce the velocity-space asymmetry of the magnetic drift. By noting that, in the absence of temperature anisotropy, the equipartition theorem predicts that $\langle{v_\parallel^2\rangle}=\langle{v_\perp^2\rangle}/2$, we can consider two approximations to the magnetic drift\cite{Terry1982KineticMode},
\begin{align}
    \omega_{ds} \approx 
    \begin{cases}  
        \left(\omega_{s,\kappa}+\omega_{s,\grad{B}}\right) \frac{v_\parallel^2}{v_{Tj}^2} & \textrm{curvature model}\\
        \left(\omega_{s,\kappa}+ \omega_{s,\grad{B}}\right) \frac{v_\perp^2}{2v_{Tj}^2} & \grad{B}\textrm{ model}
    \end{cases}.
    \label{eq:reduced-drift-models}
\end{align}
Henceforth, we shall refer to the unapproximated drift $\omega_{ds} = \omega_{s,\kappa} (v_{\parallel}/v_{Ts})^2 + \omega_{s,\grad{B}} \left(v_{\perp}/v_{Ts}\right)^2 / 2$ as the ``full-drift'' model. Comparing the ``full drift'' case to \Cref{eq:reduced-drift-models}, we observe that these drift models effectively reduce to a pure curvature or $\grad{B}$ drift, but at an augmented magnitude, which in turn reduces the resonance condition $\omega \sim \omega_{ds}$ from a curved surface in velocity space to a simple plane\cite{Cheng1981ElectrostaticTokamaks}. Those approximations alleviate the complexity of the denominators in \Cref{eq:ion-1Dres-integrals}, as the integer power terms are attributable to the $\grad{B}$ drift, while the half-integer powers are attributable to the curvature drift. In \Cref{app:curvCheng-derivation}, we show that with the curvature model, it is possible to obtain analytical expressions for \Cref{eq:ion-1Dres-integrals} as
\begin{align}
    \begin{split}
        J_{j}^{\mathrm{curv},0} = & - \Gamma_{0}(b_j) \sigma_{\gamma} \sigma_{d,j} \sqrt{\frac{a_{d,j}}{2}} Z\left(\sigma_{\gamma}\sigma_{d,j} \sqrt{\frac{a_{d,j}}{2}}\right), \\
        J_{j,\perp}^{\mathrm{curv},2} =& -2\dv{\left(b_j\Gamma_{0}(b_j)\right)}{b_j} \sigma_{\gamma} \sigma_{d,j} \sqrt{\frac{a_{d,j}}{2}} Z\left(\sigma_{\gamma}\sigma_{d,j} \sqrt{\frac{a_{d,j}}{2}}\right), \\
        J_{j,\parallel}^{\mathrm{curv},2} = & -\Gamma_0(b_j) a_{d,j} \left[1+\sigma_{\gamma} \sigma_{d,j} \sqrt{\frac{a_{d,j}}{2}} Z\left(\sigma_{\gamma} \sigma_{d,j} \sqrt{\frac{a_{d,j}}{2}}\right)\right],
    \end{split}
\end{align}
where we have introduced $a_{d,j}=\omega/(\omega_{j,\kappa}+\omega_{j,\grad{B}})$ and $\sigma_{d,j}=\sgn{(\omega_{j,\kappa}+\omega_{j,\grad{B}})}$. This analytical simplification is facilitated by a decoupling of the FLR and drift physics, since by virtue of the curvature model we have $\hat{b_j}(\xi)\rightarrow b_j$. With these approximations, the ion-density kernel in \Cref{eq:ion-el-glob-kernels} reduces to
\begin{align}
    h_{\mathrm{ion},j}^{\mathrm{curv}} = & \frac{\omega_{\grad{T_j}}}{2\omega}  \Gamma_0(b_j) 
    a_{d} \left(1+\sigma_{\gamma}\sigma_{d,j}\sqrt{\frac{a_{d,j}}{2}} Z\left(\sigma_{\gamma}\sigma_{d,j}\sqrt{\frac{a_{d,j}}{2}}\right)\right) \nonumber \\ 
    &-\sigma_{\gamma}\sigma_{d,j}\sqrt{\frac{a_{d,j}}{2}} Z\left[\sigma_{\gamma}\sigma_{d,j}\sqrt{\frac{a_{d,j}}{2}}\right]\biggl(\Gamma_0(b_j)\times  \\
    & \left[1-\frac{\omega_{\grad{n_j}}}{\omega}+\frac{\omega_{\grad{T_j}}}{2\omega}\right]-\frac{\omega_{\grad{T_j}}}{\omega}b_j\left[\Gamma_1(b_j)-\Gamma_0(b_j)\right]\biggr), \nonumber
    \label{eq:curv-model-ion-kernel}
\end{align}
which under the conventional assumptions of $\sigma_{\gamma}=\sigma_{d,j}=+1$ reduces to the form obtained by Refs.~\onlinecite{Cheng1981ElectrostaticTokamaks,Cheng1982KineticModes} in Eqns.~(11) and (18), respectively. \par
The $\grad{B}$ model does not allow for similar simplifications of \Cref{eq:ion-1Dres-integrals}, as the coupling between FLR and drift physics remains in $\hat{b_j}(\xi)$. Nonetheless, the $\grad{B}$ model does provide a reduction in the numerical cost of calculating the resonant ion integrals, as it reduces $J_{j,\parallel}^{\grad{B},2} \rightarrow J_{j,\parallel}^{\grad{B},0}$, thereby making one of the integrals redundant. It has been shown that the $\grad{B}$ model provides a better approximation to the full-drift model than the curvature model\cite{Terry1982KineticMode} (at small wavenumbers), whilst simultaneously providing reasonable agreement with the full-drift model even at intermediate wavenumbers $b\approx 0.6$ where FLR effects become more relevant. Although this latter observation is attributable to the coupling between FLR and drift physics, which is lost in the curvature model, the former observation at low wavenumbers $b\ll1$ is due to the fact that the $v_\perp^2$ resonance of the $\grad{B}$ model is accessible in a larger region of velocity space than the $v_\parallel^2$ resonance of the curvature model, and thus more closely approximates the true resonant surface of the full-drift model. By additionally invoking the Padé approximation for the Bessel function, we show in \Cref{app:gradBTerry-derivation} that, again, the resonant ion integrals become analytically tractable,
\begin{align}
    J_{j}^{\textrm{$\grad{B}$+Padé},0} &= - a_{d,j} e^{-a_{d,j}\left(1+b_j\right)} E_1\biggl(-a_{d,j}\left(1+b_j\right)\biggr), \nonumber \\
    J_{j,\perp}^{\textrm{$\grad{B}$+Padé},2} &=-2a_{d.j}\biggl[\frac{1}{1+b_j} + a_{d,j} e^{-a_{d,j}\left(1+b_j\right)} \times  \nonumber \\ 
    &E_1\biggl(-a_{d,j}\left(1+b_j\right)\biggr)\biggr], \nonumber \\
    J_{j,\parallel}^{\textrm{$\grad{B}$+Padé},2} &= J_{j}^{\textrm{$\grad{B}$+Padé},0},
\end{align}
where $E_1(z)=\int_{z}^{\infty} \dd{t} e^{-t}/t$ is the exponential integral. 
For this double approximation, the ion-density kernel in \Cref{eq:ion-el-glob-kernels} becomes
\begin{align}
    h_{\mathrm{ion},j}^{\textrm{$\grad{B}$+Padé}} =& \left(\frac{\omega_{\grad{n_j}}}{\omega}-1+\frac{\omega_{\grad{T_j}}}{\omega} \left(a_{d,j}-1\right)\right) a_{d,j} \times \nonumber \\
    & e^{-a_{d,j}\left(1+b_j\right)} E_1\left(-a_{d,j}\left(1+b_j\right)\right) + \frac{\omega_{\grad{T_j}}}{\omega} \frac{a_{d,j}}{1+b_j},
    \label{eq:gradB-Padé-model-ion-kernel}
\end{align}
which in the DK limit of $b_j\rightarrow 0$ agrees with Eqn.~(19) from Ref.~\onlinecite{Romanelli1989IonTokamaks}.
We note, however, that while we focus on the explicit simplification of the ion-density kernel in these reduced models, the trapped-electron pitch-angle density kernel $K_{\textrm{tr-el}}$ in \Cref{eq:ion-el-glob-kernels} is also implicitly modified when invoking the curvature- or $\grad{B}$-model, as this affects the computation of the bounce-averaged drift accordingly,
\begin{align}
    \overline{\omega_{de}^T} \approx
    \begin{cases}
        \overline{\left(\omega_{e,\kappa}+\omega_{e,\grad{B}}\right)(1-\lambda B)} & \textrm{curvature model} \\ 
        \\
        \overline{\left(\omega_{e,\kappa}+\omega_{e,\grad{B}}\right)\lambda B} & \grad{B}\textrm{ model}
    \end{cases}.
    \label{eq:reduced-bav-drift-models}
\end{align}
Making simplifications to the FLR term leaves $K_{\textrm{tr-el}}$ unaffected, as the trapped electrons have been treated drift-kinetically. \par
\subsubsection*{Eigenmode proxies}
Additionally, though the global dispersion model is not fully self-consistent, its variational nature allows us to use simplified proxy functions for the eigenmode structure to obtain a close approximation to the eigenmode frequency, provided the $\hat{\phi}$ proxy is reasonably accurate. Here we briefly and non-exhaustively consider two choices for an eigenmode guess: the so-called flute mode $\hat{\phi}\approx\phi_0$ with $\phi_0$ constant or a Gaussian $\hat{\phi}(z)=e^{-z^2/2}$.
Although a relatively severe approximation, flute modes are considered in TEM stability analysis to study geometric effects\cite{Adam1976,connor2006,Morren2024InfluenceStellarators}, which, unlike the eigenfunction, are known a priori. Meanwhile, Gaussian trial functions can be justified from ITG theory, where the lowest-order eigenfunction in a quadratic drift well is Gaussian\cite{Horton1981ToroidalGradients,Plunk2014,Rodriguez2025TheLocalisation}, though with a width parameter that needs to be determined self-consistently. In reduced modelling efforts for tokamaks, a similar basis-function expansion as for the quadratic well is applied to the eigenmode, where the (Gaussian) eigenmode width is an exploitable variational parameter\cite{Rewoldt1982ElectromagneticEquilibria,Bourdelle2002StabilityTEXTOR,Citrin2017,Najlaoui2025VerifyingPlasmas}. Our main goal of adopting these eigenmode proxies is to probe the model's sensitivity to the eigenfunction shape. The results presented here therefore mainly serve as a baseline for the model's performance with poor proxy functions when compared with the exact \textsc{Gene} eigenfunctions, rather than providing any degree of fidelity. The flute mode is expected to be a particularly poor eigenfunction, as it fails to localise the instability to geometric features, while the fixed Gaussian-width approach captures localisation to the outboard magnetic well, though it fails to capture the increasing mode localisation with $k_y$ due to FLR effects. As there is only one magnetic well in a tokamak to which the eigenmode can be localised, with curvature typically being worst at the outboard, we expect better performance of this proxy in axisymmetric configurations than in low-shear stellarators, where the exact mode may be highly extended at low $k_y$ or localised to other magnetic wells at high $k_y$ when FLR damping increases in importance. Ultimately, the variational nature of the model would most benefit from more intermediate-fidelity proxies that are better informed about the underlying instability physics. \par
In case of the Gaussian trial function we adopt the same ballooning-space domain for the eigenmode as considered in the \textsc{Gene} simulations, i.e. we use the same $n_{kx}$ as in \Cref{tab:GENE_sim_resolutions} to construct the geometry beyond the $2\pi N_{\mathrm{pol}}$ extent of the flux tube. The flute mode proxy, however, violates the boundary condition of the ballooning formalism (see \Cref{sec:Dglob-method}), and would consequently result in spurious unbounded contributions to the mode drive in the global dispersion relation [\Cref{eq:glob-disp}] in an (infinitely) extended geometry. Hence, to accommodate this artificial proxy, we assume an unsheared magnetic field (consistent with periodicity of the eigenmode proxy), allowing us to limit the effects of geometry on the eigenfrequency to within the original $2\pi N_{\mathrm{pol}}$ extent of the flux tube.  \par

In order to assess the effect of these approximations on the eigenfrequency, as well as whether they affect our ability to maintain the fundamental physics for toroidal ITGs and TEMs, we apply each set of these reduced models to one adiabatic-electron ITG case at gradients of $a/L_n = 2, \ a/L_{T\mathrm{i}} = 4$, and one purely density-gradient-driven kinetic electron case at $a/L_n =3$ in the absence of temperature gradients. For simplicity, the adiabatic-electron case does not contain impurities, though the chosen gradients correspond to the nominal values considered for the $Z_{\mathrm{eff}}$ scan discussed in \Cref{sec:ae}. Likewise, the density gradient considered for the kinetic-electron case corresponds to the nominal value considered in both the $a/L_{T\mathrm{i}}$ and $a/L_{T\mathrm{e}}$ scans discussed in \Cref{sec:ke}, where the absence of temperature gradients prompts the instabilities to be either a trapped-particle mode or, if the shear is sufficiently low, a UI. 

\subsection{Application to DIII-D}
\begin{figure*}
    \centering
    \begin{subfigure}{.45\linewidth}
        \includegraphics[width=\linewidth]{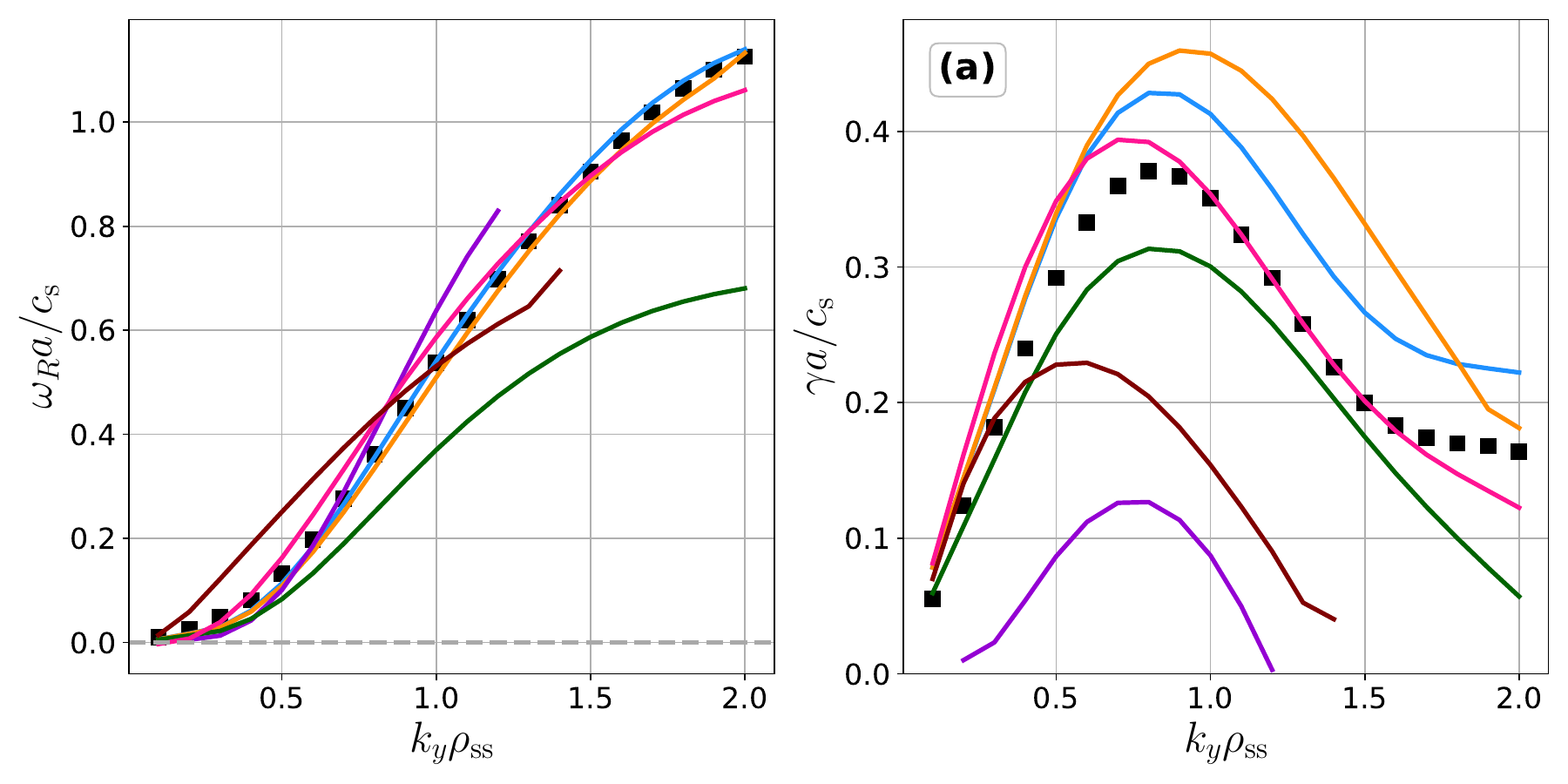}
        \phantomcaption
        \label{fig:D3D_redmod_ae_ITG}
    \end{subfigure}
    \begin{subfigure}{.45\linewidth}
        \includegraphics[width=\linewidth]{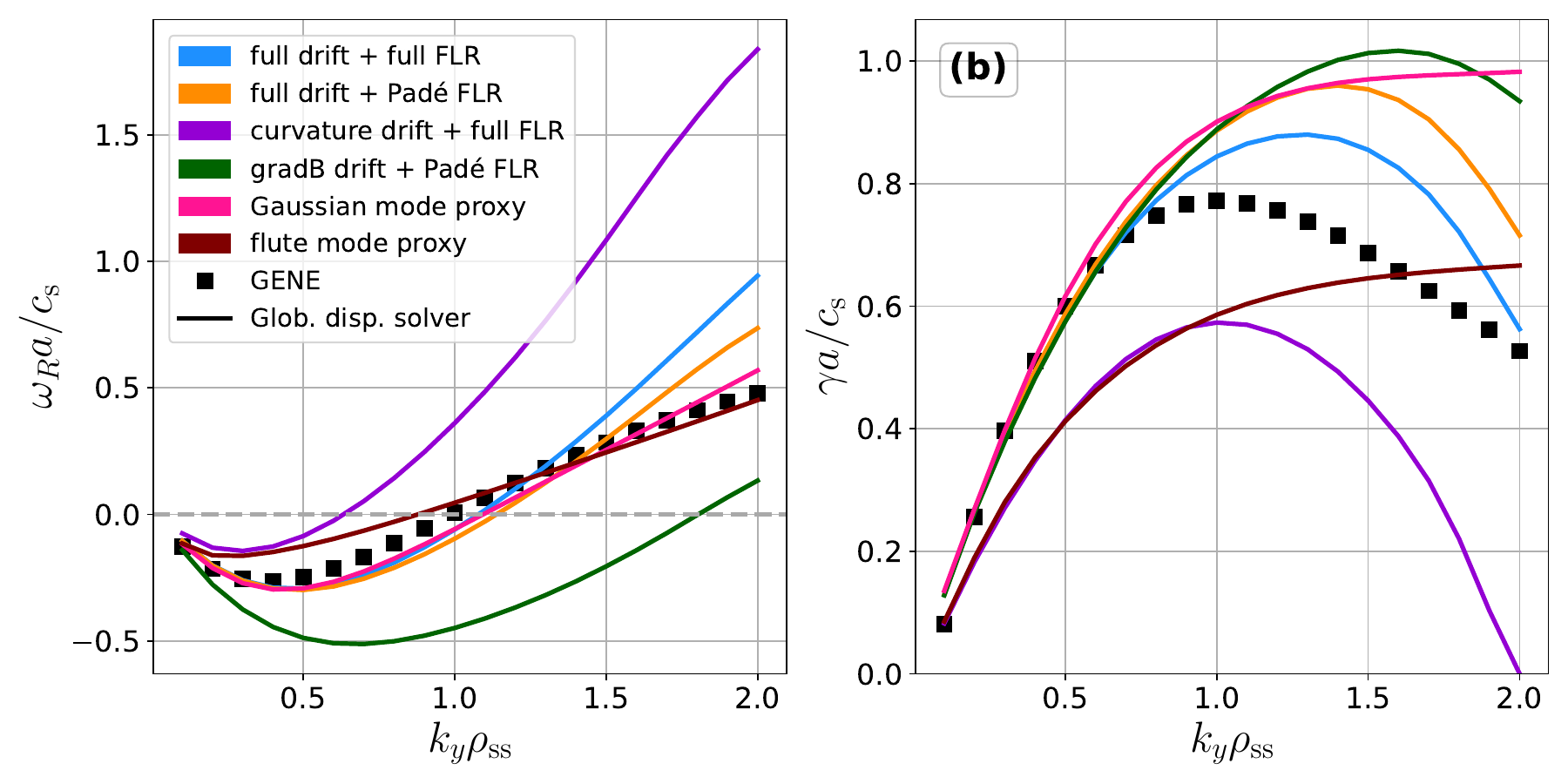}
        \phantomcaption
        \label{fig:D3D_redmod_ke_TEM}
    \end{subfigure}
    \caption{Influence of the different reduced models for the FLR damping, drift approximation, and eigenfunction proxies on the eigenfrequency solutions obtained by the global dispersion relation model (solid lines) contrasted with \textsc{Gene} simulations (symbols) in DIII-D geometry. Shown are (a) an adiabatic-electron ITG case with $a/L_n = 2, \ a/L_{T\mathrm{i}} =4$, and (b) adensity-gradient-driven TEM case with $a/L_{n} = 3$. For comparison, the solution of the baseline global dispersion model without any of the aforementioned approximations is added in blue. Note that the \textsc{Gene} eigenfunctions for $\hat{\phi}$ are used for the cases with drift and FLR approximations, while the eigenfunction proxy cases retain the full FLR and drift physics.}
    \label{fig:D3D_reduced_models}
\end{figure*}

The results for eigenfrequencies obtained by the global dispersion model in DIII-D geometry using the various reduced models are shown in \Cref{fig:D3D_reduced_models}. The results obtained from the \textsc{Gene} simulations as well as from the baseline global dispersion model in absence of further approximations are also shown to provide a reference. Focussing on the adiabatic-electron ITG scenario (see \Cref{fig:D3D_redmod_ae_ITG}) -- where we expect the model reductions to have most significant impact as they directly modify the functional properties of the ion resonant integrals [\Cref{eq:ion-1Dres-integrals}], whilst the mapping properties of the trapped-electron resonant integrals [\Cref{eq:trel-genplasmaZ-integrals}] into the complex plane remain unaltered aside from a change in the value of $\overline{\omega_{de}^{T}}$ -- we observe that, in general, the introduction of the Padé approximation (orange lines) yields the smallest discrepancy with respect to the baseline model (blue lines), with noticeable differences emerging particularly in the growth rate from $k_y \rho_{\mathrm{ss}}\approx0.8$ onwards; the higher-$k_y$ growth rates are systematically overpredicted up to $k_y\rho_{\mathrm{ss}} \approx 1.7$ compared with when full FLR effects are retained, and thereafter underpredicted at higher wavenumber. This observation is consistent with the fact that the Padé approximation $\Gamma_0(\hat{b}_j)\approx 1/(1+\hat{b}_j)$ decays less rapidly than $\Gamma_0(\hat{b}_j)$ itself from $\abs{\hat{b}_j}\leq 3$, while thereafter decaying more rapidly for larger $\abs{\hat{b}_j}$. With respect to the drift models, the $\grad{B}$-model (in combination with the Padé approximation) better captures the growth rate than the curvature model does, which is consistent with the observation from Ref.~\onlinecite{Terry1982KineticMode} about the relative areas of the resonant surfaces in velocity space between these two models; the opposite trend between drift models occurs for the propagation frequencies. Note, however, that with the curvature model approximation, the global dispersion relation model could not find any solutions for $k_y\rho_{\mathrm{ss}} = 0.1$ and $k_y\rho_{\mathrm{ss}} > 1.3$ satisfying $D=0$, with root finding methods converging onto modes with vanishingly small numerical growth/damping rates, for which the validity condition for \Cref{eq:resonce-denom-trick} used to derive \Cref{eq:curv-model-ion-kernel} is stretched to its limits, and are hence omitted from \Cref{fig:D3D_redmod_ae_ITG}. 
Lastly, with respect to the proxy functions for the electrostatic potential, the Gaussian proxy yields significantly better results than the flute mode proxy, as would be expected, thereby further cementing the variational aspect of the dispersion relation. Additionally, when using the flute mode proxy the model is only capable of finding solutions for $k_y\rho_{\mathrm{ss}} \leq 1.4$. At higher wavenumbers the initial guesses obtained from the high-frequency limit of the dispersion relation [\Cref{eq:glob-nonres-quadratic-disprel}] only yield purely real-valued frequencies for the flute mode proxy, for which the validity of \Cref{eq:resonce-denom-trick} is violated, hence such points are also omitted from \Cref{fig:D3D_redmod_ae_ITG}. With regard to the Gaussian trial function, it should be noted, however, that the closer agreement (compared with the baseline model using high-fidelity eigenfunction) between growth rates predicted by the dispersion model and \textsc{Gene} for intermediate wavenumbers should be considered as coincidental, as the model is still lacking the physics associated with finite-transit effects that would be required to obtain a one-to-one match. \par
Many of the observations regarding the influence of approximations in the model for adiabatic-electron ITG case are also directly reflected in the TEM scenario, with the discrepancy between the Padé and full-FLR model being quantitatively comparable to the adiabatic-electron ITG case, though qualitatively less severe as the discrepancy is consistent at high $k_y$. This is to be expected as the mode is predominantly driven by the trapped-electron precession resonance and thus less strongly affected by FLR damping. Similarly as in the ITG case, the curvature model does yield consistently lower growth rates than the $\grad{B}$ model, though there is a stronger discrepancy in $\omega_{R}$ with respect to the baseline model in comparison to the adiabatic-electron ITG scenario. Remarkably, the $\grad{B}$ model exhibits only a moderate discrepancy in growth rate (up to $k_y \rho_{\mathrm{ss}} \approx 1$) though a similar mismatch in $\omega_{R}$ is observed as in the ITG case. The lower sensitivity of the growth rate to the chosen drift model can be explained from \Cref{eq:reduced-bav-drift-models}, as the bounce-averaged drift is dominated by contributions near the bounce points ($\lambda B \approx 1$) where trapped-particles spend most of their time, for which the curvature component to the magnetic drift is nearly vanishing, such that the $\grad{B}$ model does not significantly affect which trapped-particle population is in resonance. However, as these particles now experience an augmented drift frequency, the mode frequency $\omega_R$ must be equivalently increased to remain in resonance, explaining the stronger discrepancy in $\omega_{R}$ compared with the adiabatic-electron ITG case. Likewise, the Gaussian proxy yields eigenfrequencies close to those obtained with the \textsc{Gene} eigenfunction with the baseline model at low wavenumber, though significant deviations occur in the growth rate beyond $k_y \rho_{\mathrm{ss}} \approx 1$. This may be attributed to the widening and flattening observed in the \textsc{Gene} UM eigenmode structures compared with the \textsc{Gene} TEM mode structures, while the proxies' width is fixed. Additionally, the flute mode proxy, despite its relentless simplicity, manages to provide reasonably accurate predictions for the propagation freuqencies -- outperforming even both options for reduced drift models when high-fidelity eigenfunctions are used -- though provides a less accurate predictions for the growth rate. Nonetheless, the growth rate predictions are comparable to those obtained with the curvature model up to $k_y \rho_{\mathrm{ss}} \approx 1$ with errors further decreasing for higher wavenumber, while with respect to the $\grad{B}$ model the opposite trend is observed. This further exemplifies the variational nature of the model when kinetic electron physics is included.

\subsection{Application to W7-X}
\begin{figure*}
    \centering
    \begin{subfigure}{.45\linewidth}
        \includegraphics[width=\linewidth]{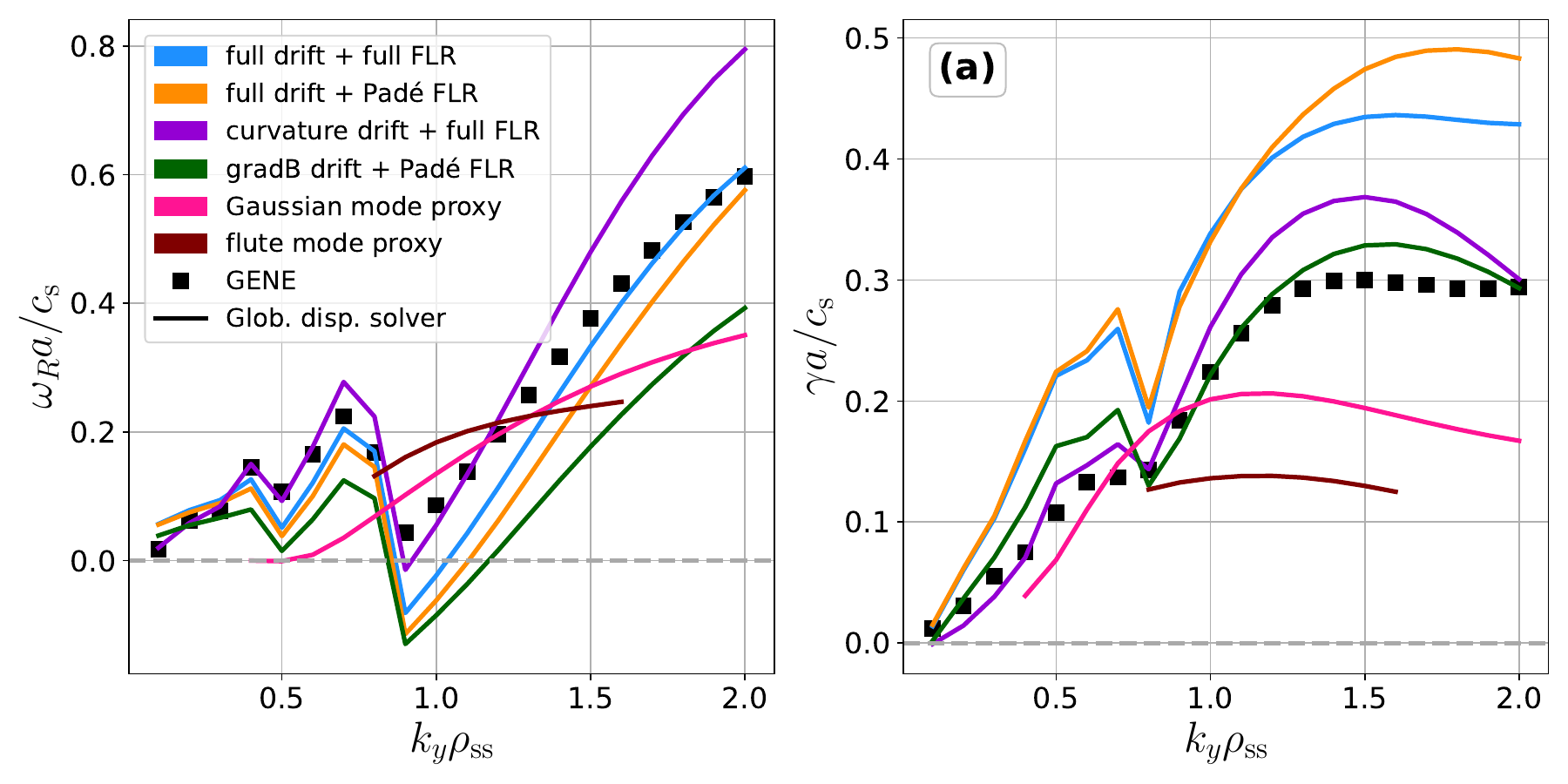}
        \phantomcaption
        \label{fig:W7Xhm_redmod_ae_ITG}
    \end{subfigure}
    \begin{subfigure}{.45\linewidth}
        \includegraphics[width=\linewidth]{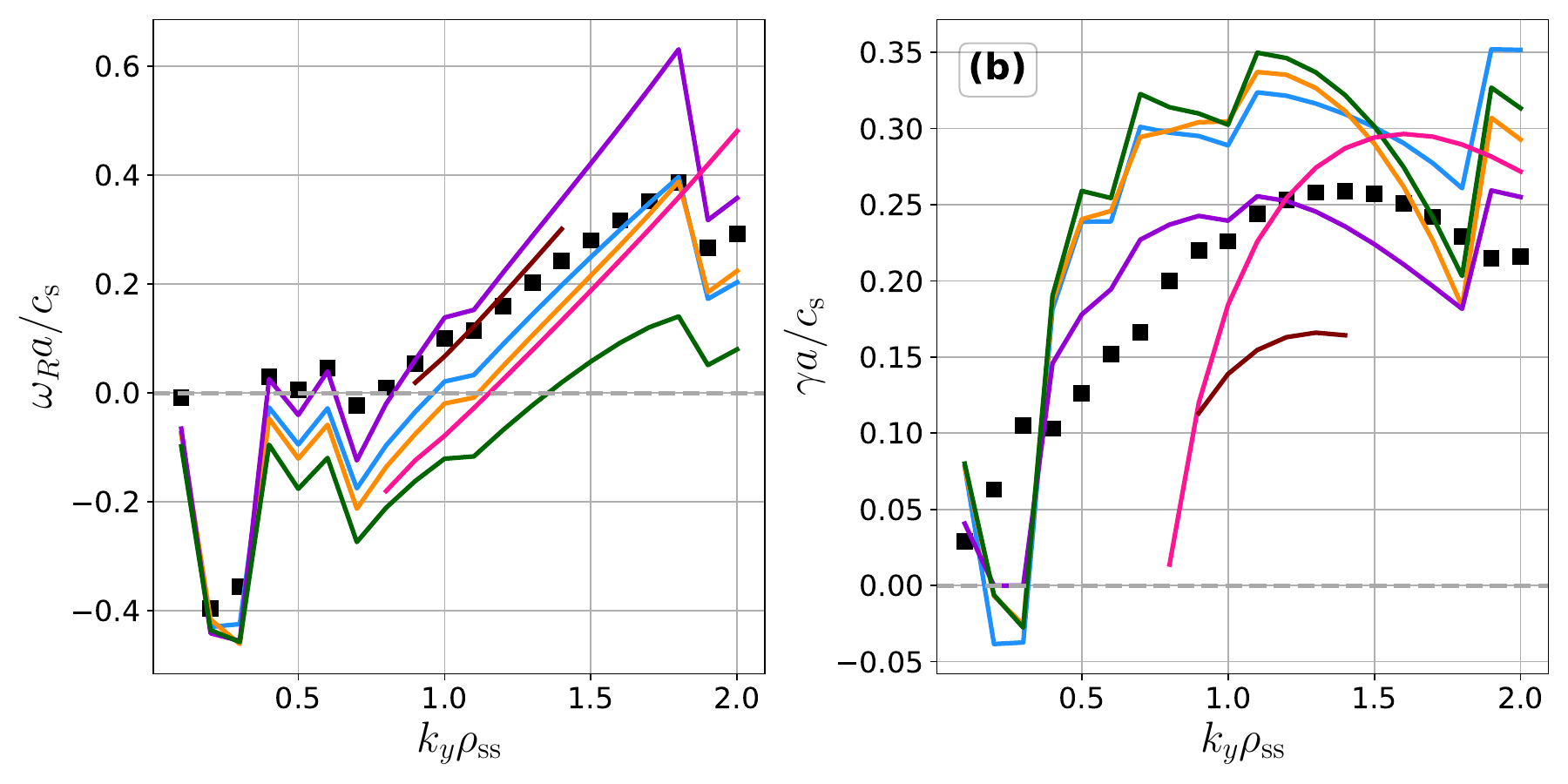}
        \phantomcaption
        \label{fig:W7Xhm_redmod_ke_TEM}
    \end{subfigure}
    \caption{Influence of the different reduced models for the FLR damping, drift approximation, and eigenfunction proxies on the eigenfrequency solutions obtained by the global dispersion relation model (solid lines) contrasted with \textsc{Gene} simulations (symbols) in the geometry of the high-mirror configuration in W7-X. Shown are (a) an adiabatic-electron ITG case with $a/L_n = 2, \ a/L_{T\mathrm{i}} =4$, and (b) adensity-gradient-driven TEM case with $a/L_{n} = 3$. For comparison, the solution of the baseline global dispersion model without any of the aforementioned approximations is added in blue. Note that the \textsc{Gene} eigenfunctions for $\hat{\phi}$ are used for the cases with drift and FLR approximations, while the eigenfunction proxy cases retain the full FLR and drift physics.}
    \label{fig:W7Xhm_reduced_models}
\end{figure*}

Results for the use of reduced models on the eigenfrequency of adiabatic-electron ITG and density-gradient-driven TEM in the high-mirror configuration of W7-X are shown in \Cref{fig:W7Xhm_redmod_ae_ITG,fig:W7Xhm_redmod_ke_TEM}, respectively.
Many of the pertinent features observed in DIII-D carry over to the high-mirror configuration of W7-X, with in particular the Padé approximation continuing to yield the best match to the baseline model. The most notable differences are the lack of mode transitions occurring when using the eigenmode proxies, which is to be expected as these are facilitated by changes in $\hat{\phi}$ obtained from \textsc{Gene}, as well as the worsening of the discrepancies between the use of proxies and the baseline model using exact eigenfunctions. \par
In particular both proxies suffer considerably from the existence of a finite range of wavenumbers for which the high-frequency limit of the dispersion relation [\Cref{eq:glob-nonres-quadratic-disprel}] predicts complex-valued rather than real-valued eigenfrequencies, which renders the applicability of the model invalid outside of those ranges, with this issue significantly exacerbated in W7-X compared with DIII-D. Remarkably, the range of wavenumbers where the model's validity is not violated is slightly smaller for the kinetic-electron TEM case compared with the adiabatic-electron ITG case. The poorer performance of the proxies in W7-X indicates that mode localisation is more complicated in (low-shear) stellarators like W7-X compared with tokamaks, necessitating the development of more realistic mode proxies than considered in this work. The fact that the kinetic-electron TEM scenario is more susceptible to this issue compared with the adiabatic-electron ITG scenario -- despite the proper variational nature of the model in the former compared with the ad-hoc applicability in the latter -- may be explained from the variation of the magnetic field strength along the field line, see \Cref{fig:FTgeo_stell}. The trapped-electron populations are strongly affected by $\norm{\bm{B}}$ through the distribution of the magnetic wells and corresponding bounce-averages. In the tokamak case, there is only one of those wells (though infinitely repeated in ballooning space), spanning the extent of a single poloidal turn, which is similar to the extent of both eigenfunction proxies. In W7-X, by contrast, there are multiple non-equivalent wells within a single poloidal turn, resulting in a mismatch between the broad extents of the eigenfunction proxies and the more narrow extents along the field line that are accessible to different trapped-particle populations, which are crucial to mode localisation for TEMs. By comparison, the variation of $\norm{\bm{B}}$ only affects the ions by slightly modifying the argument of the Bessel function and the amplitude of the magnetic drifts (see \Cref{tab:normalised-geo-quantities}), though this variation is negligible compared with the change in $\norm{\bm{k_\perp}}$ and the drift-operator $\mathcal{L}_y$, which have broader extents along the field line compared with the magnetic wells.  \par
With regard to the use of the drift approximations, the most striking difference compared with DIII-D occurs in the adiabatic-electron ITG scenario. In W7-X we observe that the $\grad{B}$ model no longer universally yields larger (and more accurate) growth rates compared with the curvature model. In fact, for the adiabatic-electron ITG case at high $k_y$, the curvature model yields better $\omega_R$ and $\gamma$ (with respect to the baseline model). This may be explained in terms of the significantly weaker FLR damping due to the lower shear, which quenches instability drive in DIII-D. Hence, it must be concluded that as far as the ITG is concerned, FLR damping is a more crucial component to determining the overall stability than the total area of the resonant surface in velocity space, otherwise the performance of the reduced model would have been universal across the two geometries. This is in line with the observation that mode localisation of adiabatic-electron ITG is most strongly correlated with $\norm{\bm{k_\perp}}$, while growth rates are subsequently most strongly correlated with the regions of curvature that are sampled within the extent of the mode\cite{Duff2022EffectTurbulence}. However, no substantial qualitative differences between W7-X and DIII-D occur for the behaviour of the reduced drift models in the TEM scenario.

\section{Summary} \label{sec:summary}
We have developed a field-line global dispersion relation for electrostatic (toroidal) ITG and TEM in arbitrary toroidal geometry, retaining FLR damping for ions and drift resonances for both ionic and electron species (semi-)analytically. This model is not only shown to satisfy a variational property, but is valid for arbitrary sign of the growth rate and magnetic curvature, and is shown to agree with conventional treatments for unstable modes with matching sign of curvature, for which the dispersion relation is more straightforward to obtain based on the integral formulation of the plasma dispersion function [\Cref{eq:plasmaZ-int}]. Such a sign generalisation is crucial for the variational formulation, as the latter is obtained by integrating the dispersion relation along the helical magnetic field line, on which the curvature drift is bound to change sign. Although this variational formulation only applies when the effect of kinetic electrons is considered, the same global dispersion relation is considered as an ad-hoc model for including non-local effects along the field line for adiabatic-electron ITG. The applicability of this model in predicting ITG-TEM eigenfrequencies in both its ad-hoc and variational states is studied by comparing against \textsc{Gene} simulations in both tokamak and stellarator geometries, where the exact \textsc{Gene} eigenfunctions for the electrostatic potential are used as trial functions. The complex eigenfrequencies produced by the model are generally in quantitative agreement with the gyrokinetic simulations, especially at transport-relevant long wavelengths. Significant deviations in that regime are predominantly observed for low-shear stellarator geometries when kinetic electrons are considered, and are attributable to the lack of non-adiabatic passing-electron physics as required to describe the Universal Instability, which is observed at these low wavenumbers in high-fidelity simulations. Additionally, when kinetic electrons are considered, the model significantly overestimates the growth rate at short wavelenghts at large ion temperature gradient in DIII-D and the high-mirror configuration of W7-X, though not in HSX. These discrepancies are attributed to higher-excitation state of ITG modes that are found in \textsc{Gene} simulations in the former set of configurations, which are more sensitive to the parallel ion dynamics that have been neglected in the model. Barring these exceptional cases, in general, the agreement between the model and the high-fidelity simulations is worst for the high-mirror configuration of W7-X. This, however, is caused by peculiarities in the geometric features of a maximum-$J$ configuration, as the model yields eigenfrequencies with similar agreement as in DIII-D and HSX for the negative-mirror configuration of W7-X -- which is distinctively not maximum-$J$ -- as shown in the Supplementary Material. Additionally, we have investigated a set of common approximations for their suitability to enhance the analytical malleability of the ion resonant integral, which may be challenging to numerically resolve as a result of coupling between the FLR damping term and the drift resonances. Such numerical limitations can particularly gain in prominence when the global magnetic shear is high, resulting in $\norm{\bm{k_\perp}\rho_s}\gg1$ even at moderate values of the bi-normal wavenumber $k_y\rho_{\mathrm{ss}}\sim\order{1}$, as has been observed in DIII-D. It is found that invoking a Padé approximation $\Gamma_0(\hat{b}_j)\approx (1+\hat{b}_j)^{-1}$ enables analytical solutions of this integral whilst consistently yielding the smallest deviations in the eigenfrequency with respect to the baseline model. \par 

Several improvements of the global dispersion model are still necessary. First and foremost, the electrostatic-potential eigenfunction needs to be determined self-consistently, rather than provided as input from linear gyrokinetic simulations. Such an approach is, however, beyond the scope of this work, where we emphasize the validation of the relevant geometry-based driving and damping physics of the model. If a self-consistent determination is not feasible, then more suitable eigenfunction proxies than the rudimentary fixed-width Gaussian and flute mode considered in \Cref{sec:redmod} will need to be developed, such that the variational nature of the dispersion relation can be fully exploited. For future work we consider furthermore including the effect of a finite-transit-frequency correction to the ion response, as explicitly done in the non-resonant limit ($\omega_{d\mathrm{i}}/\omega \ll 1$) by e.g.~Refs.~\onlinecite{Plunk2014,Romanelli1989IonTokamaks,Guo1989IonTransport}, which removes the ad-hoc nature of the model when applied to adiabatic-electron ITG scenario and may reduce the overprediction of growth rates due to missing Landau damping associated with particle streaming. Furthermore, we shall strive to include kinetic effects for passing electrons to extend the applicability of the model towards UI and, at shorter wavelengths than those considered here, ETG modes. 
Lastly we note that the analytical approaches we have considered for the global dispersion model are not limited in their applicability to studying microinstabilities, for similar velocity-space integrals over toroidal resonances in the non-adiabatic distribution are found in the evaluation of quasilinear weights for the calculation of ion and electron fluxes\cite{Casati2009ValidatingSimulations,Dudding2022ATransport,Xie2020}. As the calculation of such quasilinear weights also involve a field-line integral over the resonant velocity moments, similar to $D_{\mathrm{glob}}$, they may benefit from the sign generalisations developed in this work to correctly account for the contributions of regions of opposing curvature drift.

\begin{acknowledgments}
We would like to thank P. Costello, G.G. Plunk, and P. Helander for insightful physics discussions, as well as the \textsc{Gene} Development Team for providing support. Additionally we express our gratitude towards C.D. Stephens, J. Citrin and A.G. Goodman for their advice on the numerical framework used to solve our dispersion relation model, as well as T. Romba and F. Reimold for providing insights into experimentally accessible impurity parameters. Figures in \Cref{sec:GENE-application} have been made possible by using the colourmaps available in the open-source  {\fontfamily{cmtt}\selectfont CMasher} package \cite{vanderVelden2020CMasher:Plots}. Simulations were performed on the HPC system Viper at the Max Planck Computing and Data Facility and the HPC system Marconi at the CINECA consortium. \\
This work has been carried out within the framework of the EUROfusion Consortium, funded by the 
European Union via the Euratom Research and Training Programme (Grant Agreement No 101052200 
-- EUROfusion). Views and opinions expressed are however those of the author(s) only and do not 
necessarily reflect those of the European Union or the European Commission. Neither the European 
Union nor the European Commission can be held responsible for them. \par
\end{acknowledgments}

\appendix

\section{Derivation of the resonant ion- and electron dispersion integrals} \label{app:disp-integral-deriv-general}
\subsection{Ion dispersion integral} \label{app:tITG-derivation}
As within the model the ion population is described locally and trapping effects are neglected, one may use $v_{\parallel},v_{\perp}$ as independent velocity coordinates\cite{Romanelli1989IonTokamaks}, making the ion integral most suitable to treat with normalised cylindrical velocity coordinates $\{x_\parallel,x_\perp\}=\{v_\parallel,v_\perp\}/v_{Ts}$. In those coordinates, the single-particle Maxwellian takes the convenient parametrised form $\mathcal{F}_{Ms} \dd^3{\bm{v_s}}= \mathcal{N}_{x_\parallel}(0,1) \mathcal{N}_{x_\perp}(0,1) x_\perp \dd{x_\perp}\dd{x_\parallel} \dd{\vartheta}/\sqrt{2\pi}$ where 
\begin{align}
    \mathcal{N}_{z}(\mu,\sigma)=\frac{1}{\sqrt{2\pi \sigma^2}} \exp(-\frac{(z-\mu)^2}{2\sigma^2})
    \label{eq:param-gauss-distr}
\end{align}
is the parametrised normal distribution, where $\vartheta=[0,2\pi)$ is the gyrophase angle. In what follows we primarily use the subscript $i$ instead of $s$, though the result also holds for the other (impurity) species. \par
The ion resonant integral from \Cref{eq:Dloc} is most straightforwardly treated by simplifying the frequency fraction as $(\omega-\omega_{\ast \mathrm{i}}^{T})/(\omega-\omega_{d\mathrm{i}}) = 1+(\omega_{d\mathrm{i}}-\omega_{\ast \mathrm{i}}^T)/(\omega-\omega_{d\mathrm{i}})$, yielding
\begin{widetext}
    \begin{align}
    \int \frac{\omega - \omega_{\ast \mathrm{i}}^{T}}{\omega - \omega_{d\mathrm{i}}} \mathcal{F}_{Mi} J_0(k_\perp \rho_i)^2 \dd^3{\bm{v_i}} = \Gamma_{0}(b_i) + \underbrace{\sqrt{2\pi} \int\limits_{-\infty}^{\infty} \dd{x_\parallel} \int\limits_{0}^{\infty} \dd{x_\perp} \frac{\omega_{d\mathrm{i}}-\omega_{\ast \mathrm{i}}^{T}}{\omega-\omega_{d\mathrm{i}}} \mathcal{N}_{x_\parallel}(0,1) \mathcal{N}_{x_\perp}(0,1) x_\perp J_0(\sqrt{b_i}x_\perp)^2}_{=I_i^{\mathrm{tor}}} 
    \label{eq:ion-int-split}
\end{align}
\end{widetext}
where we introduced $b_i = \left(k_\perp \rho_{Ti}\right)^2$, the frequency-independent integral was solved in terms of Weber integrals\cite{Plunk2022EnergeticGrowth}, and $\Gamma_n(z) = \exp(-z) I_n(z)$ denote the exponentially scaled modified Bessel functions of the first kind. Here we will focus on the integral $I_i^{\mathrm{tor}}$ containing the toroidal drift resonance, which, after applying \Cref{eq:resonce-denom-trick} and using the explicit velocity dependence of the diamagnetic drift $\omega_{\ast \mathrm{i}}^{T} = \omega_{\grad{n_i}} + \omega_{\grad{T_i}} \left((x_\parallel^2+x_\perp^2)/2-3/2\right)$ and magnetic drift $\omega_{d\mathrm{i}} = \omega_{i,\kappa} x_\parallel^2 + \omega_{i,\grad{B}} x_\perp^2/2 $, reduces to
\begin{align}
    I_i^{\mathrm{tor}} =& - \frac{\omega_{\grad{n_i}}-\frac{3}{2} \omega_{\grad{T_i}}}{\omega} J_{\parallel,\perp}^{0,0} + \frac{\omega_{i,\grad{B}}-\omega_{\grad{T_i}}}{2\omega} J^{0,2}_{\parallel,\perp} \nonumber \\ 
    &+ \frac{\omega_{i,\kappa}-\frac{1}{2}\omega_{\grad{T_i}}}{\omega} J^{2,0}_{\parallel,\perp}
    \label{eq:Ii-tor-def}
\end{align}
where we introduced the following generic integrals
\begin{align}
    J_{\parallel,\perp}^{n,m} =\frac{\omega}{i\sigma_\gamma} \sqrt{2\pi}\int\limits_{0}^{\infty}  \dd{\xi} e^{i\sigma_\gamma \omega \xi} \int\limits_{-\infty}^{\infty} \dd{x_\parallel}  \mathcal{N}_{x_\parallel}(0,1) x_\parallel^n e^{-i \sigma_\gamma \omega_{i,\kappa} x_\parallel^2 \xi}\times \nonumber \\ \int\limits_{0}^{\infty} \dd{x_\perp} \mathcal{N}_{x_\perp}(0,1) x_\perp^{m+1} J_0(\sqrt{b_i}x_\perp)^2 e^{-i \sigma_{\gamma} \frac{\omega_{i,\grad{B}}}{2} x_\perp^2 \xi}.
    \label{eq:Jnm_parperp_def}
\end{align}
The velocity integrals in \Cref{eq:Jnm_parperp_def} may be performed analytically by virtue of coordinate transforms $u_\parallel = x_\parallel \sqrt{1+2i\sigma_\gamma \omega_{i,\kappa}\xi}$ and $u_\perp = x_\perp\sqrt{1+i\sigma_{\gamma}\omega_{i,\grad{B}}\xi}$, such that the exponential terms can be consolidated into a standard normal distribution in the new coordinates, yielding
\begin{widetext}
    \begin{align}
    \hspace*{-1.25cm}
        J_{\parallel,\perp}^{n,m} =\frac{\omega \sqrt{2\pi}}{i\sigma_\gamma}\int\limits_{0}^{\infty}  \dd{\xi} \frac{e^{i\sigma_\gamma \omega \xi}}{\left(1+2i\sigma_{\gamma} \omega_{i,\kappa}\xi\right)^{\frac{n+1}{2}} \left(1+i\sigma_{\gamma} \omega_{i,\grad{B}}\xi\right)^{\frac{m+2}{2}}} \int\limits_{-\infty \sqrt{1+2i\sigma_\gamma \omega_{i,\kappa}\xi}}^{\infty \sqrt{1+2i\sigma_\gamma \omega_{i,\kappa}\xi}} \dd{u_\parallel}  \mathcal{N}_{u_\parallel}(0,1) u_\parallel^n \int\limits_{0}^{\infty \sqrt{1+i\sigma_{\gamma}\omega_{i,\grad{B}}\xi}} \dd{u_\perp} \mathcal{N}_{u_\perp}(0,1) u_\perp^{m+1} J_0\left(\sqrt{\hat{b_i}(\xi)}u_\perp\right)^2
    \label{eq:Jnm_parperp_usub}
    \end{align}
\end{widetext}
where we abbreviated $\hat{b_i}(\xi) = b_i/\sqrt{1+i\sigma_{\gamma}\omega_{i,\grad{B}} \xi}$. As neither the $u_\parallel$ nor the $u_\perp$ integral contain poles (provided $n>-1,\ m>-2$) or singularities, Cauchy's theorem (applied to a pizza-slice-shaped contour) may be used to exchange the (semi) infinite line integrals in $\mathbb{C}$ with phase $-\pi/4 \leq\arg[\sqrt{1+i\sigma_\gamma \omega_{i,\grad{B}} \xi}]\leq\pi/4$, with (semi) infinite line integrals along $\mathbb{R}$, with the latter reducing to standard Gaussian and Weber integrals for $u_{\parallel}$ and $u_\perp$ respectively, yielding the three integrals of interest as
\begin{align}
    \begin{split}
        J_{\parallel,\perp}^{0,0} = \frac{\omega}{i \sigma_{\gamma}} \int\limits_{0}^{\infty} \dd{\xi} \frac{e^{i\sigma_\gamma \omega \xi} \Gamma_0(\hat{b}(\xi))}{\sqrt{1+2i\sigma_\gamma \omega_{i,\kappa}\xi}\left(1+i\sigma_{\gamma}\omega_{i,\grad{B}} \xi\right)} \\
        J_{\parallel,\perp}^{0,2} = \frac{2\omega}{i \sigma_{\gamma}} \int\limits_{0}^{\infty} \dd{\xi} \frac{e^{i\sigma_\gamma \omega \xi} \dv{}{\hat{b_i}(\xi)}\left(\hat{b_i}(\xi)\Gamma_0(\hat{b_i}(\xi))\right)}{\sqrt{1+2i\sigma_\gamma \omega_{i,\kappa}\xi}\left(1+i\sigma_{\gamma}\omega_{i,\grad{B}} \xi\right)^2}\\
        J_{\parallel,\perp}^{2,0} = \frac{\omega}{i \sigma_{\gamma}} \int\limits_{0}^{\infty} \dd{\xi} \frac{e^{i\sigma_\gamma \omega \xi} \Gamma_0(\hat{b}(\xi))}{\left(1+2i\sigma_\gamma \omega_{i,\kappa}\xi\right)^{3/2}\left(1+i\sigma_{\gamma}\omega_{i,\grad{B}} \xi\right)}
    \end{split}
    \label{eq:Jnm_final}
\end{align}
where the Bessel function expression $\Gamma_0(\hat{b}_i)+\hat{b}_i(\Gamma_1(\hat{b}_i)-\Gamma_0(\hat{b}_i))$ has been abbreviated through $\dv*{\left(\Gamma_0(\hat{b}_i)\hat{b}_i\right)}{\hat{b}_i}$. \par 
Combining \Cref{eq:ion-int-split,eq:Ii-tor-def} matches with \Cref{eq:loc-disp-solve} from the main text where we have compressed the notation $\{J_{\parallel,\perp}^{0,0}, J_{\parallel,\perp}^{2,0}, J_{\parallel,\perp}^{0,2} \}\rightarrow \{J^0,J_{\parallel}^{2}, J_{\perp}^{2}\}$, whilst \Cref{eq:Jnm_final} matches with \Cref{eq:Jnm_final} under this notation compression. 

\begin{figure}
    \centering
    \includegraphics[width=\columnwidth]{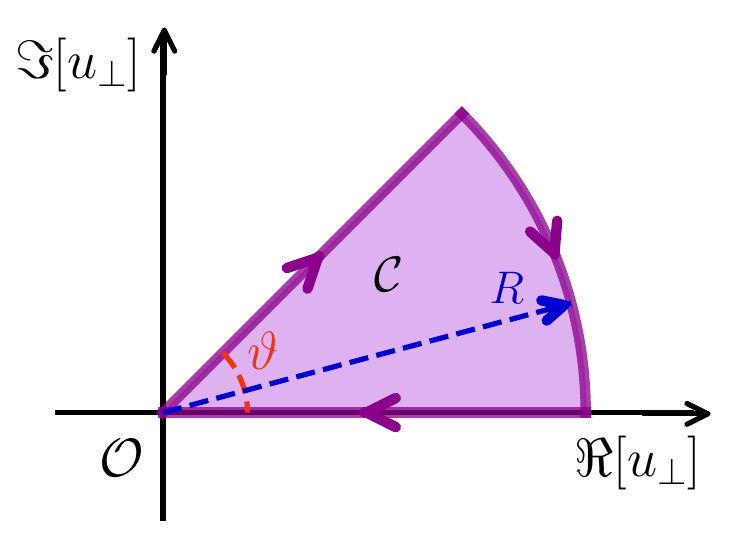}
    \caption{Sketch of the finite pizza-slice-shaped contour $\mathcal{C}$, where $\vartheta = \arg[\sqrt{1+i\sigma_\gamma \omega_{i,\grad{B}}\xi}]$ shown for a case where $\sigma_{\gamma}=\sgn{\omega_{i,\grad{B}}}=+1$, arrows indicate direction along which the contour integral $\oint_{\mathcal{C}} \dd{u_\perp}$ is evaluated. Upon taking the limit of $R\rightarrow \infty$, the contribution along the arclength vanishes asymptotically, while the contribution along the inclined segment reduces to the $u_\perp$ integral of \Cref{eq:Jnm_final}. To perform the $u_{\parallel}$ integral, we consider the contour $\mathcal{C}+\mathcal{C'}$, where $\mathcal{C'}$ is the pizza-slice-shaped contour obtained by reflecting $\mathcal{C}$ through the origin whilst preserving continuity of the integration path.}
    \label{fig:contour-sketch-vanilla}
\end{figure}

\subsection{Electron dispersion integral} \label{app:TEM-derivation}
With the kinetic contribution to the model being dominated by trapping effects, the variation of $v_{\parallel},v_{\perp}$ along the unperturbed bounce-orbits need be taken into account, hence it is not adequate to consider the velocity-space coordinates as independent. As the modulation of $v_{\parallel},v_{\perp}$ by the magnetic geometry occurs to (approximately) conserve the magnetic moment $\mu = mv_\perp^2/(2B)$ as a (lowest-order order) adiabatic invariant and is energy conserving\cite{Northrop1966}, we describe the electron velocity-space using energy and pitch angle coordinates $\varepsilon=E/T_{\mathrm{e}}=\norm{\bm{v}}^2/(2v_{T\mathrm{e}}^2)$, $\lambda = \mu/E = v_\perp^2/(v^2 B)$ such that
\begin{align}
    \mathcal{F}_{Me} \dd^3{\bm{v_e}} = \sum_{\sigma_{\parallel}=\pm1} \frac{\exp(-\varepsilon)}{\left(2\pi\right)^{3/2}} \frac{B \sqrt{2\varepsilon}}{2\sqrt{1-\lambda B}}\dd{\varepsilon}\dd{\lambda}\dd{\vartheta} 
\end{align}
where the spatial variation of $\norm{\bm{B}}$ has been suppressed, $\sigma_{\parallel} = \sgn{v_\parallel}$ accounts for the direction of the longitudinal motion and the pitch angle $\lambda \in [0,1/B]$ accounts for the the modulation of $v_\parallel,v_\perp$ across the bounce-orbits, with the boundaries of pitch angle space corresponding to strongly circulating- and (locally) deflecting trapped-particles, corresponding to the root of $\sqrt{1-\lambda B}$, respectively. \par
Hence for the electron resonant integral from \Cref{eq:Dloc} yields 
\begin{widetext}
    \begin{align}
        \int\limits_{\mathrm{trap}} \frac{\omega-\omega_{\ast \mathrm{e}}^{T}}{\omega-\overline{\omega_{d\mathrm{e}}}} \overline{\hat{\phi}} \mathcal{F}_{Me} \dd^3{\bm{v_e}} = \frac{B}{\sqrt{\pi}} \int\limits_{0}^{1/B} \dd{\lambda} H(\lambda) \frac{\overline{\hat{\phi}}(\lambda)}{\sqrt{1-\lambda B}} \left(\Gamma\left(\frac{3}{2}\right)+\underbrace{\int\limits_{0}^{\infty}\dd{\varepsilon}  \sqrt{\varepsilon}e^{-\varepsilon}\frac{\overline{\omega_{d\mathrm{e}}}-\omega_{\ast \mathrm{e}}^T}{\omega-\overline{\omega_{d\mathrm{e}}}}}_{I_{\textrm{tr-el}}} \right)
        \label{eq:electron-int-split}
    \end{align} 
\end{widetext}
where $H(\lambda)$ is the Heaviside function to account for the trapped-particle region of velocity space being exclusive to pitch angles $\lambda > 1/B_\mathrm{max}$, where $B_\mathrm{max}$ is the global maximum of the magnetic field strength on the flux surface, the frequency fraction was simplified analogously to the ion case, $\Gamma(z)=\int_{0}^{\infty}t^{z-1}e^{-t}\dd{t}$ denotes the Euler gamma function, and $\sum_{\sigma_{\parallel}=\pm} \rightarrow 2$ as the integrand depends only on $\abs{v_\parallel}$. Applying \Cref{eq:resonce-denom-trick} to the integral $I_{\textrm{tr-el}}$ containing the precession drift resonance yields
\begin{align}
    I_{\textrm{tr-el}} = - \frac{\omega_{\grad{n_e}}-\frac{3}{2}\omega_{\grad{T_e}}}{\omega}J_{\textrm{tr-el}}^{(1)} + \frac{\overline{\omega_{de}^T}(\lambda)-\omega_{\grad{T_e}}}{\omega} J_{\textrm{tr-el}}^{(2)}
    \label{eq:Itrel-def}
\end{align}
where the explicit energy dependence of the diamagnetic drift $\omega_{\ast \mathrm{e}}^{T} = \omega_{\grad{n_e}} + \omega_{\grad{T_e}} \left(\varepsilon-3/2\right)$ and precession drift $\overline{\omega_{d\mathrm{e}}} = \varepsilon \overline{\left(2 \omega_{e,\kappa} \left(1-\lambda B\right)+\omega_{e,\grad{B}} \lambda B\right)} \equiv \epsilon \overline{\omega_{de}^T}(\lambda)$ where the bounce-averaged quantity conveniently represents the precession drift of a trapped-particle at thermal energy, in terms of the following generic integral
\begin{align}
    J_{\textrm{tr-el}}^{m} = \frac{\omega}{i\sigma_{\gamma}}\int\limits_{0}^{\infty} \dd{\xi} e^{i\sigma_\gamma \omega\xi} \int\limits_{0}^{\infty} \dd{\varepsilon} \exp(-\varepsilon) \epsilon^{\frac{2m-1}{2}} e^{-i\sigma_\gamma \varepsilon \overline{\omega_{de}^T}\xi}.
    \label{eq:Jm_TEM_def}
\end{align}

\begin{figure}[ht!]
    \centering
    \includegraphics[width=\columnwidth]{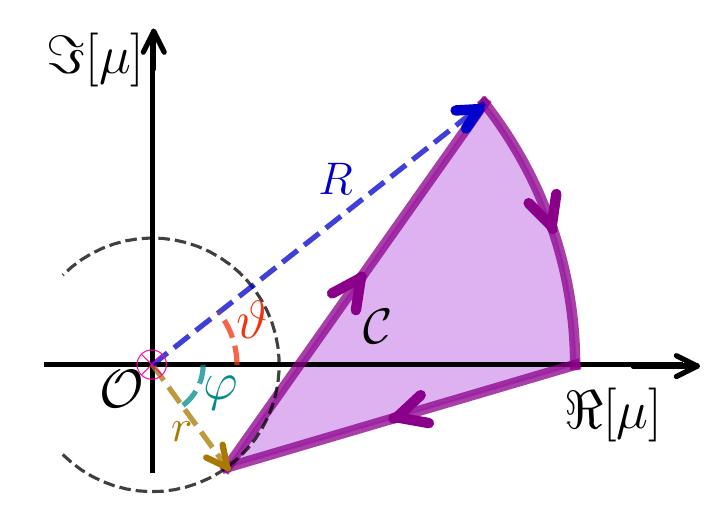}
    \caption{Sketch of the finite skewed pizza-slice-shaped contour $\mathcal{C}$, where $\vartheta = \arg\left[\sqrt{\omega/\overline{\omega_{de}^T}}\right]-\pi \sigma_{\gamma}\sigma_{d,e}/4$, $\varphi = -\arg\left[\sqrt{\omega/\overline{\omega_{de}^T}}\right]+\pi \sigma_{\gamma}\sigma_{d,e}/2$ and $r=\abs{\sqrt{\omega/\overline{\omega_{de}^T}}}$, shown for a case where $\sigma_{\gamma}\sigma_{d,e}+1$. The arrows indicate the direction along which the contour integral $\oint_{\mathcal{C}} \dd{\mu}$ is evaluated, and $\otimes$ indicates the (degenerate) pole at the origin from $1/\mu^{2m}$ for $m>0$. Upon taking the limit of $R\rightarrow \infty$, the contribution along the curved segment vanishes asymptotically, while the straight-line contribution represented by the vector $\bm{R}-\bm{r}$ reduces to the integral of \Cref{eq:Im-TEM-def}. }
    \label{fig:contour-sketch-shifted}
\end{figure}

For $m>-1/2$, the energy integral in \Cref{eq:Jm_TEM_def} can be performed analytically, facilitated by the coordinate transform $\varphi=\varepsilon\left(1+i\sigma_{\gamma}\overline{\omega_{de}^T} \xi\right)$, which, upon applying Cauchy's theorem to the pizza-slice-shaped contour from \Cref{fig:contour-sketch-vanilla} to exchange the line integral in $\mathbb{C}$ to a line integral on $\mathbb{R}$, yields
\begin{align}
    J_{\textrm{tr-el}}^{m} = 
    & \Gamma\left(\frac{2m+1}{2}\right) \frac{\omega}{i\sigma_{\gamma}} \int\limits_{0}^{\infty} \dd{\xi} \frac{e^{i\sigma_\gamma \omega\xi}}{\left(1+i\sigma_{\gamma}\overline{\omega_{de}^T}\xi\right)^{\frac{2m+1}{2}}}.
    \label{eq:Jm_epsilon_sub}
\end{align}
The remaining $\xi$ integral may also be obtained analytically, by applying a coordinate transform to simplify the denominator. This transform, however, needs be carefully considered such that it conserves the absolute convergence properties of \Cref{eq:Jm_TEM_def,eq:Jm_epsilon_sub}, regardless of parameters for the eigenfrequency or magnetic drift. The appropriate transform with the desired properties of analyticity and absolute convergence may be shown to be $\mu = - i\sigma_{\gamma}\sigma_{d,e} \sqrt{\omega/\overline{\omega_{de}^T}} \sqrt{1+i\sigma_{\gamma}\overline{\omega_{de}^T}\xi}$, where $\sigma_{d,e}=\sgn{\overline{\omega_{de}^{T}}}$, reducing the $\xi$ integral from \Cref{eq:Jm_epsilon_sub} to 
\begin{widetext}
    \begin{align}
        \frac{J_{\textrm{tr-el}}^{m}}{\Gamma\left(\frac{2m+1}{2}\right)}\equiv I^m = \left(-1\right)^{m+1} 2i\sigma_{\gamma}\sigma_{d,e} \left(\frac{\omega}{\overline{\omega_{de}^{T}}}\right)^{\frac{2m+1}{2}} e^{-\frac{\omega}{\overline{\omega_{de}^{T}}}} 
        \int\limits_{-i\sigma_{\gamma}\sigma_{d,e} \sqrt{\frac{\omega}{\overline{\omega_{de}^T}}}}^{\infty \exp(i\arg\left[\sqrt{\frac{\omega}{\overline{\omega_{de}^T}}}\right]-i\frac{\pi}{4}\sigma_{\gamma}\sigma_{d,e})} \dd{\mu} \frac{e^{-\mu^2}}{\mu^{2m}}.
    \label{eq:Im-TEM-def}
    \end{align} 
\end{widetext}
Regardless of frequency or precession drift, the phase of the upper bound is constrained to $-\pi/4\leq\arg{\mu}\leq\pi/4$, identical to the ion case, such that integrals over the Gaussian are asymptotically vanishing, and by integrating along the shifted and skewed pizza-slice-shaped contour from \Cref{fig:contour-sketch-shifted} we may exchange the integral to be performed along the path $\mu \in \left[-i\sigma_{\gamma}\sigma_{d,e} \sqrt{\omega/\overline{\omega_{de}^T}},\infty\right)$. For $m\neq 1/2$, integration by parts leads to the recurrence relation
\begin{align}
    I^{m} = - \frac{2}{2m-1} \frac{\omega}{\overline{\omega_{de}^{T}}} \left(1-I^{m-1}\right)
    \label{eq:TEM-recursive-rel}
\end{align}
where, for the integrals of interest with $m\in\mathbb{N}$, the generating integral $I^0$ of the sequence is found as
\begin{align}
    I^0 = - i \sqrt{\pi} \sigma_{\gamma}\sigma_{d,e}\sqrt{\frac{\omega}{\overline{\omega_{de}^T}}} e^{-\frac{\omega}{\overline{\omega_{de}^T}}} \erfc\left[-i\sigma_{\gamma}\sigma_{d,e} \sqrt{\frac{\omega}{\overline{\omega_{de}^T}}}\right]
    \label{eq:I0-TEM-generator}
\end{align}
which can be written succinctly in terms of the Faddeeva function\cite{Abramowitz1968HandbookTables}
% Eqn 7.1.3.
\begin{align}
    \mathcal{W}(\zeta)=\exp(-\zeta^2) \erfc[-i\zeta]
    \label{eq:Faddeeva}
\end{align}
as $I^{0}=-i\sqrt{\pi}  \sigma_{\gamma}\sigma_{d,e}\sqrt{\omega/\overline{\omega_{de}^T}} \mathcal{W}\left(\sigma_{\gamma}\sigma_{d,e} \sqrt{\omega/\overline{\omega_{de}^T}} \right)$. Using \Cref{eq:Im-TEM-def,eq:TEM-recursive-rel,eq:I0-TEM-generator,eq:Faddeeva} then yields the integrals of interest in \Cref{eq:Itrel-def} as
\begin{align}
        \frac{J_{\textrm{tr-el}}^{(1)}}{\Gamma\left(\frac{3}{2}\right)} =& -2 \frac{\omega}{\overline{\omega_{de}^T}} \left(1+ i\sqrt{\pi}\sigma_{\gamma}\sigma_{d,e}\sqrt{\frac{\omega}{\overline{\omega_{de}^T}}} \mathcal{W}\left(\sigma_{\gamma}\sigma_{d,e} \sqrt{\frac{\omega}{\overline{\omega_{de}^T}}} \right)\right) \nonumber \\
        \frac{J_{\textrm{tr-el}}^{(2)}}{ \Gamma\left(\frac{5}{2}\right)} =& -\frac{2}{3}\frac{\omega}{\overline{\omega_{de}^T}} \left(1-\frac{J_{\textrm{tr-el}}^{(1)}}{\Gamma(3/2)}\right).
        \label{eq:TEM-ana}
\end{align}
Inserting \Cref{eq:TEM-ana} into \Cref{eq:Itrel-def,eq:electron-int-split} matches with \Cref{eq:loc-disp-solve} in the main text, where the recurrence relation between Gamma functions\cite{Abramowitz1968HandbookTables} %Eqn 6.1.15. 
$\Gamma(n+1)=n\Gamma(n)$ and $\Gamma(1/2)=\sqrt{\pi}$ have been used to reduce the prefactors occurring in the pitch-angle integral, the effect of $H(\lambda)$ has been accounted for in the integration boundaries of $\lambda$, the integrals $I^{m}$ have been relabelled as $J^{m}_{\textrm{tr-el}}$ for increased consistency in notation with the ion treatment, and the plasma dispersion function $Z(\zeta)=i\sqrt{\pi}\mathcal{W}(\zeta)$ has been favoured over the Faddeeva function for historic reasons.

\section{Derivation of the slab-branch dispersion integral} \label{app:generalised-slab-derivation}
In the simplified scenario of a uniform and straight magnetic field, the magnetic drift vanishes ($\omega_{ds} \rightarrow 0$). Additionally, since the magnetic geometry does not vary along magnetic field lines, a description of the long-scale parallel mode structure in terms of Fourier modes $\hat{g}_s(l) \sim \exp(ik_\parallel l)$ (and analogously for $\hat{\phi}(l)$) is lucid\cite{Plunk2014,Helander2015b}. In this parallel-wavenumber description, a solution to the gyrokinetic Vlasov equation [\Cref{eq:lin-GKE}] is readily obtained, and the quasi-neutrality constraint [\Cref{eq:lin-QN}] reduces to 
\begin{align}
    \sum_{s} \frac{q_s^2n_s}{T_s} \left(1-\int \frac{\omega-\omega_{\ast s}^{T}}{\omega-k_\parallel v_\parallel} J_0(k_\perp \rho_s)^2 \mathcal{F}_{Ms}\dd^3{\bm{v_s}}\right) = 0.
    \label{eq:slab-QN}
\end{align}
In order to solve for the kinetic integral appearing in the slab-branch dispersion relation of \Cref{eq:slab-QN}, we proceed analogously to the derivation of the resonant ion integral in the toroidal branch presented in \Cref{app:tITG-derivation}. Using normalised cylindrical velocity coordinates the non-adiabatic density fluctuation may be written as 
\begin{widetext}
    \begin{align}
        \int \frac{\omega-\omega_{\ast s}^{T}}{\omega-k_\parallel v_\parallel} J_0(k_\perp \rho_i)^2 \mathcal{F}_{Mi}\dd^3{\bm{v_i}} = \Gamma_0(b_i) + \underbrace{\sqrt{2\pi} \int\limits_{-\infty}^{\infty} \dd{x_\parallel} \int\limits_{0}^{\infty} \dd{x_\perp} \frac{k_\parallel v_{Ts} x_\parallel-\omega_{\ast s}^{T}}{\omega-k_\parallel v_{Ts} x_\parallel} \mathcal{N}_{x_\parallel}(0,1) \mathcal{N}_{x_\perp}(0,1) x_\perp J_0(\sqrt{b_i}x_\perp)^2}_{=I_i^{\mathrm{slab}}}.
        \label{eq:slab-int-split}
    \end{align}
\end{widetext}
As the Landau resonance condition in the integral $I_s^\mathrm{slab}$ only depends on $x_\parallel$, the perpendicular velocity may be integrated over analytically making once more use of Weber integrals\cite{Plunk2022EnergeticGrowth}, and upon using  \Cref{eq:resonce-denom-trick} (with $\omega_{ds}\rightarrow k_{\parallel}v_{Ts}$) to write the resonant denominators in terms of an auxiliary integral we obtain
\begin{align}
    I_i^{\mathrm{slab}} =& -\left[\left(\frac{\omega_{\grad{n_i}}-\frac{3}{2}\omega_{\grad{T_i}}}{\omega}\right)\Gamma_{0}(b_i)+\frac{\omega_{\grad{T_i}}}{\omega}\dv{\left(b_i\Gamma_0(b_i)\right)}{b_i}\right]J_{\parallel}^{0} \nonumber \\ 
    &+\frac{k_{\parallel}v_{Ts}}{\omega} \Gamma_{0}(b_i)J_{\parallel}^1-\frac{\omega_{\grad{T_i}}}{2\omega} \Gamma_{0}(b_i) J_{\parallel}^2
    \label{eq:parameterised-slab}
\end{align}
having introduced the parametrised integral
\begin{align}
    J_{\parallel}^{n} = \frac{\omega}{i\sigma_{\gamma}} \int\limits_{0}^{\infty}  \dd{\xi} e^{i\sigma_\gamma \omega \xi} \int\limits_{-\infty}^{\infty} \dd{x_\parallel}  \mathcal{N}_{x_\parallel}(0,1) x_\parallel^n e^{-i\sigma_\gamma k_{\parallel}v_{Ts} x_\parallel\xi}.
    \label{eq:slab-J-def}
\end{align}
We note that in case of $\gamma/k_\parallel>0$, \Cref{eq:slab-J-def} may straightforwardly be expressed in terms of plasma dispersion functions through \Cref{eq:plasmaZ-int} if the resonant denominator is reintroduced by exchanging the order of integration between $\xi$ and $x_\parallel$, i.e. by reverting \Cref{eq:resonce-denom-trick}. However, in order to obtain generally valid results (lest $\gamma=0 \lor k_\parallel=0$), we proceed with \Cref{eq:slab-J-def} in its current form.
Unlike the toroidal-resonance scenario, where the Landau resonance for curvature drive depends on $x_\parallel^2$, the slab branch resonance depends linearly on $x_\parallel$, hence the exponential factor $e^{-i\sigma_\gamma k_{\parallel}v_{Ts} x_\parallel\xi}$ cannot be consolidated into a re-scaled normal distribution for just $x_\parallel$. Instead using \Cref{eq:param-gauss-distr}, it can be shown that 
\begin{widetext}
    \begin{align}
        e^{i\sigma_\gamma \omega \xi} \mathcal{N}_{x_\parallel}(0,1)  e^{-i\sigma_\gamma k_{\parallel}v_{Ts} x_\parallel\xi} = 2\pi \mathcal{N}_{x_\parallel}\left(-i\sigma_{\gamma}k_{\parallel}v_{Ts}\xi,1\right)\mathcal{N}_{\xi}\left(\frac{i\sigma_{\gamma}\omega}{k_\parallel^2v_{Ts}^2},\frac{1}{k_\parallel v_{Ts}}\right) \mathcal{N}_{\omega}\left(0,k_\parallel v_{Ts}\right)
    \end{align}
\end{widetext}
hence the velocity integral in \Cref{eq:slab-J-def} can be performed analytically after letting $u_{\parallel} = x_{\parallel} + i \sigma_{\gamma} k_\parallel v_{Ts}\xi$, thereby reducing to standard Gaussian moments in $u_{\parallel}$ once integrated along the contour in \Cref{fig:contour-sketch-vanilla} (extending the shown contour into the fourth quadrant). The three integrals of interest in \Cref{eq:parameterised-slab} thus reduce to
\begin{align}
    \begin{split}
        J_{\parallel}^0 &=  2\pi \mathcal{N}_{\omega}\left(0,k_\parallel v_{Ts}\right)  J_{\xi}^{0}  \\
        J_{\parallel}^1 &=   - i \sigma_\gamma 2\pi k_\parallel v_{Ts}\mathcal{N}_{\omega}\left(0,k_\parallel v_{Ts}\right)  J_{\xi}^{1} \\
        J_{\parallel}^2 &=  2\pi \mathcal{N}_{\omega}\left(0,k_\parallel v_{Ts}\right)  \left[J_{\xi}^0 - \left(k_\parallel v_{Ts}\right)^2 J_{\xi}^{2}\right]
        \label{eq:Jn_par_def}
    \end{split}
\end{align}
where the generic integral
\begin{align}
    J_{\xi}^{m} = \frac{\omega}{i\sigma_\gamma}  \int\limits_{0}^{\infty} \dd{\xi} \mathcal{N}_{\xi}\left(\frac{i\sigma_\gamma\omega}{k_\parallel^2 v_{Ts}^2},\frac{1}{k_\parallel v_{Ts}}\right) \xi^{m}
    \label{eq:slab-xi-integral}
\end{align}
has been introduced. Analytical solutions to \Cref{eq:slab-xi-integral} may be obtained by properly rescaling the normal distribution, however, extra care need be taken that this rescaling does not violate the convergence properties of both \Cref{eq:slab-J-def,eq:slab-xi-integral} on the semi-infinite interval. The appropriate transform may be found to be
$\eta_{\parallel} = \sigma_{k_\parallel} k_{\parallel}v_{Ts} \left(\xi-i\sigma_\gamma \omega/(k_\parallel v_{Ts})^2\right)$, such that \Cref{eq:slab-xi-integral} maps to
\begin{align*}
    J_{\xi}^{m} = \frac{\omega}{i\sigma_{\gamma} \sigma_{k_\parallel}k_{\parallel}v_{Ts}} \int\limits_{-i\sigma_\gamma \sigma_{k_\parallel} \frac{\omega}{k_{\parallel}v_{Ts}}}^{\infty} \dd{\eta_{\parallel}}& \mathcal{N}_{\eta_{\parallel}}(0,1) \times  \nonumber \\
   & \left[\frac{\eta_{\parallel}}{\sigma_{k_\parallel}k_{\parallel}v_{Ts}}+i\sigma_{\gamma}\frac{\omega}{\left(k_{\parallel} v_{Ts}\right)^2}\right]^m.
\end{align*}
Consequently the problem is reduced to finding the integrals
\begin{align*}
    I^{m} = \int\limits_{-i\sigma_\gamma \sigma_{k_\parallel} \frac{\omega}{k_{\parallel}v_{Ts}}}^{\infty} \dd{\eta_{\parallel}} \mathcal{N}_{\eta_{\parallel}}(0,1) \eta_{\parallel}^{m}
\end{align*}
which through integration by parts for $m>0$ may be shown that satisfy the recurrence relation
\begin{align}
    I^{m} = \frac{e^{\frac{\omega^2}{2\left(k_\parallel v_{Ts}\right)^2}}}{\sqrt{2\pi}} \left(-i\sigma_{\gamma}\sigma_{k_\parallel} \frac{\omega}{k_{\parallel}v_{Ts}}\right)^{m-1} + (m-1)I^{m-2}
    \label{eq:slab-recurvise-rel}
\end{align}
while $I^{0}=\erfc\left[-i\sigma_{\gamma}\sigma_{k_{\parallel}} \frac{\omega}{\sqrt{2}k_{\parallel}v_{Ts}}\right]/2$ is directly yielded in terms of the complementary error function. Combining \Cref{eq:Jn_par_def,eq:slab-xi-integral,eq:slab-recurvise-rel} then yields after a little bookkeeping the resonant slab integrals as
\begin{align}
    \begin{split}
        J_{\parallel}^{0}  &= - i \sqrt{\pi} \sigma_{\gamma} \sigma_{k_\parallel} \frac{\omega}{k_\parallel v_{Ts}\sqrt{2}} \mathcal{W}\left(\sigma_{\gamma}\sigma_{k_{\parallel}}\frac{\omega}{k_\parallel v_{Ts} \sqrt{2}}\right) \\
        J_{\parallel}^{1} &= \frac{\omega}{k_\parallel v_{Ts}} \left(J_{\parallel}^{0}-1\right) \\
        J_{\parallel}^{2} &= \frac{\omega^2}{(k_\parallel v_{Ts})^2} \left(J_{\parallel}^0-1\right)    
    \end{split}
    \label{eq:slab-J-ana}
\end{align}
where akin to the trapped-electron treatment from \Cref{app:TEM-derivation}, the Faddeeva function \Cref{eq:Faddeeva} has been introduced. Substituting  \Cref{eq:slab-J-ana} into \Cref{eq:slab-int-split,eq:parameterised-slab,eq:slab-QN} then yields the dispersion relation for slab-ITG modes which is generally valid unless $\gamma=0 \lor k_\parallel=0$.
\begin{align}
    1 + \sum_{j} Z_j^2 \frac{n_j T_e}{n_eT_j} &\left[1 - \Gamma_0(b_j) \left(\left[\frac{\omega_{\grad{n_j}}}{\omega}-1\right] \xi_{j,\parallel} Z\left(\xi_{j,\parallel}\right) \right. \right. \nonumber  \\
    &\left. \left.+\frac{\omega_{\grad{T_i}}}{\omega} \left[\xi_{j,\parallel}Z(\xi_{j,\parallel})\left(\xi_{j,\parallel}^2-\frac{1}{2}\right)+\xi_{j,\parallel}^2\right]\right) \right. \nonumber \\
    &\left.- \frac{\omega_{\grad{T_i}}}{\omega} \xi_{j,\parallel}Z\left(\xi_{j,\parallel}\right) b_j\left(\Gamma_1(b_j)-\Gamma_0(b_j)\right)\right] = 0
    \label{eq:slab-dispersion}
\end{align}
where the plasma dispersion function $Z(\zeta)=i\sqrt{\pi}\mathcal{W}(\zeta)$ has been introduced in favour of the Faddeeva function to match with typical conventions, the contribution from the kinetic electrons has been neglected since it is smaller by a factor of $\sqrt{m_\mathrm{e}/m_\mathrm{i}}$ than that from the ions (see below), and $\xi_{j,\parallel} = \sigma_\gamma \sigma_{k_\parallel} \omega/(k_\parallel v_{T_j}\sqrt{2})$ with $\sigma_{k_\parallel} = \sgn{k_\parallel}$. The fact that the sign of the parallel wavenumber $k_\parallel$, which effectively determines the mode structure along the field line, should be considered to determine the proper validity regime beyond which analytical continuation is required for the plasma dispersion function as expressed through \Cref{eq:plasmaZ-int} has also been considered in e.g.~Ref.~\onlinecite{Xie2019BO:Analysis}. In the usual limit of a pure hydrogen plasma, \Cref{eq:slab-dispersion} reduces to the well-known result by Kadomtsev \& Pogutse\cite{Kadomtsev1970TurbulenceSystems} for $\sigma_{k_\parallel}=\sigma_{\gamma}=+1$. 

\subsection*{Asymptotic forms}
The slab-branch dispersion derivation made no assumptions with regard to parameter values, and is thus valid for all species, including electrons. Although trapping effects vanish in the homogeneous plasma slab limit, the non-adiabatic electron response does not, and from \Cref{eq:lin-GKE} with an eigenmode ansatz of $\hat{g}_s(l)\sim \exp(ik_\parallel l)$, it follows the electron- and ion non-adiabatic distribution function are identical under exchange of subscripts $i\rightarrow e$. What remains identical to the toroidal branch, however, is the timescale separation between ion- and electron motion along the field line since $ v_{T\mathrm{i}}/v_{T\mathrm{e}} \ll 1$. Hence, depending on the assumption of the eigenfrequency $\omega$ we could be in different regimes. \par
If we assume $\omega/(k_\parallel v_{T\mathrm{i}}) \sim \mathcal{O}(1) $, the ions will be strongly in resonance and we have the slab-ITG branch. In this case the argument of the Faddeeva function in \Cref{eq:slab-J-ana} is ostensibly small and an expansion of \Cref{eq:Faddeeva} for $\zeta \ll 1$ is warranted, resulting in\cite{Abramowitz1968HandbookTables}
%Eqn 7.1.8.
\begin{align}
    J_{\parallel}^{0} = -i\sqrt{\pi} \sigma_{\gamma}\sigma_{k_\parallel} \frac{\omega}{k_\parallel v_{T\mathrm{e}} \sqrt{2}} \sum_{n=0}^{\infty} \frac{\left(i\sigma_{\gamma}\sigma_{k_\parallel} \frac{\omega}{k_\parallel v_{T\mathrm{e}} \sqrt{2}}\right)^n}{\Gamma\left(\frac{n}{2}+1\right)}.
    \label{eq:J-slab-el}
\end{align}
If additionally we assume that the spatial extent of the mode to be $b_i \sim O(1)$, then the Bessel function argument $b_e = b_i Z_i T_em_e/(T_im_i) \ll 1$, and the effect of FLR damping on the electrons can be neglected. Inserting \Cref{eq:J-slab-el} into \Cref{eq:slab-int-split,eq:parameterised-slab,eq:slab-recurvise-rel}, then yields to lowest order in $v_{T\mathrm{i}}/v_{T\mathrm{e}}$ to a kinetic-electron density fluctuation of
\begin{align}
    \hspace*{-0.5cm}
    \int \frac{\omega-\omega_{\ast \mathrm{e}}^{T}}{\omega-k_\parallel v_\parallel} \mathcal{F}_{Me}\dd^3{\bm{v_e}} \approx & -i\sqrt{\pi} \sqrt{\frac{T_im_e}{T_em_i}} \frac{\omega}{k_\parallel v_{T\mathrm{i}}\sqrt{2}} \times \nonumber \\
    &\left(1-\frac{\omega_{\grad{n_e}}}{\omega}+\frac{\omega_{\grad{T_e}}}{2\omega}\right)
    \label{eq:asymptotic-slabITG-electrons}
\end{align}
which is $\mathcal{O}(\sqrt{m_e/m_i})$ smaller than the kinetic-ion density fluctuation and the adiabatic electron response, and has therefore been neglected in \Cref{eq:slab-dispersion}. \par
If by contrast we assume $\omega/(k_{\parallel}v_{T\mathrm{e}}) \sim \mathcal{O}(1)$, the electrons will strongly resonate with the mode and we have the slab-ETG branch. For the ions, the argument of the Faddeeva function will be large and instead an asymptotic expansion of \Cref{eq:Faddeeva} for $\zeta \gg 1$ can be made, resulting in\cite{Abramowitz1968HandbookTables}
% Eqn 7.2.14.
\begin{align}
    J_{\parallel}^{0}\sim \sum_{m=0}^{\infty} \frac{(2m)!}{m!2^m} \left(\frac{k_\parallel v_{T\mathrm{i}}}{\omega}\right)^{2m},
    \label{eq:J-slabETG-ion}
\end{align}
which inserted into \Cref{eq:slab-int-split,eq:parameterised-slab,eq:slab-recurvise-rel} gives the kinetic-ion density fluctuation to lowest order in $v_{T\mathrm{i}}/v_{T\mathrm{e}}$ as
\begin{align}
    \int \frac{\omega-\omega_{\ast \mathrm{i}}^{T}}{\omega-k_\parallel v_\parallel} J_0(k_\perp \rho_i)^2\mathcal{F}_{Mi}\dd^3{\bm{v_i}} \approx& \Gamma_0(b_i)\left(1-\frac{\omega_{\grad{n_i}}}{\omega}\right) \nonumber \\
    &- \frac{\omega_{\grad{T_i}}}{\omega}b_i \dv{\Gamma_0(b_i)}{b_i}.
    \label{eq:asymptotic-slabETG-ions}
\end{align}
In obtaining \Cref{eq:J-slabETG-ion} we made use of the asymptotic expansion for the complementary error function $\erfc(z)$, whose validity is restricted to $\abs{\arg{z}}<3\pi/4$. In our case as $z=-i\sigma_{\gamma}\sigma_{k_\parallel} \frac{\omega}{k_\parallel v_{T\mathrm{i}}}$, by virtue of the sign generalisation, $\abs{\arg{z}}< \pi/2$, such that this expansion is always lucid except for $k_\parallel=0 \lor \gamma=0$.

\section{Derivation of reduced-fidelity models}
\subsection{Padé approximation to FLR effects} \label{app:PadéZocco-derivation}
Applying the Padé approximation of $\Gamma_0(\hat{b}_j)\approx 1/(1+\hat{b}_j)$ to the integrals of \Cref{eq:ion-1Dres-integrals} yields
\begin{align}
    \begin{split}
        J_{j}^{\textrm{Padé},0} & = \frac{\omega}{i\sigma_{\gamma}} \int\limits_{0}^{\infty} \dd{\xi} \frac{\exp(i\sigma_{\gamma}\omega\xi)}{\sqrt{1+2i\sigma_{\gamma}\omega_{j,\bm{\kappa}}\xi}\left(1+b_j+i\sigma_{\gamma}\omega_{j,\grad{B}}\xi\right)} \\
        J_{j,\perp}^{\textrm{Padé},2} & = \frac{2\omega}{i\sigma_{\gamma}} \int\limits_{0}^{\infty} \dd{\xi} \frac{\exp(i\sigma_{\gamma}\omega\xi)}{\sqrt{1+2i\sigma_{\gamma}\omega_{j,\bm{\kappa}}\xi}\left(1+b_j+i\sigma_{\gamma}\omega_{j,\grad{B}}\xi\right)^2} \\
        J_{j,\parallel}^{\textrm{Padé},2} & = \frac{\omega}{i\sigma_{\gamma}} \int\limits_{0}^{\infty} \dd{\xi} \frac{\exp(i\sigma_{\gamma}\omega\xi)}{\left(1+2i\sigma_{\gamma}\omega_{j,\bm{\kappa}}\xi\right)^{3/2}\left(1+b_j+i\sigma_{\gamma}\omega_{j,\grad{B}}\xi\right)}
    \end{split}
    \label{eq:ion-Padé-integrals}
\end{align}
where in $J_{j,\perp}^{(2)}$ we applied the Padé approximation for $\Gamma_{0}$ before evaluating its derivative. Akin to the treatment from Ref.~\onlinecite{Zocco2018ThresholdPlasmas}, we consider generalised parametrised versions of these integrals
\begin{align}
        J_{j}^{0,\nu,\lambda} & = \frac{\omega}{i\sigma_{\gamma}} \int\limits_{0}^{\infty} \dd{\xi} \frac{\exp(i\sigma_{\gamma}\omega\xi)}{\sqrt{\nu+2i\sigma_{\gamma}\omega_{j,\bm{\kappa}}\xi}\left(\lambda+i\sigma_{\gamma}\omega_{j,\grad{B}}\xi\right)} \nonumber \\
        J_{j,\perp}^{2,\nu,\lambda} &=\frac{2\omega}{i\sigma_{\gamma}} \int\limits_{0}^{\infty} \dd{\xi} \frac{\exp(i\sigma_{\gamma}\omega\xi)}{\sqrt{\nu+2i\sigma_{\gamma}\omega_{j,\bm{\kappa}}\xi}\left(\lambda+i\sigma_{\gamma}\omega_{j,\grad{B}}\xi\right)^2} \nonumber \\
        &= - 2\dv{}{\lambda} J_{j}^{0,\nu,\lambda} \nonumber \\
        J_{j,\parallel}^{2,\nu,\lambda} &= \frac{\omega}{i\sigma_{\gamma}} \int\limits_{0}^{\infty} \dd{\xi} \frac{\exp(i\sigma_{\gamma}\omega\xi)}{\left(\nu+2i\sigma_{\gamma}\omega_{j,\bm{\kappa}}\xi\right)^{3/2}\left(\lambda+i\sigma_{\gamma}\omega_{j,\grad{B}}\xi\right)} \nonumber \\
        &= - 2\dv{}{\nu} J_{j}^{0,\nu,\lambda}
    \label{eq:ion-Padé-param_integrals}
\end{align}
such that \Cref{eq:ion-Padé-integrals} is recovered in the limits $\nu\rightarrow 1,\lambda \rightarrow 1+b_j$, and the problem is in principle reduced to finding a solution of $J_{j}^{0,\nu,\lambda}$, with the remaining integrals being obtained as derivatives, much like the BDR treatment of the drift-kinetic case\cite{Biglari1998b}. \par
As the curvature-related term is similar to the precession-resonance appearing in the TEM approach \Cref{eq:Jm_TEM_def}, we make a similar coordinate transform $\eta_{\kappa} =-i \sigma_{\gamma}\sigma_{\kappa} \sqrt{\omega/(2\omega_{j,\kappa})} \sqrt{\nu+2i\sigma_{\gamma}\omega_{j,\kappa}\xi}$, with $\sigma_{\kappa} = \sgn{\omega_{j,\kappa}}$, yielding
\begin{widetext}
    \begin{align}
        J_{j}^{0,\nu,\lambda} = -2i \sigma_{\gamma}\sigma_{\kappa} \frac{\omega}{\omega_{j,\grad{B}}} \sqrt{\frac{\omega}{2\omega_{j,\kappa}}} e^{-\frac{\nu \omega}{2\omega_{j,\kappa}}} \int\limits_{-i\sigma_{\gamma}\sigma_{\kappa}\sqrt{\frac{\nu \omega}{2\omega_{j,\kappa}}}}^{\infty\exp(i\left(\arg\left[\sqrt{\frac{\nu \omega}{2\omega_{j,\kappa}}}\right]-\frac{\pi}{4}\sigma_{\gamma}\sigma_{\kappa}\right))} \frac{e^{-\eta_{\kappa}^2}}{\left(\lambda \frac{\omega_{j,\kappa}}{\omega_{j,\grad{B}}}-\frac{\nu}{2}\right) \frac{\omega}{\omega_{j,\kappa}}-\eta_{\kappa}^2}.
        \label{eq:J0Padé_etasub}
    \end{align}
\end{widetext}
By taking the $\nu$ and $\lambda$ derivatives of \Cref{eq:J0Padé_etasub}, we arrive at the following interrelation between the three integrals from \Cref{eq:ion-Padé-param_integrals}
\begin{align}
    J_{j,\parallel}^{2,\nu,\lambda} = \frac{\omega}{\omega_{j,\kappa}} \left(J^{0,\nu,\lambda}_{j}-\frac{1}{\lambda\sqrt{\nu}}\right) - \frac{\omega_{j,\grad{B}}}{2\omega_{j,\kappa}} J_{j,\perp}^{2,\nu,\lambda},
    \label{eq:Padé-integrals-interrel}
\end{align}
thus making one of the integrals redundant. For our purpose, we choose this redundant integral to be $J_{j,\perp}^{2,\nu,\lambda}$, and proceed to find an alternative expression for $J_{j,\parallel}^{2,\nu,\lambda}$. Returning to the original expressions from \Cref{eq:ion-Padé-param_integrals}, and performing partial fraction decomposition on the denominators of $J_{j}^{0,\nu,\lambda}, \ J_{j,\parallel}^{2,\nu,\lambda}$, we find that
\begin{align}
    J_{j,\parallel}^{2,\nu,\lambda} = B(\nu,\lambda) J_{j}^{0,\nu,\lambda} + A(\nu,\lambda)\frac{\omega}{i\sigma_{\gamma}} \int\limits_{0}^{\infty} \dd{\xi} \frac{e^{i\sigma_{\gamma}\omega\xi}}{\left(\nu+2i\sigma_{\gamma}\omega_{j,\kappa}\xi\right)^{3/2}}
    \label{eq:J2par_Padé-fracdecomp}
\end{align}
where $A(\nu,\lambda) = (\omega_{j,\kappa}/\omega_{j,\grad{B}})/(\lambda \omega_{j,\kappa}/\omega_{j,\grad{B}}-\nu/2)$ and $B(\nu,\lambda)=-(\lambda \omega_{j,\kappa}/\omega_{j,\grad{B}}-\nu/2))^{-1}/2$ are the expansion coefficients occurring in partial fraction decomposition of $J_{j}^{0,\nu,\lambda}$. The remaining integral over the curvature resonance in \Cref{eq:J2par_Padé-fracdecomp} is equivalent to the precession resonance integral \Cref{eq:Jm_TEM_def} (for $m=1$) under the substitutions $\omega_{j,\kappa} \rightarrow \overline{\omega_{de}^T}/2$ and $\nu\rightarrow 1$, and thus by employing the coordinate transform $\eta_{\kappa} =-i \sigma_{\gamma}\sigma_{\kappa} \sqrt{\omega/(2\omega_{j,\kappa})} \sqrt{\nu+2i\sigma_{\gamma}\omega_{j,\kappa}\xi}$, may be similarly simplified into \Cref{eq:Im-TEM-def} with the substitutions of $\mu \rightarrow \eta_{\kappa}$ and $\omega/\overline{\omega_{de}^T} \rightarrow \omega\nu/2\omega_{j,\kappa}$ in the integral bounds and exponential. It is readily verified, using the same methods as outlined in \Cref{app:TEM-derivation}, that these changes slightly modify the recurrence relation (\Cref{eq:TEM-recursive-rel}) and generating function $I^0$ (\Cref{eq:I0-TEM-generator}) to $I^{m} = - \omega/((2m-1)\omega_{j,\kappa}) \left(1/\nu^{m-1/2}-I^{m-1}\right)$ and $I^0 = -i\sqrt{\pi} \sigma_{\gamma}\sigma_{\kappa} \sqrt{\omega/(2\omega_{j,\kappa})} \mathcal{W}\left(\sigma_{\gamma}\sigma_{\kappa} \sqrt{\omega\nu/(2\omega_{j,\kappa})}\right) $ respectively. Applying to the case $m=1$ then gives 
\begin{align}
    J_{j,\parallel}^{2,\nu,\lambda} =&  -\frac{1}{2\lambda \frac{\omega_{j,\kappa}}{\omega_{j,\grad{B}}}-\nu} J_{j}^{0,\nu,\lambda} -  \frac{1}{\lambda \frac{\omega_{j,\kappa}}{\omega_{j,\grad{B}}}-\frac{\nu}{2}} \frac{\omega}{\omega_{j,\grad{B}}} \times \nonumber \\ 
    &\left[\frac{1}{\sqrt{\nu}} +i\sqrt{\pi}\sigma_{\gamma}\sigma_{\kappa} \sqrt{\frac{\omega}{2\omega_{j,\kappa}}}\mathcal{W}\left(\sigma_{\gamma}\sigma_{\kappa} \sqrt{\frac{\nu\omega}{2\omega_{j,\kappa}}}\right)\right],
    \label{eq:J2par-param-sol}
\end{align}
thereby rendering $J_{j}^{0,\nu,\lambda}$ the last remaining unknown integral. \par
Returning to \Cref{eq:J0Padé_etasub}, the upper-bound of the integral may be reduced from complex infinity to $+\infty$ by realising that for all cases of interest $\nu >0$, and the upper-bound is constraint to lie in the region $\abs{\arg{\eta_{\kappa}}}<\pi/4$ in which the Gaussian asymptotically vanishes as $\abs{\eta_{\kappa}}\rightarrow \infty$, and we integrate \Cref{eq:J0Padé_etasub} along the shifted and skewed pizza-slice-shaped-contour from \Cref{fig:contour-sketch-shifted}, under the mappings of $(\mu,\omega/\overline{\omega_{de}^T},\sigma_{d,e}) \rightarrow (\eta_{\kappa},\omega\nu/(2\omega_{j,\kappa}),\sigma_{\kappa})$. The poles at $\eta_{\kappa} = \pm \sqrt{\left(\lambda \frac{\omega_{j,\kappa}}{\omega_{j,\grad{B}}}-\frac{\nu}{2}\right) \frac{\omega}{\omega_{j,\kappa}}}$ (not shown in \Cref{fig:contour-sketch-shifted}) can be shown to lie outside of the contour whenever $\lambda \omega_{j,\kappa}/\nu\omega_{j,\grad{B}} > 0$ is satisfied, or as both $\nu,\lambda \in \mathbb{R}^+$ for the application of interest to \Cref{eq:ion-Padé-integrals}, this requires only that $\omega_{j,\kappa}/\omega_{j,\grad{B}} > 0$, which is the typical case for low $\beta$ plasmas\footnote{In regions where either $\omega_{j,\kappa}$ or $\omega_{j,\grad{B}}$ is close to vanishing, there may even at modest $\beta$ be a mismatch between the signs of the curvature and $\grad{B}$ drifts. In such cases, however, the curvature or $\grad{B}$ model (depending on which of the components is closest to vanishing) from \Cref{eq:reduced-drift-models} is expected to provide an adequate approximation to the case of full toroidal drive by both components.}. Under the assumption that $\lambda \omega_{j,\kappa}/\nu\omega_{j,\grad{B}} > 0$, it follows that along the full integration of $\eta \in [-i\sigma_{\gamma}\sigma_{\kappa}\sqrt{\nu\omega/2\omega_{j,\kappa}},\infty)$ we have that the $\sgn{\Im[(\lambda \omega_{j,\kappa}/\omega_{j,\grad{B}}-\nu/2)\omega/\omega_{j,\kappa}-\eta_\kappa^2]}=\sigma_{\gamma}\sigma_{\kappa}$, and hence by writing $1/z = (1/i\sgn\{\Im[z]\}) \int_{0}^{\infty} \dd{\xi} e^{i\sgn\{Im[z]\}z\xi}$, similar to how \Cref{eq:resonce-denom-trick} has been applied to the resonant denominator in the dispersion relation, we obtain
\begin{widetext}
    \begin{align}
        J_{j}^{0,\nu,\lambda} = -2 \sqrt{\frac{\omega}{2\omega_{j,\kappa}}} e^{-\frac{\omega\nu}{2\omega_{j,\kappa}}} \frac{\omega}{\omega_{j,\grad{B}}} \int\limits_{0}^{\infty} \dd{\xi} e^{i\sigma_\gamma\sigma_{\kappa} \left(\lambda \frac{\omega_{j,\kappa}}{\omega_{j,\grad{B}}}-\frac{\nu}{2}\right) \frac{\omega}{\omega_{j,\kappa}}\xi} \int\limits_{-i\sigma_{\gamma}\sigma_\kappa \sqrt{\frac{\omega\nu}{2\omega_{j,\kappa}}}}^{\infty}\dd{\eta_\kappa} e^{-\eta_\kappa^2 \left(1+i\sigma_{\gamma}\sigma_{\kappa}\xi\right)}.
    \end{align} 
\end{widetext}
We proceed by performing two consecutive coordinate substitutions; $u=\eta_\kappa \sqrt{1+i\sigma_{\gamma}\sigma_{\kappa}\xi}$ and $p=-i\sigma_{\gamma}\sigma_{\kappa}\sqrt{1+i\sigma_{\gamma}\sigma_{\kappa}\xi}\sqrt{\omega/2\omega_{j,\kappa}} \sqrt{2\lambda \omega_{j,\kappa}/\omega_{j,\grad{B}}-\nu}$, similar to how \Cref{eq:Jnm_parperp_def,eq:Jm_TEM_def} were simplified into Gaussian integrals for the toroidal ion- and trapped-electron resonances respectively, yielding
\begin{widetext}
    \begin{align}
        J_{j}^{0,\nu,\lambda} = - 2 \sqrt{\pi} e^{-\lambda\frac{\omega}{\omega_{j,\grad{B}}}} \frac{\omega}{\omega_{j,\grad{B}} \sqrt{2\lambda \frac{\omega_{j,\kappa}}{\omega_{j,\grad{B}}}-\nu}} \int\limits_{-i\sigma_{\gamma}\sigma_{\kappa}\sqrt{\frac{\omega}{2\omega_{j,\kappa}}}\sqrt{2\lambda \frac{\omega_{j,\kappa}}{\omega_{j,\grad{B}}}-\nu}}^{\infty} \dd{p} e^{-p^2} \erfc\left[\frac{p}{\sqrt{2\frac{\lambda\omega_{j,\kappa}}{\nu\omega_{j,\grad{B}}}-1}}\right]
        \label{eq:J0Pade_upsub}
    \end{align} 
\end{widetext}
where after each coordinate transform, we performed integration along the shifted and skewed pizza-slice-shaped contour \Cref{fig:contour-sketch-shifted} effectively allowing an exchange of the integral upper-bound from complex infinity (as by virtue of the sign inclusions in the transformations, the upper-bounds for both variables are constraint to $\abs{\arg{\{u,p\}}}<\pi/4$) to $+\infty$, and the $u$ integral has been succinctly written as an error function. The integral appearing \Cref{eq:J0Pade_upsub} may then finally be solved in terms of elementary integrals, by writing $\erfc{z}=1-\erf{z}$, we obtain 
\begin{widetext}
    \begin{align}
    \int\limits_{p_0}^{\infty} \dd{p}e^{-p^2} \erfc\left[\frac{p}{\sqrt{2\frac{\lambda\omega_{j,\kappa}}{\nu\omega_{j,\grad{B}}}-1}}\right] = \frac{\sqrt{\pi}}{2}\erfc[p_0] +2 \sqrt{\pi} \int\limits_{p_0 \sqrt{2}}^{\infty} \dd{\chi}\pdv{}{\chi} T\left[\chi,\frac{1}{\sqrt{2\frac{\lambda \omega_{j,\kappa}}{\nu\omega_{j,\grad{B}}}-1}}\right]
\end{align}
\end{widetext}
where we introduced the short-hand notation $p_0 = -i\sigma_{\gamma}\sigma_{\kappa}\sqrt{\frac{\omega}{2\omega_{j,\kappa}}}\sqrt{2\lambda \frac{\omega_{j,\kappa}}{\omega_{j,\grad{B}}}-\nu}$, rescaled the second integration variable $\chi=p\sqrt{2}$, and introduced the Owen's T-function\cite{Owen1956TablesProbabilities}
\begin{align}
    T[z,a] = \frac{1}{2\pi} \int\limits_{0}^{a} \dd{t} \frac{e^{-z^2\left(1+t^2\right)}}{1+t^2}.
    \label{eq:OwensT-def}
\end{align}
Evaluating the $\chi$ integral using the Fundamental Theorem of Calculus, then, at last, yields the admittedly wielding expression
\begin{align}
        J_{j}^{0,\nu,\lambda} =& -\frac{\pi\omega}{\omega_{j,\grad{B}}\sqrt{2\lambda \frac{\omega_{j,\kappa}}{\omega_{j,\grad{B}}}-\nu}} \left(e^{-\frac{\nu \omega}{2\omega_{j,\kappa}}} \times \right.  \nonumber \\ 
        &\left. \mathcal{W}\left[\sigma_{\gamma}\sigma_{\kappa} \sqrt{\frac{\omega}{2\omega_{j,\kappa}}}\sqrt{2\lambda \frac{\omega_{j,\kappa}}{\omega_{j,\grad{B}}}-\nu}\right] - 4e^{-\frac{\lambda \omega}{\omega_{j,\grad{B}}}} \times \right. \nonumber \\ 
        &\left. T\left[-i\sigma_{\gamma}\sigma_{\kappa} \sqrt{\frac{\omega}{\omega_{j,\kappa}}}\sqrt{2\lambda \frac{\omega_{j,\kappa}}{\omega_{j,\grad{B}}}-\nu},\frac{1}{\sqrt{2\frac{\lambda \omega_{j,\kappa}}{\nu \omega_{j,\grad{B}}}-1}}\right]\right)
        \label{eq:J0Padé-anasol}
\end{align}
where we have used that $\lim_{z\rightarrow\infty} T[z,a] \rightarrow 0$ provided that $a\in\mathbb{R}$, as is the case under our assumption of $\lambda \omega_{j,\kappa}/\nu\omega_{j,\grad{B}} > 0$, and favoured the introduction of the Faddeeva function \Cref{eq:Faddeeva} over the error function. Upon setting $\nu=1$ and $\lambda=1+b_j$ such that we recover \Cref{eq:ion-Padé-integrals} from \Cref{eq:ion-Padé-param_integrals}, we note that \Cref{eq:J2par_Padé-fracdecomp,eq:Padé-integrals-interrel,eq:J0Padé-anasol} reduce to \Cref{eq:Padé-integrals} from the main text where the plasma dispersion function $Z(\zeta)=i\sqrt{\pi}\mathcal{W}(\zeta)$ was written in favour of the Faddeeva function.

\subsection{Zero-$\beta$ drift-kinetic limit} \label{app:DK-derivation}
The drift-kinetic limit is straightforwardly obtained from \Cref{eq:Padé-integrals} by letting $b_j \rightarrow 0$. Alas, aside from simplifying $g_{\kappa,\grad{B},b}$ and the exponential multiplying the Owen's T-function, this does not yield further analytical reduction in simpler functions. However, upon further letting $\omega_{j,\kappa} \approx \omega_{j,\grad{B}}$, equivalent to $\beta \approx 0$, a significant simplification for $J^{0}$ can be achieved as $g_{\kappa,\grad{B},b}$ further reduces to unity in that limit, yielding
\begin{align}
    \lim_{b_j\rightarrow0}\lim_{\beta\rightarrow 0} J_{j}^{\textrm{Padé},0} = &\frac{\omega}{\omega_{j,\kappa}} \left(i\sqrt{\pi}e^{-\frac{\omega}{2\omega_{j,\kappa}}}Z\left[\sigma_{\gamma}\sigma_{\kappa} \sqrt{\frac{\omega}{2\omega_{j,\kappa}}}\right] \right. \nonumber \\ 
    & \left. +4\pi e^{-\frac{\omega}{\omega_{j,\kappa}}}T\left[-i\sigma_{\gamma}\sigma_{\kappa} \sqrt{\frac{\omega}{\omega_{j,\kappa}}},1\right]\right).
    \label{eq:DKlimJ0}
\end{align}
To obtain the Owen's-T function with unit upper bound, we rewrite \Cref{eq:OwensT-def} as
\begin{align}
    T[z,a] = \frac{e^{-z^2}}{2\pi} \int\limits_{0}^{\infty} \dd{\xi} e^{-\xi} \int\limits_{0}^{a} \dd{t} e^{-(\xi+\frac{z^2}{2})t^2}
    \label{eq:OwensT-expint}
\end{align}
where the denominator has been written as an exponential integral $1/\alpha = \int_{0}^{\infty} \dd{x} e^{-\alpha x}, \ (\Re{\alpha}>0)$, where for $a\in\mathbb{R}$ we have that $1+t^2 \in \mathbb{R}^+$ along the full integration domain. To further simply \Cref{eq:OwensT-expint} we perform the successive change of variables $u=t\sqrt{\xi+z^2/2}$ and $\mu=\sqrt{\xi+z^2/2}$ such that
$T[z,a] = \frac{1}{2\sqrt{\pi}} \int_{\sqrt{z^2/2}}^{\infty} \dd{\mu} e^{-\mu^2} \erf(a \mu)$. As a special case, $T[z,1]$ is then straightforwardly obtained from integration by parts as $T[z,1]=(1-\erf[\sqrt{z^2/2}]^2)/8 = \erfc[\sqrt{z^2/2}] \left(2-\erfc[\sqrt{z^2/2}]\right)/8$, which when substituted back into \Cref{eq:DKlimJ0} yields
\begin{align}
    \lim_{b_j\rightarrow0}\lim_{\beta\rightarrow 0} J_{j}^{\textrm{Padé},0} =& -\frac{\pi}{2} \frac{\omega}{\omega_{j,\kappa}} e^{-\frac{\omega}{\omega_{j,\kappa}}} \erfc\left[-i\sigma_{\gamma}\sigma_{\kappa}\sqrt{\frac{\omega}{2\omega_{j,\kappa}}}\right]^2 \nonumber \\
    =& -\frac{\pi}{2} \frac{\omega}{\omega_{j,\kappa}} \mathcal{W}\left[\sigma_{\gamma}\sigma_{\kappa}\sqrt{\frac{\omega}{2\omega_{j,\kappa}}}\right]^2
    \label{eq:BDRJ0}
\end{align}
where we used that $\sqrt{z^2/2} \rightarrow z/\sqrt{2}$ when $\abs{\arg{z}}\leq\pi/2$, which is always the case for $z=-i\sigma_{\gamma}\sigma_{\kappa}\sqrt{\omega/\omega_{j,\kappa}}$ unless the growth rate or curvature vanishes. Iterating \Cref{eq:BDRJ0} through \Cref{eq:J2par-param-sol,eq:Padé-integrals-interrel} and setting $\nu=1,\lambda=1,\omega_{j,\kappa}=\omega_{j,\grad{B}}$ to obtain the $\beta\approx0$ drift-kinetic limits for $J_{j,\parallel}^{(2)}, J_{j,\perp}^{(2)}$, and substituting the result into \Cref{eq:ion-el-glob-kernels} then yields after some rearranging
\begin{align}
    \hspace*{-0.5cm}
    \lim_{\beta\rightarrow0}\lim_{b\rightarrow0} h_{\mathrm{ion},j} &\rightarrow \frac{\omega}{2\omega_{j,\kappa}} \left(1-\frac{\omega_{\grad{n_j}}}{\omega}\right) Z^2\left(\sigma_{\gamma}\sigma_{\kappa} \sqrt{\frac{\omega}{2\omega_{j,\kappa}}}\right) \nonumber \\
    &+ \frac{\omega_{\grad{T_j}}}{\omega} \left(Z^2\left(\sigma_{\gamma}\sigma_{\kappa} \sqrt{\frac{\omega}{2\omega_{j,\kappa}}}\right)\frac{\omega}{2\omega_{j,\kappa}}\left(1-\frac{\omega}{\omega_{j,\kappa}}\right) \right. \nonumber \\ 
    &\left. -\sigma_{\gamma}\sigma_{\kappa}\sqrt{\frac{\omega}{2\omega_{\kappa}}} \frac{\omega}{\omega_{j,\kappa}} Z\left(\sigma_{\gamma}\sigma_{\kappa} \sqrt{\frac{\omega}{2\omega_{j,\kappa}}}\right)\right)
    \label{eq:DK-Biglari-dispersion}
\end{align}
which agrees with Eqn~(3) of Ref.~\onlinecite{Biglari1998b} and Eqn~(2.5) of Ref.~\onlinecite{Zocco2018ThresholdPlasmas} when considering $\sigma_{\gamma}=\sigma_{\kappa}=+1$.

\subsection{Curvature model} \label{app:curvCheng-derivation}
We straightforwardly obtain the curvature-model from \Cref{eq:ion-1Dres-integrals} by the substitutions of $\omega_{j,\grad{B}}\rightarrow 0$ and $\omega_{j,\kappa} \rightarrow \omega_{j,\grad{B}} + \omega_{j,\kappa}$, yielding
\begin{align}
    \begin{split}
        J_{j}^{\mathrm{curv},0} &= \Gamma_{0}(b_j) \frac{\omega}{i\sigma_{\gamma}} \int\limits_{0}^{\infty} \frac{e^{i\sigma_{\gamma}\omega\xi}}{\sqrt{1+2i\sigma_{\gamma}\omega_{j,\kappa}^{\mathrm{eff}}\xi}} \\
        J_{j,\perp}^{\mathrm{curv},2} &= 2\dv{\left(b_j\Gamma_{0}(b_j)\right)}{b_j} \frac{\omega}{i\sigma_{\gamma}} \int\limits_{0}^{\infty} \frac{e^{i\sigma_{\gamma}\omega\xi}}{\sqrt{1+2i\sigma_{\gamma}\omega_{j,\kappa}^{\mathrm{eff}}\xi}} \\
        J_{j,\parallel}^{\mathrm{curv},2} &= \Gamma_{0}(b_j) \frac{\omega}{i\sigma_{\gamma}} \int\limits_{0}^{\infty} \frac{e^{i\sigma_{\gamma}\omega\xi}}{\left(1+2i\sigma_{\gamma}\omega_{j,\kappa}^{\mathrm{eff}}\xi\right)^{3/2}}
        \label{eq:curv-model-start-sub}
    \end{split}
\end{align}
where we have $\hat{b_j}(\xi) = b_j/(1+i\sigma_{\gamma}\omega_{j,\grad{B}}\xi) \rightarrow b_j$ under these substitutions and introduced $\omega_{j,\kappa}^{\mathrm{eff}}=\omega_{j,\kappa}+\omega_{j,\grad{B}}$ as a short-hand notation. Note that within the curvature model there are only two unique resonant integrals to compute, as $J_{j,\perp}^{\mathrm{curv},2}$ can be obtained from $J_{j}^{\mathrm{curv},0}$ through simple rescaling by the appropriate Bessel functions. The remaining integrals over $\xi$ for $J_{j}^{\mathrm{curv},0}, J_{j,2}^{\mathrm{curv},\parallel}$ are identical to those encountered in the derivation of the TEM resonance \Cref{eq:Jm_TEM_def} (for $m=0,1$) under the mapping $\overline{\omega_{de}^T}\rightarrow 2 \omega_{j,\kappa}^{\mathrm{eff}}$, such that applying this mapping to the generating function \Cref{eq:I0-TEM-generator} and the recurrence relation \Cref{eq:TEM-recursive-rel} yields
\begin{align}
    \begin{split}
        J_{j}^{\mathrm{curv},0} &= - \Gamma_{0}(b_j) i \sqrt{\pi}\sigma_{\gamma} \sigma_{\kappa}^{\mathrm{eff}} \sqrt{\frac{\omega}{2\omega_{j,\kappa}^{\mathrm{eff}}}} \mathcal{W}\left(\sigma_{\gamma}\sigma_{\kappa}^{\mathrm{eff}} \sqrt{\frac{\omega}{2\omega_{j,\kappa}^{\mathrm{eff}}}}\right)  \\
        J_{j,\perp}^{\mathrm{curv},2} &= 2 \left(1+b_j\left(\frac{I_{1}(b_j)}{I_{0}(b_j)}-1\right)\right) J_{j}^{\mathrm{curv},0} \\
        J_{j,\parallel}^{\mathrm{curv},2} &= - \frac{\omega}{\omega_{j,\kappa}^{\mathrm{eff}}} \left( \Gamma_{0}(b_j) - J_{j}^{\mathrm{curv},0}\right)
    \end{split}
    \label{eq:curv-model-ana}
\end{align}
where the derivative of the Bessel function in $J_{j,\perp}^{\mathrm{curv},2}$ has been expanded, and we introduced $\sigma_{\kappa}^{\mathrm{eff}} = \sgn{\omega_{j,\kappa}^{\mathrm{eff}}}$. Substituting \Cref{eq:curv-model-ana} into \Cref{eq:ion-el-glob-kernels} and applying $\omega_{j,\grad{B}}\rightarrow 0, \ \omega_{j,\kappa}\rightarrow \omega_{j,\grad{B}}+\omega_{j,\kappa}$ mapping then gives the ion-density kernel in the curvature model as given by \Cref{eq:curv-model-ion-kernel} in the main text where the plasma dispersion function $Z(\zeta)=i\sqrt{\pi}\mathcal{W}(\zeta)$ was favoured over the Faddeeva function. \par
As the FLR damping terms appear explicitly in \Cref{eq:curv-model-ion-kernel}, it is straightforward to apply the reduced FLR models in \Cref{sec:redmod}, for the purpose of creating hybrid reduced models. Using the Padé approximation $\Gamma_0(b_j)\approx1/(1+b_j)$ yields
\begin{align}
    h_{\mathrm{ion},j}^{\mathrm{curv}} =& \frac{1}{1+b_j} \left[ -i\sqrt{\pi}\sigma_{\gamma}\sigma_{\kappa}\sqrt{\frac{\omega}{2\omega_{j,\kappa}^{\mathrm{eff}}}} \mathcal{W}\left(\sigma_{\gamma}\sigma_{\kappa}\sqrt{\frac{\omega}{2\omega_{j,\kappa}^{\mathrm{eff}}}}\right) \right. \nonumber \\ 
    &\left. \left(\left[1-\frac{\omega_{\grad{n_j}}}{\omega}+\frac{\omega_{\grad{T_j}}}{2\omega}\right]+\frac{\omega_{\grad{T_j}}}{\omega}\frac{b_j}{1+b_j}\right) \right. +\frac{\omega_{\grad{T_j}}}{2\omega_{j,\kappa}^{\mathrm{eff}}} \times \nonumber \\ 
    & \left. \left(1+i\sqrt{\pi}\sigma_{\gamma}\sigma_{\kappa}\sqrt{\frac{\omega}{2\omega_{j,\kappa}^{\mathrm{eff}}}} \mathcal{W}\left(\sigma_{\gamma}\sigma_{\kappa}\sqrt{\frac{\omega}{2\omega_{j,\kappa}^{\mathrm{eff}}}}\right)\right) \right]
    \label{eq:curvPadéions}
\end{align}
where we used $\Gamma_1(b_j)-\Gamma_0(b_j)=\dv*{\Gamma_0(b_j)}{b_j}$, with the drift-kinetic limit being straightforwardly obtainable from \Cref{eq:curvPadéions} by letting $b_j\rightarrow 0$.

\subsection{$\grad{B}$ + Padé model} \label{app:gradBTerry-derivation}
The $\grad{B}$ model can be mapped onto \Cref{eq:ion-1Dres-integrals} through $\omega_{j,\kappa}\rightarrow0$ and $\omega_{j,\grad{B}} \rightarrow \omega_{j,\grad{B}} + \omega_{j,\kappa}$, yielding
\begin{align}
    \begin{split}
        J_{j}^{\textrm{$\grad{B}$},0} &= \frac{\omega}{i\sigma_{\gamma}} \int\limits_{0}^{\infty} \frac{e^{i\sigma_{\gamma}\omega\xi}\Gamma_0\left(\frac{b_j}{1+i\sigma_{\gamma}\omega_{j,\grad{B}}^{\mathrm{eff}}\xi}\right)}{1+i\sigma_{\gamma}\omega_{j,\grad{B}}^{\mathrm{eff}}\xi} \\
        J_{j,\perp}^{\textrm{$\grad{B}$},2} &= \frac{2\omega}{i\sigma_{\gamma}} \int\limits_{0}^{\infty} \frac{e^{i\sigma_{\gamma}\omega\xi} \left.\left(\dv{z\Gamma_0(z)}{z}\right)\right|_{z=\frac{b_j}{1+i\sigma_{\gamma}\omega_{j,\grad{B}}^{\mathrm{eff}}\xi}}}{\left(1+i\sigma_{\gamma}\omega_{j,\grad{B}}^{\mathrm{eff}}\xi\right)^{2}}  \\
        J_{j,\parallel}^{\textrm{$\grad{B}$},2} &= \frac{\omega}{i\sigma_{\gamma}} \int\limits_{0}^{\infty} \frac{e^{i\sigma_{\gamma}\omega\xi}\Gamma_0\left(\frac{b_j}{1+i\sigma_{\gamma}\omega_{j,\grad{B}}^{\mathrm{eff}}\xi}\right)}{1+i\sigma_{\gamma}\omega_{j,\grad{B}}^{\mathrm{eff}}\xi} \\
    \end{split}
    \label{eq:gradB-model-integrals}
\end{align}
where the explicit representation of $\hat{b_j}(\xi)$ was used to highlight its modification under the $\grad{B}$ model, and $\omega_{j,\grad{B}}^{\mathrm{eff}}=\omega_{j,\kappa}+\omega_{j,\grad{B}}$ is introduced as short-hand notation. We immediately note from \Cref{eq:gradB-model-integrals} that $J_{j}^{\textrm{$\grad{B}$},0} = J_{j,\parallel}^{\textrm{$\grad{B}$},2}$, thus rendering one of the integrals redundant, and reducing the ion-density kernel \Cref{eq:ion-el-glob-kernels} in the $\grad{B}$ model to
\begin{align}
    h_{\mathrm{ion},j}^{\grad{B}} = \Gamma_0(b_j)-\frac{\omega_{\grad{n_j}}-\omega_{\grad{T_j}}}{\omega} J_{j}^{\textrm{$\grad{B}$},0} + \frac{\omega_{j,\grad{B}}^{\mathrm{eff}}-\omega_{\grad{T_j}}}{2\omega} J_{j,\perp}^{\textrm{$\grad{B}$},2}.
    \label{eq:gradB-vanilla-kernel}
\end{align}
Alas, further analytical is not possible as the (default) $\grad{B}$ model does not alleviate the coupling between the FLR and drift physics, but rather modifies $\hat{b_j}(\xi)$ by the augmented $\grad{B}$-drift. \par
This difficulty can be alleviated by invoking the Padé approximation $\Gamma_0(\hat{b}_j)\approx1/(1+\hat{b}_j)$ for the Bessel function, where in case of $J_{j,\perp}^{\grad{B},2}$ we apply the approximation prior to evaluating the derivative, yielding the mapping $\left\{J_{j}^{\grad{B},0},J_{j,\perp}^{\grad{B},2} \right\}\rightarrow \{J_j^{\textrm{$\grad{B}$+Padé},0},J_j^{\textrm{$\grad{B}$+Padé},1}\} $ 
in terms of the following generic integral
\begin{align}
    J_{j}^{\textrm{$\grad{B}$+Padé},m} = (1+m)\frac{\omega}{i\sigma_{\gamma}} \int\limits_{0}^{\infty} \frac{e^{i\sigma_\gamma\omega\xi}}{\left(1+b_j+i\sigma_{\gamma}\omega_{j,\grad{B}}^{\mathrm{eff}}\right)^{m+1}}.
    \label{eq:JgradB-Padé-m-def}
\end{align}
Using the coordinate transform $\eta_{\grad{B}}=- (\omega/\omega_{j,\grad{B}}^{\mathrm{eff}}) (1+b_j+i\sigma_{\gamma}\omega_{j,\grad{B}}^{\mathrm{eff}}\xi)$, \Cref{eq:JgradB-Padé-m-def} can be solved analytically in terms of 
\begin{align}
    J_{j}^{\textrm{$\grad{B}$+Padé},m} =& (1+m) \left(-\frac{\omega}{\omega_{j,\grad{B}}^{\mathrm{eff}}}\right)^{m+1} e^{-\frac{\omega\left(1+b_j\right)}{\omega_{j,\grad{B}}^{\mathrm{eff}}}} \times \nonumber \\ 
    &\int\limits_{-\frac{\omega\left(1+b_j\right)}{\omega_{j,\grad{B}}^{\mathrm{eff}}}}^{\infty e^{i\arg[-i\sigma_{\gamma}\omega]}} \frac{e^{-\eta_{\grad{B}}}}{\eta_{\grad{B}}^{m+1}}.
    \label{eq:JgradB-m-etasub}
\end{align}
By virtue of the transform, the phase of the upper-bound is constrained to within $-\pi/2\leq\arg{\eta_{\grad{B}}}\leq \pi/2$ regardless whether the mode is unstable or damped, such that integrals over the exponential are asymptotically vanishing, hence (for $m>-2$) by integrating \Cref{eq:JgradB-m-etasub} along the shifted and skewed pizza-slice-shaped contour from \Cref{fig:contour-sketch-shifted}, we may perform the integral along the path $\eta_{\grad{B}}\in[-\omega(1+b)/\omega_{j,\grad{B}}^{\mathrm{eff}},\infty)$ instead. For $m\neq0$ integration by parts yields the recurrence relation
\begin{align}
    J_{j}^{\textrm{$\grad{B}$+Padé},m} = - \frac{\omega}{\omega_{j,\grad{B}}^{\mathrm{eff}}} \frac{1+m}{m} \left(\frac{1}{(1+b_j)^m}-\frac{J_{j}^{\textrm{$\grad{B}$+Padé},{m-1}}}{m}\right)
    \label{eq:gradBPadé-recursive-rel}
\end{align}
with generating integral of the sequence being $m=0$
\begin{align}
    J_{j}^{\textrm{$\grad{B}$+Padé},0} = - \frac{\omega}{\omega_{j,\grad{B}}^{\mathrm{eff}}} e^{-\frac{\omega\left(1+b_j\right)}{\omega_{j,\grad{B}}^{\mathrm{eff}}}}E_1\left(-\frac{\omega\left(1+b_j\right)}{\omega_{j,\grad{B}}^{\mathrm{eff}}}\right)
    \label{eq:J0-gradBPadé-generator}
\end{align}
where $E_1(z)=\int_{z}^{\infty} \dd{t} e^{-t}/t$ is the exponential integral. Note that alternative to using \Cref{eq:gradBPadé-recursive-rel}, one could also find $J_{j}^{\textrm{$\grad{B}$+Padé},1}=-2\dv*{J_{j}^{\textrm{$\grad{B}$+Padé},0}}{b_j}$, which using the properties of the generalised exponential integral\cite{Abramowitz1968HandbookTables} %Eqn 5.1.4., 5.1.24, 5.1.26.
$E_{n}(z) = \int_{1}^{\infty} \dd{t} e^{-zt}/t^{n}$, can be shown to match with \Cref{eq:gradBPadé-recursive-rel} for $m=1$, in case of $\omega_R/\omega_{j,\grad{B}}^{\mathrm{eff}}<0$. It should be noted that the recurrence relation \Cref{eq:gradBPadé-recursive-rel}, however, is only limited in validity to $(\omega_{j,\grad{B}}^{\mathrm{eff}},\gamma)\neq(0,0)$, and is otherwise an exact result. Combining \Cref{eq:J0-gradBPadé-generator,eq:gradBPadé-recursive-rel} to obtain \Cref{eq:JgradB-Padé-m-def} and inserting into \Cref{eq:gradB-vanilla-kernel} along with setting $\Gamma_0(b_j)\rightarrow 1/(1+b_j)$ to consistently apply the Padé approximation then yields \Cref{eq:gradB-Padé-model-ion-kernel} from the main text. The DK limit is then straightforwardly obtained by setting $b_j \rightarrow0$ in \Cref{eq:gradB-Padé-model-ion-kernel}.

\section{Strongly-driven limit of charge-density kernels} \label{app:fluid-limit-derivs}
In typical drift-wave ordering, the mode frequency scales as\cite{Kadomtsev1970TurbulenceSystems,Helander2013,Smolyakov2002ShortPlasmas,Horton1999DriftTransport} $\omega \sim \omega_{\grad{n_e}}$, hence as the gradients are increased, the mode frequency will satisfy $\omega_{ds}/\omega \ll 1$ in the thermal bulk, and consequently resonances with the toroidal drift are weak as they are driven by particles in the tails of the distribution function. \par 
In such scenarios, it is lucid to expand the integrands in the ion-resonance integrals \Cref{eq:ion-1Dres-integrals} for $\omega_{j,\kappa}/\omega \sim \omega_{j,\grad{B}}/\omega\ll1$. Such frequency ratios can explicitly be enforced, to provide a formal basis for these expansions, by the coordinate transform $\zeta = - i\sigma_{\gamma}\omega$, yielding
\begin{align}
    \begin{split}
        J_{j}^{(0)} &=  \int\limits_{0}^{\infty e^{i\arg[-i\sigma_{\gamma\omega}]}} \dd{\zeta} \frac{e^{-\zeta}\Gamma_0\left(\frac{b_j}{1-\frac{\omega_{j,\grad{B}}}{\omega}\zeta}\right)}{\sqrt{1-2\frac{\omega_{j,\kappa}}{\omega} \zeta}(1- \frac{\omega_{j,\grad{B}}}{\omega}\zeta)} \\
        J_{j,\perp}^{(2)} &=2 \int\limits_{0}^{\infty e^{i\arg[-i\sigma_{\gamma\omega}]}} \dd{\zeta} \frac{e^{-\zeta} \left.\dv{z\Gamma_0(z)}{z}\right|_{z=\frac{b_j}{1-\frac{\omega_{j,\grad{B}}}{\omega}\zeta}}}{\sqrt{1-2\frac{\omega_{j,\kappa}}{\omega} \zeta}(1- \frac{\omega_{j,\grad{B}}}{\omega}\zeta)^2} \\
        J_{j,\parallel}^{(2)} &= \int\limits_{0}^{\infty e^{i\arg[-i\sigma_{\gamma}\omega]}} \dd{\zeta} \frac{e^{-\zeta}\Gamma_0\left(\frac{b_j}{1-\frac{\omega_{j,\grad{B}}}{\omega}\zeta}\right)}{\left(1-2\frac{\omega_{j,\kappa}}{\omega} \zeta\right)^{3/2}(1- \frac{\omega_{j,\grad{B}}}{\omega}\zeta)}.
    \end{split}
    \label{eq:ion-1Dres-zetasub}
\end{align}
In \Cref{eq:ion-1Dres-zetasub} the singularities at $\zeta^{\ast}_{\grad{B}}=\omega/\omega_{j,\grad{B}}$ and $\zeta^{\ast}_{\kappa}=\omega/(2\omega_{j,\kappa})$ have a fixed phase of $\arg{\zeta^\ast_{\{\kappa,\grad{B}\}}} = \arg{\omega} + \pi(1-\sigma_{\{\kappa,\grad{B}\}})/2$ where $\sigma_{\{\kappa,\grad{B}\}}=\sgn{\omega_{j,\{\kappa,\grad{B}\}}}$ is the sign of drift associated with the singularity, which do not intersect with the straight-integration path $\zeta \in [0,\infty e^{\arg[-i\sigma_{\gamma}\omega]})$, at constant phase of $\arg{\zeta}=\omega-\pi \sigma_{\gamma}/2$. Consequently, as this path is further constraint to lie within the region $\abs{\arg{\zeta}}<\pi/2$ where the exponential function is asymptotically vanishing, an evaluation of \Cref{eq:ion-1Dres-zetasub} along the pizza-slice-shaped contour of \Cref{fig:contour-sketch-vanilla} allows to perform the integrals along the path $\zeta\in[0,\infty)$ instead by virtue of Cauchy's theorem. By additionally replacing each term in \Cref{eq:ion-1Dres-zetasub} containing the small parameters $\omega_{j,\grad{B}}/\omega\ll 1$ and $\omega_{j,\kappa}/\omega\ll 1$ by its first-order Taylor expansion we obtain
\begin{widetext}
    \begin{align}
        J_{j}^{0}\approx & \int\limits_{0}^{\infty} \dd{\zeta} e^{-\zeta} \left(1+\frac{\omega_{j,\grad{B}}}{\omega}\zeta\right)\left(1+\frac{\omega_{j,\kappa}}{\omega}\zeta\right) \left(\Gamma_0(b_j)+b_j\dv{\Gamma_{0}(b_j)}{b_j} \frac{\omega_{j,\grad{B}}}{\omega}\zeta\right) + \order{\left(\frac{\omega_{j,\grad{B}}}{\omega}\right)^2,\left(\frac{\omega_{j,\kappa}}{\omega}\right)^2}\nonumber \\
        \approx & \Gamma_0(b_j) \left(1+\frac{\omega_{j,\grad{B}}+\omega_{j,\kappa}}{\omega}\right) + b_j\left[\Gamma_{1}(b_j)-\Gamma_0(b_j)\right] \frac{\omega_{j,\grad{B}}}{\omega} + \order{\left(\frac{\omega_{j,\grad{B}}}{\omega}\right)^2,\frac{\omega_{j,\grad{B}}\omega_{j,\kappa}}{\omega^2},\left(\frac{\omega_{j,\kappa}}{\omega}\right)^2} \nonumber\\
        J_{j,\perp}^{(2)} \approx& 2 \int\limits_{0}^{\infty} \dd{\zeta} e^{-\zeta} \left(1+2\frac{\omega_{j,\grad{B}}}{\omega}\zeta\right)\left(1+\frac{\omega_{j,\kappa}}{\omega}\zeta\right) \left(\dv{\left(b_j\Gamma_0(b_j)\right)}{b_j} + b_j \dv[2]{\left(b_j\Gamma_0(b_j)\right)}{b_j} \frac{\omega_{j,\grad{B}}}{\omega}\zeta\right)  + \order{\left(\frac{\omega_{j,\grad{B}}}{\omega}\right)^2,\left(\frac{\omega_{j,\kappa}}{\omega}\right)^2}\nonumber \\
        \approx & 2 \left(\left(\Gamma_0(b_j)+b_j\left[\Gamma_1(b_j)-\Gamma_0(b_j)\right]\right)\left(1+\frac{\omega_{j,\kappa}}{\omega}\right)+\frac{\omega_{j,\grad{B}}}{\omega} \left(b_j \left[\Gamma_{1}(b_j)-\Gamma_0(b_j)\right]\left(3-2b_j\right)+\Gamma_0(b_j)\left(2-b_j\right)\right) \right) + \nonumber \\
        &\order{\left(\frac{\omega_{j,\grad{B}}}{\omega}\right)^2,\frac{\omega_{j,\grad{B}}\omega_{j,\kappa}}{\omega^2},\left(\frac{\omega_{j,\kappa}}{\omega}\right)^2} \nonumber \\
        J_{j,\parallel}^{(2)} \approx &\int\limits_{0}^{\infty} \dd{\zeta} e^{-\zeta} \left(1+\frac{\omega_{j,\grad{B}}}{\omega}\zeta\right)\left(1+3\frac{\omega_{j,\kappa}}{\omega}\zeta\right) \left(\Gamma_0(b_j)+b_j\dv{\Gamma_{0}(b_j)}{b_j} \frac{\omega_{j,\grad{B}}}{\omega}\zeta\right) + \order{\left(\frac{\omega_{j,\grad{B}}}{\omega}\right)^2,\left(\frac{\omega_{j,\kappa}}{\omega}\right)^2} \nonumber \\
        \approx & \Gamma_0(b_j) \left(1+\frac{\omega_{j,\grad{B}}+3\omega_{j,\kappa}}{\omega}\right) + b_j\left[\Gamma_{1}(b_j)-\Gamma_0(b_j)\right] \frac{\omega_{j,\grad{B}}}{\omega} + \order{\left(\frac{\omega_{j,\grad{B}}}{\omega}\right)^2,\frac{\omega_{j,\grad{B}}\omega_{j,\kappa}}{\omega^2},\left(\frac{\omega_{j,\kappa}}{\omega}\right)^2}
        \label{eq:ion-res-int-fluid-expand}
    \end{align}
\end{widetext}
where all $\zeta$ integrals are evaluated using the Euler gamma function $\Gamma(n+1)=n!$, and the recurrence relation between the modified Bessel functions of first-kind $I_n$\cite{Abramowitz1968HandbookTables} were used to evaluate the derivatives of $\Gamma_0(b_j)$. In \Cref{eq:ion-res-int-fluid-expand} only first-order effects in $\omega_{j,\kappa}/\omega$ and $\omega_{j,\grad{B}}/\omega$ were retained, since the quadratic and higher-order terms caused by the products of the expansions are spurious, resulting in erroneous coefficients as they lack the proper contributions from the second- and higher-order Taylor expansions of the integrand terms. \par
Substituting \Cref{eq:ion-res-int-fluid-expand} into \Cref{eq:ion-el-glob-kernels} then gives the strongly-driven limit for the ion-density kernel after some rearrangement as 
\begin{widetext}
    \begin{align}
        \label{eq:fluid-ion-kernel}
        h_{\mathrm{ion},j}^{\mathrm{str. drive}} \approx& \Gamma_0(b_j) \left[1-\frac{\omega_{\grad{n_j}}}{\omega} +\frac{\omega_{j,\kappa}+\omega_{j,\grad{B}}}{\omega} -\frac{\left(\omega_{\grad{n_j}}+\omega_{\grad{T_j}}\right)\left(\omega_{j,\grad{B}}+\omega_{j,\kappa}\right)}{\omega^2}+b_j\frac{\omega_{\grad{T_j}}\omega_{j,\grad{B}}}{\omega^2}\right] \\ \nonumber 
        &+ b_j \left(\Gamma_1(b_j)-\Gamma_{0}(b_j)\right) \left[\frac{\omega_{j,\grad{B}}}{\omega}-\frac{\omega_{\grad{T_j}}}{\omega}-\frac{\omega_{j,\grad{B}}\left(\omega_{\grad{n_j}}+\omega_{\grad{T_j}}\right)}{\omega^2}-\frac{\omega_{\grad{T_j}}\left(\omega_{j,\kappa}+\omega_{j,\grad{B}}\right)}{\omega^2}+2b_j \frac{\omega_{\grad{T_j}}\omega_{j,\grad{B}}}{\omega^2}\right]
    \end{align}
\end{widetext}
where quadratic terms in $\omega_{j,\grad{B}}/\omega \sim \omega_{j,\kappa}/\omega \ll 1$ arising from multiplication with frequency prefactors have been neglected, as such higher-order terms would be spurious for the aforementioned reason. \par
With regards to the trapped-electrons, the strongly driven-limit of $\omega_{e,\kappa}/\omega \sim \omega_{e,\grad{B}}/\omega \ll 1$ is equivalent to $\overline{\omega_{de}^T}/\omega \ll 1$, hence one could one could take the proper asymptotic limit for the plasma dispersion function in the trapped-electron resonant integrals \Cref{eq:trel-genplasmaZ-integrals}. However, extra care must be taken when the bounce-averaged drift is marginally close to vanishing rather than finite. In the limit that $\overline{\omega_{de}^T}/\omega\rightarrow0$, whilst $\sqrt{\omega/\overline{\omega_{de}^{T}}}$ will approach complex infinity, simultaneously $\mathcal{W}(\sigma_{\gamma}\sigma_{d}\sqrt{\omega/\overline{\omega_{de}^{T}}})$ vanishes asymptotically. When the bounce-averaged drift becomes vanishingly small, (which shall always be achieved by some populations of trapped-particles for the drift-wells must have roots in order for the passing particle drift to vanish on a transit-average), the trapped-electron energy integral $I_{\textrm{tr-el}}$ \Cref{eq:Itrel-def} has an almost trivially well defined limit. The intricacy of complex infinities in the analytical solution \Cref{eq:trel-genplasmaZ-integrals} is an unphysical aftermath from the coordinate transform $\mu=-i\sigma_{\gamma}\sigma_{d}\sqrt{\omega/\overline{\omega_{de}^{T}}}\sqrt{1+i\sigma_{\gamma}\overline{\omega_{de}^T}\xi}$, considered in \Cref{app:TEM-derivation} to simplify the emergent integrals of \Cref{eq:Jm_TEM_def}, breaking down when the bounce-averaged drift vanishes. \par 
If instead we invoke the same $\zeta=-i\sigma_{\gamma}\omega\xi$ coordinate transform considered for the ions, these integrals are rendered as
\begin{align}
    J_{\textrm{tr-el}}^{m} = \Gamma\left(\frac{2m+1}{2}\right) \int\limits_{0}^{\infty} \dd{\zeta} \frac{e^{-\zeta}}{\left(1-\frac{\overline{\omega_{de}^T}}{\omega}\zeta\right)^{\frac{2m+1}{2}}}
    \label{eq:Jtrel-zetasub}
\end{align}
where we have performed integration along the pizza-slice-shaped contour from \Cref{fig:contour-sketch-vanilla} to reroute the integration path along the real-line, valid for similar arguments as in the ion case. As \Cref{eq:Jtrel-zetasub} is continuous in $\overline{\omega_{de}^{T}}$, the strongly-driven limit may be used to smoothly connect the solutions to \Cref{eq:Jm_TEM_def} between asymptotically small and completely vanishing drifts, where upon applying the binominal expansion to the denominator we have
\begin{widetext}
    \begin{align}
        \frac{J_{\textrm{tr-el}}^{m}}{\Gamma\left(\frac{2m+1}{2}\right)} =I^{m} &\approx \int\limits_{0}^{\infty} e^{-\zeta} \left(1+\frac{2m+1}{2}\frac{\overline{\omega_{de}^T}}{\omega}\zeta+\frac{\left(2m+1\right)\left(2m-1\right)}{8} \zeta^2\right)  + \order{\left(\frac{\overline{\omega_{de}^T}}{\omega}\right)^3} \nonumber \\
        & \approx 1 + \frac{2m+1}{2} \frac{\overline{\omega_{de}^T}}{\omega}+\frac{(2m+1)(2m+3)}{4} \left(\frac{\overline{\omega_{de}^T}}{\omega}\right)^2 + \order{\left(\frac{\overline{\omega_{de}^T}}{\omega}\right)^3}
    \label{eq:Im-TEM-fluidexpand}
    \end{align}
\end{widetext}
where we have defined $I^m$ analogous to \Cref{eq:Im-TEM-def}, and maintained the expansion up to second order, as it insightfully reveals that the recursive relation \Cref{eq:TEM-recursive-rel} remains maintained up to first order in $\overline{\omega_{de}^{T}}/\omega$ for all $m\neq1/2$ if one delays truncation. In fact, it is straightforward to show that \Cref{eq:TEM-recursive-rel} remains valid up to order $\order{\left(\overline{\omega_{de}^{T}}/\omega\right)^N}$ if one truncates the binomial expansion at $(N+1)^{\mathrm{th}}$ order. Note that \Cref{eq:Im-TEM-fluidexpand} is always well-defined, including when $\overline{\omega_{de}^T}\rightarrow 0$. When the bounce-averaged drift is finite, but still sufficiently small for the strongly-driven limit $\overline{\omega_{de}^T}/\omega\ll1$ to be applicable, an asymptotic expansion of the Faddeeva function \Cref{eq:Faddeeva} for $\zeta \gg 1$ in the analytical solutions \Cref{eq:trel-genplasmaZ-integrals} yields\cite{Abramowitz1968HandbookTables} 
%Eqn 7.1.8.
\begin{align}
    I^{1} \sim& \sum_{m=0}^{\infty} \frac{(2(m+1))!}{(m+1)! 2^{2m+1}} \left(\frac{\overline{\omega_{de}^T}}{\omega}\right)^{m} \nonumber \\
    I^{2} \sim& \frac{1}{3} \sum_{m=0}^{\infty} \frac{(2(m+2))!}{(m+2)! 2^{2m+2}} \left(\frac{\overline{\omega_{de}^T}}{\omega}\right)^{m}
    \label{eq:TEM-asymp-int}
\end{align}
where we accounted for the notational difference that the $J_{\textrm{tr-el}}^{m}$ as appear in the main text are actually the integrals $I^{m}$ as mentioned in \Cref{app:TEM-derivation} for consistency, and note that the first three terms of \Cref{eq:TEM-asymp-int} agree with \Cref{eq:Im-TEM-fluidexpand} for both $m=1,2$, showing consistency between the analytical solution and the strongly-driven limit for trapped-particles not subject to a marginally vanishing bounce-averaged drift. In obtaining \Cref{eq:TEM-asymp-int}, the large-argument asymptotic expansion of the $\erfc(z)$ was used, which is valid only when $\abs{\arg{z}}<3\pi/4$. For the resonant TEM integrals $z=-i\sigma_{\gamma}\sigma_{de} \sqrt{\omega/\overline{\omega_{de}^T}}$, whose argument is always constrained within $\abs{\arg{z}}\in(0,\pi)$ and hence always lies in applicable range by virtue of the sign generalisation gained from the chain of integral transforms used to obtain the plasma dispersion function. \par  
Inserting \Cref{eq:Im-TEM-fluidexpand} into \Cref{eq:ion-el-glob-kernels}, and taking into account that the $J_{\textrm{tr-el}}^{m}$ as appear in the main text are actually the integrals $I^{m}$ as mentioned in \Cref{app:TEM-derivation}, we obtain the strongly-driven limit for the trapped-electron pitch-angle density kernel as
\begin{align}
    K_{\textrm{tr-el}}^{\mathrm{str. drive}} \approx \frac{1}{2} \left(1-\frac{\omega_{\grad{n_e}}}{\omega}\right) + \frac{3}{4} \frac{\overline{\omega_{de}^{T}}}{\omega} \left(1-\frac{\omega_{\grad{n_e}}+\omega_{\grad{T_e}}}{\omega}\right)
    \label{eq:fluid-TEM-kernel}
\end{align}
where quadratic and higher-order terms in $\overline{\omega_{de}^T}/\omega$ have been neglected to keep the expansion consistent with \Cref{eq:fluid-ion-kernel} for the ions.

\subsection*{Application to reduced models}
Here we will only focus on the effect of reduced models on the ion-density kernel, since the trapped-electron pitch-angle density kernel \Cref{eq:fluid-TEM-kernel} is only moderately affected by the choice of the drift model through modification of the bounce-averaged drift \Cref{eq:reduced-bav-drift-models}, with the trapped-electrons response already assumed to be drift-kinetic. \par 
As the curvature- and $\grad{B}$ models of the magnetic drift can be mapped onto the ``full drift'' scenario by setting the name-sake drift-component equal to the full drift $\omega_{j,\alpha} \rightarrow \omega_{j,\kappa}+\omega_{j,\grad{B}}$ whilst making the other component vanish $\omega_{j,\neq\alpha} \rightarrow 0$, where $\alpha=\{\kappa,\grad{B}\}$, the appropriate strongly-driven limit for the reduced drift models whilst retaining the full FLR effects are readily obtained by making the suitable substitutions to \Cref{eq:fluid-ion-kernel}. This substitution ensures that the relative qualitative aspect of the toroidal drift-driven instability mechanism with respect to the full resonant integrals remains conserved across the reduced models. Perhaps enlightening, the explicit form of \Cref{eq:fluid-ion-kernel} reveals that much of the physics in the strongly-driven limit does not depend on a particular component, but rather the combined magnetic drift $\omega_{j,\kappa}+\omega_{j,\grad{B}}$, hence leaving those terms unaffected. Rather, as the few symmetry breaking terms between the drift components are all proportional to $\omega_{j,\grad{B}}$, these terms are either completely suppressed or enhanced, by adopting the reduced drift model. \par
For the suitable strongly-driven limit of the reduced FLR models, we reconsider \Cref{eq:ion-Padé-param_integrals} considered for the Padé approximation of the Bessel functions, as for the reduced FLR models the resonant integrands are directly modified and thus cannot be straightforwardly mapped onto the original ``full FLR'' model by making an appropriate substitution to \Cref{eq:fluid-ion-kernel}. For the the integrals of \Cref{eq:ion-Padé-param_integrals} we proceed analogous to the TEM resonant integral by applying the change of coordinates $\zeta = -i\sigma_{\gamma}\omega\xi$, integrating along the pizza-slice-shaped contour from \Cref{fig:contour-sketch-vanilla} to reroute the integration path along the real-line as the emerging singularities $\zeta^{\ast,\textrm{Padé}}_{\kappa}=\nu\omega/(2\omega_{j,\kappa}$ and $\zeta^{\ast,\textrm{Padé}}_{\grad{B}}=\lambda \omega/(\omega_{j,\grad{B}})$ lie outside of the contour given $\lambda,\nu>0$, resulting in
\begin{widetext}
    \begin{align}
        J_{j}^{0,\nu,\lambda} =& \int\limits_{0}^{\infty} \dd{\zeta} \frac{e^{-\zeta}}{\sqrt{\nu-2\frac{\omega_{j,\kappa}}{\omega}\zeta}\left(\lambda-\frac{\omega_{j,\grad{B}}}{\omega}\zeta\right)} \nonumber \\
        \approx& \frac{1}{\lambda \sqrt{\nu}} \left(1+\frac{\nu \omega_{j,\grad{B}}+\lambda\omega_{j,\kappa}}{\lambda\nu\omega}+\frac{2(\nu\omega_{j,\grad{B}})^2+2\lambda\nu\omega_{j,\grad{B}}\omega_{j,\kappa}+3(\lambda\omega_{j,\kappa})^2}{(\lambda\nu\omega)^2}\right) + \order{\delta^3} & \nonumber \\
        J_{j,\perp}^{2,\nu,\lambda} =& 2\int\limits_{0}^{\infty} \dd{\zeta} \frac{e^{-\zeta}}{\sqrt{\nu-2\frac{\omega_{j,\kappa}}{\omega}\zeta}\left(\lambda-\frac{\omega_{j,\grad{B}}}{\omega}\zeta\right)^2} \nonumber \\
         \approx& \frac{2}{\lambda^2 \sqrt{\nu}} \left(1+\frac{2\nu \omega_{j,\grad{B}}+\lambda\omega_{j,\kappa}}{\lambda\nu\omega}+\frac{6(\nu\omega_{j,\grad{B}})^2+4\lambda\nu\omega_{j,\grad{B}}\omega_{j,\kappa}+3(\lambda\omega_{j,\kappa})^2}{(\lambda\nu\omega)^2}\right) + \order{\delta^3} \nonumber \\
        J_{j,\parallel}^{2,\nu,\lambda} =& \int\limits_{0}^{\infty} \dd{\zeta} \frac{e^{-\zeta}}{\left(\nu-2\frac{\omega_{j,\kappa}}{\omega}\zeta\right)^{3/2}\left(\lambda-\frac{\omega_{j,\grad{B}}}{\omega}\zeta\right)} \nonumber \\
        \approx& \frac{1}{\lambda \nu^{3/2}} \left(1+\frac{\nu \omega_{j,\grad{B}}+3\lambda\omega_{j,\kappa}}{\lambda\nu\omega}+\frac{2(\nu\omega_{j,\grad{B}})^2+6\lambda\nu\omega_{j,\grad{B}}\omega_{j,\kappa}+15(\lambda\omega_{j,\kappa})^2}{(\lambda\nu\omega)^2}\right) + \order{\delta^3}
        \label{eq:Padé-fluid-expand}
    \end{align}
\end{widetext}
where for each integral in the subsequent step we performed a binomial expansion of the terms in the denominator, retaining all terms up to second order in $\omega_{j,\grad{B}}/\omega \sim \omega_{j,\kappa}/\omega \sim \delta \ll 1$ and $\order{\delta^3}$ is a shorthand notation for $\order{\left(\frac{\omega_{j,\grad{B}}}{\omega}\right)^3,\left(\frac{\omega_{j,\kappa}}{\omega}\right)^3,\frac{\omega_{j,\grad{B}}\omega_{j,\kappa}^2}{\omega^3},\frac{\omega_{j,\grad{B}}^2\omega_{j,\kappa}}{\omega^3}}$ accounting for all combinations of cubic order terms. In particular by retaining second order effects in $J_{j}^{0,\nu,\lambda}$ but truncating both $J_{j,\perp}^{2,\nu,\lambda},J_{j,\parallel}^{2,\nu,\lambda}$ at first order, the interrelation \Cref{eq:Padé-integrals-interrel} between these integrals holds up to first order in $\omega_{j,\kappa}/\omega\sim\omega_{j,\grad{B}}/\omega\sim\delta$ as well, and like the TEM case, one may show that this interrelation continues being valid up to $\order{\delta^N}$ if the binomial expansions for $J_{j}^{0,\nu,\lambda},J_{j,\perp}^{2,\nu,\lambda}, J_{j,\parallel}^{2,\nu,\lambda}$ are truncated at $(N+1)^{\mathrm{th}},N^{\mathrm{th}},N^{\mathrm{th}}$ order respectively. \par
Upon setting $\nu \rightarrow 1, \lambda \rightarrow 1+b_j$ we note that the expressions for $J_{j}^{0,\nu,\lambda}$ and $J_{j,\parallel}^{2,\nu,\lambda}$ from \Cref{eq:Padé-fluid-expand} match with the equivalent expressions of $J_{j}^{0}$ and $J_{j,\parallel}^{(2)}$ from \Cref{eq:ion-res-int-fluid-expand} if the Bessel functions are Padé expanded (recalling that $\Gamma_{1}(b_j)-\Gamma_0(b_j) = \dv*{\Gamma_0(b_j)}{b_j}\approx -b_j/(1+b_j)$). The expression for $J_{j,\perp}^{2,\nu,\lambda}$, however, slightly deviates from the equivalent Padé expansion of $J_{j,\perp}^{(2)}$ from \Cref{eq:ion-res-int-fluid-expand} by 
\begin{align}
    \Delta^{\textrm{Padé,str. drive}} \equiv J_{j,\perp}^{(2)} - J_{j,\perp}^{2,1,1+b_j} = -\frac{2b_j^2(1-b_j)}{(1+b_j)^3} \frac{\omega_{j,\grad{B}}}{\omega}
    \label{eq:Padé-fluid-J2perp-mismatch}
\end{align} 
where the expression is to be understood to be only valid in the strongly-driven limit and when considering the Padé approximation to be applied to the Bessel function in $J_{j,\perp}^{(2)}$. Using \Cref{eq:Padé-fluid-J2perp-mismatch}, the strongly-driven limit for the ion-density kernel under Padé approximation is then straightforwardly obtained as
\begin{widetext}
    \begin{align}
    h_{\mathrm{ion},j}^{\textrm{Padé, str. drive}} &\approx \frac{1}{1+b_j} \left[1-\frac{\omega_{\grad{n_j}}}{\omega} +\frac{\omega_{j,\kappa}+\omega_{j,\grad{B}}}{\omega} -\frac{\left(\omega_{\grad{n_j}}+\omega_{\grad{T_j}}\right)\left(\omega_{j,\grad{B}}+\omega_{j,\kappa}\right)}{\omega^2}+b_j\frac{\omega_{\grad{T_j}}\omega_{j,\grad{B}}}{\omega^2}\right] \nonumber \\
    &-\frac{b_j}{(1+b_j)^2} \left[\frac{\omega_{j,\grad{B}}}{\omega}-\frac{\omega_{\grad{T_j}}}{\omega}-\frac{\omega_{j,\grad{B}}\left(\omega_{\grad{n_j}}+\omega_{\grad{T_j}}\right)}{\omega^2}-\frac{\omega_{\grad{T_j}}\left(\omega_{j,\kappa}+\omega_{j,\grad{B}}\right)}{\omega^2}+2b_j \frac{\omega_{\grad{T_j}}\omega_{j,\grad{B}}}{\omega^2}\right] + \Delta^{\textrm{Padé,str. drive}}\frac{\omega_{\grad{T_j}}}{2\omega} \nonumber \\
    &=\frac{1}{1+b_j} \left[1-\frac{\omega_{\grad{n_j}}}{\omega} +\frac{\omega_{j,\kappa}+\omega_{j,\grad{B}}}{\omega} -\frac{\omega_{\grad{n_j}}\left(\omega_{j,\grad{B}}+\omega_{j,\kappa}\right)}{\omega^2}-\frac{\omega_{\grad{T_j}}\left(\omega_{j,\kappa}-\omega_{j,\grad{B}}\right)}{\omega^2}\right] \nonumber \\
    &-\frac{b_j}{(1+b_j)^2} \left[\frac{\omega_{j,\grad{B}}}{\omega}-\frac{\omega_{\grad{T_j}}}{\omega}-\frac{\omega_{j,\grad{B}}\left(\omega_{\grad{n_j}}-\omega_{\grad{T_j}}\right)}{\omega^2}-\frac{\omega_{\grad{T_j}}\omega_{j,\kappa}}{\omega^2}\right] - \frac{2}{(1+b_j)^3} \frac{\omega_{j,\grad{B}}\omega_{\grad{T_j}}}{\omega^2}
    \label{eq:Padé-fluid-ion-kernel}
    \end{align}
\end{widetext}
where only terms up to first order in $\delta$ were retained for consistency across the various strongly-driven models. \par
The fact that the discrepancy \Cref{eq:Padé-fluid-J2perp-mismatch} applies only to $J_{j,\perp}^{(2)}$ can be traced back to the fact that within the strongly-driven limit $J_{j}^{0}$ and $J_{j,\parallel}^{(2)}$ are fully characterised by just $\Gamma_0$ and its first derivative, which are closely approximated by the Padé approximation since $\Gamma_0(b_j)\approx 1/(1+b_j)$ has an identical first-order Maclaurin series as $\Gamma_0(b_j)$\cite{Adler1994SeriesExpansions}, whilst information about the second derivative is embedded into $J_{j,\perp}^{(2)}$, whose behaviour is thus not closely followed by the Padé approximation. As this second-derivative term is multiplied by $\omega_{j,\grad{B}}/\omega$, explaining the proportionality from \Cref{eq:Padé-fluid-J2perp-mismatch}, any discrepancy between the Padé approximation of the strongly-driven limit and the strongly-driven limit of the Padé approximation vanishes for the curvature model. This agrees with the above observation, since the underlying equations of the curvature model \Cref{eq:curv-model-start-sub} only involve $\Gamma_0(b_j)$ and its first derivative. \par
Lastly, in the drift-kinetic limit of $b_j \rightarrow 0$ we obtain from both \Cref{eq:fluid-ion-kernel,eq:Padé-fluid-ion-kernel} 
\begin{align}
    h_{\mathrm{ion},j}^{\mathrm{DK, str. drive}} \approx&  1-\frac{\omega_{\grad{n_j}}}{\omega} +\frac{\omega_{j,\kappa}+\omega_{j,\grad{B}}}{\omega} \nonumber \\ 
    &-\frac{\left(\omega_{\grad{n_j}}+\omega_{\grad{T_j}}\right)\left(\omega_{j,\grad{B}}+\omega_{j,\kappa}\right)}{\omega^2}
    \label{eq:DK-fluid-limit}
\end{align}
such that the discrepancy between the Padé approximation and the full-FLR model vanishes. This re-matching between Padé and full-FLR solution can be traced back to the vanishing of the $\dv*[2]{\left(b_j\Gamma_0(b_j)\right)}{b_j}$ term in the expression for $J_{j,\perp}^{(2)}$ in \Cref{eq:ion-res-int-fluid-expand} in this limit, such that only $\Gamma_0(b_j)$ and its first-derivative, evaluated at $b=0$, make an explicit appearance in the drift-kinetic limit of the ion-density kernel. Meanwhile, as the first-order Maclaurin series of $1/(1+b_j)$ matches that of $\Gamma_0(b_j)$, the Padé approximation accurately both predicts $\Gamma_0(b_j)$ and its first derivative at $b_j=0$, causing the coefficients of the surviving terms to match. Additionally we note that any further discrepancy between the various approximations for the drift models \Cref{eq:reduced-drift-models} also vanishes, as \Cref{eq:DK-fluid-limit} only depends on the sum of the drift components $\omega_{j,\grad{B}}+\omega_{j,\kappa}$, which through the mappings $\omega_{j,\alpha} \rightarrow \omega_{j,\kappa}+\omega_{j,\grad{B}}$ and $\omega_{j,\neq\alpha} \rightarrow 0$ for $\alpha=\{\kappa,\grad{B}\}$ remains invariant, conform the findings from Ref.~\onlinecite{Terry1982KineticMode}. \par

\putbib[references,mainNotes]

\end{bibunit}

\UseRawInputEncoding 

\makeatletter
\def\@email#1#2{%
 \endgroup
 \patchcmd{\titleblock@produce}
  {\frontmatter@RRAPformat}
  {\frontmatter@RRAPformat{\produce@RRAP{*#1\href{mailto:#2}{#2}}}\frontmatter@RRAPformat}
  {}{}
}%
\makeatother

\setcounter{section}{0}
\setcounter{equation}{0}
\setcounter{figure}{0}
\setcounter{table}{0}
\setcounter{page}{1}
\setcounter{affil}{0}
\renewcommand{\theequation}{S\arabic{equation}}
\renewcommand{\thefigure}{S\arabic{figure}}
\renewcommand{\thesection}{S\arabic{section}}
\renewcommand{\bibnumfmt}[1]{[S#1]}
\renewcommand{\citenumfont}[1]{S#1}

\crefname{equation}{eqn.}{eqns.}
\Crefname{equation}{Eqn.}{Eqns.}

\let\bmog\bm 
\renewcommand{\bm}[1]{{\mathbf{#1}}}

\draft 

\begin{titlepage}

\title{Supplementary Material for ``Fast electrostatic microinstability evaluation in arbitrary toroidal magnetic geometry using a variational approach''}

\author{M.C.L. Morren}
    \email{m.c.l.morren@tue.nl}

\author{P. Mulholland}
\affiliation{Department of Applied Physics and Science Education, Eindhoven University of Technology, 5600 MB Eindhoven, The Netherlands}

\author{J.H.E. Proll}
\affiliation{Department of Applied Physics and Science Education, Eindhoven University of Technology, 5600 MB Eindhoven, The Netherlands}
\affiliation{Max-Planck-Institut für Plasmaphysik, 17491 Greifswald, Germany}

\author{M.J. Pueschel}
\affiliation{Department of Applied Physics and Science Education, Eindhoven University of Technology, 5600 MB Eindhoven, The Netherlands}
\affiliation{Dutch Institute for Fundamental Energy Research, 5612 AJ Eindhoven, The Netherlands}
\affiliation{Department of Physics \& Astronomy, Ruhr-Universität Bochum, 44780 Bochum, Germany}

\author{L. Podavini}
\affiliation{Max-Planck-Institut für Plasmaphysik, 17491 Greifswald, Germany}

\author{D.D. Kiszkiel}

\author{J.A. Schuurmans}
\affiliation{Department of Applied Physics and Science Education, Eindhoven University of Technology, 5600 MB Eindhoven, The Netherlands}

\author{A. Zocco}
\affiliation{Max-Planck-Institut für Plasmaphysik, 17491 Greifswald, Germany}

\date{5 December 2025}

\maketitle

\end{titlepage}

\begin{bibunit}

\section{Extending the flux-tube geometry into ballooning space}
When expressed in (normalised) Boozer coordinates $(s,\theta,\zeta)$, the metric tensor $\underline{\underline{g_{B}}}$ obeys the two-fold $2\pi$ physical periodicity of the torus, i.e. $g^{nm}_{B}(s,\theta,\zeta+2\pi)=g^{nm}_{B}(s,\theta+2\pi,\zeta)=g^{nm}_{B}(s,\theta,\zeta)$ for all components of the metric tensor. Using \Cref{eq:flux-tube-coo} from the main text, the metric tensor for the flux-tube coordinate system $\underline{\underline{g_{\textrm{FT}}}}$ are obtained in terms of the Boozer system as
\begin{align}
    g^{xx}_{\textrm{FT}} = \grad{x}\vdot\grad{x} &\approx \frac{L_{\textrm{ref}}^2}{4s_0} g^{ss}_{B} \nonumber \\
    g^{xy}_{\textrm{FT}} = \grad{x}\vdot\grad{y} &\approx \frac{L_{\textrm{ref}}^2}{2} \left(g_{B}^{s\theta}-\frac{g_{B}^{s\zeta}}{q_0}+\frac{\theta g^{ss}_{B}}{q_0} \left.\dv{q}{s}\right|_{s=s_0} \right) \nonumber \\
    g^{xz}_{\textrm{FT}} = \grad{x}\vdot\grad{z} &\approx \frac{L_{\textrm{ref}}}{2\sqrt{s_0}} g^{s\theta}_{B} \nonumber \\
    g^{yy}_{\textrm{FT}} = \grad{y}\vdot\grad{y} &\approx L_{\textrm{ref}}^2 s_0 \left(g^{\theta\theta}_{B} + \frac{g_{B}^{\zeta\zeta}-2q_0 g^{s\zeta}_{B}}{q_0^2} + \right. \nonumber \\
    \left. \left(\frac{\theta}{q_0}\left.\dv{q}{s}\right|_{s=s_0}\right)^2 \right. & \left. g^{ss}_{B} +  \frac{2\theta \left(q_0g^{s\theta}_{B}-q^{s\zeta}_{B}\right)}{q_0^2} \left.\dv{q}{s}\right|_{s=s_0}\right) \nonumber \\ 
    g^{yz}_{\textrm{FT}} = \grad{y}\vdot\grad{z} &\approx L_{\textrm{ref}} \sqrt{s_0} \left(g^{\theta\theta}_{B} - \frac{g^{\zeta\theta}}{q_0} + \frac{\theta g^{s\theta}_{B}}{q_0} \left.\dv{q}{s}\right|_{s=s_0}\right) \nonumber \\
    g^{zz}_{\textrm{FT}} = \grad{z} \vdot\grad{z} &= g^{\theta\theta}_{B}
    \label{eq:FT-Boozer-metric-rel}
\end{align}
where the thin flux-tube approximation was invoked to neglect the small variation in $s$ after evaluating the gradients, as the equilibrium quantities vary on radial scales comparable to the minor radius. Note that unlike the Boozer coordinate system, the metric tensor in the flux-tube coordinates no longer satisfies the two-fold $2\pi$ physical symmetry in the torus due to the secular terms in $\theta$ arising in $g^{xy}_{\textrm{FT}}, g^{yy}_{\textrm{FT}},g^{yz}_{\textrm{FT}}$.  As the secularity only arises in the field-aligned coordinate $z=\theta$, it follows that $2\pi$ periodicity is still preserved in the toroidal direction, i.e. $g_{\textrm{FT}}^{mn}(s_0,\theta,\zeta+2\pi) =g_{\textrm{FT}}^{mn}(s_0,\theta,\zeta+2\pi)$, though a $\theta \rightarrow \theta +2\pi N_{\textrm{pol}}$ revolution now results into an interrelation between the metric components
\begin{align}
    g^{xy}_{\textrm{FT}}(s_0,\theta+2\pi N_{\textrm{pol}},\zeta) =& g^{xy}_{\textrm{FT}}(s_0,\theta,\zeta) \nonumber \\ 
    &+ \frac{2s_0}{q_0} \left. \dv{q}{s} \right|_{s=s_0} 2\pi N_{\textrm{pol}} g^{xx}_{\textrm{FT}}(s_0,\theta,\zeta) \nonumber \\
    g^{yy}_{\textrm{FT}}(s_0,\theta+2\pi N_{\textrm{pol}},\zeta) =& g^{yy}_{\textrm{FT}}(s_0,\theta,\zeta) \nonumber \\ 
    & +\frac{4s_0}{q_0} \left. \dv{q}{s} \right|_{s=s_0} 2\pi N_{\textrm{pol}} g^{xy}_{\textrm{FT}}(s_0,\theta,\zeta) \nonumber \\
    &+ \left(\frac{2s_0}{q_0} \left.\dv{q}{s}\right|_{s=s_0}\right)^2 \left(2\pi N_{\textrm{pol}}\right)^2 g^{xx}_{\textrm{FT}}(s_0,\theta,\zeta) \nonumber \\
    g^{yz}_{\textrm{FT}}(s_0,\theta+2\pi N_{\textrm{pol}},\zeta) =& g^{yz}_{\textrm{FT}}(s_0,\theta,\zeta) \nonumber \\ 
    &+ \frac{2s_0}{q_0} \left.\dv{q}{s}\right|_{s=s_0} 2\pi N_{\textrm{pol}} g^{xz}_{\textrm{FT}}(s_0,\theta,\zeta) 
    \label{eq:secular-FT-metric-rels}
\end{align}
which causes the secular terms to be experience a non-periodic upshift due to finite magnetic shear, whilst other metric components remain periodic in $\theta$. \par 
Since the remaining flux tube geometric quantities outlined in \Cref{sec:fluxtubes} are all derived from $\underline{\underline{g_{\textrm{FT}}}}$, they are also affected by the secularity. As shall be clear momentarily, it will suffice to investigate how the secularity of the metric seeps into the metric quantities $\gamma^{1}=g_{\textrm{FT}}^{xx}g^{yy}_{\textrm{FT}} - (g^{xy}_{\textrm{FT}})^2, \gamma^{2} = g^{yz}_{\textrm{FT}} g^{xx}_{\textrm{FT}}-g^{xz}_{\textrm{FT}} g^{xy}_{\textrm{FT}}, \ \gamma^{3} = g^{xz}_{\textrm{FT}}g^{yy}_{\textrm{FT}} - g^{xy}_{\textrm{FT}} g^{yz}_{\textrm{FT}}$. Using \Cref{eq:secular-FT-metric-rels} a little algebra yields
\begin{align}
    \label{eq:metric-gamma-secularity}
    \gamma^{1}(s_0,\theta+2\pi N_{\textrm{pol}},\zeta) =& \gamma^{1}(s_0,\theta,\zeta) \nonumber \\
    \gamma^{2}(s_0,\theta+2\pi N_{\textrm{pol}},\zeta) =& \gamma^{2}(s_0,\theta,\zeta)  \\
    \gamma^{3}(s_0,\theta+2\pi N_{\textrm{pol}},\zeta) =& \gamma^{3}(s_0,\theta,\zeta) \nonumber \\ 
    & - \frac{2s_0}{q_0} \left. \dv{q}{s} \right|_{s=s_0} 2\pi N_{\textrm{pol}} \gamma^2(s_0,\theta,\zeta) \nonumber
\end{align}
where by virtue of the similarity in the interrelations between secular and non-secular components, most of the secular contributions cancel out in these combinations. As within the flux tube the normalised magnetic field strength is given by $B_N(s_0,\theta,\zeta) = \sqrt{\gamma^{1}(s_0,\theta,\zeta)}$ it follows that the field strength retains the two-fold $2\pi$ symmetry of the torus, like it should since it is a physical quantity. Corollary, with regard to magnetic drift-operators [\Cref{eq:gradB-operators-GENE} from the main text], it follows that the derivatives of $B_N$ also satisfy this symmetry, and hence from \Cref{eq:metric-gamma-secularity} it follows that
\begin{align*}
    \mathcal{L}_{x}(s_0,\theta &+ 2\pi N_{\textrm{pol}},\zeta) = \mathcal{L}_{x}(s_0,\theta,\zeta)  \\
    \mathcal{L}_{y}(s_0,\theta &+ 2\pi N_{\textrm{pol}},\zeta) = \mathcal{L}_{y}(s_0,\theta + 2\pi N_{\textrm{pol}},\zeta)  \\ 
    &-\frac{\gamma^2(s_0,\theta,\zeta)}{\gamma^{1}(s_0,\theta,\zeta)} \left.\pdv{B_N(s_0,\theta,\zeta)}{z}\right|_{x,y} \frac{4\pi N_{\textrm{pol}}s_0}{q_0} \left. \dv{q}{s} \right|_{s=s_0}
\end{align*}
We may write these relations more succinctly in interrelated forms mimicking those of \Cref{eq:metric-gamma-secularity,eq:secular-FT-metric-rels}, by invoking the definition of the drift-operators from \Cref{eq:gradB-operators-GENE} in the main text as
\begin{align}
\label{eq:drift-operators-secularity}
    \mathcal{L}_{x}(s_0,\theta + 2\pi N_{\textrm{pol}},\zeta) =& \mathcal{L}_{x}(s_0,\theta,\zeta) \\
    \mathcal{L}_{y}(s_0,\theta + 2\pi N_{\textrm{pol}},\zeta) =& \mathcal{L}_{y}(s_0,\theta,\zeta) + \frac{4\pi N_{\textrm{pol}} s_0}{q_0} \left.\dv{q}{s}\right|_{s=s_0}  \times  \nonumber \\ 
    & \left(\mathcal{L}_{x}(s_0,\theta,\zeta) + \left.\pdv{B_N(s_0,\theta,\zeta)}{y}\right|_{x,z}\right) \nonumber
\end{align}
with identical relations also holding for the curvature drift-operators $\mathcal{K}_x = \mathcal{L}_{x}$ and $\mathcal{K}_{y} = \mathcal{L}_{y} + \dv*{\beta_{\textrm{ref}}}{x} / (2B_N)$ as the pressure correction term in the latter is strictly non-secular. Lastly, the determinant of the (contravariant) metric tensor may be directly given by $\det[\underline{\underline{g_{\textrm{FT}}}}] = g^{zz}_{\textrm{FT}}\gamma^1-g^{yz}_{\textrm{FT}}\gamma^{2}-g^{xz}_{\textrm{FT}}\gamma^{3}$, such that by combining \Cref{eq:secular-FT-metric-rels,eq:metric-gamma-secularity} it follows that $\sqrt{g}(s_0,\theta+2\pi N_{\textrm{pol}},\zeta)=\sqrt{g}(s_0,\theta,\zeta)$, where $\sqrt{g}=1/\det[\underline{\underline{g_{\textrm{FT}}}}]$. \par 
The secular mappings from \Cref{eq:secular-FT-metric-rels,eq:drift-operators-secularity} apply to a $2\pi N_{\textrm{pol}}$ poloidal revolution on the flux surface at a fixed toroidal angle $\zeta$, thus corresponding to the same physical location on the torus. The ends of a flux tube are, however, toroidally displaced by $\Delta\zeta=q_0\Delta\theta$ to maintain a fixed field-line label $\alpha$. Applying \Cref{eq:secular-FT-metric-rels,eq:drift-operators-secularity} in general toroidal configuration will result in discrepancies as the metric tensor $g_{B}^{nm}(s,\theta,\zeta)\neq g_{B}^{nm}(s,\theta+2\pi N_{\textrm{pol}},\zeta + 2 q_0 \pi N_{\textrm{pol}})$, lest the system be axisymmetric or the flux tube is located on a rational surface where $q_0 N_{\textrm{pol}} \in \mathbb{Z}$. To alleviate this issue we proceed to invoke a \textit{pseudo-axisymmetric} approximation, whereby assuming that $q_0\Delta\theta/L_\zeta\ll1$, where $L_\zeta$ is the smallest characteristic toroidal gradient scale length for the metric tensor $1/L_\zeta = \max\pdv*{\ln g^{mn}}{\zeta}$, the variation of the geometry along the toroidal direction can be considered negligible with respect to the variation along the poloidal direction. This \textit{pseudo-axisymmetric} approximation implies that $\left.\pdv*{B_N}{y}\right|_{x,z}\approx 0$, as under this assumption we have $B_N(s_0,\theta,\zeta)\rightarrow\sqrt{\gamma^1(s_0,\theta)}$, and from \Cref{eq:flux-tube-coo} of the main text it follows that a variation in $y$ (at fixed $\theta$) can only be due to a variation in $\zeta$. Hence, by additionally neglecting $\left.\pdv*{B_N}{y}\right|_{x,z}$  \Cref{eq:drift-operators-secularity,eq:FT-Boozer-metric-rel,eq:secular-FT-metric-rels} agree with \Cref{eq:geometry-extend-formulae} from the main text, where we additionally used \Cref{eq:flux-tube-coo} from the main text to write $2(s_0/q_0) \left.\dv*{q}{s}\right|_{s=s_0} = \hat{s}$, and by induction the formulae where generalised to making $2\pi pN_{\textrm{pol}}$ poloidal revolutions where $p\in\mathbb{Z}$ \par
\begin{figure*}
    \centering
    \begin{subfigure}{.85\linewidth}
        \includegraphics[width=\linewidth]{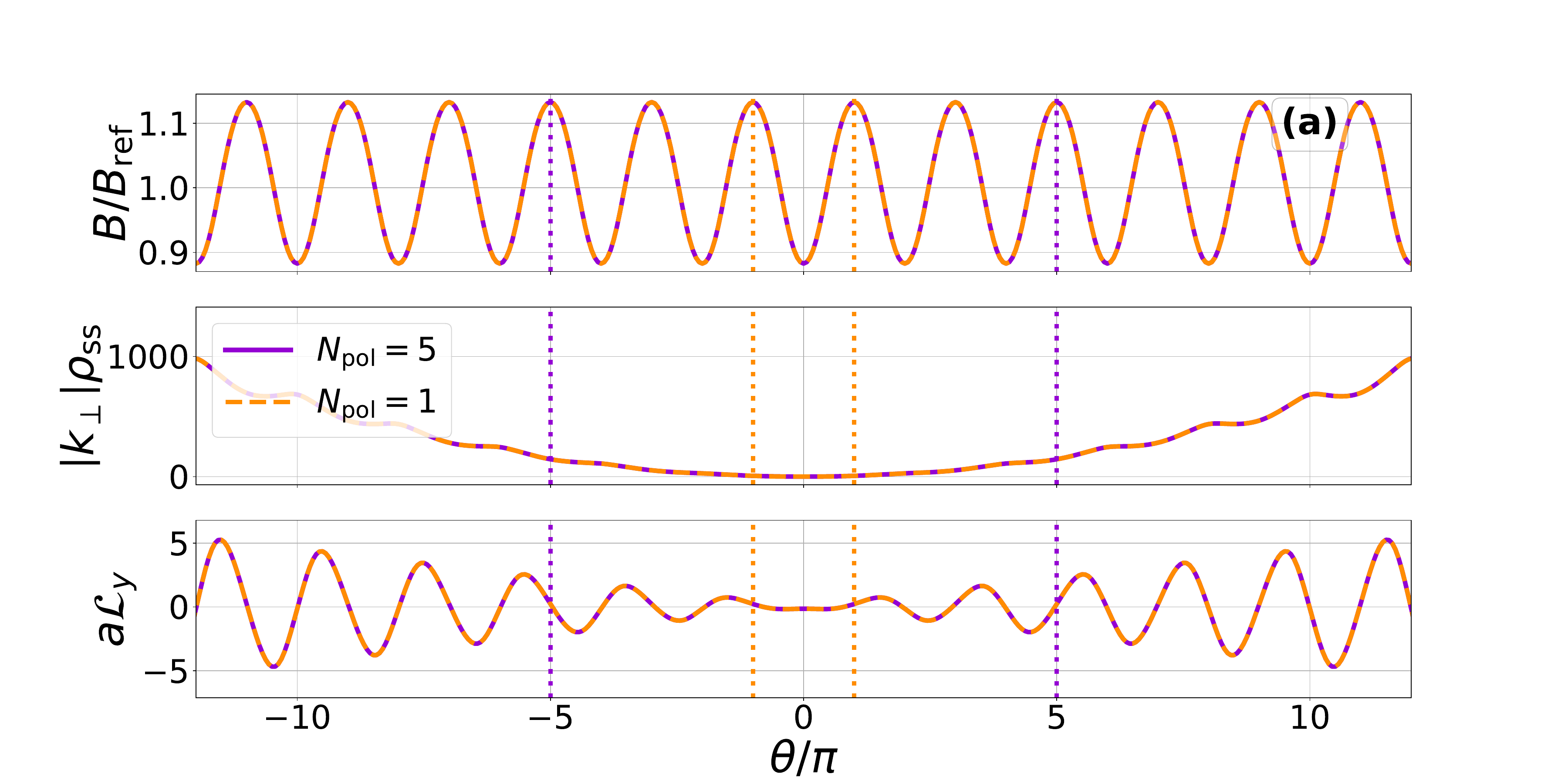}
    \end{subfigure}
    \begin{subfigure}{.85\linewidth}
        \includegraphics[width=\linewidth]{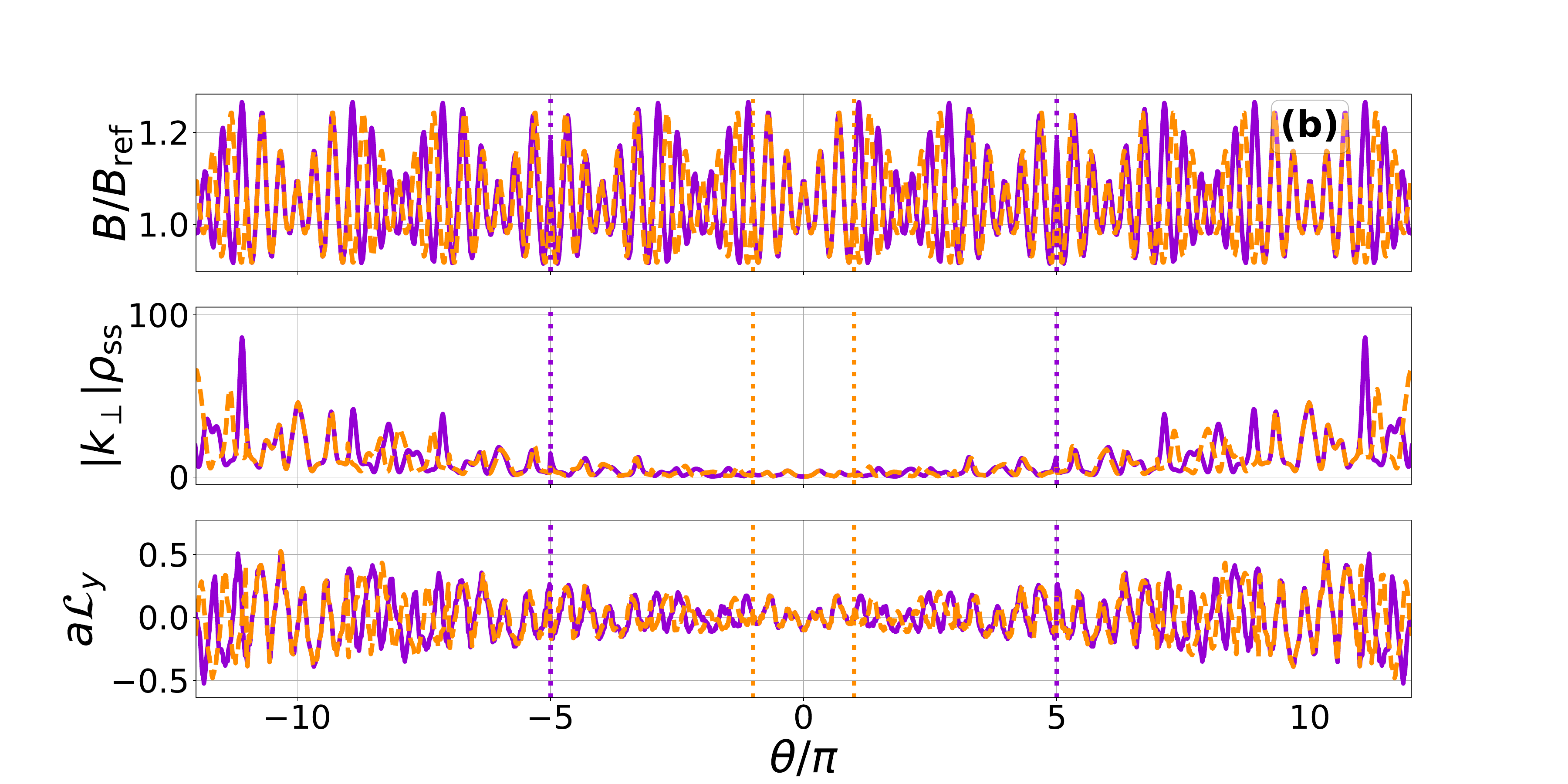}
    \end{subfigure}
    \caption{Extended flux tube geometries into ballooning-space reconstructed from $N_{\textrm{pol}}=1$ (dashed orange) and $N_{\textrm{pol}}=5$ (solid purple) geometric data using \Cref{eq:geometry-extend-formulae} from the main text under the \textit{pseudo-axisymmetric} approximation for (a) a tokamak equilibrium and (b) the high-mirror configuration of W7-X. For both configurations, an $\alpha=0$ flux tube is considered at the $s_0 = 0.5$ flux surface, where the global magnetic shear is given by $\hat{s}\approx0.8$ and $\hat{s}\approx-0.11$ for (a) and (b), respectively. Vertical dotted lines indicate the extent of the original ``raw'' geometric data. Here we focus on most important geometric features for the ITG-TEM dispersion relation, being the arrangement of trapping regions through $B$, the magnetic drift wells through the bi-normal component of the $\grad{B}$ drift-operator $\mathcal{L}_y$, and the FLR damping through $\norm{\bm{k_\perp}}$, where we taken a wavenumber of $k_x \rho_{\textrm{ref}}= 0, \ k_y \rho_{s} = 1$ for visualisation purposes.}
    \label{fig:FT-extension-rule-comparison}
\end{figure*}

The impact of the \textit{pseudo-axisymmetric} approximation on the reconstructed geometry beyond the original $2\pi N_{\textrm{pol}}$ extend of the flux tube is visualised in \Cref{fig:FT-extension-rule-comparison}, where we compare the extended geometry in ballooning space $\theta \in [-12\pi,12\pi]$ for an $N_{\textrm{pol}}=1$ and $N_{\textrm{pol}}=5$ flux tube generated from an axisymmetric and stellarator equilibrium. For the tokamak case, the extended geometry from the $N_{\textrm{pol}}=1$ flux tube generated by \Cref{eq:geometry-extend-formulae} from the main text perfectly agrees with the $N_{\textrm{pol}}=5$ flux-tube geometry (and its extension), which is to be expected due to the exact axisymmetry of the equilibrium. In the stellarator case, however, deviations between the $N_{\textrm{pol}}=1$ and $N_{\textrm{pol}} = 5$ geometry occur beyond the center $2\pi$ range, which is due to the violation of the \textit{pseudo-axisymmetric} approximation. The extended geometry generated by \Cref{eq:geometry-extend-formulae} from the main text then again overlaps at the $2\pi$ interval at $\theta = \pm 10\pi$, where the extensions from $N_{\textrm{pol}}=1$ and $N_{\textrm{pol}}=5$ synchronise (at $p=\pm 5$ and $p=\pm 1$, respectively) since the global shear $\hat{s}$ determining the secular shifts is identical for both flux tubes. Furthermore, the role of the secularity is clearly observable from $\norm{\bm{k_\perp}}$, the variation of which under the assumption of $k_x =0$ is determined by $k_y \sqrt{g^{yy}(s_0,\theta,\zeta)}$, and the drift-operator $\mathcal{L}_{y}$, whose extreme values increase with every $2\pi N_{\textrm{pol}}$ iteration due to global shear, with the effect being noticeably stronger in the axisymmetric case due to the higher shear. The violation of the \textit{pseudo-axisymmetric} approximation in stellarator geometry results in discontinuities in e.g. the drift operators and $\norm{\bm{k_\perp}}$ at every successive $2\pi N_{\textrm{pol}}$, with the discontinuity proportional to the global shear\cite{Martin2018}. 

\section{Result for additional gradient-drive scenarios} \label{sec:addl-cases-results}
Here we present additional results obtained with the global dispersion model for both adiabatic ($a/L_{T\mathrm{i}}$ scan without impurities and $a/L_{n\mathrm{C}}$ scan with $\mathrm{C}^{6+}$ impurity at concentration of $Z_{\textrm{eff}}=1.4$) and kinetic-electron scenarios ($a/L_{n}$ scan for pure density-gradient driven TEM and simultaneous $a/L_{T\mathrm{i}},a/L_{T\mathrm{e}}$ variation with fixed density gradient) in the flux-tube geometries DIII-D tokamak, quasi-symmetric configuration of the HSX stellarator and high-mirror configuration of the W7-X stellarator not discussed in the main text. The relevant information about the range of considered gradients and numerical resolutions used for the \textsc{Gene} simulations are discussed in \Cref{sec:num_settings} and \Cref{tab:GENE_sim_resolutions} of the main text, respectively.

\subsection{Adiabatic-electron scenarios}
\begin{figure}
    \centering
    \begin{subfigure}{\linewidth}
        \centering
        \includegraphics[width=\linewidth]{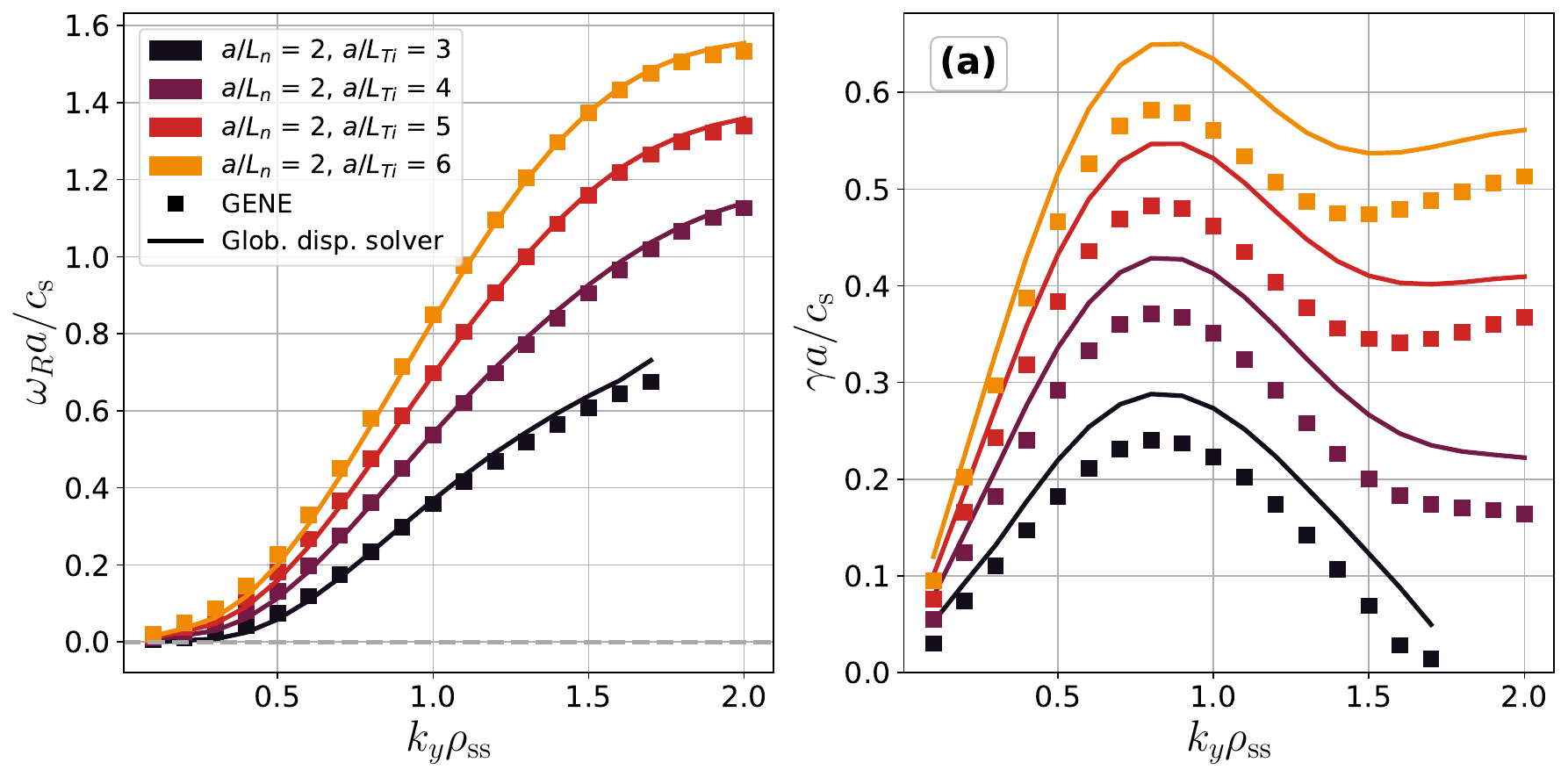}
    \end{subfigure}
    \begin{subfigure}{\linewidth}
        \centering
        \includegraphics[width=\linewidth]{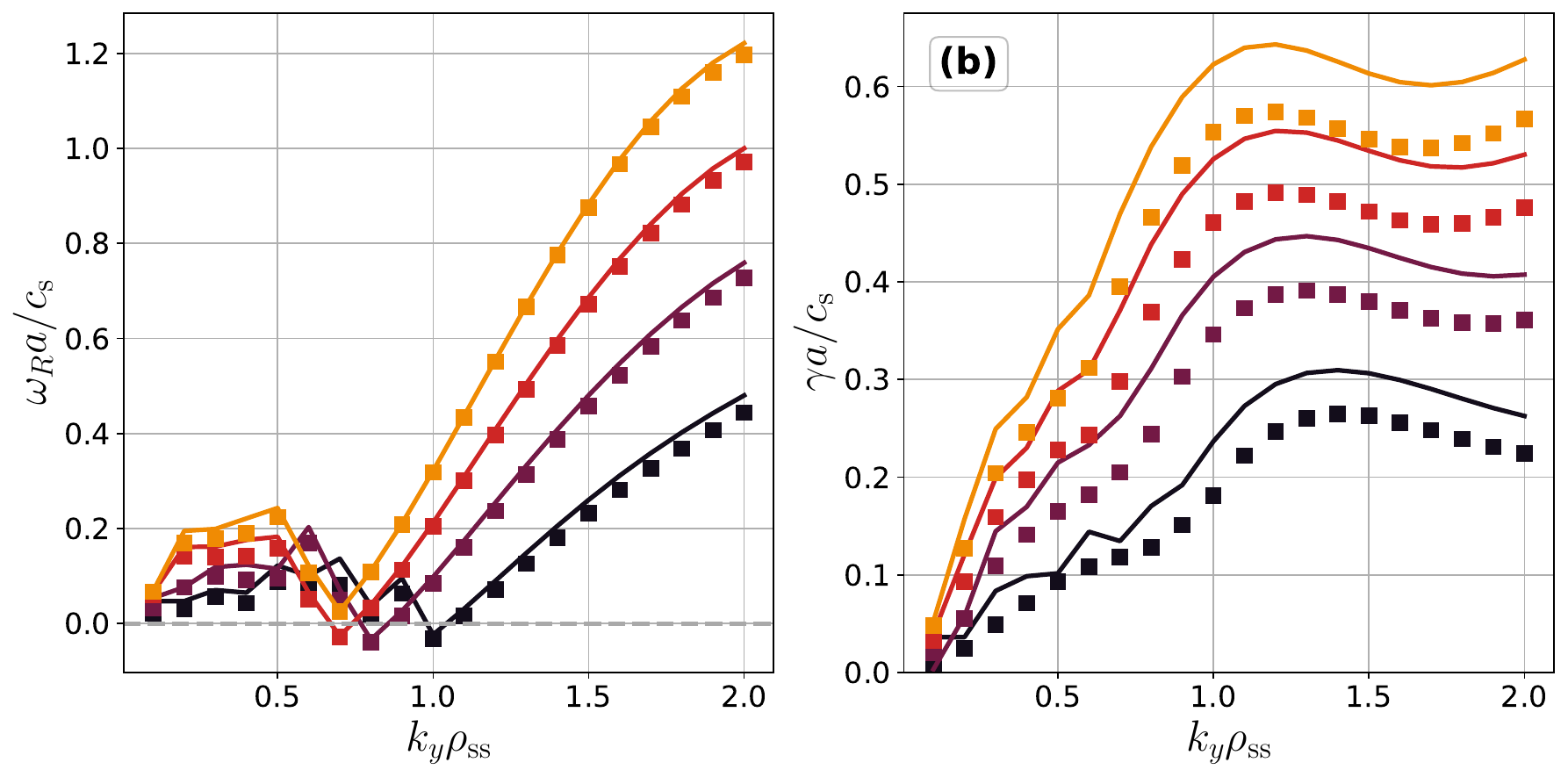}
    \end{subfigure}
    \begin{subfigure}{\linewidth}
        \centering
        \includegraphics[width=\linewidth]{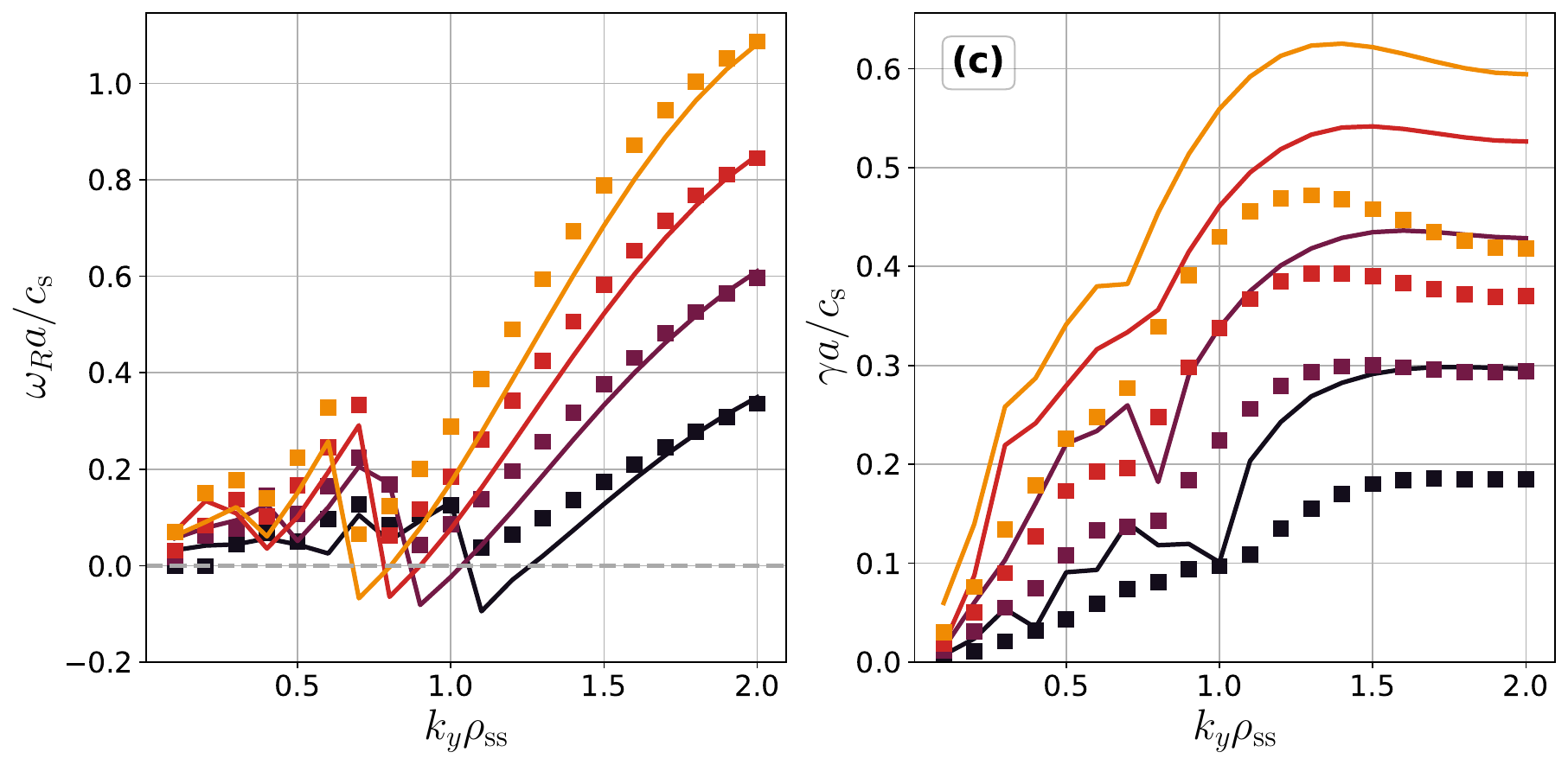}
    \end{subfigure}
    \caption{Eigenfrequency solutions obtained by the global dispersion model (solid lines) contrasted with \textsc{Gene} simulations (symbols) for adiabatic-electron ITG while varying the ion-temperature gradient $a/L_{T\mathrm{i}}$ (lighter colours indicate a stronger temperature gradient). Shown are results for (a) the DIII-D tokamak, (b) the HSX stellarator and (c) the high-mirror configuration of the W7-X stellarator. In all cases the density gradient is fixed at $a/L_{n}=2$ and no $C^{6+}$ impurities are considered.}
    \label{fig:ae-base-itg}
\end{figure}

\Cref{fig:ae-base-itg} shows results in the case of driving-gradient scan of the ion-temperature gradient in absence of impurities. As expected from the ITG instability, growth rates monotonically increase and propagation frequency monotonically shifts in the direction of the ion diamagnetic direction across all length scales with increasing ion temperature gradient. This is most apparent in DIII-D, where for the lowest temperature gradient of $a/L_{T\mathrm{i}}=3$ modes no instabilities are observed in \textsc{Gene} simulations beyond a wavenumber of $k_y \rho_\mathrm{ss}\geq 1.6$, whereas an increase in temperature gradient of $a/L_{T\mathrm{i}}=4$ subsequently destabilises these high wavenumber modes, as also reproduced by the global dispersion model. In both stellarator configurations, unstable modes occur at all length scales, even at the smallest gradient of $a/L_{T\mathrm{i}}=3$, with several mode transitions occurring at lower wavenumber range, until the eigenmodes become strongly ballooned around $\theta=0$ for intermediate-high wavenumbers, with the transition point shifting to smaller $k_y$ as temperature gradient is increased. Regardless of the eigenmode, propagation frequencies shift further into the ion diamagnetic direction and growth rates increase as $a/L_{T\mathrm{i}}$ is increased, like in DIII-D, which is reproduced by the dispersion model. The dispersion model performs quantitatively better in HSX, with a smaller growth rate overprediction and tighter match of propagation frequencies. At the transition wavenumber, the global dispersion model consistently estimates a propagation frequency in the electron diamagnetic direction, while \text{GENE} simulations only result in modes traversing in the ion diamagnetic direction. In HSX a few modes are also found to propagate in the electron diamagnetic direction at the transition wavenumber, though there this matches the behaviour observed from \textsc{Gene} simulations. The model, however, does correctly reproduce the stronger dispersion of growth rates at high wavenumber in HSX, compared to virtually constant growth rates (with exception of $a/L_{T\mathrm{i}}=6$) in W7-X.

\begin{figure}
    \centering
    \begin{subfigure}{\linewidth}
        \centering
        \includegraphics[width=\linewidth]{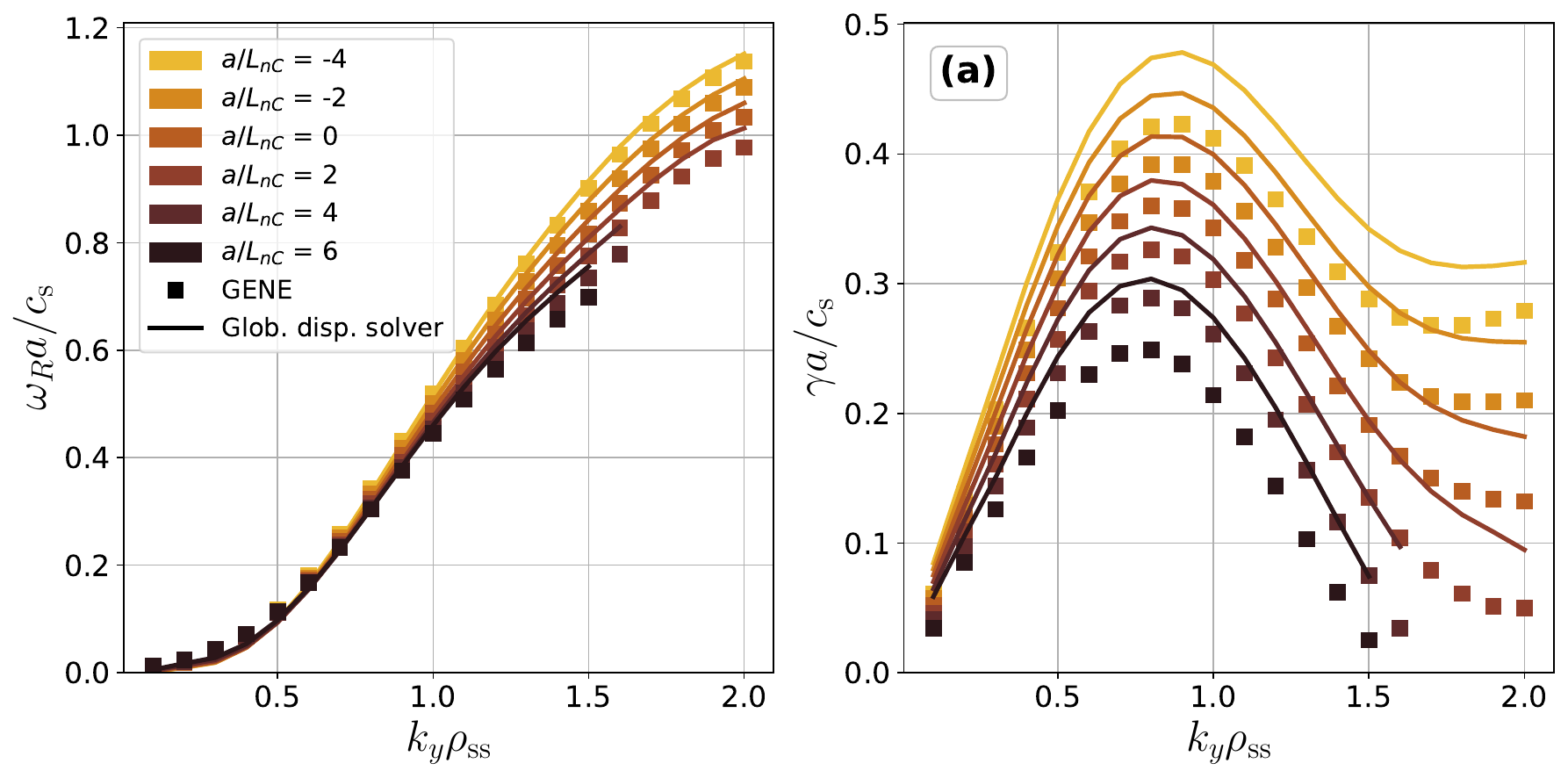}
    \end{subfigure}
    \begin{subfigure}{\linewidth}
        \centering
        \includegraphics[width=\linewidth]{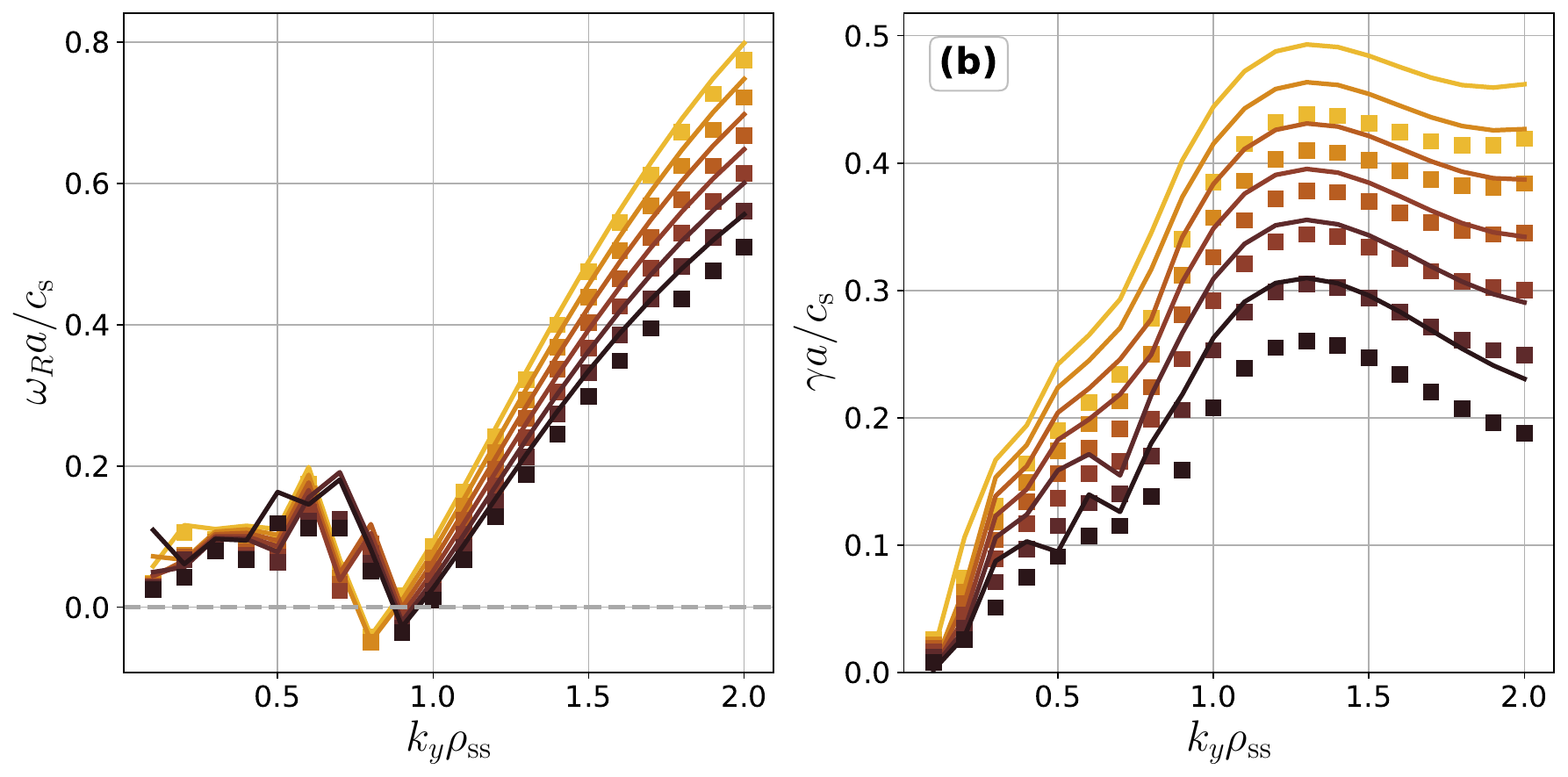}
    \end{subfigure}
    \begin{subfigure}{\linewidth}
        \centering
        \includegraphics[width=\linewidth]{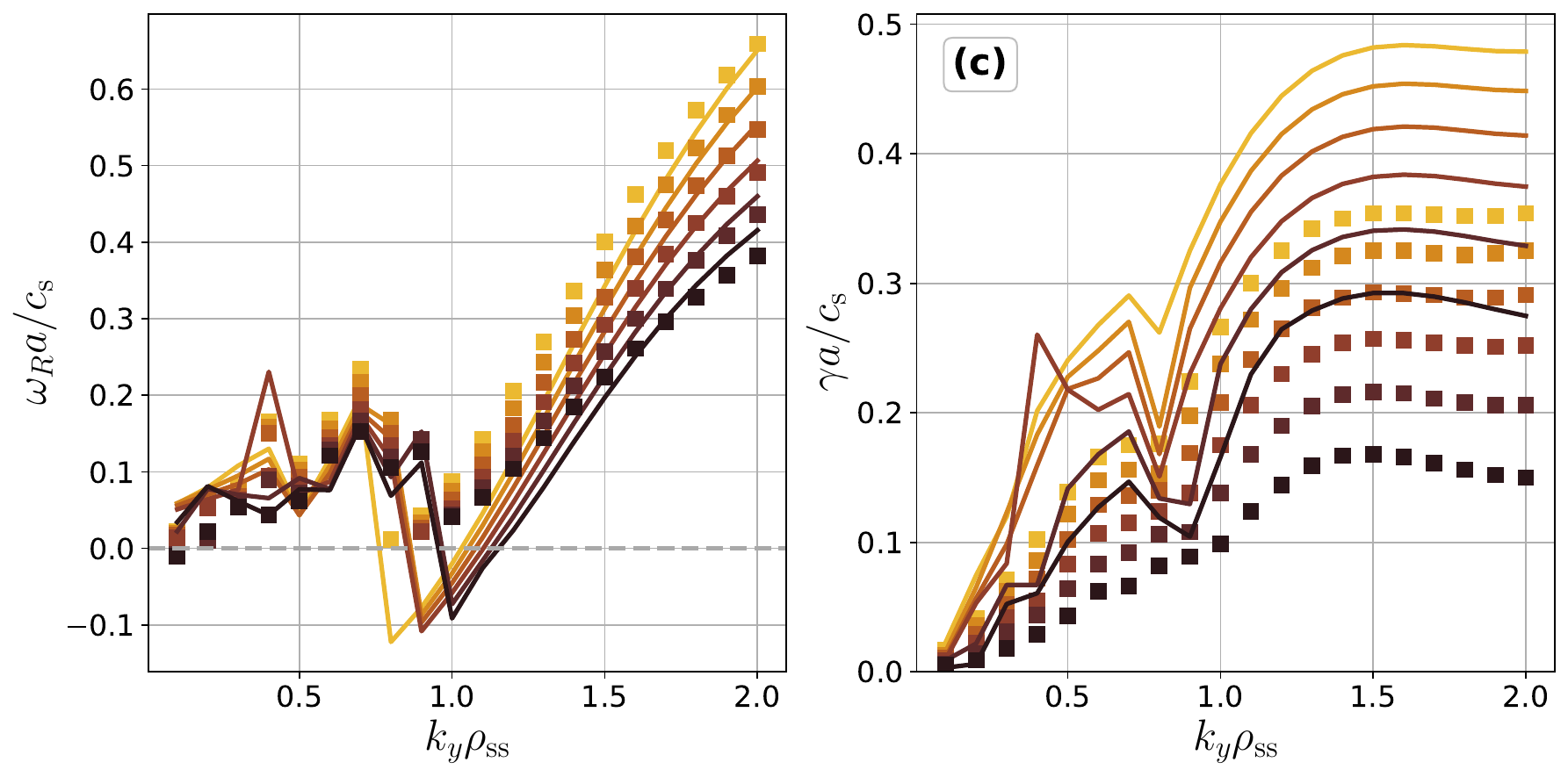}
    \end{subfigure}
    \caption{Eigenfrequency solutions obtained by the global dispersion model (solid lines) contrasted with \textsc{Gene} simulations (symbols) for adiabatic-electron ITG while varying the impurity density gradient $a/L_{n\mathrm{C}}$ (darker colours indicate a stronger density gradient). Shown are results for (a) the DIII-D tokamak, (b) the HSX stellarator and (c) the high-mirror configuration of the W7-X stellarator. In all cases the electron density gradient is fixed at $a/L_{ne}=2$ while the ion density gradient $a/L_{ni}$ is modified to respect ambipolarity, the temperature gradient is fixed at $a/L_{Ts}=4$ for both deuterium ions and carbon impurity, while the $C^{6+}$ impurity content is fixed to maintain a value of $Z_{\textrm{eff}}=1.4$.}
    \label{fig:imp-drive-dillution-itg}
\end{figure}

When impurities are considered, a second dilution effect beyond the reduction of the (main) ion concentration, as considered in the main text, may occur due to the shape of the impurity density profile\cite{Angioni2021ImpurityExperiments}. To maintain ambipolarity, the profile of the (main) ion density has to adjust itself such that the net positive charge gradient equals the electron density gradient (see \Cref{eq:impurity-ambipolarity} in the main text). Unlike the density-dilution effect, this gradient dilution may result in either reduction/enhancement of ITG drive by the main ion species if the impurity profile is more/less steep than the electron density gradient. The results of a scan in $a/L_{n\mathrm{C}}$ probing the two-sided effect of this drive dilution are shown in \Cref{fig:imp-drive-dillution-itg}, where we consider both peaked impurity density profiles -- over a range where the impurity density gradient both exceeds and is eclipsed by $a/L_{ne}$ -- and hollow profiles (indicated by $a/L_{n\mathrm{C}}<0$), at a fixed impurity concentration. Unlike in \Cref{fig:ae-base-itg} where the ion-temperature gradient is modified, the shift of propagation frequency towards ion diamagnetic direction as $a/L_{n\mathrm{C}}$ is diminished (corresponding to an increase in $a/L_{ni}$) is less pronounced than the destabilisation of the growth rates. This may be attributed to the fact that the changes in the ion density gradient are comparatively moderate (ranging from $a/L_{ni} \in [1.652,2.522] $), such that the overall diamagnetic frequency is not significantly impacted. In general both the growth rate and propagation frequency exhibit similar trends in model performance when $a/L_{n\mathrm{C}}$ is varied as when $a/L_{T\mathrm{i}}$ was varied in absence of impurities, thus showing that the global dispersion model correctly captures the physics associated with the density dillution effect on adiabatic-electron ITG modes.

\subsection{Kinetic-electron scenarios}
\begin{figure}
    \centering
    \begin{subfigure}{\linewidth}
        \centering
        \includegraphics[width=\linewidth]{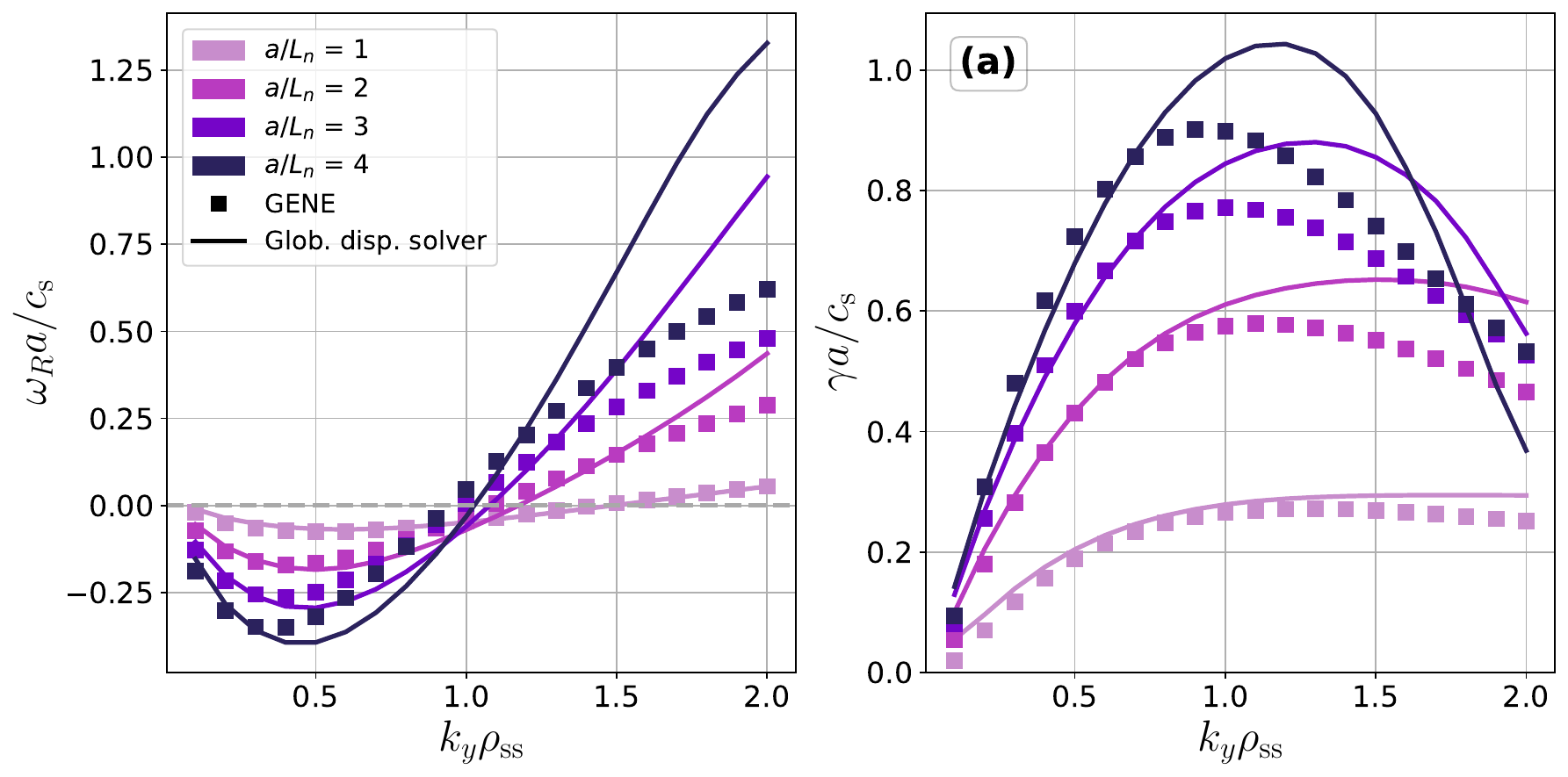}
    \end{subfigure}
    \begin{subfigure}{\linewidth}
        \centering
        \includegraphics[width=\linewidth]{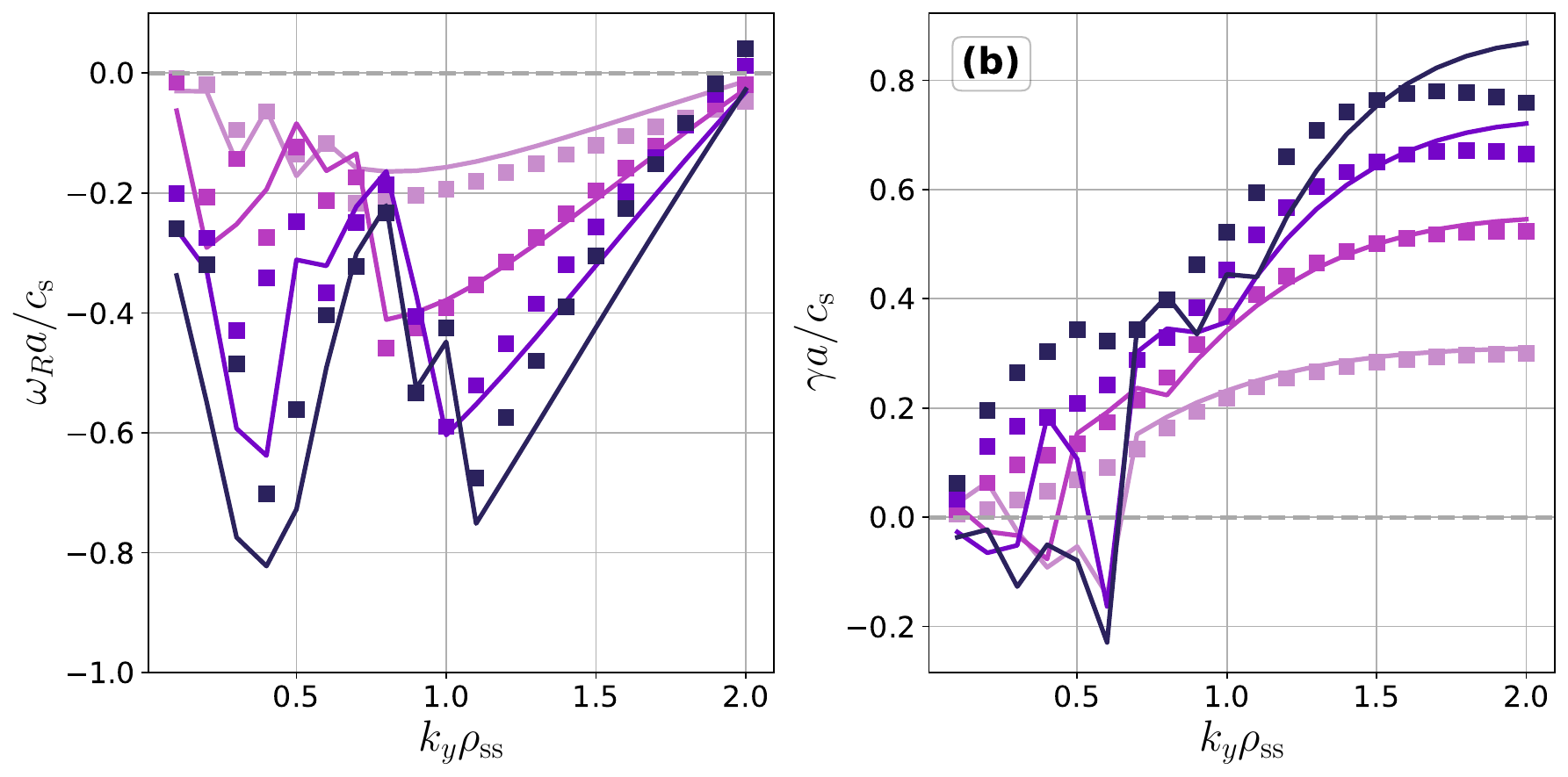}
    \end{subfigure}
    \begin{subfigure}{\linewidth}
        \centering
        \includegraphics[width=\linewidth]{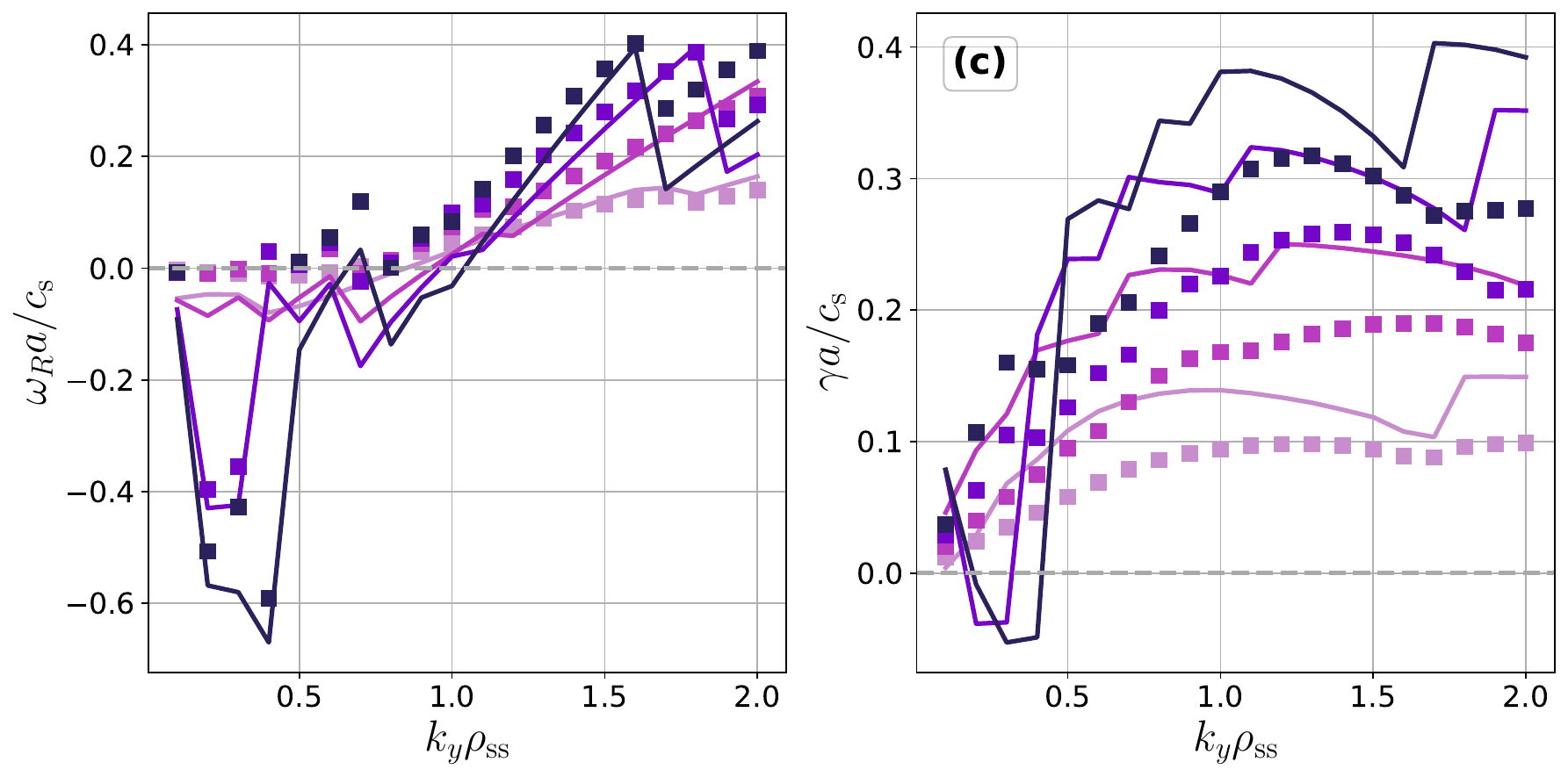}
    \end{subfigure}
    \caption{Eigenfrequency solutions obtained by the global dispersion model (solid lines) contrasted with \textsc{Gene} simulations (symbols) including kinetic-electrons while varying the density gradient $a/L_{n}$ (darker colours indicate a stronger density gradient). Shown are results for (a) the DIII-D tokamak, (b) the HSX stellarator and (c) the high-mirror configuration of the W7-X stellarator. In all cases both temperature gradients are suppressed ($a/L_{T\mathrm{i}}=a/L_{T\mathrm{e}}=0$).}
    \label{fig:ke-gradn-TEM}
\end{figure}

In \Cref{fig:ke-gradn-TEM} we present results for the density-gradient drive scan including kinetic-electron effects, where neither ITG nor $\grad{T_e}$-TEM can occur, and thus predominantly density-gradient-driven TEMs are expected. Indeed, in DIII-D only TEMs are found below $k_y \rho_{s}\approx 1.0$, while for larger wavenumbers the ubiquitous mode (UM) occurs, with the transition point where the frequency crosses into the ion diamagnetic direction slightly shifting towards lower wavenumber as the density gradient is increased. Growth rates increase monotonically with wavenumber, though the largest increase is obtained between $a/L_n=1$ and $a/L_{n}=2$, as also reproduced by the model. The global dispersion relation model, however, makes a larger error at high $k_y$ where the UMs occur compared to low wavenumber where TEMs occur. In HSX, nearly all eigenmodes are found in the electron diamamgnetic direction, indicating they correspond to different branches of TEMs. As the density gradient is increased, a new cluster of modes appear around $0.2 \leq k_y \rho_{s} \leq 0.6$ in a separate frequency band, which are distinctly stronger destabilised as $a/L_{n}$ is increased compared to the TEMs. These instabilities are characterised by significantly more extended eigenmodes without strong preferential localisation to any of the magnetic or drift wells and correspond to cases of the universal instability (UI)\cite{Landreman2015,Helander2015b}. Similar modes also occur in the high-mirror configuration of W7-X, though only for the higher density gradients of $a/L_n = 3,4$ in the smaller wavenumber region of $0.2 \leq k_y \rho_\mathrm{ss} \leq 0.4$, with the remainder of the instabilities corresponding to ion-driven trapped electron modes (iTEMs)\cite{Plunk2014}. Both of these observations regarding the UI may be explained in terms of the shear-dependent threshold density gradient which is necessary for this instability to thrive\cite{Landreman2015}, with the shear in W7-X being larger by an order of magnitude, however, simultaneously the instability drive for (i)TEMs being smaller by an order of the trapped-electron fraction because of the lack of unfavourable bounce-average curvature due to the maximum-$J$ property of the configuration. Aside from the UIs -- for which the dispersion relation model fails to predict accurate growth rate attributable to lack of kinetic response for passing electrons -- the global dispersion relation does reproduce the fact that growth rates are significantly smaller in W7-X compared to either DIII-D or HSX for a given wavenumber and density gradient, indicating the weaker drive of iTEMs compared to conventional TEMs. However, the frequency of these iTEMs is not as accurately predicted by the model compared to frequency of conventional TEMs, with low wavenumber modes being assigned a frequency in the electron diamagnetic direction, opposed iTEMs being characterised by propagation frequencies in the ion diamagnetic direction. Comparing the output of the model to eigenfrequencies obtained by \textsc{Gene} simulations, there is an overall downshift of propagation frequencies into the electron diamagnetic direction and an overprediction of growth rates, which are significantly larger compared to the errors made for TEMs in DIII-D and HSX. This may be attributed to the diminished influence of trapped electrons, which are strictly necessary to make the model variational, and the increased importance of passing electrons, which are only accounted for adiabatically. This hypothesis is further confirmed in \Cref{sec:config-effects} where the negative mirror configuration is examined, which is distinctively not maximum-$J$. Most notably, with the exception of the UI modes, the accuracy of the model to reproduce \textsc{Gene} eigenfrequencies decreases with increasing $a/L_n$ universally across all three geometries.

\begin{figure}
    \centering
    \begin{subfigure}{\linewidth}
        \centering
        \includegraphics[width=\linewidth]{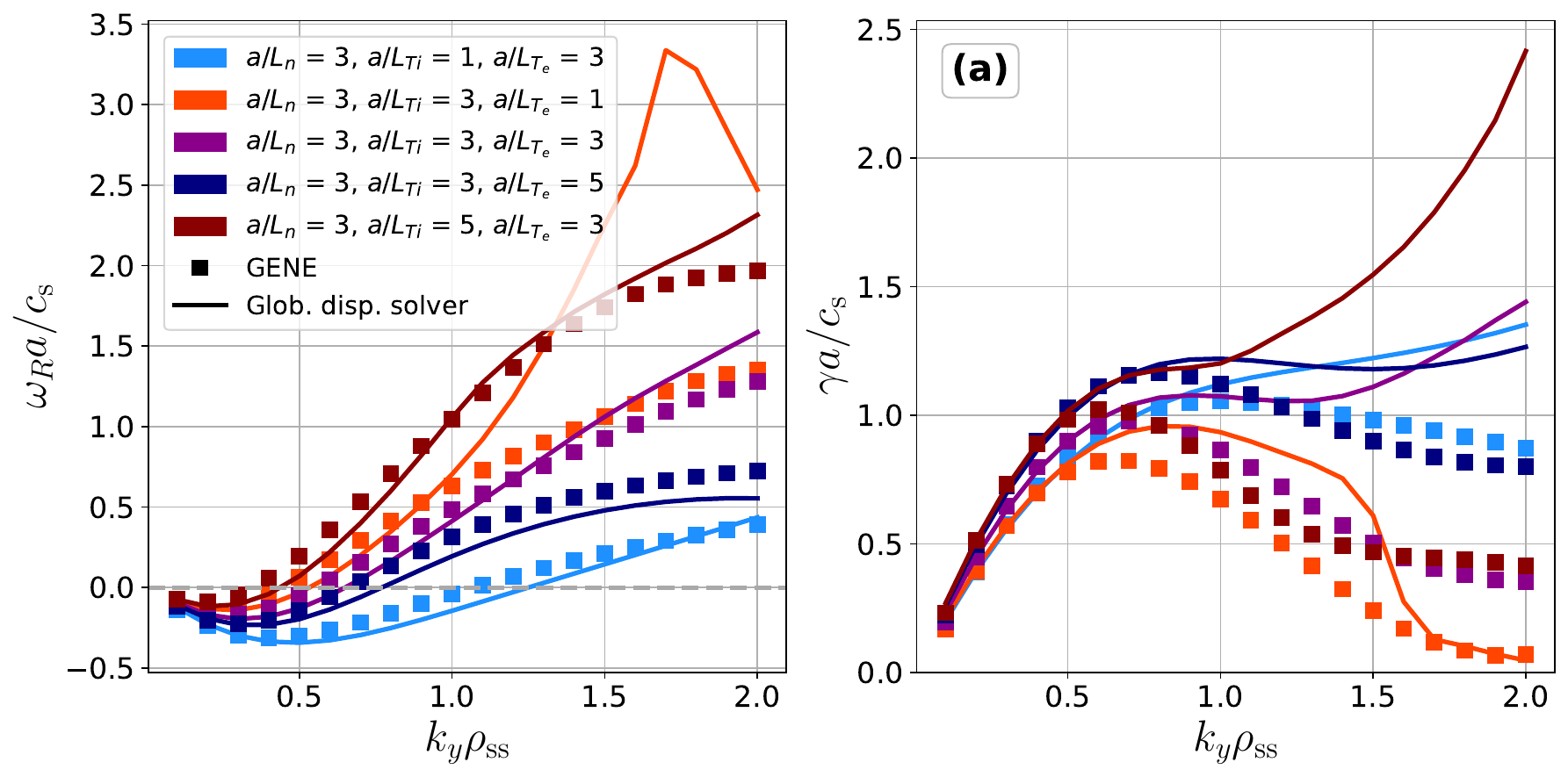}
    \end{subfigure}
    \begin{subfigure}{\linewidth}
        \centering
        \includegraphics[width=\linewidth]{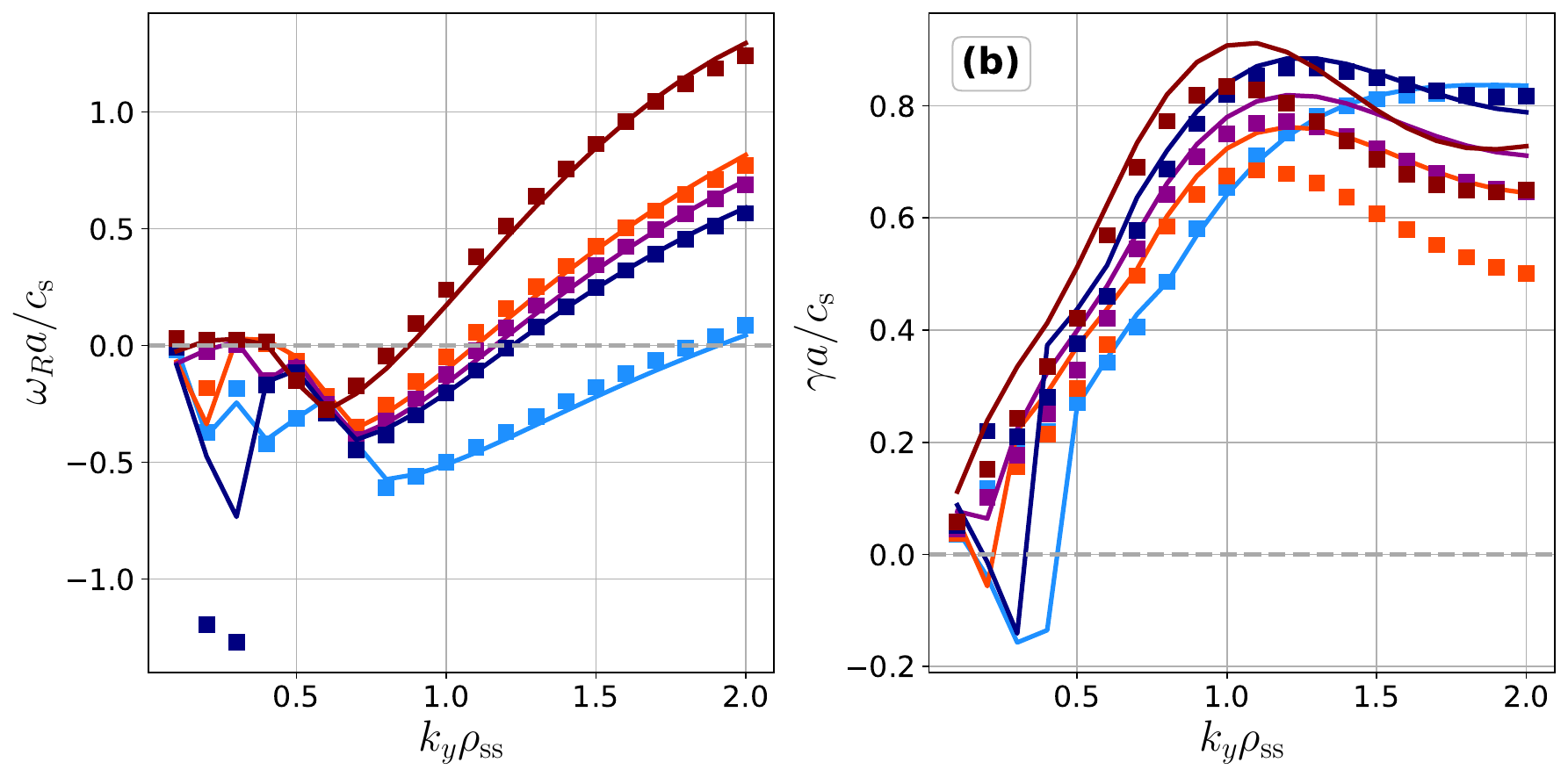}
    \end{subfigure}
    \begin{subfigure}{\linewidth}
        \centering
        \includegraphics[width=\linewidth]{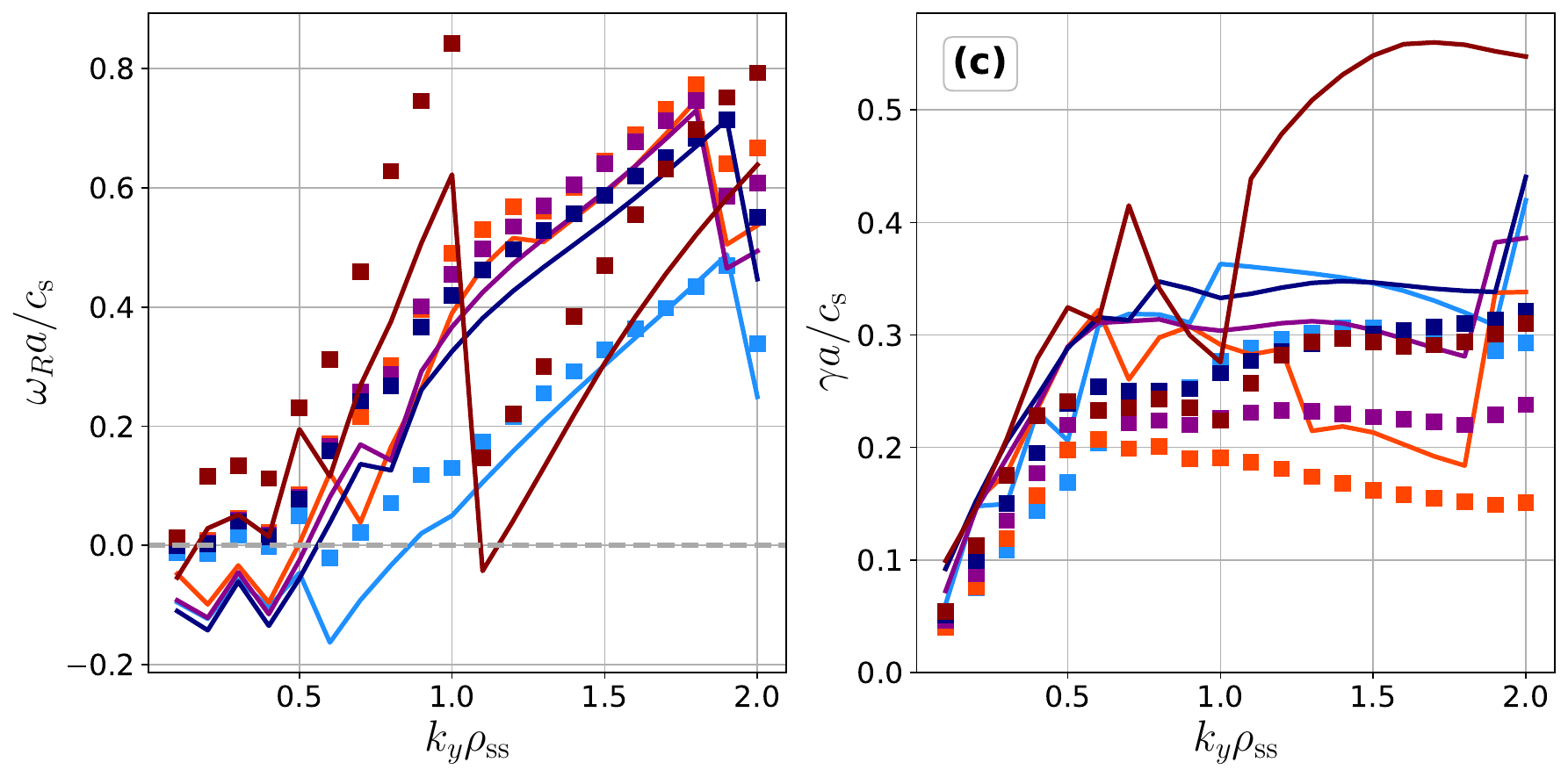}
    \end{subfigure}
    \caption{Eigenfrequency solutions obtained by the global dispersion model (solid lines) contrasted with \textsc{Gene} simulations (symbols) including kinetic-electrons for realistic non-flat kinetic profiles of both species. Both the ion- and electron-temperature gradient are varied simultaneously, with blue and red hues indicating cases of dominant electron-temperature gradient ($\eta_e > \eta_i$) and dominant ion-temperature gradient ($\eta_i>\eta_e$), respectively, with darker hues indicating a stronger (dominant) temperature gradient. Shown are results for (a) the DIII-D tokamak, (b) the HSX stellarator and (c) the high-mirror configuration of the W7-X stellarator. In all cases the density gradient is fixed at $a/L_{n}=3$.}
    \label{fig:ke-fullgrad}
\end{figure}

Lastly in \Cref{fig:ke-fullgrad}, we consider the scenario of realistic profiles where finite ion- and electron-temperature gradient coexist, where both situations with a dominant effect of ion temperature gradient ($a/L_{T\mathrm{i}}>a/L_{T\mathrm{e}}$, indicated by reddish hues) and dominant electron temperature gradient ($a/L_{T\mathrm{e}}>a/L_{T\mathrm{i}}$, indicated by blueish hues) are considered around a nominal scenario with balanced gradients ($a/L_{T\mathrm{i}}=a/L_{T\mathrm{e}}=a/L_n$, indicated in purple). For each of the two dominant scenarios, we consider a case with low temperature gradient ($a/L_{Ts}=a/L_n$, indicated by lighter colour) and high temperature gradient ($a/L_{Ts}>a/L_n$, indicated by darker colour) to also probe for $\eta_s = L_{Ts}/L_{n}$ effects, where $s=i,e$. Across the three geometries, both the model and \textsc{Gene} simulations indicate that the ion temperature gradient plays a more determining role on the type of instabilities that occur, with the most significant changes in propagation frequency occurring when $a/L_{T\mathrm{i}}$ is changed from its nominal value of $3$ to $5$ or $1$. When the ion temperature gradient is kept fixed, both the model and the \textsc{Gene} simulations predict an increase in the growth rate as $a/L_{T\mathrm{e}}$ increases, particularly at high wavenumbers. In particular, the inclusion of finite temperature gradient for both species is sufficient for UI modes to no longer appear in the high-mirror configuration of W7-X (though they may persist as subdominant unstable mode), resulting in improved agreement between the model and \textsc{Gene} simulations at low wavenumber. For the $a/L_{T\mathrm{i}}=5,a/L_{T\mathrm{e}}=3$ case, growth rates in W7-X are however significantly overestimated by the model for $k_y \rho_\mathrm{ss} >1 $  compared to other gradients. Like with high ion temperature gradient case discussed in the main text, this corresponds to a scenario of iTEM transitioning into a higher-excitation state of ITG, for which, as explained in the main text, any error made in the growth rate as a result from the lack of Landau damping from ion parallel streaming is subsequently enhanced. Similar observations can also be made about the ion-temperature-gradient dominated cases in DIII-D, where the high-$k_y$ large-$a/L_{T\mathrm{i}}$ mode growth rates (maroon curve) obtained by the model continue to grow without bound, while the \textsc{Gene} simulations clearly show that those modes are stabilised as $k_y$ increases as a result of FLR damping. This behaviour of the model is robust to changes in numerical resolution used to calculate the resonant ion integrals as well as change of root finding algorithm from the Broyden method\cite{Broyden1965AEquations} to a conventional Newton method, and is deemed a numerical artifact as a result of geometric details of the DIII-D flux-tube -- presumably the high global shear -- rather than a physical aspect predicted by the model in axisymmetric configurations. Indeed, as will be shown in \Cref{sec:config-effects}, the dispersion relation model accurately predicts both growth rates and propagation frequencies for identical gradients in the TCV tokamak -- where the shear is reduced by more than a factor of two compared to DIII-D -- which closely follow the trends observed in the \textsc{Gene} simulations. Unlike the pure $a/L_{T\mathrm{i}}$ case, however, these high-$k_y$ high-$a/L_{T\mathrm{i}}$ modes do not correspond to a higher excitation-state ITG, but are some form of hybrid UM-ITG as there are no mode transitions observed in the propagation frequency. It should also be remarked that only in the $a/L_{T\mathrm{i}}=3, a/L_{T\mathrm{e}}=1$ case we observe the issue for intermediate-to-high-$k_y$ UMs identified in the main text -- being that the model predicts simultaneously large propagation frequency and small growth rates -- while this issue is resolved in the $a/L_{T\mathrm{i}}=3, a/L_{T\mathrm{e}}=3$ case. This indicates that as the electron temperature gradient is increased the trapped-electron contribution to the dispersion relation gains more prominence, which, containing the regularisation physics through the bounce-average of the electrostatic potential, aids to alleviate the issue arising from the ion contribution. The model's sensitivity to high global shear would also explain why the global dispersion relation model does not suffer from similar divergences (aside from the aforementioned higher excitation state ITG mode in W7-X) in both stellarator configurations, where the shear is substantially lower.

\section{Investigation of configuration effects} \label{sec:config-effects}

\begin{figure*}
    \centering
    \begin{subfigure}{0.45\linewidth}
        \centering
        \includegraphics[width=\linewidth]{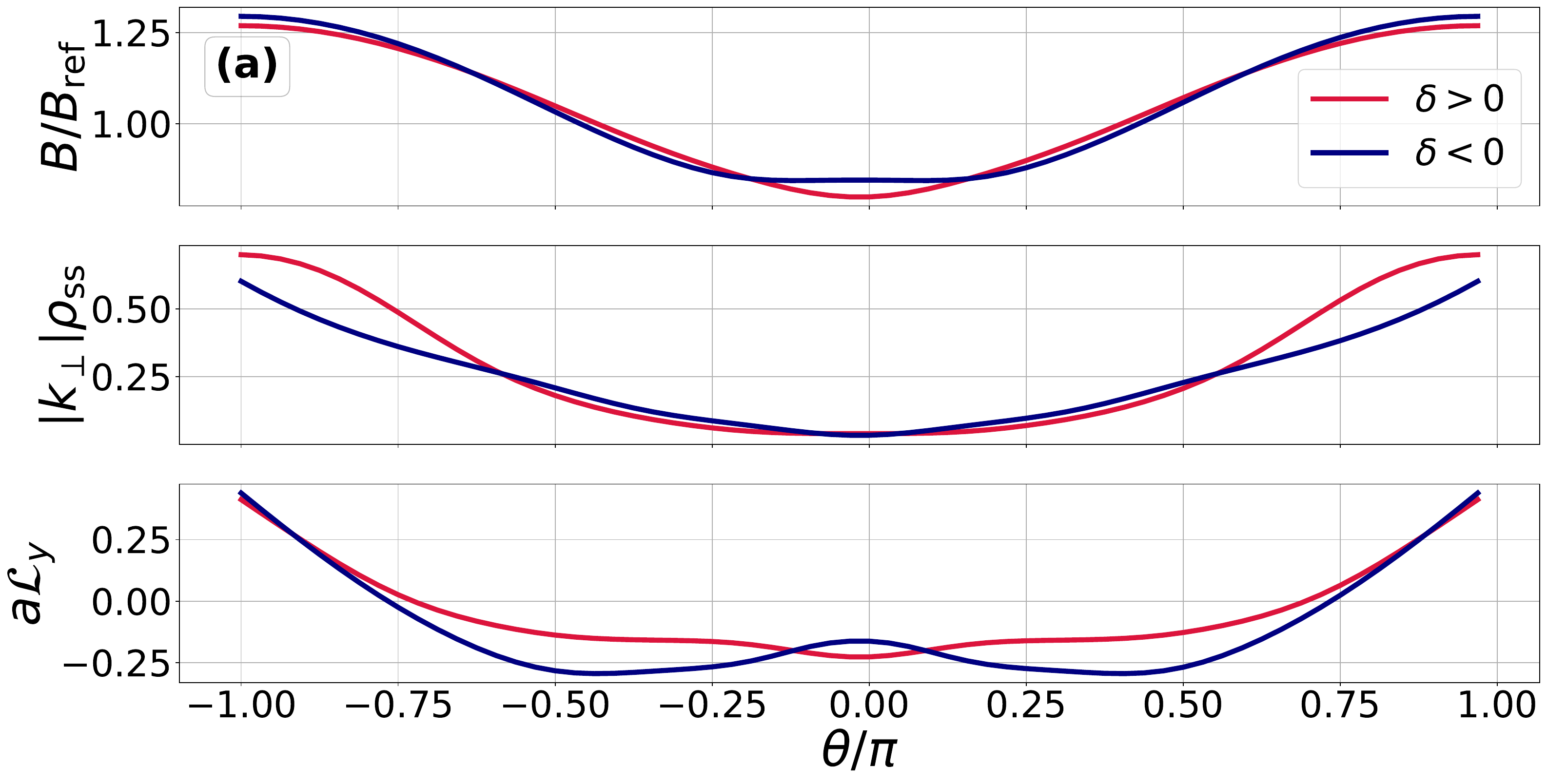}
    \end{subfigure}
    \begin{subfigure}{.45\linewidth}
        \centering
        \includegraphics[width=\linewidth]{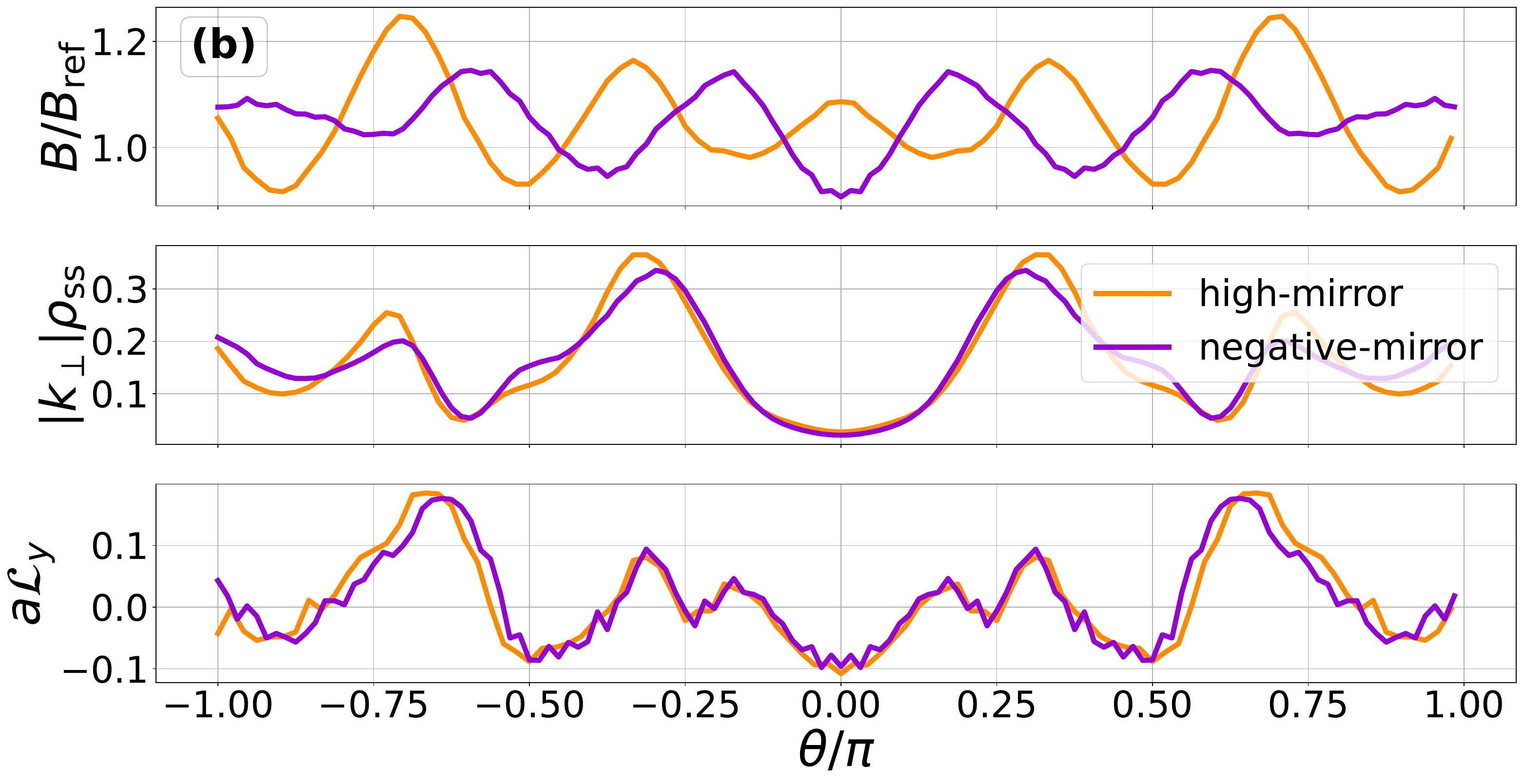}
    \end{subfigure}
    \caption{Comparison of flux-tube geometric quantities between (a) positive- (red) and negative-triangularity (blue) configurations of the TCV tokamak and (b) high- (orange) and negative-mirror (purple) configurations of the W7-X stellarator. Shown are the variation of the magnetic field strength (top panel), magnitude of the perpendicular wavenumber (middle panel) and bi-normal component of the $\grad{B}$ drift (bottom panel). For $\norm{\bm{k_\perp}}$ we have consider a wavenumbers of $k_x \rho_\mathrm{ss} =0$ and $k_y \rho_\mathrm{ss} =0.3$, to facilitate comparisons with \Cref{fig:loc-freq-sol-visual,fig:FTgeo_stell} from the main text.}
    \label{fig:config_fluxtube_differences}
\end{figure*}

The results from \Cref{sec:GENE-application} of the main text and \Cref{sec:addl-cases-results} show that the global dispersion relation model can accurately account for inherent geometric differences between different classes of toroidal magnetic confinement devices. Consequently, the differences in geometric features between these devices are significant. To test whether the model is equally sensible to more subtle changes in the local geometric flux-tube quantities, we performed additional tests between two configurations of the same device. For this, both a positive- and negative-triangularity equilibrium of the TCV tokamak and the negative-mirror configuration of the W7-X stellarator (to be compared with the high-mirror configuration discussed in the main text) are chosen. The differences in the most important flux-tube geometric quantities between these configurations are shown in \Cref{fig:config_fluxtube_differences}. The main differences between the TCV configurations are found in the FLR term, which increases faster in the $\delta>0$ configuration, where $\delta$ denotes the triangularity, and the magnetic-drift well, which is both steeper and more stretched out in the $\delta<0$ configuration. Meanwhile, the magnetic field strength is nearly identical between the configurations, aside from a slightly shallower and flatter magnetic well observed in the $\delta<0$ configuration, corresponding to the straight section of the D-shaped flux-surface coinciding with the low-field side. By contrast, both the magnetic-drift wells and FLR term are nearly identical between the two W7-X configurations, while the magnetic field strengths differ significantly. As a consequence, the magnetic wells and regions of bad curvature are again aligned in the negative-mirror configuration, rendering the negative-mirror flux tube unstable to conventional TEMs. 

\subsubsection*{Comparing positive and negative triangularity TCV}

\begin{figure*}
    \centering
    \begin{subfigure}{.45\linewidth}
        \centering
        \includegraphics[width=\linewidth]{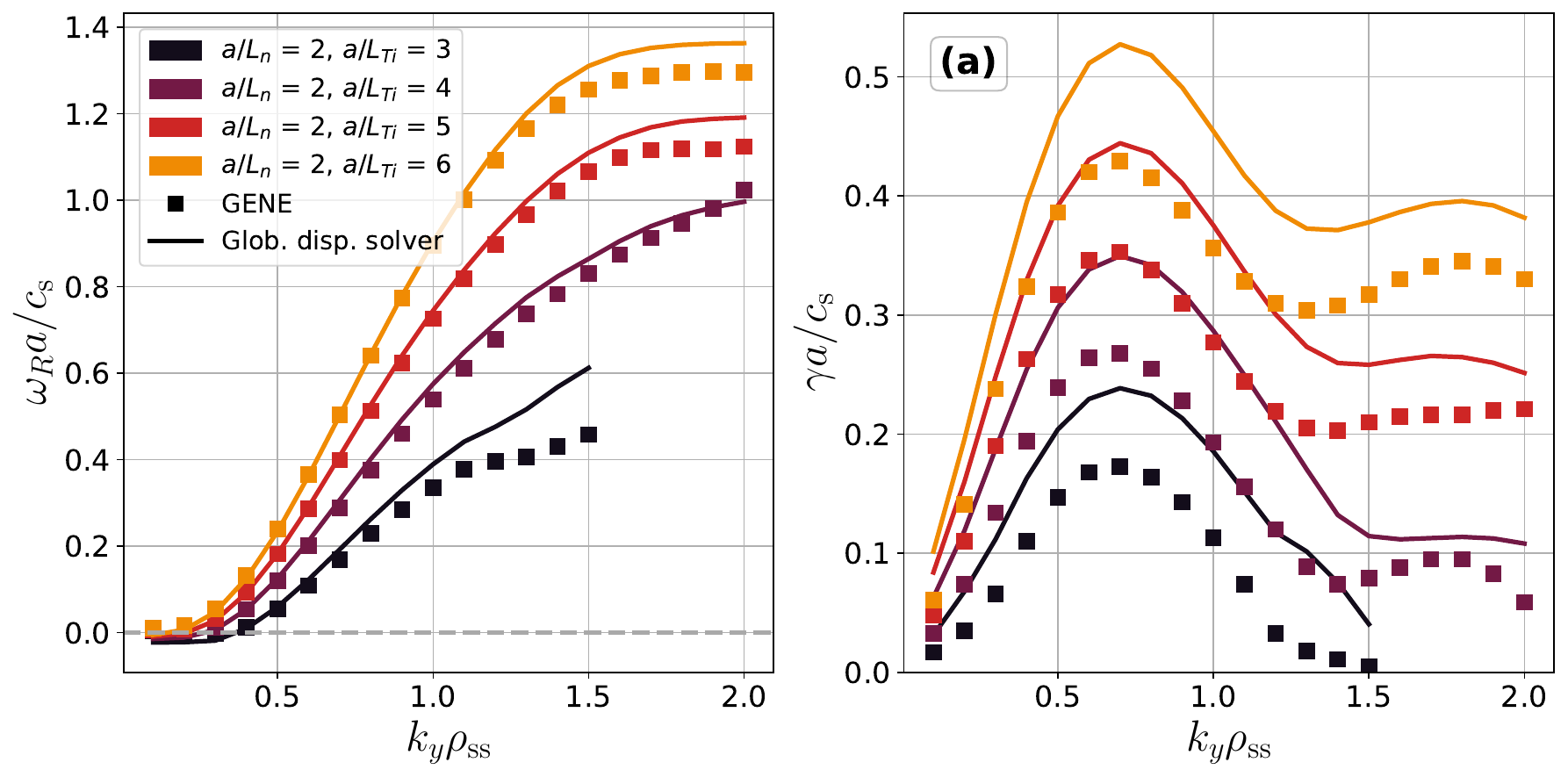}
    \end{subfigure}
    \begin{subfigure}{.45\linewidth}
        \centering
        \includegraphics[width=\linewidth]{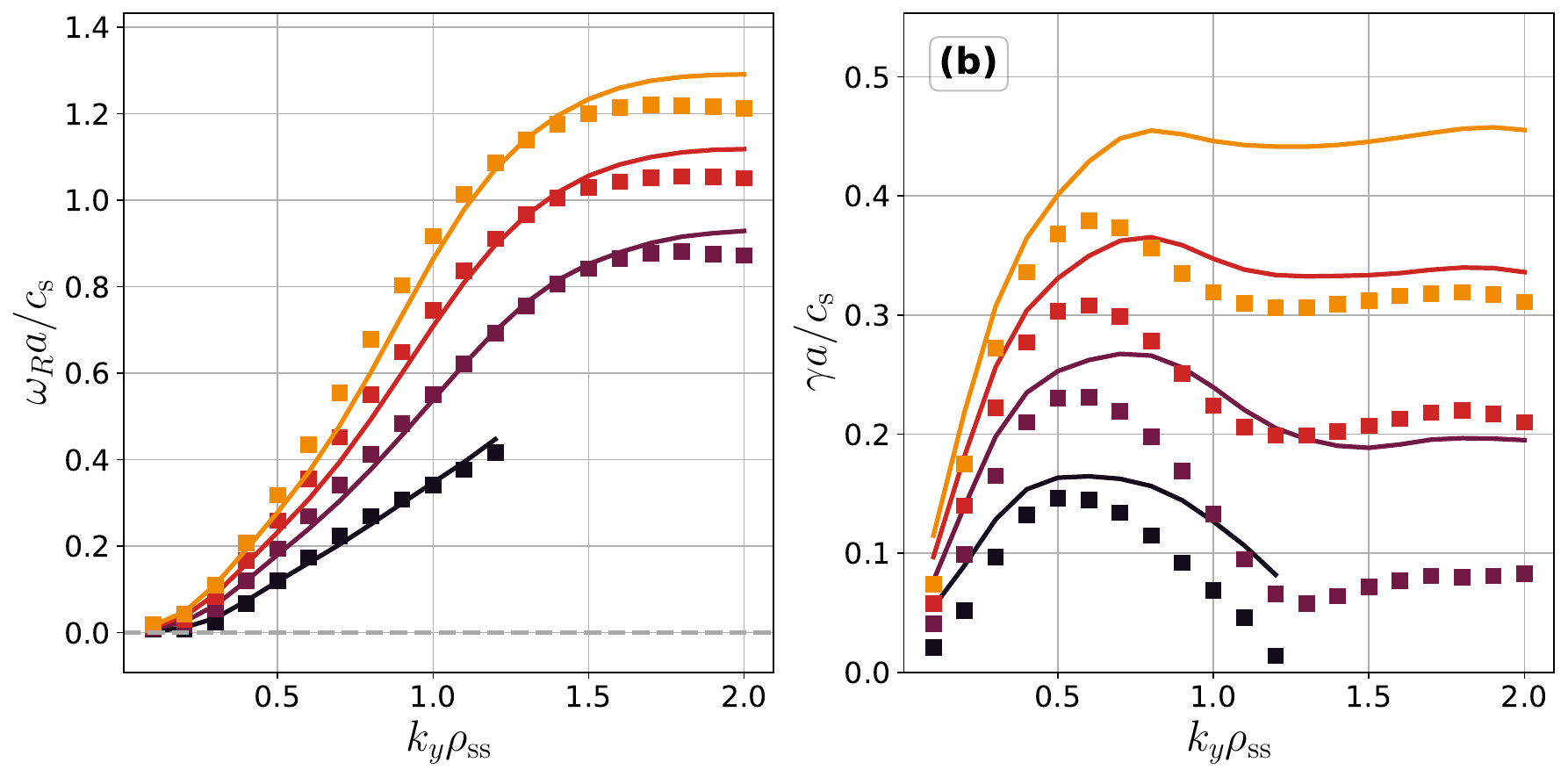}
    \end{subfigure}
    \begin{subfigure}{.45\linewidth}
        \centering
        \includegraphics[width=\linewidth]{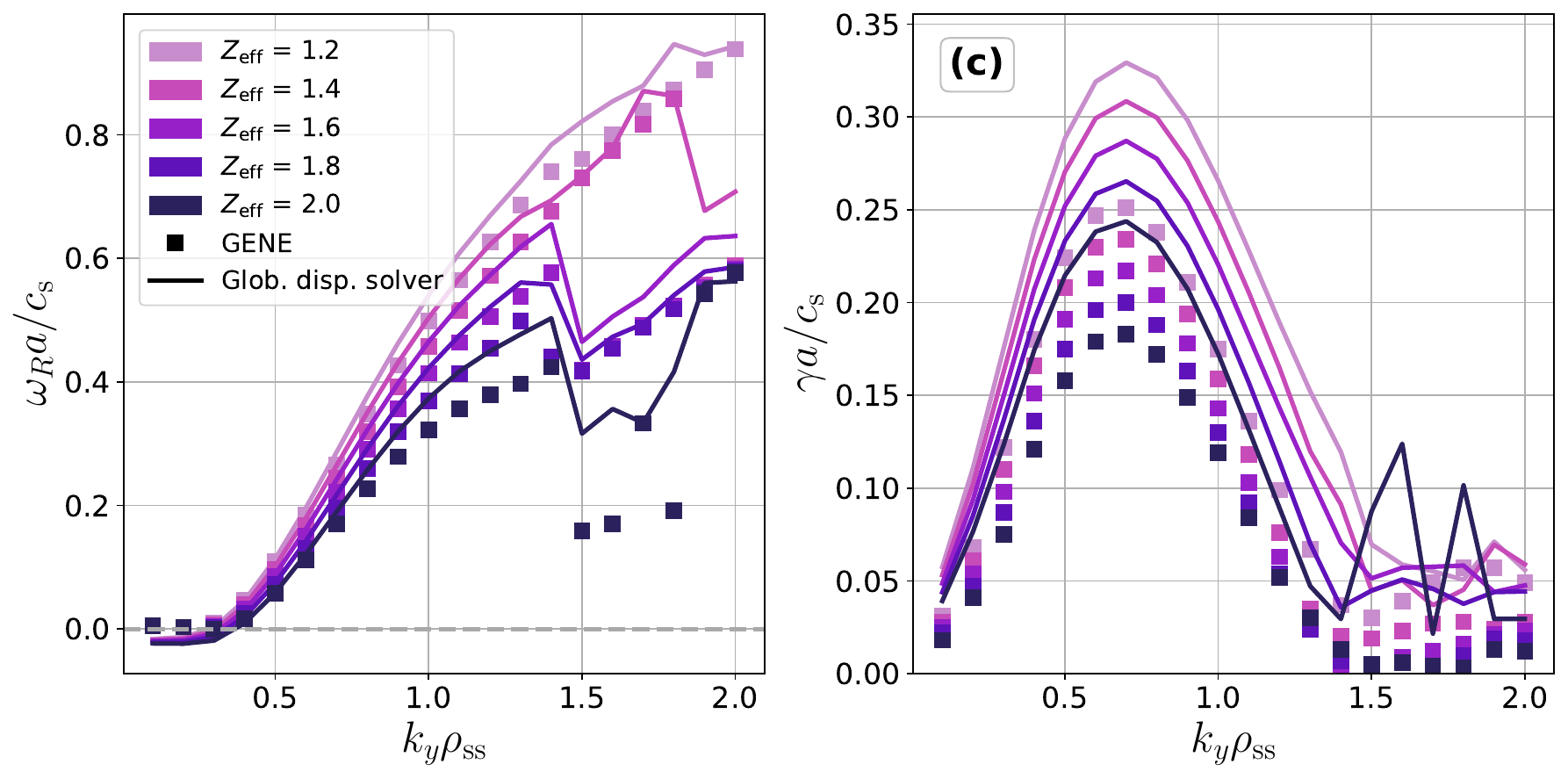}
    \end{subfigure}
    \begin{subfigure}{.45\linewidth}
        \centering
        \includegraphics[width=\linewidth]{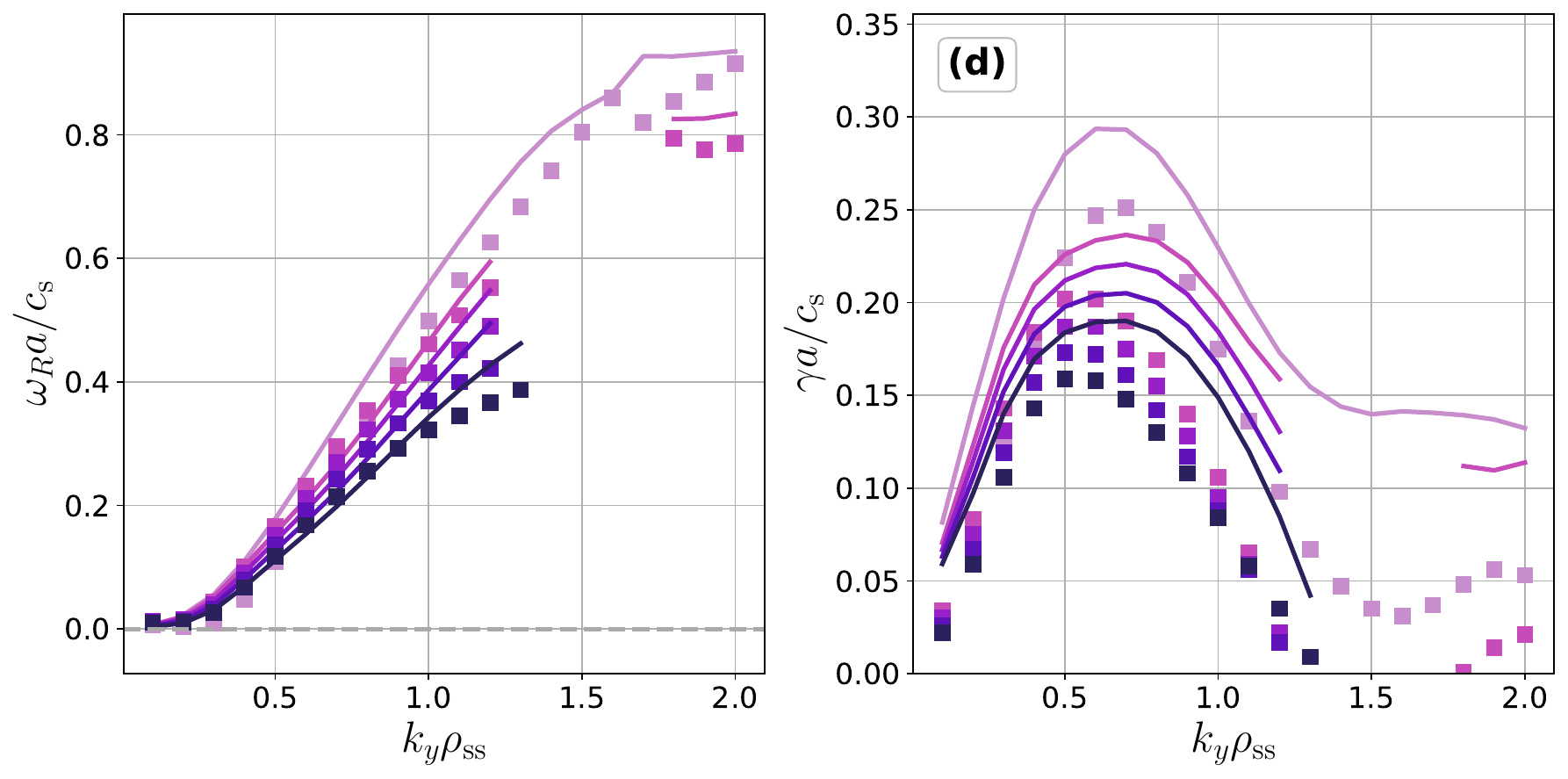}
    \end{subfigure}
    \begin{subfigure}{.45\linewidth}
        \centering
        \includegraphics[width=\linewidth]{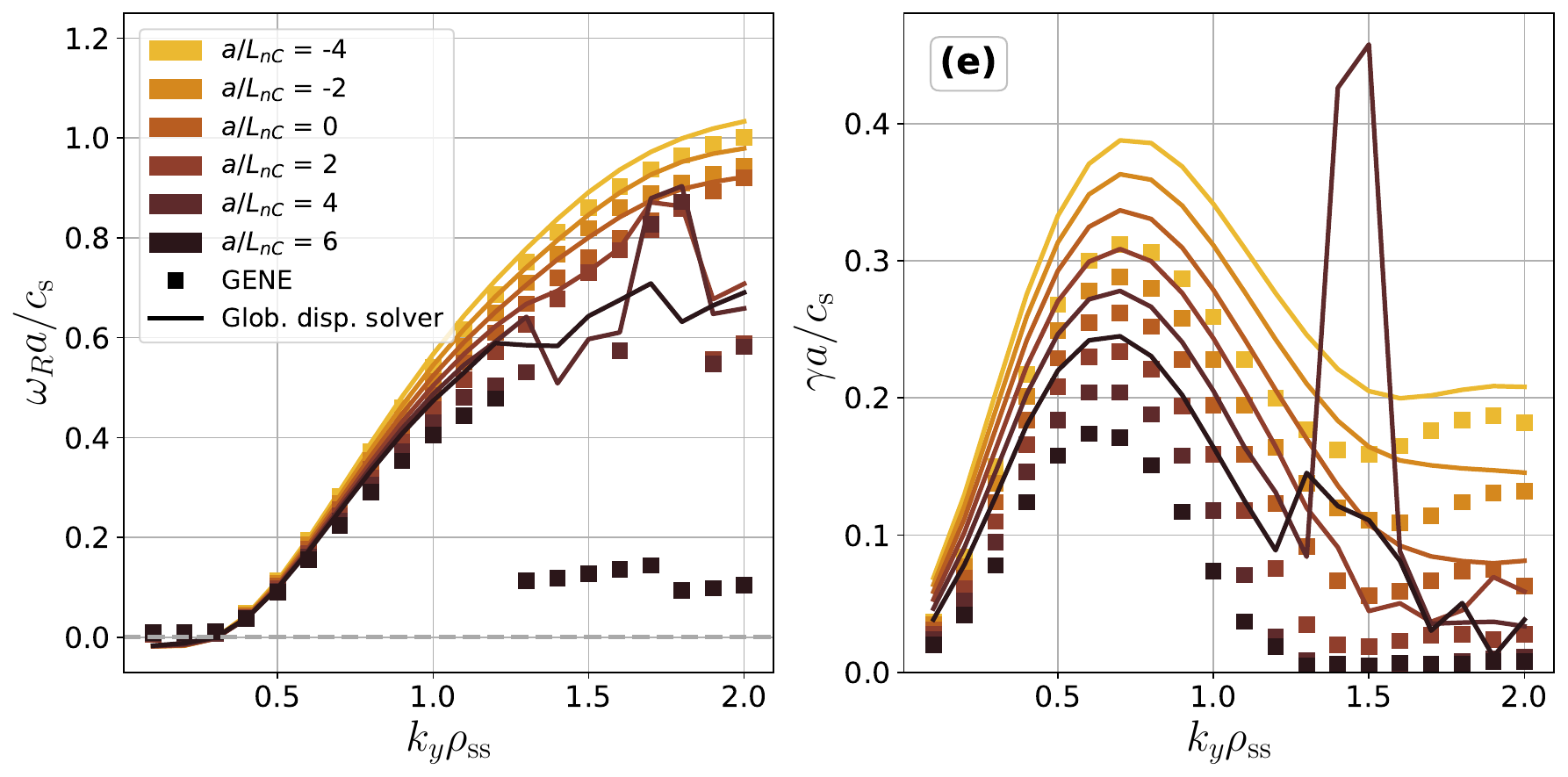}
    \end{subfigure}
    \begin{subfigure}{.45\linewidth}
        \centering
        \includegraphics[width=\linewidth]{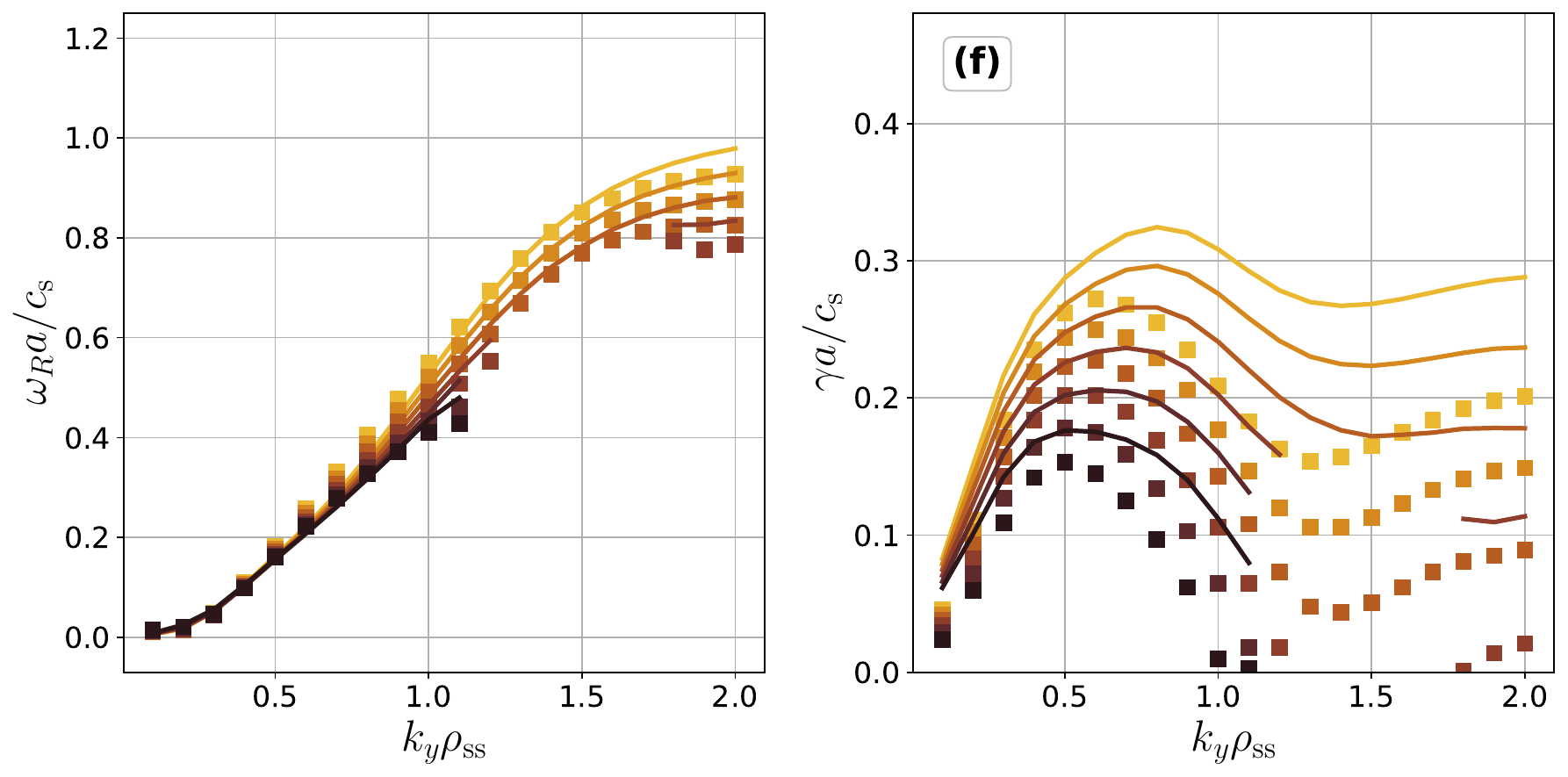}
    \end{subfigure}
    \caption{Eigenfrequency solutions obtain by the global dispersion model (solid lines) contrasted with \textsc{Gene} simulations (symbols) for adiabatic-electron scenarios in positive triangularity TCV flux tube (left column) and negative triangularity flux tube (right column) of the TCV tokamak. Shown are results while varying (a)-(b) ion-temperature gradient $a/L_{T\mathrm{i}}$ [no $C^{6+}$ impurities], (c)-(d) impurity concentration $Z_{\textrm{eff}}$ [with ambipolar gradients $a/L_{ni}=a/L_{n\mathrm{C}}=2$], and (e)-(f) impurity density gradient $a/L_{n\mathrm{C}}$ [at fixed $Z_{\textrm{eff}}=1.4$]. In all cases the electron density gradient is fixed at $a/L_{ne}=2$, with ion and impurity density (gradient) determined by ambipolarity constraints [\Cref{eq:impurity-ambipolarity} in main text]. Note that the axis limits for both $\omega_R$ and $\gamma$ are matched between left and right columns to facilitate quantitative comparison.}
    \label{fig:TCV-aeITG-compars}
\end{figure*}

In \Cref{fig:TCV-aeITG-compars} results for adiabatic-electron scenarios are presented for both TCV configurations. In general the model captures the trend that peak growth rates are both smaller and shifted to lower wavenumber in the negative triangularity flux-tube compared to its positive-triangularity counterpart. At very low wavenumbers $k_y \rho_\mathrm{ss} \leq 0.3$, ITG modes are, however, marginally more unstable in the negative-triangularity configuration as a result of the aforementioned shift in growth rate peak -- as observed in both \textsc{Gene} simulations and solutions to the global dispersion model. The model does qualitatively capture that (in absence of impurities) there is a weaker dispersion in growth rates in the negative triangularity case, with growth rates being nearly constant as a function of $k_y$ at higher wavenumber in both the \textsc{Gene} simulations and solutions to the dispersion model. Whereas the overprediction of growth rates by the model is virtually constant in the positive triangularity case, the negative triangularity case displays near-accurate growth rates for low-wavenumber while a significantly larger error is made for high-wavenumber modes, which contributes to the disparity observed between the \textsc{Gene} simulations and the global dispersion model with regards to the preferential role of negative triangularity for ITG stability. Setting aside this discrepancy in the large-wavenumber regime, the results here are in agreement with earlier investigations on the effects of triangularity in on adiabatic-electron ITG\cite{Merlo2023InterplayPlasmas,Merlo2023OnTokamaks,Duff2022EffectTurbulence}, though it should be emphasized that the beneficial role of negative triangularity is not a general property and depends on other details of the magnetic configuration under investigation, most notably the (global) magnetic shear\cite{Merlo2023InterplayPlasmas}. These observed trends about the model's performance mostly extrapolate to the cases where impurity effects are included, where both dillution effects observed in the \text{Gene} simulations are qualitatively reproduced by the model. Noticeable exceptions, however, occur for a handful of high-wavenumber modes when the dilution effects are strongest (particularly $k_y \rho_{ss} = 1.5,1.6,1.8$ for $Z_{\textrm{eff}}=2.0$, $k_y \rho_{ss} = 1.4,1.5$ for $a/L_{n\mathrm{C}}=4$ and $k_y\rho_{ss} = 1.3 \textrm{-}2$ for $a/L_{n\mathrm{C}}=6$) where both frequency and growth rate are significantly overpredict by the model, though this only occurs for the positive triangularity configuration. The eigenmodes corresponding to those cases show a significantly broader structure with weaker localisation to the outboard mid-plane, and are contrast to conventional ITG modes observed at other wavenumbers. As these discrepancies only occur for the aforementioned modes, it may be possible that this disparity between the model and high-fidelity simulations is a result of the neglect of the effect of particle streaming.

\begin{figure*}
    \centering
    \begin{subfigure}{.45\linewidth}
        \centering
        \includegraphics[width=\linewidth]{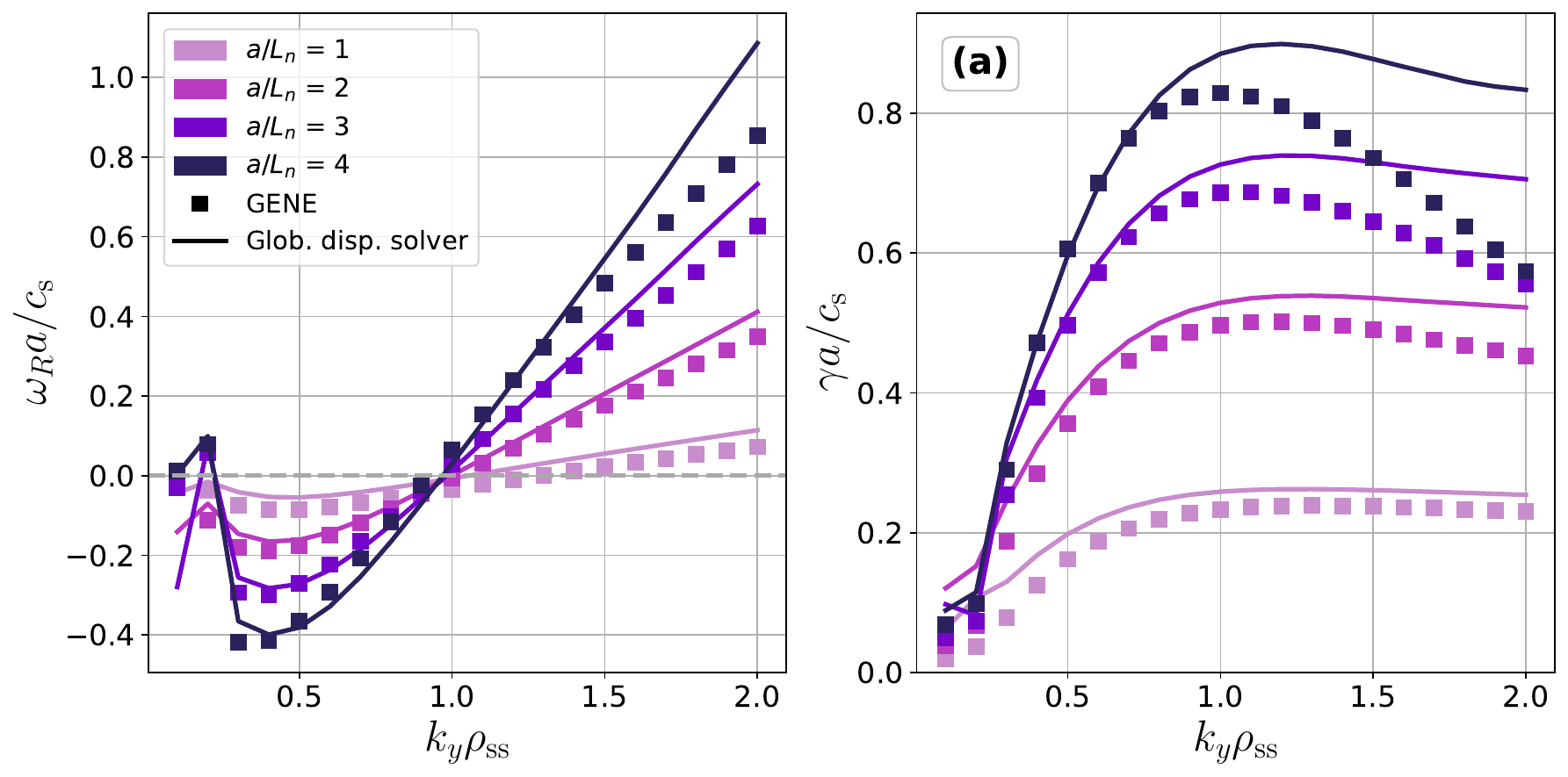}
    \end{subfigure}
    \begin{subfigure}{.45\linewidth}
        \centering
        \includegraphics[width=\linewidth]{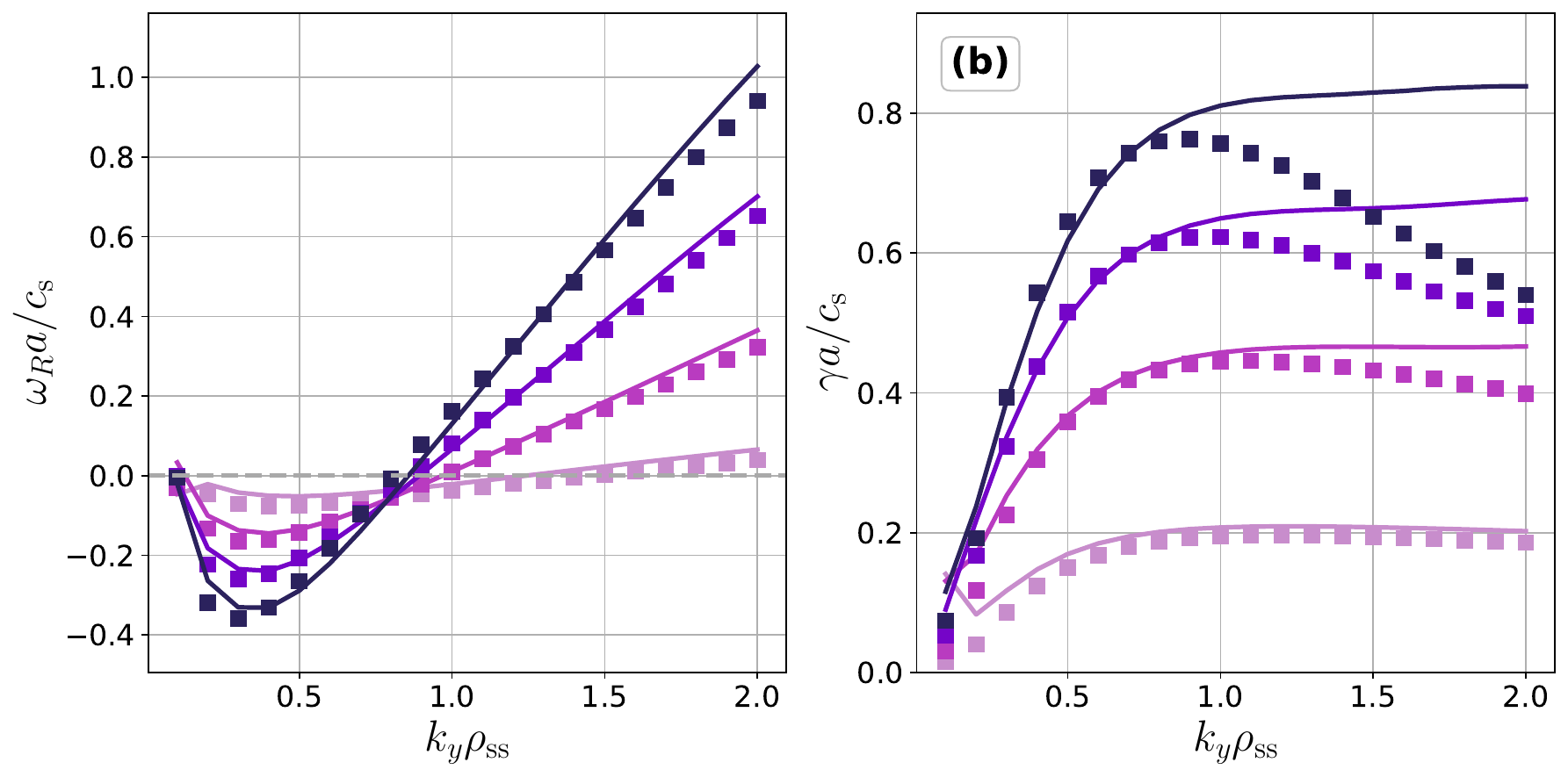}
    \end{subfigure}
    \begin{subfigure}{.45\linewidth}
        \centering
        \includegraphics[width=\linewidth]{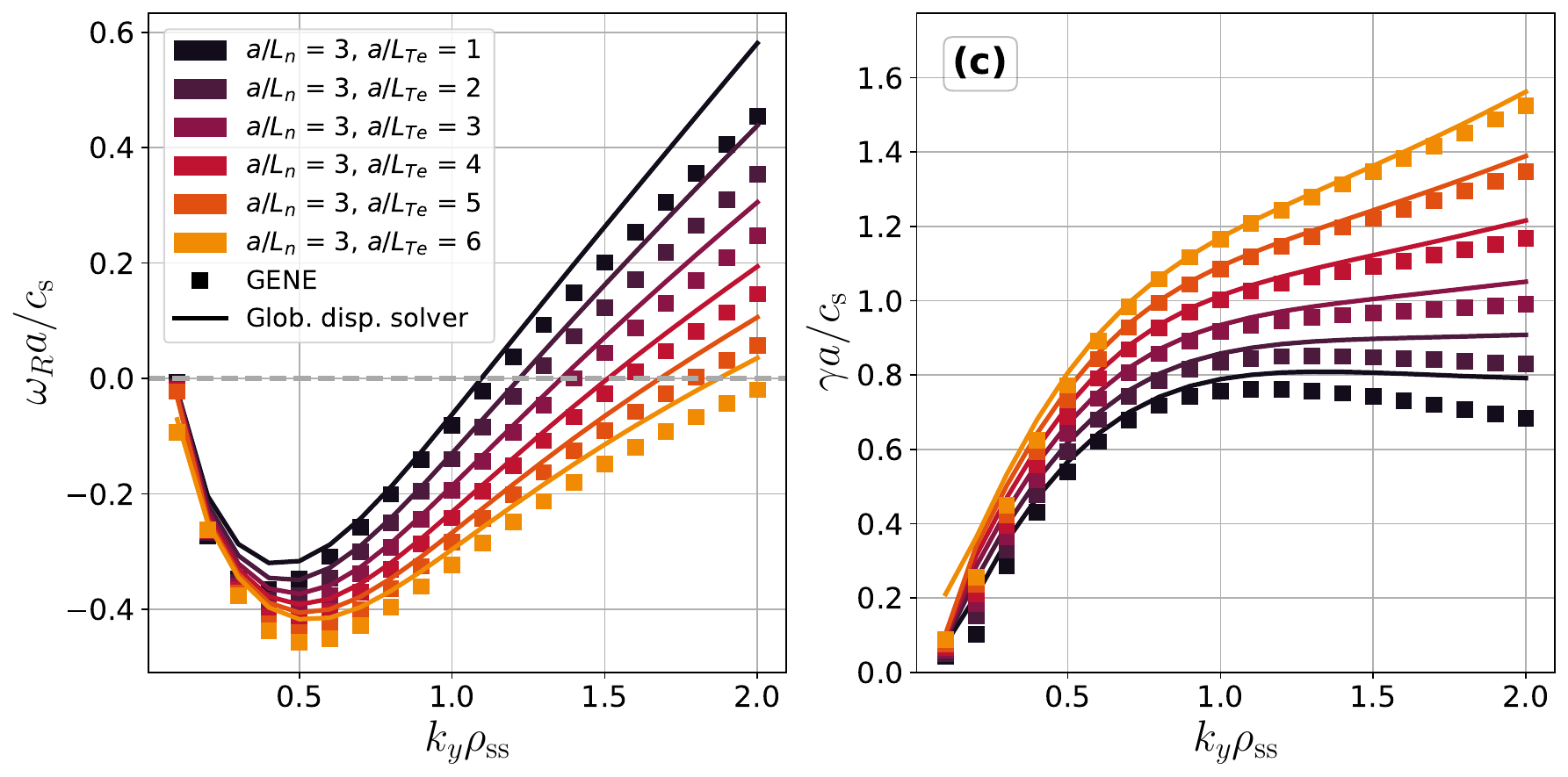}
    \end{subfigure}
    \begin{subfigure}{.45\linewidth}
        \centering
        \includegraphics[width=\linewidth]{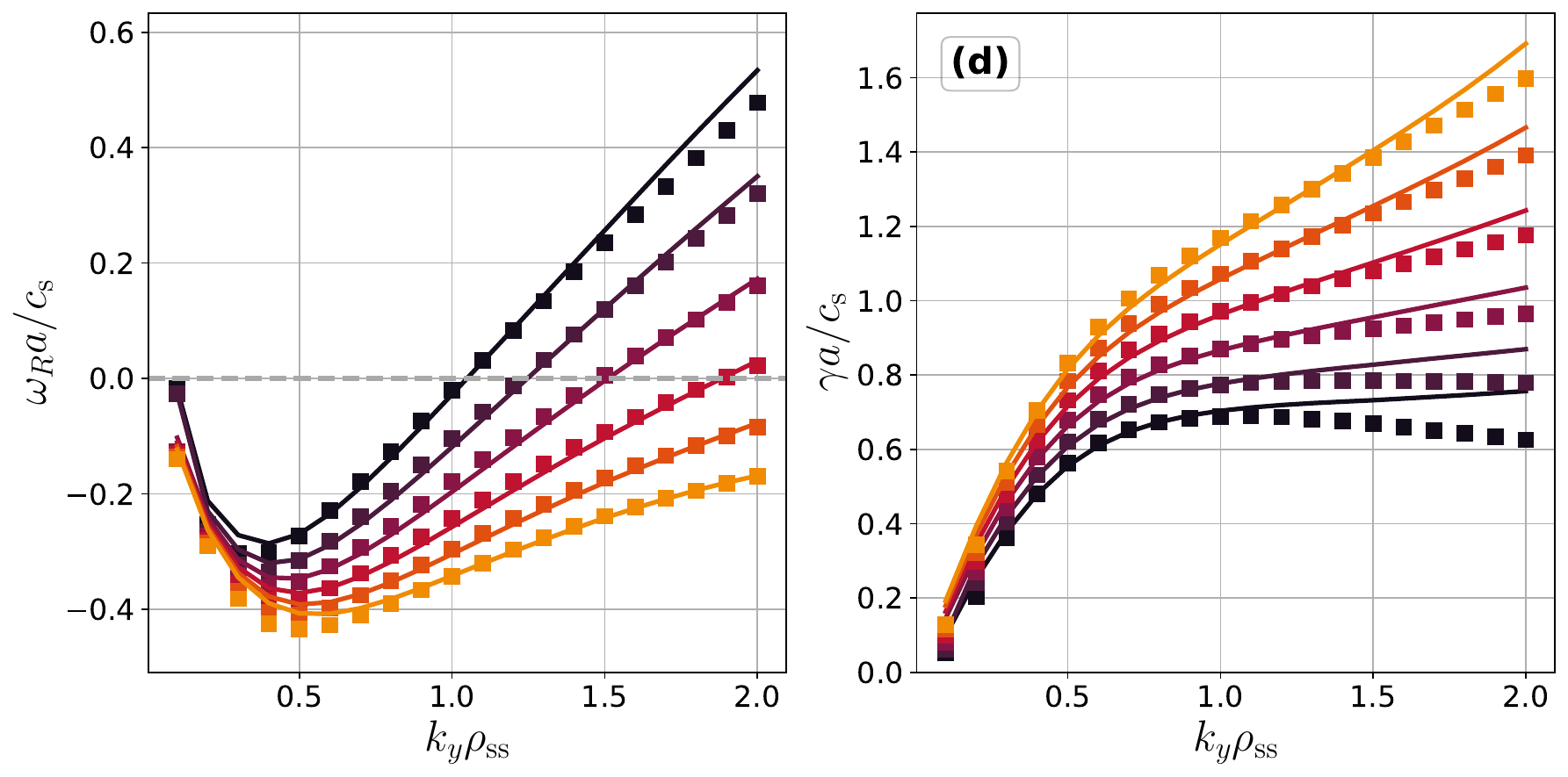}
    \end{subfigure}
    \caption{Eigenfrequency solutions obtain by the global dispersion model (solid lines) contrasted with \textsc{Gene} simulations (symbols) for kinetic-electron scenarios in response to TEM driving gradients in positive triangularity TCV flux tube (left column) and negative triangularity flux tube (right column) of the TCV tokamak. Shown are results while varying (a)-(b) the density gradient $a/L_{n}$ with suppressed electron temperature gradient ($a/L_{T\mathrm{e}}=0$), (c)-(d) the electron temperature gradient $a/L_{T\mathrm{e}}$ at a fixed density gradient of $a/L_n=3$. In all cases the ion temperature gradient is suppressed ($a/L_{T\mathrm{i}}=0$) to prevent the emergence of ITG modes. Note that the axis limits for both $\omega_R$ and $\gamma$ are matched between left and right columns to facilitate quantitative comparison.}
    \label{fig:TCV-keTEM-compars}
\end{figure*}

\begin{figure*}
    \centering
    \begin{subfigure}{.45\linewidth}
        \centering
        \includegraphics[width=\linewidth]{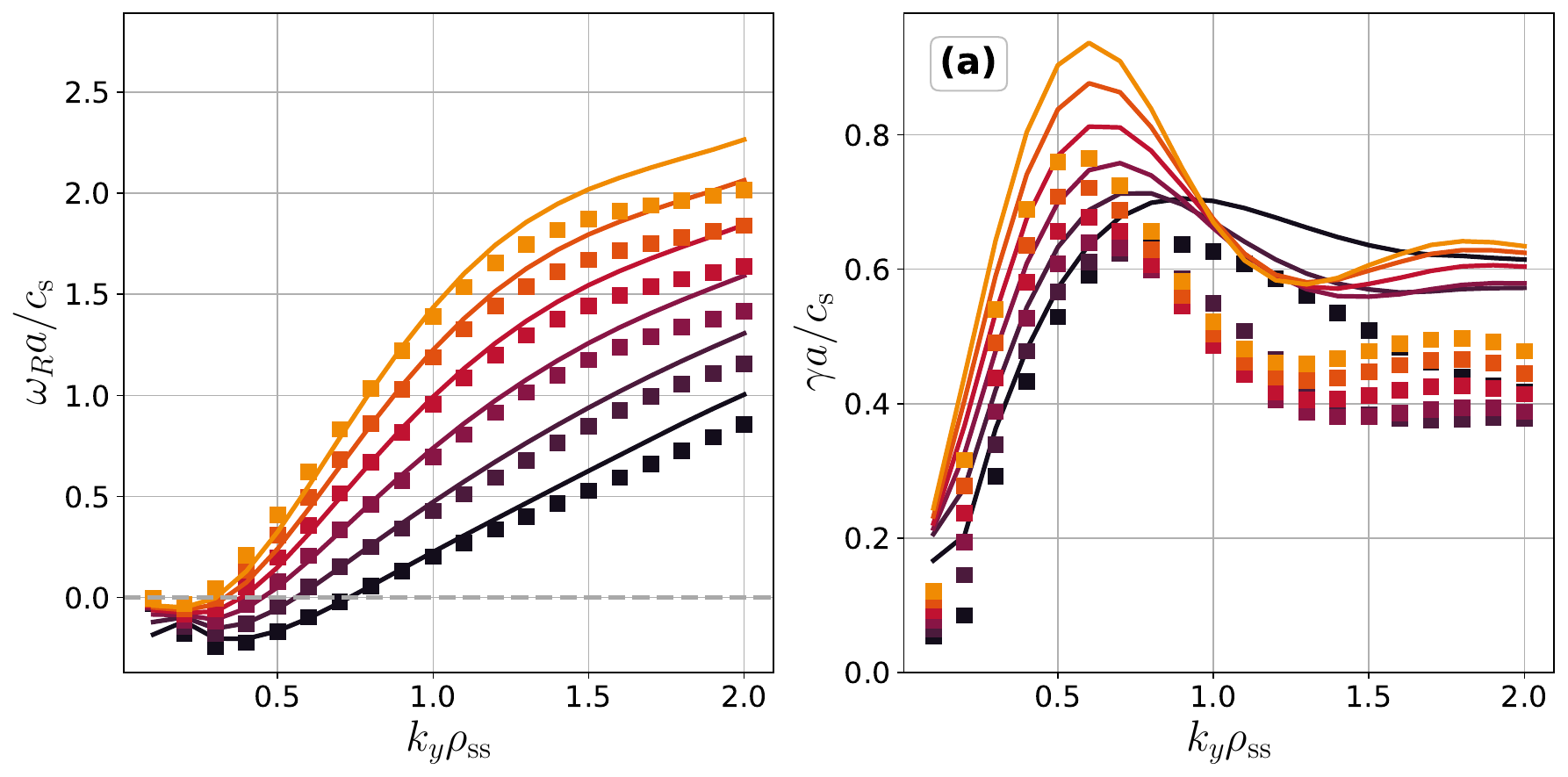}
    \end{subfigure}
    \begin{subfigure}{.45\linewidth}
        \centering
        \includegraphics[width=\linewidth]{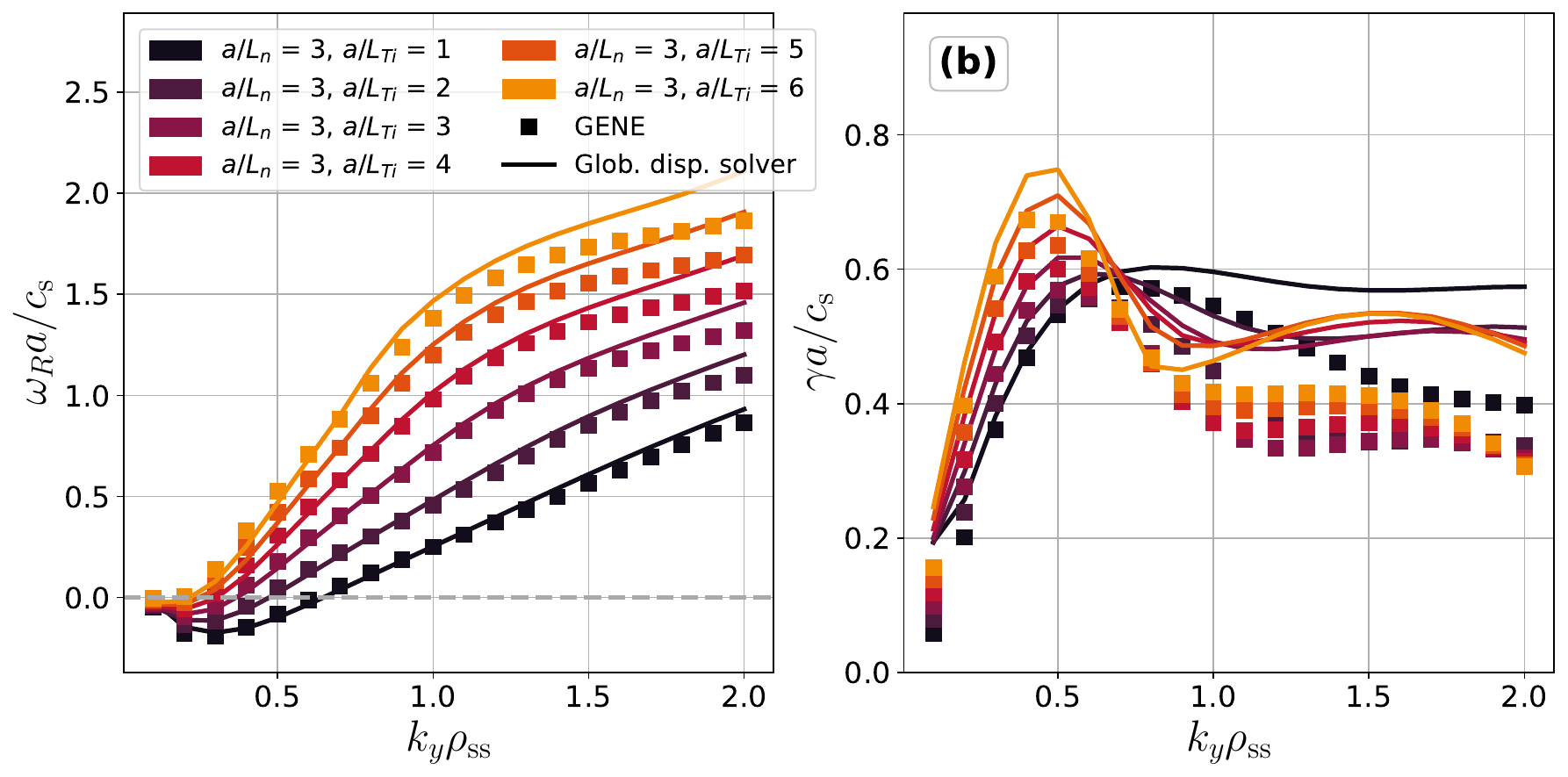}
    \end{subfigure}
    \begin{subfigure}{.45\linewidth}
        \centering
        \includegraphics[width=\linewidth]{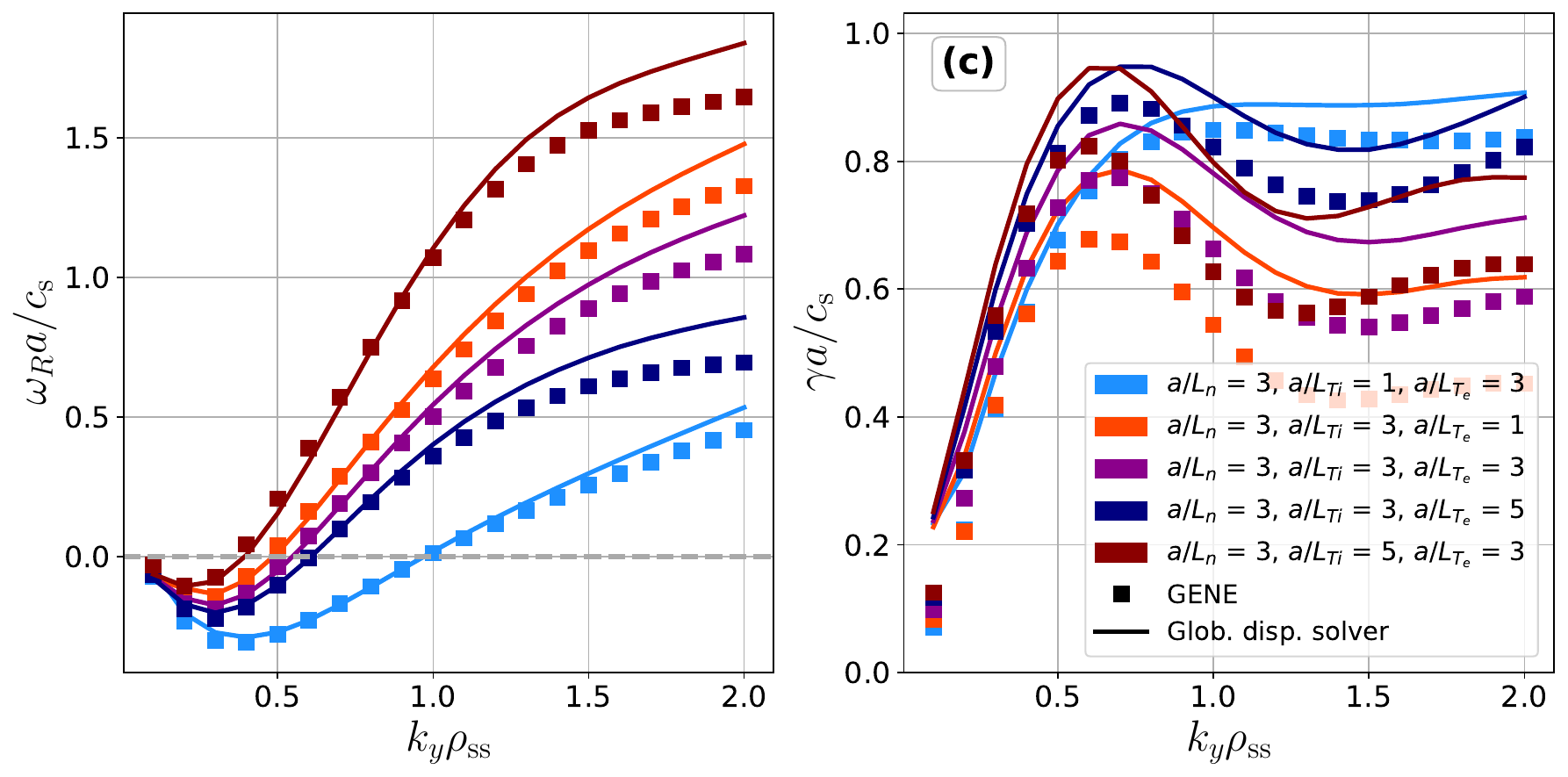}
    \end{subfigure}
    \begin{subfigure}{.45\linewidth}
        \centering
        \includegraphics[width=\linewidth]{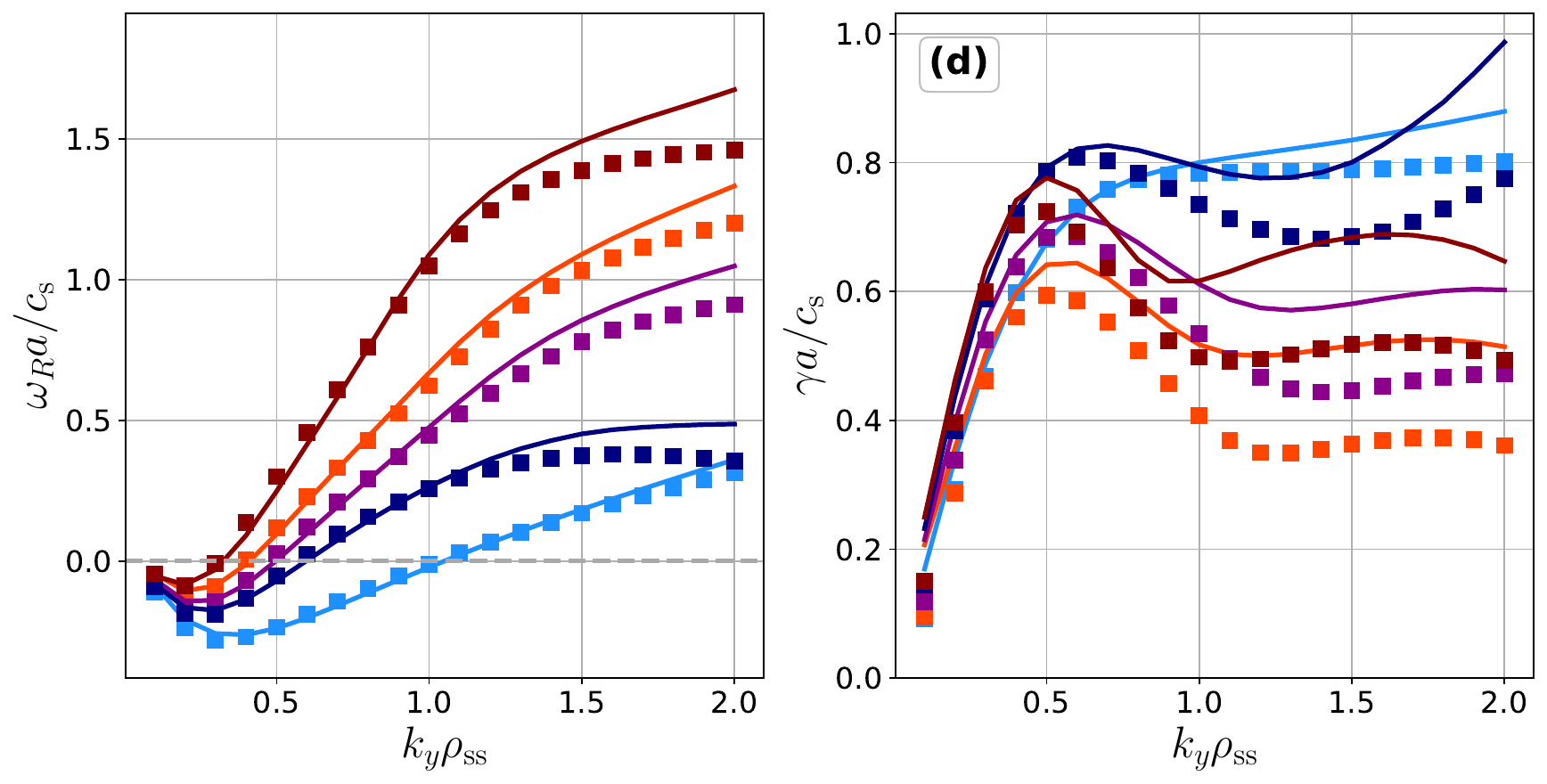}
    \end{subfigure}
    \caption{Eigenfrequency solutions obtain by the global dispersion model (solid lines) contrasted with \textsc{Gene} simulations (symbols) for gradients supporting mixed ITG-TEM modes in positive triangularity TCV flux tube (left column) and negative triangularity flux tube (right column) of the TCV tokamak. Shown are results while varying (a)-(b) the ion temperature gradient $a/L_{T\mathrm{i}}$ with suppressed electron temperature gradient ($a/L_{T\mathrm{e}}=0$), (c)-(d) both $a/L_{T\mathrm{i}},a/L_{T\mathrm{e}}$ to create realistic non-flat kinetic profiles. In all cases the density gradient is fixed at $a/L_n =3$. Note that the axis limits for both $\omega_R$ and $\gamma$ are matched between left and right columns to facilitate quantitative comparison.}
    \label{fig:TCV-kegradTi-compars}
\end{figure*}

To investigate the model's sensitivity to configuration effects from triangularity we split the kinetic-electron scenarios in two cases: those with and without a finite ion temperature gradient to focus on pure TEM scenarios and mixed ITG-TEM scenarios, with results shown in \Cref{fig:TCV-keTEM-compars,fig:TCV-kegradTi-compars}, respectively. Akin to the adiabatic-electron ITG case, both the global dispersion model and \textsc{Gene} simulations show lower peak growth rates with the peak shifted to smaller wavenumbers in the negative-triangularity configuration for density-gradient-driven TEM, with the model closely following \textsc{Gene} results.  Again, the shift of growth-rate peak to smaller $k_y$ results in a marginal destabilisation of low $k_y$ modes when going from $\delta >0$ to $\delta<0$ (most notably in the region $k_y \rho_\mathrm{ss} \leq 0.3$, though the effect persists up to $k_y \rho_\mathrm{ss} = 0.5$ at the largest gradient), though simultaneously the growth rates are erroneously predict to be constant at high $k_y$ in the negative-triangularity configuration by the model. However, this situation changes when the effects of a finite electron-temperature gradient are considered. For $\eta_e\leq 1$, where $\eta_e = L_{n}/L_{Te}$, aside from a shift in propagation frequency into the electron diamagnetic direction, the stability properties of positive and negative triangularity configurations are unchanged. By contrast, when $\eta_e >1$, a peak in growth rate spectrum eludes both \textsc{Gene} simulations and results from the global dispersion model, with both indicating that high $k_y$ modes are progressively more destabilised in the negative triangularity configuration compared to its positive triangularity counterpart as $a/L_{T\mathrm{e}}$ is increased. This effect can be attributed to the behaviour of the propagation frequency, where the shift into the electron diamagnetic direction of the propagation frequency is more pronounced in the negative triangularity configuration, making resonant excitation of temperature-gradient-driven TEMs ($\omega_R<0$) with the electron precession drift over non-resonant excitation from UMs ($\omega_R>0$) possible. Additionally, as the drift-well in the negative triangularity is both deeper and wider, all but the most shallowly trapped-electrons will have more destabilising bounce-averaged magnetic drift compared to the positive triangularity configuration\cite{Marinoni2009TheSimulations}. Unlike for the adiabatic-electron ITG scenario, there is no clear consensus on whether negative triangularity should have beneficial effects over positive triangularity for TEM -- though it would be expected based on the strong reduction in electron heat diffusivity observed in both experiments\cite{Camenen2007ImpactPlasmas} and non linear simulations\cite{Marinoni2009TheSimulations,Merlo2015InvestigatingTransport} in TCV -- with previous studies reporting both a stabilising and destabilising effect depending on wavenumber, gradients and geometric details of the chosen flux tube\cite{Marinoni2009TheSimulations,Merlo2015InvestigatingTransport,Garbet2024TheModel,Merlo2023InterplayPlasmas}. 

When the effect of a finite ion temperature gradient is included [see \Cref{fig:TCV-kegradTi-compars}] the instability classification becomes more complex as modes propagating in the ion diamagnetic direction may be either UM or ITGs, and unlike the DIII-D case discussed in \Cref{fig:ke_LTi_scan_result} of the main text and \Cref{fig:ke-fullgrad} there is an absence of clear mode transitions. Rather, in both TCV configurations we observe that as the ion temperature gradient is increased (in absence of electron temperature gradient) there is a gradual change in the dispersive behaviour, with propagation frequencies and growth rates no longer approximately linearly increasing and decreasing, respectively, at high $k_y$, but instead flattening off and attaining a second smaller instability peak, respectively, as the ion temperature gradient is increased. This would suggest the possibility of hybrid UM-ITG modes. Regardless of the nature of these modes, we observe that the global dispersion model closely follows the trends observed in the \textsc{Gene} simulations, which like the adiabatic-electron ITG case and pure TEM case show that the negative triangularity configuration has increased resilience to instability at intermediate-high $k_y$, though is marginally more unstable at low $k_y$ compared to its positive triangularity counterpart. In particular the increased stabilisation is more pronounced compared to the pure TEM cases from \Cref{fig:TCV-keTEM-compars}. When in addition to the $a/L_{T\mathrm{i}}$ also the electron temperature gradient is reintroduced, a combination of the aforementioned effect occurs, though the difference between positive and negative triangularity is most pronounced for cases where the ion temperature gradient dominates ($a/L_{T\mathrm{i}} \geq a/L_{T\mathrm{e}}$, corresponding to red and purple curves). For cases with a dominant electron temperature gradient $a/L_{T\mathrm{e}}>a/L_{T\mathrm{i}}$ high (low) wavenumber modes are only marginally (de)stabilised by negative triangularity in \textsc{Gene} simulations, though as a result of larger growth-rate estimation in $\delta<0$ configuration, the opposite effect is predicted by the global dispersion model at the highest wavenumbers.

\subsubsection*{Results for W7-X negative-mirror configuration}

\begin{figure}
    \centering
    \begin{subfigure}{\linewidth}
        \centering
        \includegraphics[width=\linewidth]{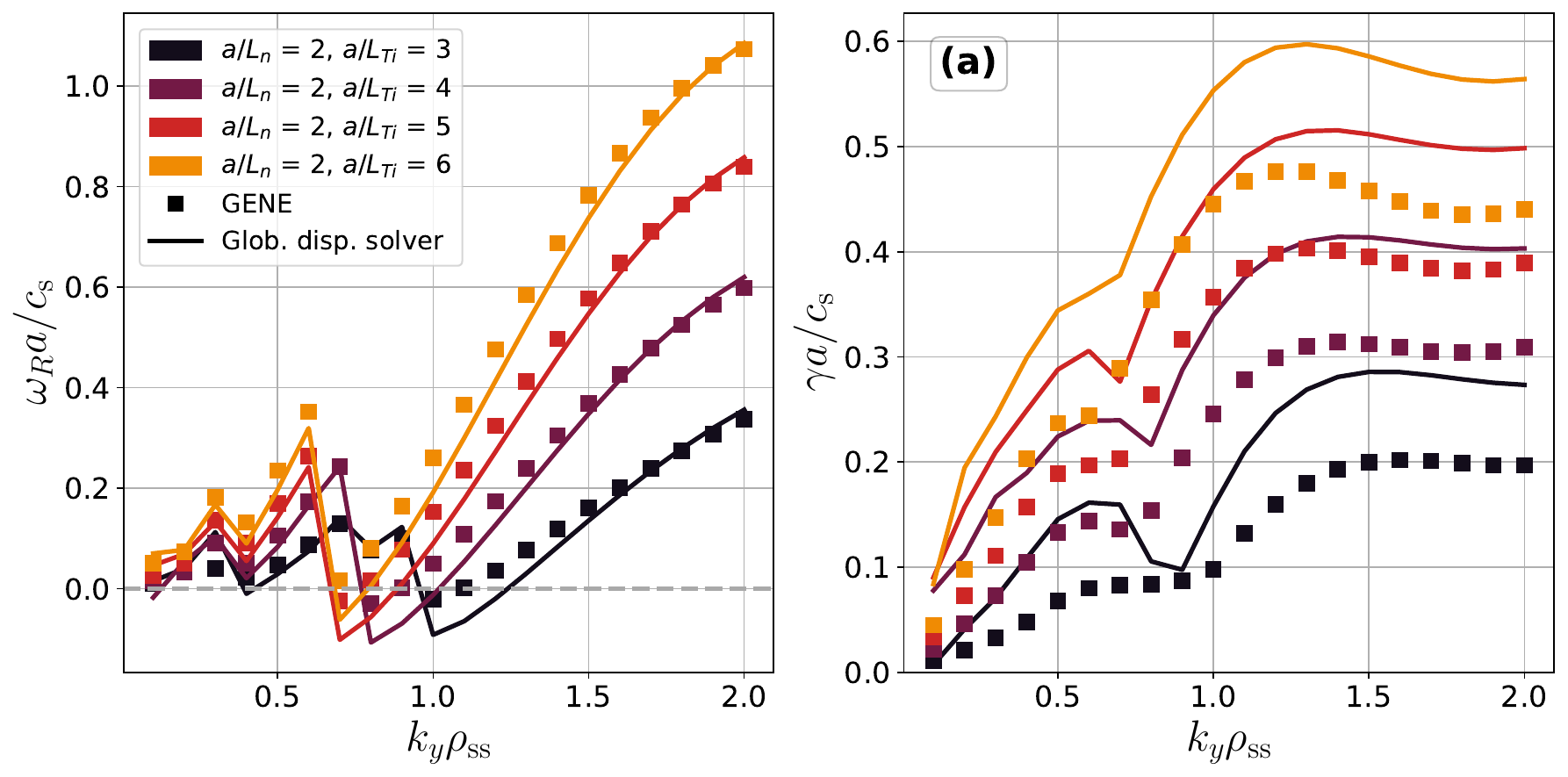}
    \end{subfigure}
    \begin{subfigure}{\linewidth}
        \centering
        \includegraphics[width=\linewidth]{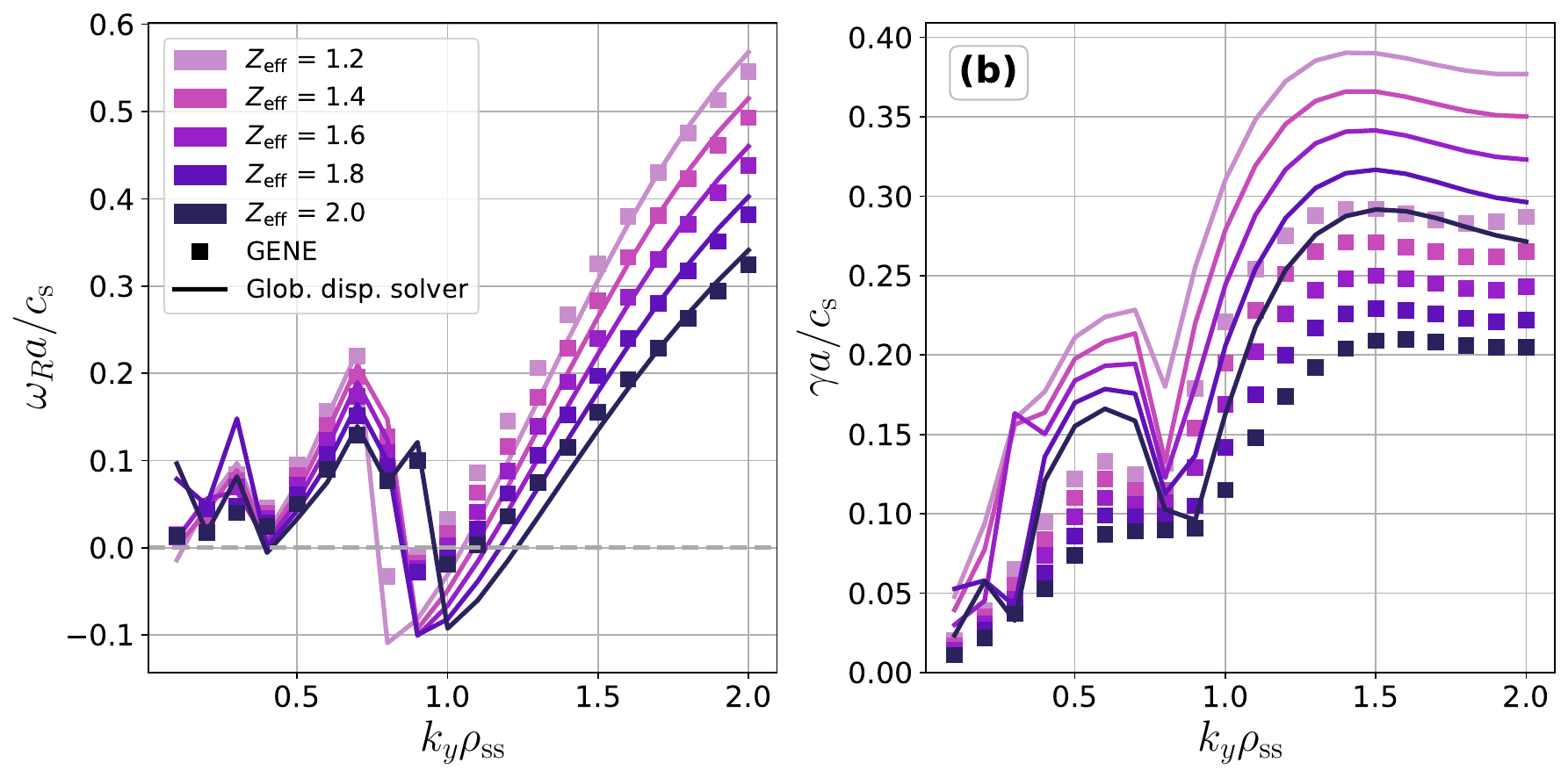}
    \end{subfigure}
    \begin{subfigure}{\linewidth}
        \centering
        \includegraphics[width=\linewidth]{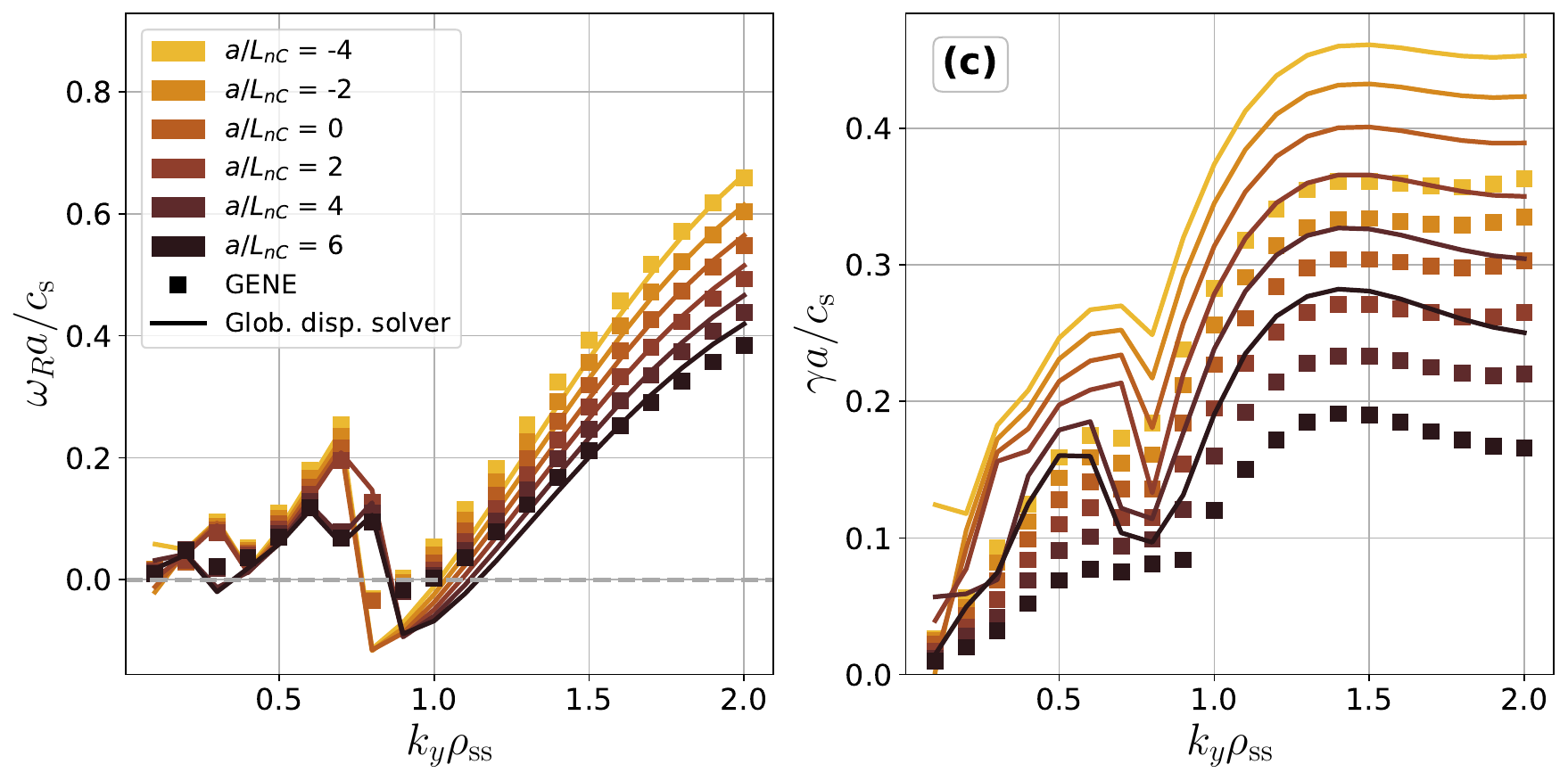}
    \end{subfigure}
    \caption{Eigenfrequency solutions obtain by the global dispersion model (solid lines) contrasted with \textsc{Gene} simulations (symbols) for adiabatic-electron scenarios in the negative mirror configuration of W7-X. Shown are results while varying (a) ion-temperature gradient $a/L_{T\mathrm{i}}$ [no $C^{6+}$ impurities], (b) impurity concentration $Z_{\textrm{eff}}$ [with ambipolar gradients $a/L_{ni}=a/L_{n\mathrm{C}}=2$], and (c) impurity density gradient $a/L_{n\mathrm{C}}$ [at fixed $Z_{\textrm{eff}}=1.4$]. In all cases the electron density gradient is fixed at $a/L_{ne}=2$, with ion and impurity density (gradient) determined by ambipolarity constraints [\Cref{eq:impurity-ambipolarity} in main text].}
    \label{fig:negativemirror-ae-plots}
\end{figure}

\begin{figure*}
    \centering
    \begin{subfigure}{.45\linewidth}
        \centering
        \includegraphics[width=\linewidth]{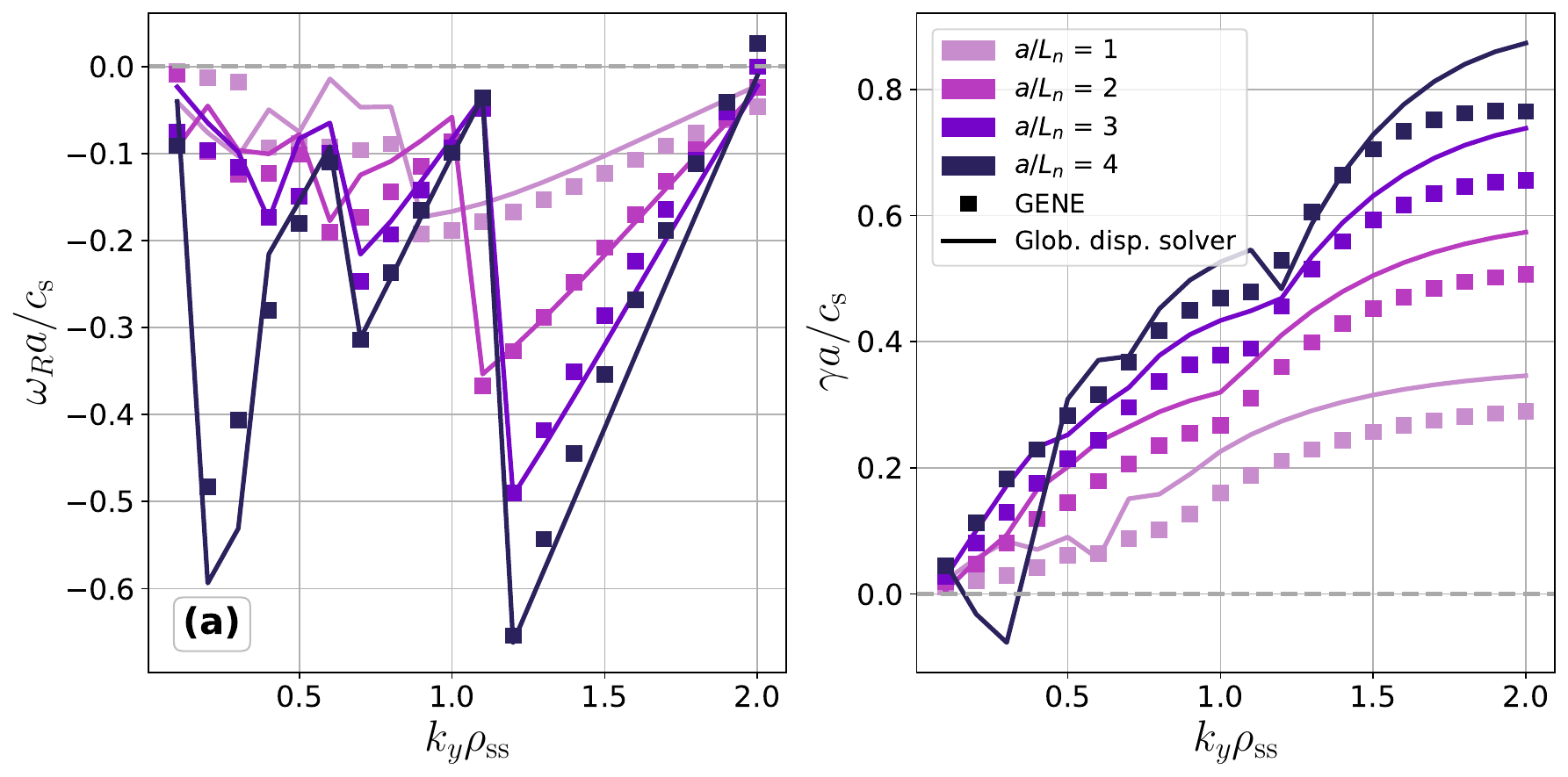}
    \end{subfigure}
    \begin{subfigure}{.45\linewidth}
        \centering
        \includegraphics[width=\linewidth]{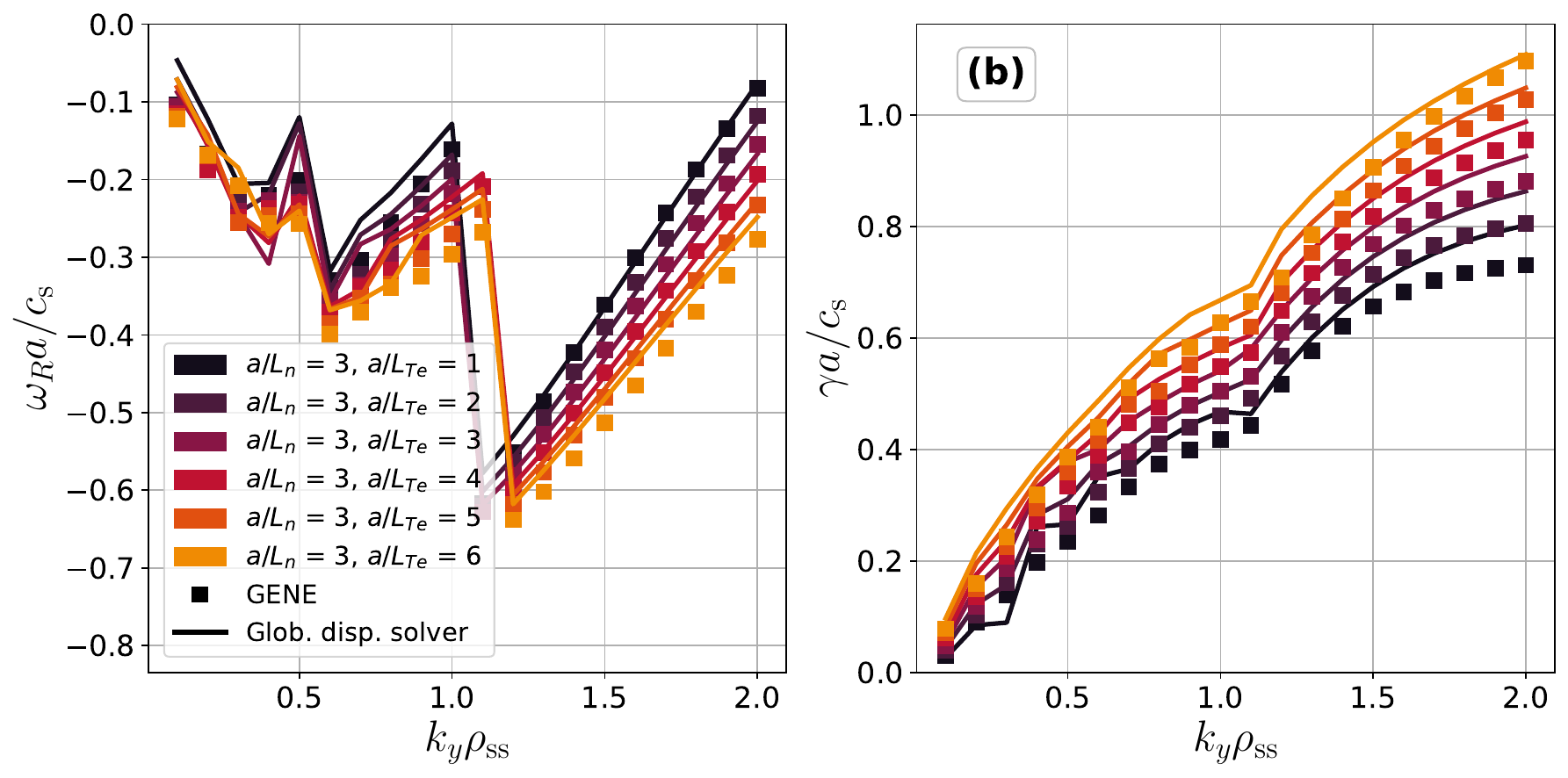}
    \end{subfigure}
    \begin{subfigure}{.45\linewidth}
        \centering
        \includegraphics[width=\linewidth]{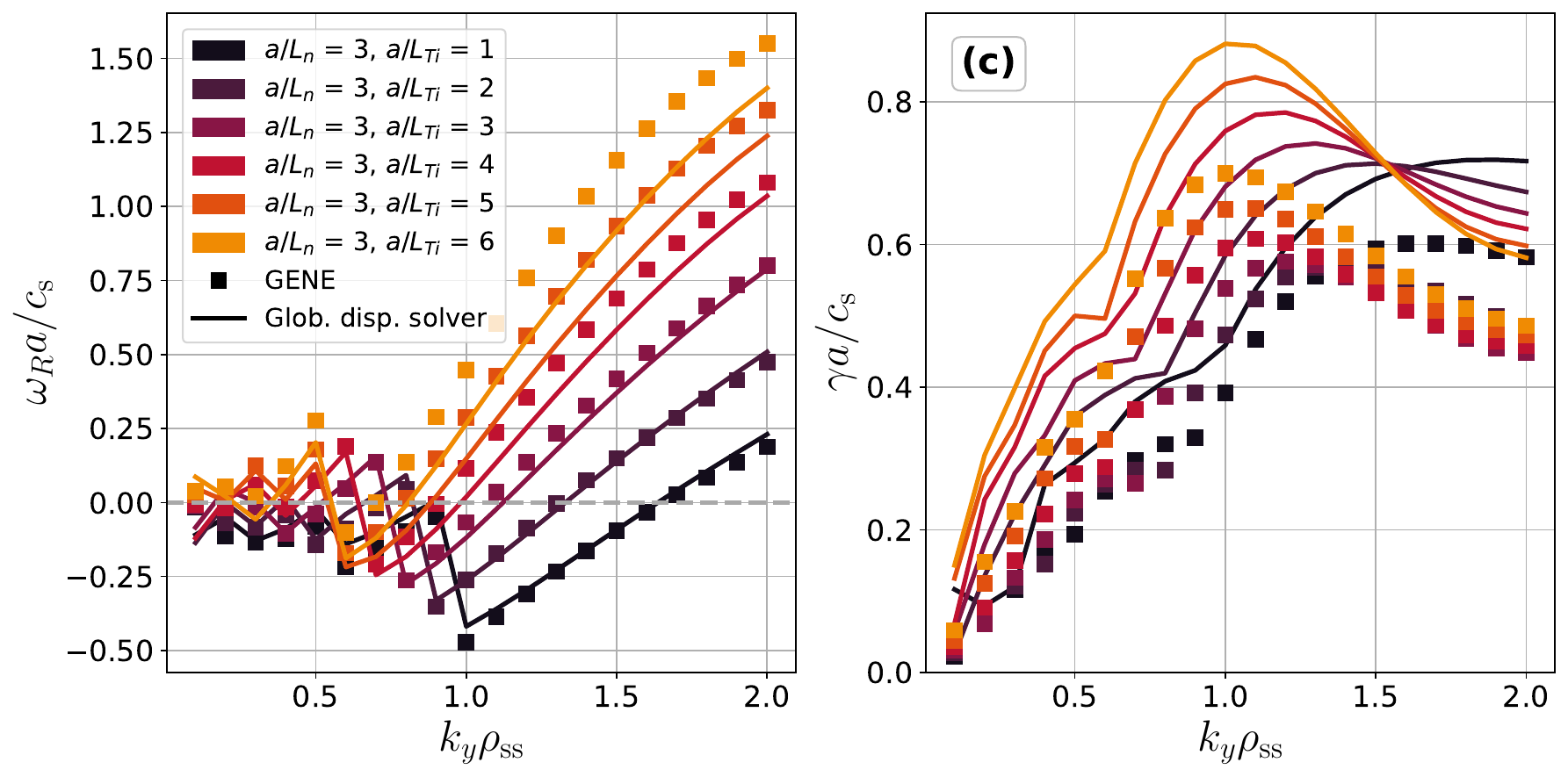}
    \end{subfigure}
    \begin{subfigure}{.45\linewidth}
        \centering
        \includegraphics[width=\linewidth]{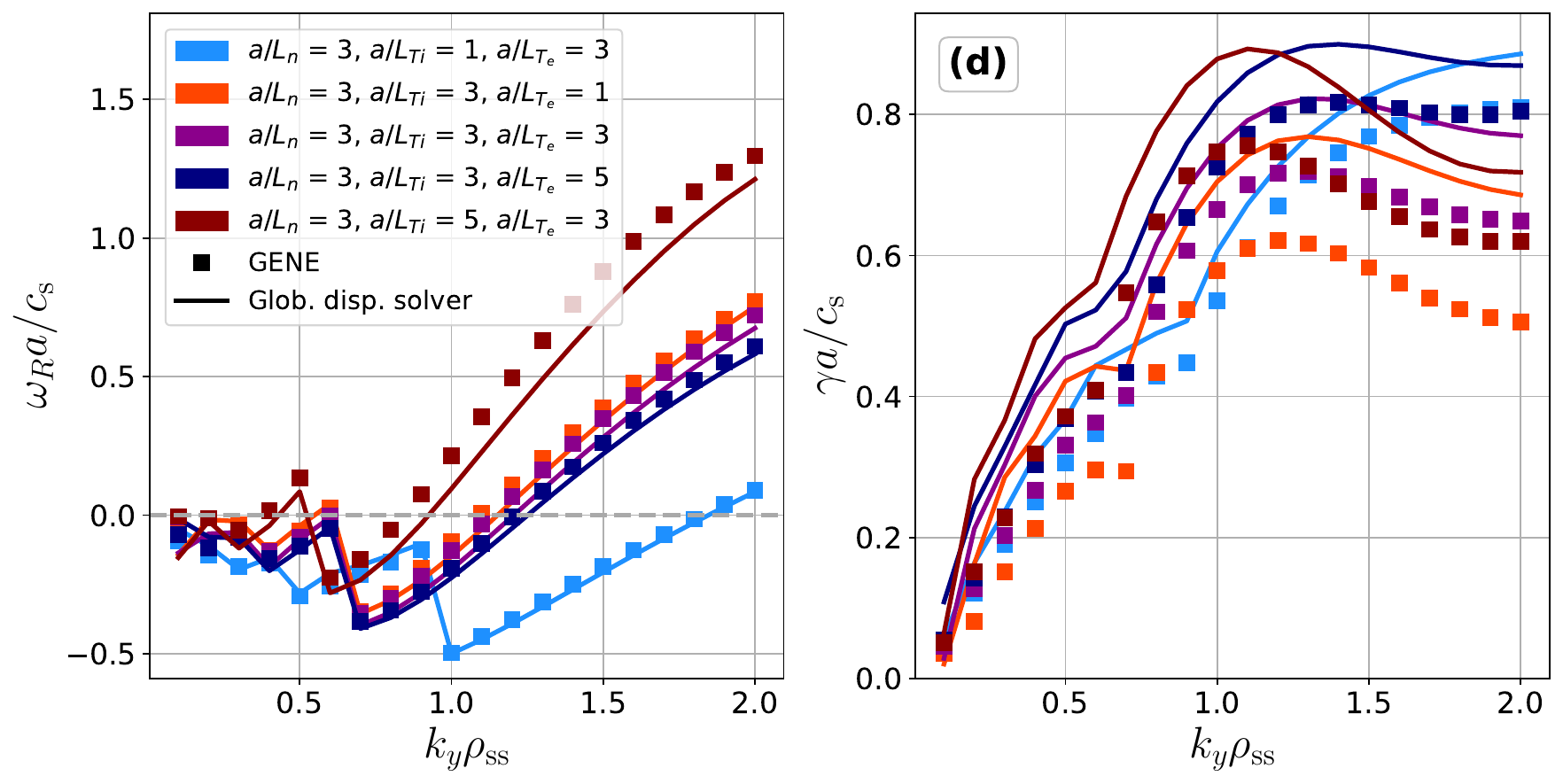}
    \end{subfigure}
    \caption{Eigenfrequency solutions obtain by the global dispersion model (solid lines) contrasted with \textsc{Gene} simulations (symbols) for kinetic-electron scenarios in the negative mirror configuration of W7-X. Shown are results while varying (a) the density gradient $a/L_n$ [both temperature gradients suppressed $a/L_{T\mathrm{i}}=a/L_{T\mathrm{e}}=0$], (b) the electron temperature gradient $a/L_{T\mathrm{e}}$ [suppressing ion temperature gradient $a/L_{T\mathrm{i}}=0$], (c) the ion temperature gradient [suppressing electron temperature gradient $a/L_{T\mathrm{e}}=0$] and (d) both $a/L_{T\mathrm{i}},a/L_{T\mathrm{e}}$ corresponding to realistic non-flat kinetic profiles. In cases (b)-(d) with non-zero temperature gradients the density gradient is fixed at $a/L_n =3$.}
    \label{fig:negativemirror-ke-plots}
\end{figure*}

Results for adiabatic-electron and kinetic-electron scenarios in the flux-tube geometry of the negative-mirror configuration of W7-X are shown in \Cref{fig:negativemirror-ae-plots,fig:negativemirror-ke-plots}, respectively. Comparing \Cref{fig:negativemirror-ae-plots} to \Cref{fig:ae-base-itg,fig:imp-drive-dillution-itg} and \Cref{fig:Zeff_result} of the main text we notice that there are only marginal differences in the dispersion of adiabatic-electron ITG between the negative-mirror and high-mirror configurations of W7-X in both the \textsc{Gene} simulations and the global dispersion model. This can be explained from the near-identicalness of both the magnetic-drift wells and perpendicular wavenumber $\norm{\bm{k_\perp}}$ between these configurations [see \Cref{fig:config_fluxtube_differences}], such that instability drive by regions of unfavourable magnetic curvature and FLR damping effects on the ITG mode are approximately equal, unlike for the TCV case. 
  
Significant differences (more pronounced than those observed between the positive- and negative-triangularity configurations of TCV in \Cref{fig:TCV-kegradTi-compars,fig:TCV-keTEM-compars}) with the high-mirror configuration are, however, observed for kinetic-electron cases when comparing \Cref{fig:negativemirror-ke-plots} with \Cref{fig:ke-gradn-TEM} and \Cref{fig:LTe_scan_result,fig:ke_LTi_scan_result} of the main text. Due to (re)alignment of magnetic wells and regions of bad-curvature conventional TEMs prevail in the negative mirror configuration in situations without an ion temperature gradient ($a/L_{T\mathrm{i}}=0$), where in the high-mirror configuration predominantly iTEMs occurred. Most notably in the aforementioned scenarios the growth rates are approximately twice as large in the negative-mirror configuration compared to the high-mirror configuration -- a feature which is also closely reproduced by the global dispersion model -- which can be attributed to the additional instability drive from trapped-electrons residing in regions of bad-bounce-averaged magnetic curvature in the former configuration. Despite being unstable against conventional TEMs, a handful of UI modes still appear in the negative-mirror configuration for $k_y \rho_\mathrm{ss} = 0.2\textrm{-}0.4$ at large density gradient $a/L_{n} = 4$, being in a similar wavenumber range where UIs are observed in the high-mirror configuration. Qualitatively different behaviour between both configurations is also observed when a finite ion temperature gradient is considered. For $k_y \rho_\mathrm{ss} \geq 0.6$ the eigenfrequency monotonously transitions from electron-diamagnetic direction to ion-diamagnetic direction as the ion temperature gradient is increased, with a simultaneous increase of peak growth rate and shift of growth-rate-maximum to lower $k_y$, analogous to what was previously observed in HSX [see \Cref{fig:ke_LTi_scan_result} of the main text], and thus by similar conventions being identified as iTEM. Unlike the high-mirror configuration, however, a transition to a second branch of higher excitation states of ITG at higher ion-temperature gradients is absent, which may also be attributed to the additional drive of trapped-electrons to iTEMs, as in the wavenumber range of $k_y \rho_\mathrm{ss} \geq 0.6$ growth rates for the negative mirror configuration exceed those found in the high-mirror configuration [see \Cref{fig:ke_LTi_scan_result}]. In the low-wavenumber region of $k_y \rho_{s} \leq 0.5$ we observe a transition from conventional TEMs to ITG for $a/L_{T\mathrm{i}}=5,6$, much akin to the iTEM-ITG transition observed in the high-mirror configuration, though unlike the latter, there is a threshold temperature gradient for this transition to occur. Analogous observations also hold for the case with realistic gradients when both ion- and electron temperature gradients are present. In general, the quantitative agreement between the global dispersion model and \textsc{Gene} simulations is significantly better in the negative-mirror configuration compared to the high-mirror configuration, which can be directly attributed to the diminished influence of trapped-electrons to the instability drive in the latter as a result of the configuration (approximately) satisfying the maximum-$J$ criterion.

\putbib[references]

\end{bibunit}

\end{document}